\documentclass[11pt, letterpaper, cleveref, autoref]{paper}

\usepackage{float,amsthm,amsfonts,amsmath,amssymb,url,tikz,circuitikz,tabularx,multirow,multicol,hhline,tablefootnote,xcolor,etoolbox,clipboard,titlesec,setspace,titletoc,tocloft,enumitem,latexsym,graphicx,algpseudocode,xparse,hyperref}
\usepackage[blocks]{authblk}
\usepackage[margin=1in]{geometry}
\usepackage[sortcites=true,maxnames=1000,style=alphabetic]{biblatex} 
\usepackage[toc]{appendix}
\usepackage[margin=0pt,font={small},labelfont={bf},labelsep=colon,format=plain]{caption}
\usepackage[left]{lineno}
\usepackage[mathscr]{eucal}
\usepackage[ruled,vlined]{algorithm2e}
\usepackage[T1]{fontenc}
\usepackage{thm-restate}
\usepackage[normalem]{ulem}
 
\makeatletter 
\let\original@algocf@latexcaption\algocf@latexcaption
\long\def\algocf@latexcaption#1[#2]{%
  \@ifundefined{NR@gettitle}{%
    \def\@currentlabelname{#2}%
  }{%
    \NR@gettitle{#2}%
  }%
  \original@algocf@latexcaption{#1}[{#2}]%
}
\makeatother 

\usetikzlibrary{fadings,patterns,shadows.blur,shapes,positioning,decorations.pathreplacing,arrows.meta,positioning,snakes}
\def\titlename{}
\def\authorsnames{}

\def\squiggylinescale{0.08}
\def\squiggylinedist{0.6}
\def\squiggylinerange{3.2}

\newcommand{\approxdist}{d}

\newcommand{\onepluseps}{1+\varepsilon}
\newcommand{\twopluseps}{2+\varepsilon}

\newcommand{\seventhirds}{\frac{7}{3}}

\newcommand{\sstar}{^*}

\newcommand{\dhztwothirdapprox}{2(\lfloor \frac{k}{3} \rfloor + 1)}

\newcommand{\singleton}[1]{\{#1\}}

\NewDocumentCommand\case{mg}
    {%
    \ensuremath{\textbf{Case (#1\IfNoValueTF{#2}{}{$_#2$})}}%
    }
\newcommand{\ccase}[1]{\underline{\case{#1}}:}
\NewDocumentCommand\notcase{mg}
    {%
    \ensuremath{\textbf{Case (\textoverline{#1}\IfNoValueTF{#2}{}{$_#2$})}}%
    }
\newcommand{\expectation}[1]{\mathbb{E}\bracke{#1}}
\NewDocumentCommand{\level}{omm}{
  \IfNoValueTF{#1}
    {
      \ell_{#2}(#3)
    }
    {
      \ell_{#2}^{(#1)}(#3)
    }
}
\NewDocumentCommand{\levelh}{omm}{
  \IfNoValueTF{#1}
    {
      {\hat \ell}_{#2}(#3)
    }
    {
      {\hat \ell}_{#2}^{\paren{#1}}(#3)
    }
}
\NewDocumentCommand{\noi}{o}{
  \IfNoValueTF{#1}
    {
      n^{o(1)}
    }
    {
      n^{#1 + o(1)}
    }
}
\NewDocumentCommand\onolog{g}{\displaystyle \tilde{O}\IfNoValueTF{#1}{}{\paren{#1}}}
\NewDocumentCommand\weight{ggg}{\displaystyle \mathbf{w} \IfValueTF{#2}{\IfValueTF{#3}{_{#2,#3}}{_{#2}}}{\IfValueTF{#3}{_{#3}}{}}\IfValueTF{#1}{\paren{#1}}{}} 
\NewDocumentCommand\dist{ggg}{\displaystyle \delta \IfValueTF{#2}{\IfValueTF{#3}{_{#2,#3}}{_{#2}}}{\IfValueTF{#3}{_{#3}}{}}\IfValueTF{#1}{\paren{#1}}{}} 
\NewDocumentCommand\hopdist{mmgg}{\displaystyle \delta^{(#2)} \IfValueTF{#3}{\IfValueTF{#4}{_{#3,#4}}{_{#3}}}{\IfValueTF{#4}{_{#4}}{}}\paren{#1}} 
\newcommand{\hops}[1]{h\paren{#1}}

\NewDocumentCommand{\auxpiv}{om}{
  \IfNoValueTF{#1}
    {
      \tilde{p}\paren{#2}
    }
    {
      \tilde{p}_{#1}\paren{#2}
    }
}
\NewDocumentCommand{\piv}{om}{
  \IfNoValueTF{#1}
    {
      p\paren{#2}
    }
    {
      p_{#1}\paren{#2}
    }
}
\NewDocumentCommand{\bunch}{mg}{
      \IfNoValueTF{#2}
        {
          B\paren{#1}
        }
        {
          B_{#2}\paren{#1}
        }
}
\NewDocumentCommand{\ball}{om}{
      \IfNoValueTF{#1}
        {
          \mathcal{B}\paren{#2}
        }
        {
          \mathcal{B}_{#1}\paren{#2}
        }
}

\NewDocumentCommand{\cluster}{mg}{
  \IfNoValueTF{#2}
    {
      C\paren{#1}
    }
    {
      C_{#2}\paren{#1}
    }
}
\NewDocumentCommand{\onediveps}{o}{
  \IfNoValueTF{#1}
    {
        \frac{1}{\varepsilon}
    }
    {
      \frac{1}{\varepsilon^{#1}}
    }
}
\newcommand{\bigo}[1]{O\paren{#1}}

\newcommand{\floor}[1]{\left\lfloor #1 \right\rfloor}
\newcommand{\ceil}[1]{\left\lceil #1 \right\rceil}
\newcommand{\paren}[1]{\left( #1 \right)}
\newcommand{\bracke}[1]{\left[ #1 \right]}
\newcommand{\bracce}[1]{\left\{ #1 \right\}}

\newcommand{\abs}[1]{\left| #1 \right|}

\newcommand{\tabref}[1]{\autoref{tab:#1}}
\newcommand{\secref}[1]{\hyperref[sec:#1]{Section~\ref*{sec:#1}}}
\newcommand{\appref}[1]{\hyperref[app:#1]{Appendix~\ref*{app:#1}}}
\newcommand{\thmref}[1]{~\protect\autoref{thm:#1}}
\newcommand{\defref}[1]{~\protect\hyperref[def:#1]{Definition~\ref*{def:#1}}}
\newcommand{\lemref}[1]{\hyperref[lem:#1]{Lemma~\ref*{lem:#1}}}
\renewcommand{\algref}[1]{\hyperref[alg:#1]{Algorithm~\ref*{alg:#1}}}

\newcommand{\cncref}[1]{~\protect\hyperref[cnc:#1]{Conclusion~\ref*{cnc:#1}}}
\newcommand{\corref}[1]{~\protect\hyperref[cor:#1]{Corollary~\ref*{cor:#1}}}
\newcommand{\propref}[1]{~\protect\hyperref[prop:#1]{Proposition~\ref*{prop:#1}}}
\newcommand{\figref}[1]{Figure~\ref*{fig:#1}}

\newcommand{\obvref}[1]{\hyperref[obv:#1]{Observation~\ref*{obv:#1}}}
\newcommand{\clmref}[1]{\hyperref[clm:#1]{Claim~\ref*{clm:#1}}}
\newcommand{\stpref}[3]{\hyperref[alg:#1_#2]{\textit{#3}}}
\newcommand{\prbref}[1]{\hyperref[prb:#1]{Problem~\ref*{prb:#1}}}
\newcommand{\queref}[1]{\hyperref[que:#1]{Question~\ref*{que:#1}}}

\NewDocumentCommand\mmin{mgg}{\displaystyle \min\IfNoValueTF{#2}{}{_{#2}}\IfNoValueTF{#3}{}{^{#3}}\bracce{#1}}
\NewDocumentCommand\mmax{mgg}{\displaystyle \max\IfNoValueTF{#2}{}{_{#2}}\IfNoValueTF{#3}{}{^{#3}}\bracce{#1}}
\NewDocumentCommand\mprod{mgg}{\displaystyle \prod\IfNoValueTF{#2}{}{_{#2}}\IfNoValueTF{#3}{}{^{#3}}{#1}}
\NewDocumentCommand\msum{mgg}{\displaystyle \sum\IfNoValueTF{#2}{}{_{#2}}\IfNoValueTF{#3}{}{^{#3}}{#1}}
\NewDocumentCommand\mcup{mgg}{\displaystyle \bigcup\IfNoValueTF{#2}{}{_{#2}}\IfNoValueTF{#3}{}{^{#3}}{#1}}
\NewDocumentCommand\mcap{mgg}{\displaystyle \bigcap\IfNoValueTF{#2}{}{_{#2}}\IfNoValueTF{#3}{}{^{#3}}{#1}}
\newcommand{\textoverline}[1]{\={#1}}
\newcommand{\midline}{\,\middle| \,\,}

\newcommand{\qoute}[1]{``#1''}
\newcommand{\subpol}{o\paren{1}}

\newcommand{\squiggly}{
\begin{tikzpicture} 
    {
    \draw[domain=-\squiggylinedist:0,smooth,variable=\x,scale=\squiggylinescale] plot({\x},{0.75+sin(4*0 r)});
    \draw[domain=0:\squiggylinerange,smooth,variable=\x,scale=\squiggylinescale] plot({\x},{0.75+sin(4*\x r)});
    \draw[domain=\squiggylinerange:\squiggylinerange+\squiggylinedist,smooth,variable=\x,scale=\squiggylinescale] plot({\x},{0.75+sin(4*\squiggylinerange r)});
    }
\end{tikzpicture}
}
\newcommand{\straight}{
\begin{tikzpicture} 
    {
    \draw[domain=-\squiggylinedist:-\squiggylinedist+0.0001,smooth,variable=\x,scale=\squiggylinescale] plot({\x},{-0.25+sin(4*\squiggylinerange r)});
    \draw[domain=-\squiggylinedist+0.0001:\squiggylinerange+\squiggylinedist,smooth,variable=\x,scale=\squiggylinescale] plot({\x},{0.75+sin(4*\squiggylinerange r)});
    }
\end{tikzpicture}
}

\ifdefined\socgtabular\else
  \let\socgtabular\tabular
\fi

\newcommand*{\settitlename}[1]{\title{#1}\def\titlename{#1}}
\newcommand*{\addauthor}[2]{\author[#2]{#1} 
\IfNoValueTF{\authorsnames}{\def\authorsnames{#1}}{\expandafter\def\expandafter\authorsnames\expandafter{\authorsnames, #1}}}
\renewcommand{\keywords}[1]{\def\keywordnames{#1}}

\newcommand{\rightabovearrow}[2]{\overset{\notin #2}{\overrightarrow{#1}}}
\newcommand{\leftabovearrow}[2]{\overset{\notin #2}{\overleftarrow{#1}}}
\newcommand{\sizeindex}[2]{\abs{#1}_{\notin #2}}

\newtheorem{theorem}{Theorem}
\newtheorem*{theorem*}{Theorem}
\newtheorem{lemma}{Lemma}[section]
\newtheorem*{lemma*}{Lemma}
\newtheorem{corollary}[lemma]{Corollary}
\newtheorem{claim}[lemma]{Claim}
\newtheorem*{claim*}{Claim}
\newtheorem*{observation*}{Observation}
\newtheorem{definition}[lemma]{Definition}
\newtheorem{observation}[lemma]{Observation}
\newtheorem{proposition}[lemma]{Proposition}

\newtheorem{remark}[lemma]{Remark}
\newtheorem*{note}{Note}

\newtheorem{question}{Question}
\newtheorem{conclusion}[lemma]{Conclusion}

\algnewcommand\algorithmicforeach{\textbf{for each}}
\algdef{S}[FOR]{ForEach}[1]{\algorithmicforeach\ #1\ \algorithmicdo}

\settitlename{First Near-Additive  All-Pairs Approximate Shortest Paths in Decremental Weighted Graphs}

\newif\ifanonymous
\anonymousfalse

\ifanonymous
\addauthor{Anonymous Author(s)}{1}
\affil[1]{Anonymous Institution}

\else
\addauthor{Eden Koveshi}{1,$\S$}
\addauthor{Liam Roditty}{1,$\propto$}
\addauthor{Ariel Sapir}{1,$\divideontimes$}

\affil[1]{Department of \href{http://cs.biu.ac.il/}{Computer Science}, Bar-Ilan University}
\affil[ ]{\small $^\S$eden.koveshi@live.biu.ac.il \quad $^\propto$liam.roditty@biu.ac.il \quad $^\divideontimes$sapirar@biu.ac.il}
\fi

\date{}


\makeatletter
\def\@maketitle{%
  \newpage
  \null
  \vskip 2em%
  \begin{center}%
  \let \footnote \thanks
    {\huge \textsf{\textbf{\@title}} \par}%
    \vskip 1.5em%
    {\large
      \lineskip .5em%
      \begin{tabular}[t]{c}%
        \@author
      \end{tabular}\par}%
    \vskip 1em
    {\large \@date}
  \end{center}
  \par
  \vskip 1.5em}
\makeatother

\makeatletter
\renewenvironment{abstract}{
    \begin{center}%
    {\bfseries\sffamily \Large\abstractname\vspace{\z@}}
      \end{center}%
      \quotation
    }
    \endquotation
\makeatother

\titleformat*{\section}{\Large\bfseries\sffamily}
\titleformat*{\subsection}{\large\bfseries\sffamily}

\dottedcontents{section}[1.5em]{}{1.3em}{.6em}

\setlist[itemize,1]{label=\scalebox{1.0}[1.0]{\textbullet}}
\setlist[itemize,2]{label=\scalebox{0.9}[1.3]{$\blacktriangleright$}}
\setlist[itemize,3]{label=\scalebox{0.9}[1.2]{--}}
\setlist[itemize,4]{label=\scalebox{0.666}[0.47]{---}}

\begin{document}

\let\originalleft\left
\let\originalright\right
\renewcommand{\left}{\mathopen{}\mathclose\bgroup\originalleft}
\renewcommand{\right}{\aftergroup\egroup\originalright}

\setlength{\textfloatsep}{8pt}


\maketitle

\begin{abstract}
\small 

Near-additive \textbf{A}ll-\textbf{P}airs \textbf{A}pproximate \textbf{S}hortest \textbf{P}aths (APASP) algorithms for decremental weighted undirected graphs have, until now, remained unknown.
We extend the  hierarchy induced edge sets of Baswana and Kavitha [FOCS'06 and SICOMP'10] from static to decremental graphs, obtaining the \textit{first} four near-additive $\left(1+\varepsilon,\beta\right)$-APASP algorithms for decremental weighted graphs, for an arbitrarily small $\varepsilon>0$. 

Our additive stretch guarantees are local to each vertex pair $u,v\in V$. At any point in time, we let $W_{i}$ denote the weight of the $i^{\textnormal{th}}$ heaviest edge on any  shortest path in the current graph. 

For weighted graphs, 
our first  result is  a  $\left(1+\varepsilon,2W_{2}\right)$-APASP  algorithm  with $\tilde O \left( n^{\frac{3}{2}+o\left(1\right)}m^{\frac{1}{2}}\right)$ total update time. This  is the same, up to a subpolynomial, as of a single computation of the static $+2W_{2}$-APASP algorithm of Cohen and Zwick  [SODA’97 and JALG’01]. 
Our algorithm also improves the stretch of the
$\left(2+\varepsilon\right)$-APASP algorithm of Dory, Forster, Nazari and
de Vos [ICALP’24].

Our second  result is a $\left(1+\varepsilon,2W_{1}+2W_{2}\right)$-APASP  algorithm 
with a total update time of 
$\tilde O \left( n^{\frac{7}{3}}\right)$. 
For $m=\Omega\left(n^{\frac{11}{6}}\right)$, this improves upon the 
$\left(3+\varepsilon\right)$-APASP algorithm of 
Chechik [FOCS'18], whose total update time is $\tilde{O} \left(mn^{\frac{1}{2}+o\left(1\right)}\right)$.

For a broader range of additive stretches, we obtain 
 two  $\left(1+\varepsilon,2\cdot\left(k-1\right)W_{1}\right)$-APASP algorithms with total update times $\tilde O \left( n^{2-\frac{1}{k}+o\left(1\right)}m^{\frac{1}{k}}\right)$ and $\tilde O \left( n^{2+\frac{2}{3k-2}+o\left(1\right)}\right)$.
 These adapt the static
$+2\sum_{i=1}^{k-1}W_{i}$-APASP algorithms of Cohen and Zwick
[SODA'97 and JALG'01] and Roditty and Sapir [FSTTCS'25] to the
decremental setting.

For unweighted graphs, 
we present 
 two near-additive algorithms:
$\left(1+\varepsilon,2\right)$-APASP in
$\tilde O\left(n^{\frac{7}{3}}\right)$ total update time and
$\left(1+\varepsilon,2\cdot \left(k-1\right)\right)$-APASP in
$\tilde O\left(n^{2+\frac{1}{3k-4}+o(1)}\right)$ total update
time. Combined with the results
of Dory, Forster, Nazari and De Vos [ICALP'24], we get total update times of 
$\tilde{O} \left(\min{\left\{n^{\frac{3}{2}+o\left(1\right)}m^{\frac{1}{2}},n^{\frac{7}{3}}\right\}}\right)$ and 
$\tilde{O} \left(\min{\left\{n^{2-\frac{1}{k}+o\left(1\right)}m^{\frac{1}{k}},n^{2+\frac{1}{3k-4}+o\left(1\right)}\right\}}\right)$, 
respectively, matching the runtimes of the currently fastest combinatorial  static $+2$-APASP algorithms and $+2\cdot\left(k-1\right)$-APASP algorithms of 
Dor, Halperin and Zwick [FOCS'96 and SICOMP'00].
\end{abstract}

\newpage


\tableofcontents
\newpage

\pagenumbering{arabic}

\section{Introduction}\label{sec:intro}
Dynamic graphs provide a natural framework for modeling networks whose
topology or edge weights change over time. A dynamic algorithm preprocesses
the initial graph and then processes updates while answering queries about
the current graph. Its aim is to reuse previous computation, rather than solve each new instance from scratch.   

Dynamic algorithms are commonly studied in three models: the
\emph{incremental} model allows edge insertions and weight decreases;
the \emph{decremental} model allows edge deletions and weight
increases; the \emph{fully dynamic} model allows both types of updates.
Our work focuses on
the decremental setting, where the performance of a decremental algorithm
is measured by its \emph{total update time}: the cumulative runtime required by an algorithm to process the entire sequence of update operations.

We study the \textbf{A}ll-\textbf{P}airs \textbf{S}hortest
\textbf{P}aths (\emph{APSP}) problem, which asks for the distances
between all vertex pairs. Let $G=\paren{V,E,w}$ be a weighted
undirected graph with $n=\abs{V}$ vertices and initially $m=\abs{E}$
edges. In the decremental setting, these distances must be maintained
throughout an update sequence. 
An \emph{oblivious adversary} fixes this sequence independently
of the algorithm's choices, while an \emph{adaptive adversary} may choose
updates based on earlier answers. Our algorithms are randomized and assume
an oblivious adversary.

Recent progress has yielded truly subcubic algorithms for
exact APSP in the static setting.
Alman and Vassilevska Williams~\cite{AlmVas2026} refuted
the APSP hypothesis ~\cite{Vas2015, Vas2018, AbbVas2023} by providing an APSP algorithm with $\tilde O\paren{n^{2.998}}$ runtime. For polynomially bounded integer weights, Saha and
Ye~\cite{SahYe2026} subsequently improved the runtime to $\tilde O\paren{n^{2.9964}}$. Nonetheless, the exponents in these bounds remain close to $3$,
motivating the search for approximation algorithms with faster runtime. As eventually an $n\times n$ matrix needs to be returned, an optimal runtime is $\tilde O\paren{n^2}$. 

In the decremental setting, exact distance maintenance faces
a separate conditional lower bound.
The \textbf{O}nline \textbf{M}atrix \textbf{V}ector
\textbf{M}ultiplication (\emph{OMVM}) Conjecture~\cite{GasJanLevLin2021, HenKriNanSara2015, LarRWill2017} concerns
the products of a fixed $n\times n$ Boolean matrix with
$n$ Boolean vectors arriving one by one, where each product
must be returned before the next vector arrives.
It asserts that this problem cannot be solved in
$\tilde O\paren{n^{3-\eta}}$ time for any constant $\eta>0$. 
 Henzinger, Krinninger, Nanongkai and
Saranurak~\cite{HenKriNanSara2015} showed that, under
this conjecture, no decremental APSP algorithm with
polynomial preprocessing and polylogarithmic query time
can achieve  $\tilde O\paren{m n^{1-\eta}}$ total update time. For dense graphs where $m=\Theta\paren{n^{2}}$ this excludes truly  
subcubic total update time.

This motivates the
\textbf{A}ll-\textbf{P}airs \textbf{A}pproximate
\textbf{S}hortest \textbf{P}aths (\emph{APASP}) problem.
For a \emph{multiplicative stretch} $\alpha\geq 1$ and
an \emph{additive stretch} $\beta\geq 0$,  
a decremental $\paren{\alpha,\beta}$-APASP algorithm maintains,
after every update at \textit{time} $t\geq 0$, a distance estimate $d_{t}\bracke{u,v}$ satisfying:
$$d_{t}\bracke{u,v} \in \bracke{\delta_{G,t}\paren{u,v}, \alpha \cdot \delta_{G,t}\paren{u,v} + \beta}$$
for all $u,v \in V$, where $\delta_{G,t}\paren{u,v}$ is the distance at time $t$. 
When $\beta=0$, the approximation is \textit{multiplicative}, denoted by $\alpha$-APASP; 
when $\alpha=1$, it is \textit{additive}, denoted by $+\beta$-APASP; when $\alpha=1+\varepsilon$ for a small $\varepsilon>0$, the approximation is  \textit{near-additive}; 
otherwise, the approximation is \textit{mixed}.

For unweighted graphs, the classical combinatorial algorithms of
Dor, Halperin and Zwick~\cite{DorHalZwi2000} obtain $+2$-APASP
and a family of $+2\paren{k-1}$-APASP tradeoffs.
Extending additive approximation to weighted graphs 
presents an issue. Any  $+\beta$-APASP for weighted graphs with a constant additive  is inherently problematic: multiplying the edge weights by a sufficiently large value reduces $+\beta$-APASP to exact APSP. 

To overcome this, the additive stretch must have a dependence of some sort on the weight function $\mathbf{w}: E \rightarrow \mathbb{R}^{+}$. A naive approach would be to 
consider the largest weight in the graph $W_{\max}$. However, as this is a global parameter of the entire graph, 
a single heavy edge elsewhere in the graph can make
$W_{\max}$ arbitrarily larger. An additive stretch that depends on $W_{\max}$ might therefore be meaningless for many vertex pairs. 
This leads us to seek additive stretch determined by a  parameter that is local for each pair of vertices.

In their seminal work, Cohen and Zwick~\cite{CohZwi1997} extended  unweighted additive
algorithms to weighted graphs, obtaining local additive stretch determined by edge weights along a shortest path. For vertices $u,v\in  V$ and a shortest path $u\squiggly v$ between  them,  they defined $W_{i}\paren{u \squiggly v}$ as the weight of the $i^{\textnormal{th}}$ heaviest edge\footnote{If $u\squiggly v$ has fewer than $i$ edges, they set $W_{i}\paren{u \squiggly v}=0$.}
along $u \squiggly v$. 
This additive is not only local for a pair of vertices, but can also be converted to multiplicative stretch, as $W_{i}\paren{u\squiggly v} \leq \frac{\delta\paren{u,v}}{i}$, which is informative for $i\geq 2$.

The algorithms of Cohen and Zwick match the same runtimes as the unweighted additive algorithms of Dor, Halperin and Zwick~\cite{DorHalZwi2000}.
Cohen and Zwick thus took the step from unweighted additive
approximation to weighted additive approximation.
We take the next step, from static weighted graphs to decremental
weighted graphs.
Despite extensive work on dynamic shortest paths within the last three decades (e.g. \cite{KingWeightedESTree,Thorup2005,Bernstein2016,ChechikADO}), 
only one previous decremental result has a local additive stretch: the $\paren{2+\varepsilon,W_{1}}$-APASP algorithm of Dory, Forster, Nazari and de Vos~\cite{Doryetal}.

This appears in contrast to the variety of usage of $W_{i}$s in the static setting, for several graph theory problems, such as  APASP algorithms \cite{BK,Kavitha2012,Ariel+kDense,DorForKirNazVasVos2023}, spanners \cite{Tzalik2026,ElkGitNei2021, ElkinGN2022SWAT,LaLe2024} and emulators \cite{ElkinGN2022SWAT,RodittySapir2026Emulators}.

\begin{figure}[H]
  \centering
\begin{tikzpicture}[
    vertex/.style={circle, draw, fill=white, minimum size=10pt, inner sep=0pt, font=\Large\bfseries},
    small vertex/.style={circle, draw, fill=gray!20, minimum size=5pt, inner sep=0pt},
    edge label/.style={font=\small, text=black, inner sep=1.5pt},
    path1 style/.style={->, >={Stealth[scale=1.2]}, ultra thick, blue!60!cyan, shorten >=1pt, shorten <=1pt},
    path2 style/.style={->, >={Stealth[scale=1.2]}, thick, red!70!orange, shorten >=1pt, shorten <=1pt},
    scale=1.2 
]

\node[vertex] (u) at (0,0) {\small$u$};
\node[vertex] (v) at (8,0) {\small$v$};

\coordinate (p1_c1) at (4, 0.9);
\node[small vertex] (p1_1) at (p1_c1) {};

\draw[path1 style] (u) to[bend left=12] node[edge label, above] {10} (p1_1);
\draw[path1 style] (p1_1) to[bend left=12] node[edge label, above] {9} (v);

\foreach \i in {1,...,3} {
    \pgfmathsetmacro{\xcoord}{8 * \i / 4}
    \pgfmathsetmacro{\ycoord}{-0.035 * \xcoord * (8 - \xcoord) - 0.2}
    \node[small vertex] (p2_\i) at (\xcoord, \ycoord) {};
}

\draw[path2 style] (u) to node[edge label, below] {5} (p2_1);
\foreach \i [evaluate=\i as \j using int(\i+1)] in {1,...,2} {
    \draw[path2 style] (p2_\i) to node[edge label, below] {5} (p2_\j);
}
\draw[path2 style] (p2_3) to node[edge label, below] {5} (v);

\end{tikzpicture}
\caption{Consider a $+2W_{2}$-APASP. The blue  path is a shortest path and has weight $19$ and $W_{2}=9$, therefore  $d_{t-1}\bracke{u,v}\in\bracke{19,37}$. The red path has weight $20$ and $W_{2}=5$. If a blue edge is erased, the red path becomes the new shortest path. The estimation guarantee at that time tightens to  $d_{t}\bracke{u,v}\in\bracke{20,30}$. } \label{fig:w+2w2:any-path-vs-shortest}
\end{figure}
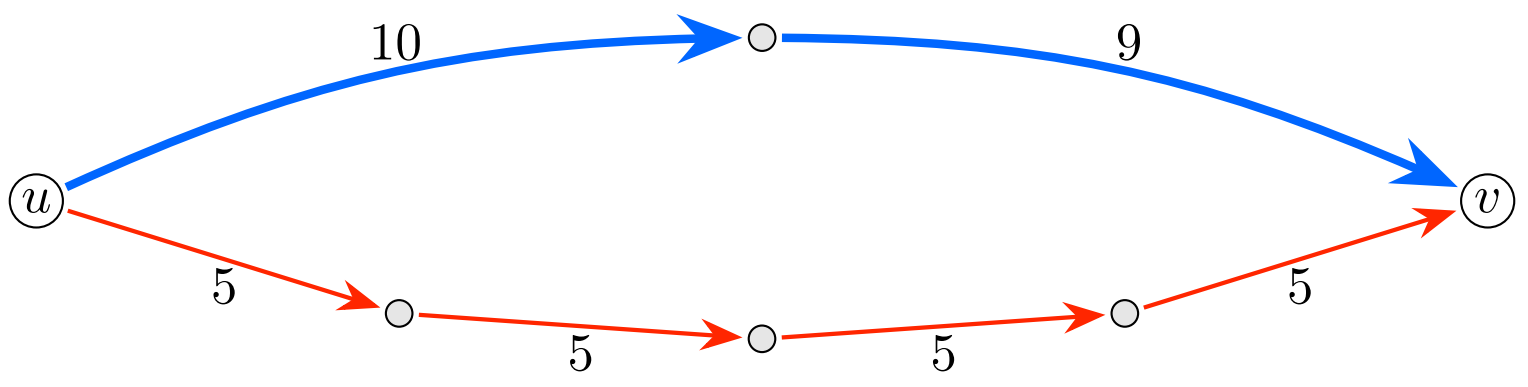

The required approximation upper bound can decrease under
decremental updates, although distances are monotonically nondecreasing.
At every time $t$ we have
$\delta_{G,t}\paren{u,v}\geq \delta_{G,t-1}\paren{u,v}$, but an update may replace the chosen shortest path by another
path of the same weight whose heaviest edges are lighter. 
Consequently, an upper bound of the form
$\alpha\cdot\delta_{G,t}\paren{u,v}
+\msum{\beta_{i}W_{i}\paren{\paren{u\squiggly v}_{t}}}{i\in I}$
with $I\subseteq\bracke{n}$, $\alpha\geq 1$ and $\beta_{i}\geq 0$
need not be nondecreasing. 
An estimate that satisfied the guarantee at time $t-1$ may
therefore exceed the upper bound at time $t$.
We refer to this as the \emph{monotonicity challenge}. For example, see \figref{w+2w2:any-path-vs-shortest}.

\begin{question}\label{que:additive-techniques}
Which decremental $\paren{\alpha,\msum{\beta_{i}W_{i}}{i\in I}}$-APASP algorithms can we find for weighted graphs, and with what total update time?
\end{question}

We answer \queref{additive-techniques} with four decremental
algorithms for weighted graphs. All four have multiplicative stretch $1+\varepsilon$ and additive stretch expressed in terms of $W_{i}$. 
These are the first near-additive APASP
 algorithms for weighted graphs in the decremental setting.

\subsection{Related Work}\label{sec:intro:known}

To properly understand the scope of our contribution, we begin by reviewing notable static APASP algorithms. We distinguish between \emph{algebraic} APASP algorithms,
which use fast matrix multiplication, and \emph{combinatorial} algorithm, which do not. 
Our algorithms are combinatorial, and our comparisons are
with static combinatorial algorithms. Algebraic APASP forms a separate research line that we do not
survey in detail.
Representative results include multiplicative approximations for
weighted and unweighted graphs~\cite{BriKunWeg2019,DorForKirNazVasVos2023},
additive approximations for unweighted
graphs~\cite{SahYe2023,JinKSW25} and near-additive approximations
for weighted graphs~\cite{SahYe2023,Ariel+kDense}.

The classical work of Dor, Halperin and
Zwick~\cite{DorHalZwi2000} established a family of combinatorial
tradeoffs between additive stretch and running time for
unweighted APASP.
For sparse graphs, their $+2$-APASP and
$+2\paren{k-1}$-APASP algorithms have running times
$\tilde O\paren{n^{\frac{3}{2}}m^{\frac{1}{2}}}$ and
$\tilde O\paren{n^{2-\frac{1}{k}}m^{\frac{1}{k}}}$,
respectively, where $k\in \mathbb{N}$. 
For dense graphs, they obtained these additive stretches in
$\tilde O\paren{n^{\frac{7}{3}}}$ and
$\tilde O\paren{n^{2+\frac{1}{3k-4}}}$ time, respectively.

Cohen and Zwick~\cite{CohZwi1997} extended the sparse
algorithms to weighted graphs, obtaining $+2W_{2}$-APASP
and $+2\msum{W_{i}}{i=1}{k-1}$-APASP within the corresponding
unweighted running times.
For dense graphs, Cohen and Zwick obtained additive stretch
$+\mmin{2W_{1},4W_{2}}$ within the running time of the unweighted
$+2$-APASP algorithm.
Only recently did Roditty and Sapir~\cite{Ariel+kDense} obtain the
dense $+2\msum{W_{i}}{i=1}{k-1}$-APASP tradeoff,
matching its unweighted counterpart.
Further static results appear in \tabref{known_static}. 

While static APASP admits a broad range of additive tradeoffs,
the corresponding decremental results remain more limited (See: \tabref{known_decremental}).
For unweighted undirected graphs, Henzinger, Krinninger and
Nanongkai~\cite{HKN14} obtained a decremental
$\paren{1+\varepsilon,2\cdot\paren{1+\frac{2}{\varepsilon}}^{k-2}}$-APASP
algorithm for every constant $\varepsilon>0$ and integer $k\in\mathbb{N}$. 
Their algorithm has $\tilde O\paren{\paren{\frac{37}{\varepsilon}}^{k-1}\cdot n^{2+\frac{1}{k}}
}$ total update time and it serves as an essential ingredient in our unweighted
results.

Another unweighted result is by Dory, Forster, Nazari and
de Vos~\cite{Doryetal} that obtained a  decremental
$\paren{1+\varepsilon,2\cdot\paren{k-1}}$-APASP algorithms.  Their  total update times match the sparse static
runtime of Dor, Halperin and Zwick \cite{DorHalZwi2000}  up to an $n^{o\paren{1}}$ factor\footnote{Here this factor may depend on the fixed value of $\varepsilon$.}.

For weighted graphs with polynomially bounded weights,
Bernstein~\cite{Bernstein2016} obtained
$\paren{1+\varepsilon}$-APASP with
$\tilde O\paren{mn}$ total update time.
Dory, Forster, Nazari and de Vos provided a partial answer to
\queref{additive-techniques} by obtaining a
$\paren{2+\varepsilon,W_{1}}$-APASP algorithm with total update time $\tilde O\paren{m^{\frac{3}{4}}\cdot n^{1+o\paren{1}}}$. 

\subsection{Statically Competitive Algorithms}
\label{sec:intro:static-competitive}

A central goal in decremental algorithms is to maintain a solution
after every update while processing the \emph{entire sequence of updates}
in essentially the runtime of one static
computation. We pursue this goal for additive APASP and show that it can be
achieved for several classical additive guarantees.

As is common in decremental shortest path
algorithms~\cite{BerRod2011,HKNLinSSSP,ChechikADO,
Bernstein2021DetSSSP,LackiNazari22,Doryetal}, 
we allow the multiplicative stretch to increase by an arbitrarily
small $+\varepsilon$ and permit an $n^{o\paren{1}}$ overhead in
total update time, compared to the static setting.
Such losses can arise from using tools such as
near-additive \textbf{S}ingle \textbf{S}ource \textbf{S}hortest \textbf{Paths} (\emph{SSSP}) algorithms~\cite{HKN14}, hopsets~\cite{LackiNazari22} and edge weight
scaling~\cite{BernsteinRounding}. The additive stretch remains unchanged, and for fixed
$\varepsilon$ there is no polynomial gap between processing the
entire update sequence and one static computation.
We formalize this comparison as follows.

\begin{definition}\label{def:intro:static-competitive}
Let $T_{\mathrm{stat}}\paren{n,m}$ be the best currently known
running time for a static combinatorial 
$\paren{\alpha,\beta}$-APASP algorithm.
A decremental combinatorial
 algorithm is \emph{statically competitive}
if, for $\varepsilon>0$, it maintains
an $\paren{\alpha+\varepsilon,\beta}$-APASP with an expected total
update time of $\tilde O_{\varepsilon} \paren{T_{\mathrm{stat}}\paren{n,m}\cdot n^{o\paren{1}}}$.
\end{definition}

\begin{question}\label{que:static-competitive}
For which $\paren{\alpha,\beta}$-APASP  can we find
statically competitive decremental algorithms?
\end{question}

Dory, Forster, Nazari and de Vos~\cite{Doryetal} provide a
partial answer to \queref{static-competitive}.
Their decremental algorithms are statically competitive with
the sparse $+2$-APASP and $+2\paren{k-1}$-APASP algorithms
of Dor, Halperin and Zwick~\cite{DorHalZwi2000}.
However, their results do not cover the corresponding dense
tradeoffs.
We provide the dense counterparts in \thmref{+2} and
\thmref{+k}. We also provide another  affirmative answer to
  \queref{static-competitive} for weighted $+2W_{2}$-APASP in \thmref{+2W2}. We elaborate on our results in the following section.

\subsection{Our Results}\label{sec:intro:ours}

We present two decremental algorithms for unweighted graphs and
four for weighted graphs. \tabref{ours_decremental} summarizes
their stretches and total update times.
We also 
extend the static hierarchy of induced edge sets used by Baswana and Kavitha~\cite{BK}  to the decremental setting. To do so, we utilize a hopset that supports edge insertions to a subgraph of a decremental graph.
We believe this extension of the weighted edge sets hierarchy is of valuable use for maintaining weighted APASP.

\begin{table}[t]
\centering
\begingroup
\footnotesize
\setlength{\tabcolsep}{2pt}
\renewcommand{\arraystretch}{1.5}
\begin{tabular}{|c|c|c|c|c|c|c|}
\hline
\multirow{2}{*}{\textbf{Weights}}
& \multirow{2}{*}{\textbf{Approximation}}
& \multirow{2}{*}{\shortstack[c]{\textbf{Expected Total}\\
                                  \textbf{Update Time}}}
& \multicolumn{2}{|c|}{\textbf{ Answers To}}
& \multirow{2}{*}{\textbf{Section}}
& \multirow{2}{*}{\textbf{Theorem}}\\
\cline{4-5}
& & & \queref{additive-techniques}
    & \queref{static-competitive} & &\\
\hline

Unweighted
& $\paren{\onepluseps,2}$
& $\onolog{n^{\frac{7}{3}}}$
& & $\checkmark$
& \secref{uw+2} & \thmref{+2}\\
\hline

Unweighted
& $\paren{\onepluseps,2\cdot\paren{k-1}}$
& $\onolog{n^{2+\frac{1}{3k-4}+o\paren{1}}}$
& & $\checkmark$
& \secref{uw+k} & \thmref{+k}\\
\hline

Weighted
& $\paren{\onepluseps,2W_{2}}\leq\paren{\twopluseps,0}$
& $\onolog{n^{\frac{3}{2}}m^{\frac{1}{2}}}^{\star}$
& $\checkmark$ & $\checkmark$
& \secref{w+2w2:unbounded} & \thmref{+2W2}\\
\hline

Weighted
& $\paren{\onepluseps,2W_{1}+2W_{2}}$
& $\onolog{n^{\frac{7}{3}}}$
& $\checkmark$ &
& \secref{w+2w1+2w2} & \thmref{+2w1+2w2}\\
\hline

\multirow{2}{*}{Weighted}
& \multirow{2}{*}{$\paren{\onepluseps,2\cdot\paren{k-1}W_{1}}$}
& $\onolog{n^{2-\frac{1}{k}+\subpol}m^{\frac{1}{k}}}^{\circledast}$
& $\checkmark$ &
& \secref{w+2kw1-sparse} & \thmref{w+2kw1}\\
\cline{3-7}
&
&
$\onolog{n^{2+\frac{2}{3k-2}+\subpol}}^{\circledast}$
& $\checkmark$ &
& \secref{w+2kw1-dense} & \thmref{w+2kw1-dense}\\
\hline
\end{tabular}
\endgroup
    \renewcommand\arraystretch{1}
    \caption{Our contributions for $\paren{\alpha,\beta}$-APASP algorithms for undirected (weighted and unweighted) graphs in the decremental setting.
    For simplicity of reading we assume that $\varepsilon$ is a constant and is absorbed into $\onolog{\cdot}$. Exact total update times are stated in the theorems.
    $^{\star}$ The bound holds when $m = n^{1+\Omega(1)}$; otherwise, an extra $n^{o(1)}$ factor is incurred. 
    $^{\circledast}$ For general $k$ and $\varepsilon$ we can replace the $n^{\subpol}$ factor with $\paren{\frac{k^2 \log n}{\varepsilon}}^{O(k)}$.}~\label{tab:ours_decremental}
\end{table}

\subsubsection{Unweighted Graphs}
\label{sec:intro:ours:unweighted}

We first consider additive stretch for  $k=2$.

\begin{restatable}{theorem}{thmunweightedplustwo}
\label{thm:+2}
For every $\varepsilon>0$, there is a decremental
$\paren{1+\varepsilon,2}$-APASP algorithm for unweighted
undirected graphs undergoing edge deletions by an oblivious adversary, with an expected
total update time of
$\tilde O\paren{\frac{1}{\varepsilon^2}\cdot n^{\frac{7}{3}}}$.
\end{restatable}

We then obtain a tradeoff for  any $k\geq 3$:

\begin{restatable}{theorem}{thmunweightedplusk}
\label{thm:+k}
For every integer $3\leq k\in o\paren{\log n}$ and every fixed
$\varepsilon>0$, there is a decremental
$\paren{1+\varepsilon,2\paren{k-1}}$-APASP algorithm for
unweighted undirected graphs undergoing edge deletions  by an oblivious adversary, with
an expected total update time of
$\tilde O\paren{n^{2+\frac{1}{3k-4}+o\paren{1}}}$.
\end{restatable}

For fixed $\varepsilon$, the algorithms in \thmref{+2} and
\thmref{+k} match the dense static bounds of Dor, Halperin
and Zwick~\cite{DorHalZwi2000} up to an $n^{o\paren{1}}$ factor.
Together with the sparse decremental algorithms of Dory,
Forster, Nazari and de Vos~\cite{Doryetal}, they answer
\queref{static-competitive} affirmatively for the classical
unweighted additive tradeoffs across all graph densities.

\subsubsection{Weighted Graphs}
\label{sec:intro:ours:weighted}

Our first result for weighted graphs maintains the local additive term
$2W_{2}$.

\begin{restatable}{theorem}{thmsecedge}
\label{thm:+2W2}
For every fixed $\varepsilon>0$, there is a decremental
$\paren{1+\varepsilon,2W_{2}}$-APASP algorithm for weighted
undirected graphs undergoing edge deletions and edge weight increases by an oblivious
adversary, with an expected total update time of
$\tilde O\paren{n^{\frac{3}{2}}m^{\frac{1}{2}}}$
when $m=n^{1+\Omega\paren{1}}$, and
$\tilde O\paren{n^{\frac{3}{2}+o\paren{1}}m^{\frac{1}{2}}}$
otherwise.
\end{restatable}

This algorithm is statically competitive with the static
$+2W_{2}$-APASP algorithm of Cohen and Zwick~\cite{CohZwi1997}. Furthermore, 
Dory, Forster, Nazari and de Vos~\cite{Doryetal} obtained
$\paren{2+\varepsilon}$-APASP within the same total update time. Since  $2W_{2}\paren{P}\leq\weight{P}$ for any path $P$, our algorithm improves their
 stretch to $\paren{1+\varepsilon,2W_{2}}\leq \paren{2+\varepsilon,0}$ without
increasing the  total update time bound.

We next obtain an additive stretch depending on both $W_{1}$
and $W_{2}$, with a total update time independent of  the sparsity of the graph:

\begin{restatable}{theorem}{thmtwoheaviestedges}
\label{thm:+2w1+2w2}
For every $0<\varepsilon<1$, there is a decremental
$\paren{1+\varepsilon,2W_{1}+2W_{2}}$-APASP algorithm for
weighted undirected graphs undergoing edge deletions and edge weight increases by an
oblivious adversary, with an expected total update time of
$\tilde O\paren{\frac{1}{\varepsilon^{9}}\cdot n^{\frac{7}{3}}}$.
\end{restatable}

Chechik~\cite{ChechikADO} obtained a decremental
$\paren{3+\varepsilon}$-APASP algorithm with total update
time $\tilde O\paren{mn^{\frac{1}{2}+o\paren{1}}}$.
Since $W_{1}\paren{P}+W_{2}\paren{P}\leq\weight{P}$
for every path $P$, the algorithm in \thmref{+2w1+2w2}
improves this stretch to
$\paren{1+\varepsilon,2W_{1}+2W_{2}}$-APASP.
For fixed $\varepsilon$, it also improves the total update
time whenever
$m\in \Omega\paren{n^\frac{11}{6}}$.  In particular, for $m=\Theta\paren{n^{2}}$, it reduces the total update time from $\tilde O\paren{n^{\frac{5}{2}+o\paren{1}}}$ to
$\tilde O\paren{n^{\frac{7}{3}}}$.

Our remaining weighted algorithms give two tradeoffs with
additive stretch $2\paren{k-1}W_{1}$.
hey build on the sparse static algorithm of Cohen and
Zwick~\cite{CohZwi1997} and the dense static algorithm of
Roditty and Sapir~\cite{Ariel+kDense}, respectively.

\begin{restatable}{theorem}{thmtwokheaviest}
\label{thm:w+2kw1}
For every integer $k\geq2$ satisfying
$k\in O\paren{\sqrt{\log n}}$
and every fixed $\varepsilon>0$, there is a decremental
$\paren{1+\varepsilon,2\paren{k-1}W_{1}}$-APASP algorithm
for weighted undirected graphs undergoing edge deletions and edge weight increases by
an oblivious adversary, with an expected total update time of
$\onolog{\paren{\frac{k^2 \log n}{\varepsilon}}^{\bigo{k}} n^{2-\frac{1}{k}}m^{\frac{1}{k}}}$ if $m=n^{1+\Omega(1)}$
or $\tilde O\paren{n^{2-\frac{1}{k}+o\paren{1}}m^{\frac{1}{k}}}$ otherwise.
\end{restatable}

\begin{restatable}{theorem}{thmweightedkdense}
\label{thm:w+2kw1-dense}
For every integer $k\geq 2$ satisfying
$k\in O\paren{\sqrt{\log n}}$
and every fixed $\varepsilon>0$, there is a decremental
$\paren{1+\varepsilon,2\paren{k-1}W_{1}}$-APASP algorithm
for weighted undirected graphs undergoing edge deletions and edge weight increases by
an oblivious adversary, with an expected total update time of
$\onolog{\paren{\frac{k^2\log n}{\varepsilon}}^{3k} n^{2 + \frac{2}{3k-2}}}
= \onolog{n^{2 + \frac{2}{3k-2} + \subpol}}$
\end{restatable}

Together, \thmref{w+2kw1} and
\thmref{w+2kw1-dense} give  total update time
bounds for sparse and dense graphs, respectively. For fixed $k$ and $\varepsilon$, these bounds meet at
$m=\tilde\Theta\paren{n^{1+\frac{2k}{3k-2}}}$. We use the faster one.

\subsection{Organization}\label{sec:intro:org}

The remainder of this paper is organized as follows. 
\secref{preliminaries} introduces the definitions and notation,
and \secref{ov} presents an overview of our algorithms and
the main challenges we face, together with the solution for each.
\secref{unweighted} gives the unweighted algorithms.
\secref{weighted:bounded} develops the bounded depth weighted $+2W_{2}$-APASP 
algorithm.
\secref{tools} presents the hierarchy and hopset tools used
to extend this construction to unrestricted distances in
\secref{w+2w2:unbounded}.
The remaining weighted algorithms appear in
\secref{w+2w1+2w2} and \secref{w+2kw1}, first for bounded
depth and then for unrestricted distances.
Finally, \secref{discussion} examines extensions to the fully dynamic
setting and the limitations of our methods.

\section{Preliminaries}
\label{sec:preliminaries}

\paragraph{Basic Notations and Definitions}
Consider a graph $G = \paren{V, E}$ with $n = |V|$ vertices and $m = |E|$ edges. For a weighted graph $G$, let $\weight{u,v}{G}$ denote the non-negative weight of an edge $(u,v) \in E$; when $G$ is clear from context, we write $\weight{u,v}$. $W_{\max}$ denotes the maximal edge weight in $G$. We assume that weights are integers and $W_{\max} \in \bigo{n^{\bigo{1}}}$\footnote{These are common assumptions in decremental graph algorithms. If $W_{\max}$ is not polynomial, all of the presented total update time carry an additional factor of $\log ^2 W_{\max}$}. For a vertex $v \in V$, let $N(v) = \{u \in V \mid (u,v) \in E\}$ denote the neighbor set of $v$, $E(v)$ the edges incident to $v$, and $\deg(v) = \abs{N(v)}$ its degree.

For a set of edges $F$ and a graph $H$, we write $F \subseteq H$ to indicate $F \subseteq E(H)$, and $H \cup F$ for the graph with edge set $E(H) \cup F$.

A path $P$ between $u$ and $v$ is a sequence of edges connecting them; $V(P)$ denotes the vertex set of $P$. For two paths $P_1, P_2$ whose endpoints are compatible, $P_1 \circ P_2$ denotes their concatenation. The \emph{weight} of $P$ is $\weight{P} = \sum_{e \in P}\weight{e}$, and its \emph{hop-length} $\hops{P}$ is the number of edges in $P$ \footnote{For unweighted graphs, $\hops{P} = \weight{P}$.}. For a weighted path $P$ and $1 \le i \le \hops{P}$, $W_i(P)$ denotes the weight of the $i$-th heaviest edge on $P$.

For $u,v \in V$, $u \squiggly v$ denotes the shortest path between $u$ and $v$ and $\dist{u,v}{G}=\weight{u \squiggly v}{G}$ denotes the shortest-path distance in $G$. The \emph{$\gamma$-hop distance} $\hopdist{u,v}{\gamma}{G}$ is the minimum weight of any $u$-$v$ path with at most $\gamma$ hops. In both notations, we omit $G$ when clear from context. 
A distance estimate $d$ is an $\paren{\alpha,\beta}$-approximation if $\dist{u,v} \le d[u,v] \le \alpha\dist{u,v} + \beta$ for all $u,v \in V$. 

We use $\expectation{\cdot}$ to denote expectation.

\paragraph{Hierarchy, Clusters and Bunches.}

For a set $S \subseteq V$, define $\dist{u,S}=\min_{s\in S}\dist{u,s}$ and the \emph{pivot} $\piv[S]{u}=\arg\mmin{\dist{u,s}}{s\in S}$ (breaking ties arbitrarily). Define the \emph{hierarchy-induced edge set} $E_S(u)=\{(u,v)\in E \mid \weight{u,v} < \dist{u,S}\}$ and $E_S=\bigcup_{u \in V}E_S(u)$.

Given a path $P_{u,v}$, let $\sizeindex{P_{u,v}}{E_{S}}$ be the number of edges in $P_{u,v}$ not in $E_{S}$. When $\sizeindex{P_{u,v}}{E_{S}}>0$, let $\rightabovearrow{P_{u,v}}{E_{S}}$ (resp.\ $\leftabovearrow{P_{u,v}}{E_{S}}$) be the first (resp.\ last) such edge from $u$'s perspective. To compute $\piv[S]{u}$, we maintain SSSP from an artificial source $\hat{s}$ connected to every $s \in S$ by a zero-weight edge.

\begin{definition}
\label{def:prelim:hierarchy}
Given $q \in \paren{0,1}$ and integer $k \ge 1$, the \emph{hierarchy} $\mathcal{R}\paren{q,k}$ consists of sets $S_0 \supseteq \cdots \supseteq S_k$ where $S_0 = V$, $S_k = \emptyset$, and for each $0 < i < k$ ,$S_i$ is formed by independently sampling each vertex of $S_{i-1}$ with probability $q$. The \emph{priority} of $v$ is the largest $j$ with $v \in S_j$. The $i$-th \emph{pivot} of $v$ is $\piv[i]{v} \in \arg\mmin{\dist{v,s}}{s \in S_i}$, breaking ties by highest priority (and then arbitrarily); if $v \in S_i$ then $\piv[i]{v} = v$.
\end{definition}

Given the hierarchy $\mathcal{R}\paren{q,k}$, for $1 \le i < k$ and $w \in S_i \setminus S_{i+1}$, the \textbf{cluster} of $w$ is $\cluster{w} = \{ u \in V \mid \dist{u,w} < \dist{u,S_{i+1}} \}$, and the $i$-th \textbf{bunch} of $u \in V$ is $\bunch{u}{i} = \{ w \in S_i \mid \dist{u,w} < \dist{u,S_{i+1}} \}$.

\begin{observation}
    \label{obv:prelim:clusters-and-bunches}
    For any $u \in V$ and $w \in S_i \setminus S_{i+1}$, $u \in \cluster{w} \iff w \in \bunch{u}{i}$.
\end{observation}

\paragraph{Dynamic Graphs and Adversary Models.}

A \emph{dynamic graph} undergoes a sequence of updates over time, indexed by $t = 0, 1, 2, \ldots$. In the \emph{decremental setting}, only edge deletions and weight increases are permitted; in the \emph{incremental setting}, only insertions and weight decreases; All algorithms in this paper operate in the decremental setting.

We also consider an adversary that makes the updates.
An \emph{oblivious adversary} is unaware of the randomness employed in the algorithm,
whereas an \emph{adaptive adversary} can observe the algorithm's choices.
Our algorithms assume oblivious adversaries.

We measure performance by \emph{total update time}: the cumulative time to process the entire update sequence. \emph{Time} refers to the number of updates applied so far (initialization is time~$0$). We write $\dist{u,v}{G}{t}$ and $\weight{u,v}{G}{t}$ for distance and edge weight at time $t$, omitting $G$ or $t$ when clear, and use $W_{i,t}(P)$ for the $i$-th heaviest edge weight on $P$ at time $t$.

We consider a path as a \emph{fixed set of edges} present from initialization until its first edge is deleted.

\paragraph{Even-Shiloach Trees.}

The Even-Shiloach (ES) tree~\cite{ESTree,KingWeightedESTree} maintains an exact SSSP tree from a root $s$ under decremental updates. Each vertex $v$ maintains a distance label $\level{s}{v} = \dist{s,v}$; when a deletion may increase some distances, the level of the relevant vertex is updated, and its' edges are scanned to check whether any of its neighbors requires a level update as well. A \emph{distance-bounded} ES-tree caps maintained distances at a threshold $\tau$: vertices with $\dist{s,v} > \tau$ are treated as disconnected.

\begin{lemma}[\cite{KingWeightedESTree}]
    Given a weighted undirected graph $G$ undergoing edge deletions and a distance bound $\tau$, a distance-bounded ES-Tree rooted at $s$ maintains exact distances $\dist{s,v}$ for all $v$ with $\dist{s,v} \le \tau$ in $\bigo{\tau \cdot m}$ total update time.
\end{lemma}

The bound holds because the edges of each vertex $v$ are scanned at most $\tau$ times (once per level increase); summing over all vertices gives $\bigo{\tau \cdot m}$.

\paragraph{Monotone Even-Shiloach Trees.}
A \emph{Monotone ES-Tree}~\cite{HKNMonESTree} extends the ES-tree to sub-graphs of decremental graphs, which may have edges inserted to them as edges are deleted from the input graph.

Let $G = (V, E, w)$ be a decremental graph and let $H = (V, E', w)$ be an auxiliary graph subject to edge deletions, edge weight increases, edge insertions and edge weight decreases. Let $T_u(H)$ be a Monotone ES-Tree rooted at $u \in V$ on $H$. The distance estimate for $v \in V$ in $T_u(H)$ is denoted by $\level{u}{v}$, and called $u$'s \emph{level} in $v$'s tree. When an edge $\paren{x, v}$ is inserted into $E'$ (or undergoes a weight decrease) at time step $t$, the edge is inserted into $T_u(H)$, but the standard relaxation step $\level{u}{v} \leftarrow \min\{\level{u}{v}, \level{u}{x} + \weight{x,v}\}$ is \textbf{omitted}. Instead, the edge $\paren{x, v}$ is registered in $v$'s incoming adjacency list without updating $\level{u}{v}$. The distance label $\level{u}{v}$ increases only when all incoming edges $\paren{z, v} \in E'$ fail to satisfy $\level{u}{z} + \weight{z,v} \le \level{u}{v}$.

Formally, the algorithm maintains the invariant: $\level{u}{v} \ge \mmin{\level{u}{x} + \weight{x,v}}{\paren{x,v} \in E'} \quad \forall u,v \in V$.

If a newly inserted edge $\paren{x,v}$ satisfies $\level{u}{x} + \weight{x,v} < \level{u}{v}$, $\paren{x,v}$ is defined to be \emph{stretched}. As long as $v$ has stretched edges incident on it, $\level{u}{v}$ does not change. The edge $\paren{x,v}$ becomes \emph{active} at the exact future time step $t' > t$ when $\level[t'-1]{u}{v} \le \level[t']{u}{x} + \weight{x,v}{t'}$. At this point, $\paren{x,v}$ serves as a valid incoming edge for $v$, preventing $\level{u}{v}$ from incrementing further. 

Monotone ES-Trees are designed to operate on emulators, which are auxiliary graphs $H$ for a decremental graph $G$ in which for every $u,v \in V$, at any point in time, $\dist{u,v}{G} \le \dist{u,v}{H} \le \alpha \dist{u,v}{G} + \beta$. The standard correctness argument to prove an $\paren{\alpha,\beta}$-approximation is that at any time, $\level{u}{v}$ holds the value of the $\dist{u,v}{H}$, or there are stretched edges on the path - in which case one must show that the old estimate suffices for the approximation purposes.

\begin{lemma}[\cite{HKNMonESTree}]
    \label{lem:prelim:monotone-es-properties}
    The following holds for a monotone ES-tree of $H$ rooted at $s$, maintained up to distance $\tau$:
    (1). Initially, $\level{s}{v} = \dist{s,v}{H}$ for all $v$ with $\dist{s,v}{H} \leq \tau$.
    (2). $\dist{s,v}{H} \leq \level{s}{v}$ at all times.
    (3). $\level{s}{v}$ never decreases.
    (4). Whenever $\level{s}{v}$ increases to a value $\leq \tau$, $\level{s}{v} = \mmin{\level{s}{x} + \weight{x,v}{H}}{x \in N(v)}$.
\end{lemma}

\begin{lemma}[\cite{HKNMonESTree}]
    \label{lem:prelim:monotone-es-update-time}
    For $\tau \geq 1$, the total update time of a monotone ES-tree up to level $\tau$ on a graph $H$ undergoing edge deletions, insertions, and weight increases is $\onolog{\mathcal{E}(H) \cdot \tau + \mathcal{W}(H) + n}$,
where $\mathcal{E}(H)$ is the total number of edges ever contained in $H$ and $\mathcal{W}(H)$ is the total number of updates to $H$.
\end{lemma}

The bound follows the same argument of ES-trees, replacing $\deg(v)$ with all historical edges incident on $v$.

\paragraph{Decremental Hopsets.}

A \emph{$\paren{\beta, \onepluseps}$-hopset} for $G = \paren{V,E}$ is a weighted edge set $F$ on $V$ such that for every $u, v \in V$,
\[
  \dist{u,v}{G} \le \hopdist{u,v}{\beta}{G \cup F} \le (\onepluseps)\dist{u,v}{G}.
\]
Adding $F$ to $G$ allows any shortest path to be $(\onepluseps)$-approximated using at most $\beta$ hops. We elaborate on the use of hopsets in \secref{tools:hopset}.

In our algorithms, we make use of the following theorem:
\begin{theorem*}[\cite{LackiNazari22}]
Given an undirected graph $G=(V,E)$ with polynomial weights, subject to edge deletions, we can maintain a $(\beta, 1+\varepsilon)$-hopset of size $\tilde{O}(n^{1+\frac{1}{2^k -1}})$ in total expected update time  $\tilde{O}(\frac{\beta}{\epsilon} \cdot (m+n^{1+\frac{1}{2^k -1}})n^{\rho})$, where $\beta= (O(\frac{\log n}{\varepsilon} \cdot (k+\frac{1}{\rho})))^{k+\frac{1}{\rho}+1}$, $k \geq 1$ is an integer, $0 < \varepsilon <1$ and $\frac{2}{2^k-1} < \rho <1$. 
\end{theorem*}

\section{Overview and Challenges}\label{sec:ov}


We begin with our unweighted algorithms as a warmup for
the weighted setting. These results also settle the dense unweighted case left
open by Dory, Forster, Nazari and de Vos~\cite{Doryetal}. A direct weighted extension provides an additive stretch depending on $W_{\max}$, motivating local bounds in
terms of $W_{i}$.
To obtain these bounds, we develop a decremental version
of the hierarchy induced edge sets of Baswana and
Kavitha~\cite{BK}.

We first present our $+2W_{2}$-APASP algorithm for depth $\tau$, then present our tools to obtain a
$\paren{1+\varepsilon,2W_{2}}$-APASP for arbitrary distances.  Additional hierarchy levels and different sampling probabilities
give three further near-additive algorithms: one with
additive stretch $2W_{1}+2W_{2}$ and two with additive
stretch $2\cdot\paren{k-1}W_{1}$.

\subsection{Unweighted Graphs}\label{sec:ov:unweighted}

We separate vertex pairs using a distance threshold $\tau$.
For pairs at distance greater than $\tau$, we use the
$\paren{1+\varepsilon', 2\cdot \paren{1+\frac{2}{\varepsilon'}}^{k-2}}$-APASP algorithm of Henzinger, Krinninger and
Nanongkai~\cite{HKN14} as a black box.
We choose $\tau$ so that its additive stretch is at most
$\varepsilon'  \cdot \dist{u,v}$ for these pairs. Taking $\varepsilon'=\frac{\varepsilon}{2}$ gives purely
multiplicative stretch $\onepluseps$.

For pairs at distance at most $\tau$, we choose a different strategy.
We use increasing degree thresholds $s_{i}$, dominating sets
$D_{i}$ for vertices of degree at least $s_{i}$ and dominating vertices. We define two edge sets: $E_{i}$ of edges for  which at least one endpoint has degree at most $s_{i}$ and $E^{*}$ of dominating edges across all degree thresholds. 
This is a decremental adaptation of the degree hierarchy
used by Dor, Halperin and Zwick~\cite{DorHalZwi2000}.

However, 
as edges are deleted from $G$, vertex degrees decrease,
dominating vertices may need to be replaced and surviving
edges may enter $E_{i}$ and $E^{*}$.
We maintain these edge sets as in the decremental algorithms of Dory,
Forster, Nazari and de Vos~\cite{Doryetal}.

At a high view, we maintain trees rooted in
$u\in D_{i}$ that use $E_{i+1}$, $E^{*}$ and shortcut edges between $u$ and vertices with previously computed distance estimates to $u$. 
To handle insertions into the auxiliary graphs, we use
Monotone ES-Trees~\cite{HKNMonESTree}, whose levels do not
decrease even when a new edge creates a shorter route.
The total update time of a tree depends linearly on its
depth. The degree thresholds balance the number of tree roots
against the size of their auxiliary graphs. 

For \algref{uw+2:bounded}, we consider two degree thresholds $s_{1}= n^{\frac{1}{3}}$ and $s_{2}= n^{\frac{2}{3}}$.
For each pair $u,v \in V$ we show that the auxiliary edge set over which 
$u$'s tree is maintained contains a path of weight at most
$\dist{u,v}+2$. 
We then prove by induction that the same additive guarantee
holds even when the Monotone ES-Tree uses levels retained from
earlier updates.

For \algref{uw+k:bounded}, we consider $k-1$ degree thresholds  $s_{i}= n^{\frac{i}{k}}$. The general tradeoff here requires a stronger inductive invariant. 
In the static setting,
the  analysis of Dor, Halperin and Zwick~\cite{DorHalZwi2000}
uses an intermediate additive stretch bound $+e_{i_1,j,i_2}$ for the additive offset between $u \in D_{i_1}$ and $v \in D_{i_2}$ such that  
the highest degree vertex on a shortest path $u\squiggly v$ has degree in $\left[s_{j-1},s_{j}\right)$.

When an edge on this path is deleted, the new shortest path
may increase $\delta\paren{u,v}$ but at the same time contain only vertices of degree less than $s_{j-1}$, which decreases the additive term.
As a result, the value $\dist{u,v} + e_{i_1,j,i_2}$ may either increase or  decrease, with no monotonicity guarantee.

Yet, Monotone ES-Trees do not allow distance estimate decreases, and in fact rely on the assumption that the approximation error is monotone.
To handle this, we strengthen the inductive claim to cover
every surviving path of at most $\tau$ edges and allow
looser intermediate bounds when degree levels change.
These relaxed bounds also hold for levels retained from
earlier updates, and their additive stretch remains at most
the claimed additive stretch.

\subsection{Challenges for Weighted Graphs}\label{sec:ov:weighted}

As a simple idea, assume we invoke \algref{uw+2:bounded} over a weighted graph.
Bounding the two additional edges of \algref{uw+2:bounded} by
$W_{\max}$ gives additive stretch $2W_{\max}$.
Besides the loss of informativeness  in the stretch as  discussed in the introduction, 
this creates a major runtime obstacle: even for pairs at distance
at most $\tau$, the trees may require depth up to
$\tau+2W_{\max}$, introducing a runtime dependence on $W_{\max}$.
Local additive stretch can avoid this dependence.
For example, since $2W_{2}\paren{P}\leq\weight{P}$ for every
path $P$, a $+2W_{2}$ stretch on paths of weight at most
$\tau$ has a depth upper bound of at most $2\tau$. We seek  to obtain such guarantees while keeping the auxiliary
graphs sparse. While doing so, we present three challenges that we face: \emph{size}, \emph{depth} and \emph{monotonicity}. 

To obtain sparse auxiliary graphs, we use an edge hierarchy
with pivots, as in Cohen and Zwick~\cite{CohZwi1997} and
Baswana and Kavitha~\cite{BK}.
Let $p_{S}\paren{u}$ be the pivot
of $u$ from a set $S$.
The edge set $E_{S}$ contains every edge $\paren{u,v}$ whose weight is smaller than $\dist{u,p_{S}\paren{u}}$. Thus, if an  edge $\paren{x,y}\notin E_{S}$, both $\dist{x,p_{S}\paren{x}},\dist{y,p_{S}\paren{y}}\leq w\paren{x,y}$.
Therefore, a detour through $p_{S}\paren{x}$ or $p_{S}\paren{y}$  can have additive
stretch at most $2w\paren{x,y}$.

\paragraph{Size challenge.} 
The total update time of Monotone ES-Trees~\cite{HKNMonESTree}
depends on all distinct edges ever present in their
auxiliary graphs.
Cohen and Zwick deterministically retain a fixed number
of the lightest incident edges at each vertex.
An (oblivious) adversary with a strategy of always deleting the lightest edge from $G$ forces all $m$
original edges to eventually be retained  over time, even though at each time the set is small.

The approach of Baswana and Kavitha involves
sampling each vertex independently into $S$ with probability $q$.
For each vertex $u$, we choose a closest vertex of $S$ as
its pivot and retain the incident edges of weight less
than $\dist{u,S}$.
The resulting edge set $E_{S}$ has expected size
$\bigo{\frac{n}{q}}$ at any fixed time.
As distances to $S$ increase, surviving edges may enter
$E_{S}$, so this bound does not control its cumulative size.
We first maintain a bounded depth version $E_{S}^{\tau}$.
We charge insertions to maintained distance levels, with
expected charge $\bigo{\frac{1}{q}}$ per vertex and level.
Since each vertex has at most $\bigo{\tau}$ such levels,
the expected cumulative size of $E_{S}^{\tau}$ is
$\bigo{\frac{n\tau}{q}}$.
In \secref{ov:unbounded} we discuss an extension to arbitrary depths, replacing $\tau$ with a subpolynomial factor.

\paragraph{Depth challenge.} 
Even a sparse auxiliary graph may have shortest paths of weight $\Theta\paren{nW_{\max}}$. Maintaining ES-Trees up to these distances is too expensive. We would therefore like to maintain a hopset $F$ to reduce hop-length of paths. 

If $F$ is a hopset for $E$, the hopbound guarantee only applies for $E\cup F$. However, our sparsity constraints allow us to consider only $E_{S}\cup F$, for which the hopbound guarantee does not necessarily apply. 
Another attempt it to define $E_S$ over $E \cup F$. Yet, a missing 
edge from $E_{S}$ may represent an edge of $F$, which in turn represents a long path in $E$, and hence its weight cannot be
upper bounded by a single original edge from $E$.
On the other hand, constructing a decremental hopset for $E_{S}$ with the existing algorithms is problematic, as it must also 
accommodate edge insertions. 

We adapt the construction of \L\k{a}cki and
Nazari~\cite{LackiNazari22} to use the same sampled vertex hierarchy for the hopset construction 
as for our APASP algorithms. The distances to the sampled sets determine both which
edges are retained and how far the hopset trees must reach.
We use this relation to maintain a single hopset $F$
that provides approximating routes with few hops in
$E_{S}\cup F$ for every hierarchy level despite insertions.

Yet, few hops alone do not bound the weight of a route.
We therefore combine this hopset with the edge weight
scaling technique of Bernstein~\cite{BernsteinRounding}.
At the relevant distance scale, the required tree depth
depends on the hopbound and the accuracy rather than
the entire original path weights.
The exact process is described in \secref{ov:unbounded}. Scaling introduces additional multiplicative stretch,
which we absorb into the final factor $\onepluseps$.



\paragraph{Monotonicity challenge.} 
As discussed
in the introduction, this  concerns upper bounds that may decrease,
such as $\dist{u,v} + 2W_{2}\paren{u \squiggly v}$,
while Monotone ES-Tree levels cannot decrease.
Our analysis overcomes  this by 
establishing an upper bound for every surviving path. Taking as example $+2W_{2}$, this means that we establish  upper bounds of the form $d\bracke{u,v}\leq \mmin{w\paren{P}+2W_{2}\paren{P}}{P}$, where $P$ is an arbitrary path between $u$ and $v$ that survives at the current time. Since the graph is decremental, each surviving path at a current point in time already existed by the start. Such an approach was not necessary in the static regime, as in the $+2W_{2}$-APASP algorithm of Cohen and Zwick \cite{CohZwi1997}, yet is essential for the decremental setting.

\subsection{Algorithms for Weighted Bounded Depth}\label{sec:ov:bounded}
All of our algorithms share a common two-phase structure. Given a parameter $\tau$, we first design a bounded-distance algorithm that handles all pairs of vertices at distance at most~$\tau$. We then extend it to pairs at arbitrary distance using the general framework described in \secref{ov:unbounded}.

We demonstrate our tools and techniques for weighted graphs in our bounded weighted algorithms with
\algref{w+2w2:bounded}, which maintains $+2W_{2}$-APASP
for pairs at distance at most $\tau$.
We sample a pivot set $S$ and maintain $E_{S}^{\tau}$.
For each $s\in S$, we maintain an ES-Tree on $G$ up to
depth $2\tau$.
The nearest pivot of a vertex $v$ may change even when
$\dist{v,S}$ remains fixed.
Connecting $v$ directly to successive pivots can therefore
introduce distinct edges without an increase in this distance, causing a large number of updates.
For each $v\in V$, we instead introduce a fixed auxiliary
vertex $\tilde p\paren{v}$, called its \emph{auxiliary pivot}.
The edge $\paren{v,\tilde p\paren{v}}$ is weighted by the
maintained distance from $v$ to $S$.
Each source $u\in V$ also has edges to these auxiliary
pivots, with weights supplied by the pivot ES-Trees and
updated only by increases.
Pivot changes are thus handled through weight updates
on fixed edges, using only $\bigo{n}$ distinct auxiliary
pivot edges in each auxiliary graph over the entire
update sequence.

For each $u\in V$, we maintain two Monotone ES-Trees up to
depth $2\tau$.
The first uses $E_{S}^{\tau}$, the original edges incident
to $u$ and the auxiliary pivot connections.
The second uses $E_{S}^{\tau}$ and the auxiliary pivot
connections, together with a shortcut $\paren{u,v}$ to
each $v\in V$ whose weight is supplied by the first-round
estimate from $v$ to $u$.
After each deletion, we update the pivot ES-Trees, all
first-round trees and then all second-round trees.
For a query $\paren{u,v}$, we return the minimum of the
second-round estimate from $u$ to $v$ and the one from
$v$ to $u$.

These two rounds adapt the approach of Cohen and
Zwick~\cite{CohZwi1997}.
If $\paren{x,y}$ is the only edge outside $E_{S}^{\tau}$
on a path from $u$ to $v$ of weight at most $\tau$, the
first tree rooted at $y$ contains this original edge.
The first-round estimate from $y$ to $u$ supplies the
shortcut $\paren{u,y}$ in the second tree rooted at $u$,
which can then follow the remaining suffix inside
$E_{S}^{\tau}$.
With multiple edges outside $E_{S}^{\tau}$, a pivot detour
can be charged to the lighter of the first and last,
whose weight is at most $W_{2}$.
Separate trees allow  estimates from the second round  to improve
upon estimates from the first round while each tree remains
monotone over time.

For the analysis, we prove the bound
$\weight{P}+2W_{2}\paren{P}$ for every surviving path $P$
with $\weight{P}\leq\tau$, by induction on time.
Every such path was present at earlier updates, with
no larger edge weights.
We handle stretched edges by applying the induction
to retained levels.
A shortcut also requires care: $\paren{u,y}$ may inherit
stretch from $y$'s first tree even when it is not stretched
in $u$'s second tree.
Here the estimates pass between trees at the same hierarchy
level.
We apply the induction in the tree supplying the estimate
to the corresponding subpath of $P$.
Its inherited stretch is at most $2W_{2}\paren{P}$,
and following an unstretched suffix adds only its weight.

Additional hierarchy levels and different sampling probabilities
give our other bounded weighted algorithms, with additive
stretches $+2W_{1}+2W_{2}$ and $+2\cdot \paren{k-1}W_{1}$.
These algorithms use nested sampled sets $S_{i}$ and their
hierarchy induced edge sets.
Trees at one level use distance estimates from higher
levels as shortcut weights.
Across hierarchy levels, this \emph{stretch propagation
challenge} requires further control:
these shortcuts may retain estimates computed before the
current pivot was assigned.
Their additive stretch may therefore involve edges of an
earlier detour, rather than distinct edges of the comparison
path $P$ as in the static analysis.
We bound some of these contributions by $W_{1}\paren{P}$,
where the static analysis uses $W_{i}\paren{P}$ for $i>1$.
We prove the resulting guarantees by induction on both
time and hierarchy level for every surviving path $P$
with $\weight{P}\leq\tau$.

\subsection{Beyond Bounded Depth}\label{sec:ov:unbounded}

To extend our algorithms to arbitrary graph distances and remove the dependence on $\tau$,
we develop two algorithmic tools that, combined with the edge-rounding technique of Bernstein~\cite{BernsteinRounding}, generalize our bounded constructions to arbitrary, unbounded distances with only a subpolynomial total update time overhead.

For a parameter $k = \bigo{\sqrt{\log n}}$,
we maintain an unbounded variant of the hierarchy-induced edge sets $E_{S_i}$ for all $i = 1 ,\dots ,k-1$
using decremental near-linear $(\onepluseps)$-SSSP data structures.
At each vertex, we retain incident edges of weight below
its maintained approximate distance to $S_{i}$.
We show that each such estimate changes only
$n^{\subpol}$ times, or $\log^{\bigo{1}}n$ times in some
regimes.
Charging newly inserted edges to these changes replaces
the $\tau$ factor in the cumulative size bound by the
corresponding factor.
We also show that the set is maintainable within the total update time of our APASP algorithms, and we offer several size-update time tradeoffs.


Because paths within $E_{S_i}$ can still require $\Omega(n)$ hops, we construct a specialized hopset for the hierarchy-induced edge sets.
To construct such a hopset, we adapt the decremental hopset framework of \L\k{a}cki and Nazari~\cite{LackiNazari22} to use the same
sampled hierarchy as our APASP algorithms. The distances to the next sampled set determine both which
edges are retained and which vertices the hopset trees must reach.
Using the same hierarchy aligns these two conditions.
We prove that this process results in a \emph{single} set of edges $F$ 
with expected size $\onolog{\abs{E_{S_1}}}$
that simultaneously guarantees a low-hop $(\onepluseps)$-alternative for paths in $E_{S_i}$, for every $1 \le i < k$, despite continuous edge insertions into these sets.
To align with our arbitrary-path analysis, we show that our hopset guarantee applies for \emph{any} path in $E_{S_i}$ rather than only shortest paths. At last, we prove that the hopset can be maintained within the total update time of our APASP algorithms, and again, we offer some hopbound-update time tradeoffs.


Finally, we apply the edge weight rounding technique of
Bernstein~\cite{BernsteinRounding} at
$\bigo{\log\paren{nW_{\max}}}$ distance scales.
At the relevant scale, rounding and rescaling the edge
weights gives the routes supplied by the hopset a
rescaled weight bounded in terms of the hopbound and
the accuracy.
We can therefore invoke our bounded algorithms with
tree depth determined by these parameters.
Taking the minimum estimate across all scales yields
the final
$\paren{\onepluseps,\msum{\beta_{i}W_{i}}{i=1}{k-1}}$-APASP over arbitrary distances.

\section{Unweighted Graphs}
\label{sec:unweighted}
Dor, Halperin, and Zwick~\cite{DorHalZwi2000} presented two combinatorial algorithms
for additive APASP in unweighted graphs. The first achieves a $+2$ approximation in
$\tilde{O}(\min\{n^{3/2}m^{1/2}, n^{7/3}\})$ time, and the second, for every even
$k > 2$, achieves a $+k$ approximation in
$\tilde{O}(\min\{n^{2-\frac{2}{k+2}}m^{\frac{2}{k+2}}, n^{2+\frac{2}{3k-2}}\})$
time, where the two terms correspond to the sparse and dense regimes, respectively.
While improvements have been obtained via fast matrix multiplication
techniques~\cite{DenKirVicVasZho2022,Durr2023,SahYe2023,JinKSW25}, these results remain the state of the art for
combinatorial algorithms after over two decades, serving as a benchmark for
efficient combinatorial APASP algorithm design.

Recently, Dory, Forster, Nazari, and De Vos~\cite{Doryetal} extended these results to the decremental setting. They present a $\paren{\onepluseps, 2}$-APASP algorithm for
unweighted decremental graphs with total update time
$\tilde{O}(n^{\frac{3}{2}+o(1)}m^{\frac{1}{2}})$, and a
$\paren{\onepluseps, k}$-APASP algorithm with total update time
$\tilde{O}(n^{2-\frac{2}{k+2}+o(1)}m^{\frac{2}{k+2}})$:
\begin{theorem*}[\cite{Doryetal}]
\label{thm:uw:dory-apsp}
Given an undirected unweighted graph $G$, a constant $0 < \varepsilon < 1$, and an integer $2 \leq k \leq \log n$, there is a decremental data structure maintaining a $(\onepluseps,2(k-1))$-approximation for APSP with constant query time. The expected total update time is $$\mmin{\bigo{n^{2-\frac{1}{k}+o(1)} m^{\frac{1}{k}}},\onolog{\paren{n^{2-\frac{1}{k}} m^\frac{1}{k}}\bigo{\frac{1}{\varepsilon}}^{\frac{k}{\rho}}}}$$ where $m = n^{1+\rho}$.
\end{theorem*}

In this section, we present decremental algorithms for $\paren{\onepluseps, 2}$
and $\paren{\onepluseps, k}$-APASP in unweighted graphs with total update times of
$\onolog{n^{\seventhirds}}$ and
$\onolog{n^{2+\frac{2}{3k-2}+\subpol}}$, respectively. These match the dense-regime
runtimes of the static algorithms of Dor, Halperin, and Zwick~\cite{DorHalZwi2000}
up to a sub-polynomial factor and constitute the first decremental algorithms
achieving these approximation guarantees at these runtimes.

Both results follow the same bounded-then-unbounded strategy: we construct a $\tau$-bounded $+k$-APASP algorithm covering all vertex pairs at distance at most $\tau$, and combine it with the algorithm of Henzinger, Krinninger and
Nanongkai~\cite{HKN14} for the large-distance regime, which provides a $(1+\varepsilon)$-approximation for distant pairs:
\begin{lemma}[\cite{HKN14}]
\label{lem:uw:hkn14-apsp}
For any constant $0 < \epsilon < 1$ and integer $2 \le k \le \log n$, there is a $(1+\varepsilon, 2(1+\frac{2}{\varepsilon})^{k-2})$-approximation algorithm for decremental APSP on unweighted undirected graphs with constant worst-case query time and expected total update time $\onolog{n^{2+\frac{1}{k}}(\frac{37}{\varepsilon})^{k-1}}$.
\end{lemma}

We begin with the $+2$-APASP algorithm.

\subsection{$(\onepluseps,2)$-APASP}
\label{sec:uw+2}

In this section we establish the following theorem:

\thmunweightedplustwo*

\begin{algorithm}
\caption{$+2$ up to depth $\tau$}
\label{alg:uw+2:bounded}
\textbf{Input:} An unweighted, undirected graph $G = \paren{V, E}$ and a depth parameter $\tau$.

\textbf{Output:} A matrix $\{ d[u,v] \}_{u,v \in V}$ of estimated distances.

$s_1 \gets n^{1/3}$, $s_2 \gets n^{2/3}$

\For{$i = 1$ to $2$}
{
    $V_i \gets \{ v \in V \mid \deg(v) \geq s_i \}$
    
}
\For{$i = 1$ to $2$}
{
    $E_i \gets \{ \{u,v\} \in E \mid \deg(u) \leq s_i \text{ or } \deg(v) \leq s_i \}$
    
}

$D_0 \gets V$

\For{$i = 1$ to $2$}
{
    $(D_i,E_i^*) \gets $ dominating set of $G$ with parameter $s_i$
    
}

$E^* \gets E_1^* \cup E_2^*$

$E_3 \gets E$, $E_0 \gets \emptyset$

\For{every $u \in D_2$}
{
    Maintain an ES-Tree on $G$ from $u$ up to depth $\tau+2$
}
\For{every $u \in D_1$}
{
    Maintain a Monotone ES-Tree on $G_2=(V, E_2)$ from $u$ up to depth $\tau+2$
    
}
\For{every $u \in D_0$}
{
    Maintain a Monotone ES-Tree on $G_1(u)=(V, E_1 \cup E^* \cup (D_2 \times V) \cup (\singleton{u} \times D_1))$ up to depth $\tau+2$
}
\end{algorithm}

\paragraph{Algorithm Description.} Our approach adapts the framework of \cite{DorHalZwi2000} to the decremental setting. \algref{uw+2:bounded} receives an unweighted graph $G=\paren{V,E}$ and a depth parameter $\tau$. Initially, it computes hitting sets $D_{1}$ and $D_{2}$ for vertices with degrees at least $s_{1}=n^{1/3}$ and $s_{2}=n^{2/3}$, respectively. We define the corresponding edge sets as $E_{i}^{*}=\{\paren{u,v}\in E \mid u\in D_{i}, v\in V_{i}\}$ and set $E^{*}=E_{1}^{*}\cup E_{2}^{*}$. Throughout the sequence of edge deletions, \algref{uw+2:bounded} maintains three instances of Monotone ES-Trees up to bounded depth $\tau+2$.

In the first two iterations, \algref{uw+2:bounded} maintains ES-Trees from each $u \in D_2$ on $G$ and Monotone ES-Trees from each $u \in D_1$ on $G_2 = \paren{V,E_2}$.
In the third iteration, for every $u \in D_0 = V$, \algref{uw+2:bounded} creates shortcuts from each $x_2 \in D_2$ to each $y_2 \in V$ with weight $\level{x_2}{y_2}$ and from each $x_1 \in D_1$ to $u$ with weight $\level{x_1}{u}$. \algref{uw+2:bounded} then maintains a Monotone ES-Tree from $u$ on the graph $G_1(u) = \paren{V, E_1 \cup E^* \cup (D_2 \times V) \cup (\singleton{u} \times D_1)}$.

The final estimate is $d[u,v] = \mmin{\level{u}{v},\level{v}{u}}$.

\paragraph{Maintaining Distances Under Edge Deletions.}
The sets $D_0$, $D_1$ and $D_2$ are created at the beginning of the algorithm and remain fixed.
Since the graph $G$ is decremental, $E_0 = \emptyset \subseteq E_1 \subseteq E_2 \subseteq E_3 = E$ holds throughout all graph updates. When the degree of a vertex $v$ falls below $s_{i-1}$ for the first time, we insert the edges incident on $v$ into $E_i$.

Each time the distance estimate between $x \in D_2$ and a node $v \in V$ increases, we update the weight
of the corresponding shortcut edge $(x,v)$ in $G_1(u)$ for every $u\in D_0 = V$.

Whenever the distance estimate between $w \in D_1$ and $u \in V$ increases, we update the corresponding shortcut edge $\paren{w,u}$ in $G_1(u)$.

We can have at most $\tau+2$ updates per shortcut. If the distance estimate exceeds $\tau+2$, meaning $u$ leaves $x$'s tree, we remove the shortcut edge from every graph containing it.

The edge set $E^*$ is also used in the algorithm of Dory, Forster, Nazari, and De Vos \cite{Doryetal} for $+2k$-APASP, and we adapt their method for maintaining it:

When an edge $(v, x) \in E_i\sstar$ between a node $v \in V_i$ and $x \in D_i$ is deleted from the graph $G$, we distinguish two cases. If $v$ still has a neighbor $w \in D_i$, we first add a new edge $(v, w)$ to the set $E_i\sstar$ and then delete the edge $(v, x)$.
Otherwise, we add all edges incident to $v$ in $G$ to $E_{i+1}$, and then delete the edge $(v, x)$.

\paragraph{Differences from the static algorithm of Dor, Halperin, and Zwick~\cite{DorHalZwi2000}.}
The primary difference from the static algorithm is the use of Monotone ES-Trees in place of Dijkstra's algorithm.
This substitution introduces a key challenge: edge insertions into the subgraphs $G_1(u)$ and $G_2$ can cause underlying distances to decrease,
which Monotone ES-Trees ignore to preserve monotonicity.

Furthermore, our algorithm differs from the decremental algorithm of Dory, Forster, Nazari, and de Vos \cite{Doryetal} by incorporating shortcut edges in the constructed subgraphs - a distinct challenge involving out-of-sync edge weight propagation.

The use of Monotone ES-Trees creates two distinct challenges that we must address:
\begin{itemize}
     \item \textbf{Shortcut accuracy.} For every $w \in D_1$, the level $\level{w}{u}$ defines a shortcut in $G_1(u)$ and also serves as a distance estimate for $\dist{w,v}$. We must show both usages yield sufficient accuracy and prevent cascading errors. Any error in the approximation of $\dist{w,u}$ may increase the error of estimated distances in $u$'s tree.
     \item \textbf{Final estimate accuracy.} For every $u \in D_0 = V$, the level $\level{u}{v}$ computed on $G_1(u)$ serves as the estimate for $\dist{u,v}$. We must show this estimate maintains the desired approximation throughout all deletions.
\end{itemize}

We must also show that all computed paths have depth at most $\tau+2$, as this is the depth parameter given to the tree and larger paths will not be explored.

The use of Monotone ES-Trees also affects the total update time of the algorithm: since the total update time of the Monotone ES-Tree depends on the total number of edges ever added to those subgraphs, we must bound this number. Moreover, the use of ES-Trees (regular and monotone) introduces an additional $\tau$ factor to the total update time.

We begin with a total update time analysis.

We use the following claim\footnote{The original presentation in \cite{Doryetal} uses fixed values for $\{s_i\}_{i=1}^{k-1}$, but the proof itself does not rely on these fixed values and apply for variable values of $\{s_i\}_{i=1}^{k-1}$ and in particular our values.}\footnote{The original presentation in \cite{Doryetal} uses a different indexing convention where the degree thresholds are decreasing as the index grows and hence stated differently. The claim presented here is an adaptation to our indexing convention which uses an ascending order.} from~\cite{Doryetal} to bound the total number of edges in $E^*$:

\begin{claim}[\cite{Doryetal}]
    \label{clm:uw+2:doryetal-total-edges-estar}
    For every $k \ge 2$, the expected total number of edges added to the set $E\sstar$ throughout all edges updates is
    bounded by $\bigo{\frac{m \log n}{s_{k-1}} + \sum_{i=1}^{k-1}\frac{n \log n\cdot s_i}{s_{i-1}}}$.
\end{claim}

\begin{lemma}[Total Update Time]
    \label{lem:uw+2:total-update-time}
    The expected total update time of \algref{uw+2:bounded} is $\onolog{\tau n^{7/3}}$.
\end{lemma}
\begin{proof}
    We bound the total update time by evaluating the number of edge scans across all three iterations of tree maintenance.

    Since $D_1$ and $D_2$ are chosen at the beginning of the algorithm and remain fixed, we have $|D_1| = \onolog{n^{2/3}}$ and $|D_2| = \onolog{n^{1/3}}$.
    
    Using \lemref{prelim:monotone-es-update-time}, we must bound $\mathcal{E}\paren{H}$ and $\mathcal{W}\paren{H}$ for $H \in \{G_1(u),G_2\}_{u \in V}$.

    For the first iteration, we maintain exact ES-Trees up to maximum depth $\tau+2$ from each vertex in $D_2$ on the unweighted graph $G$. Since $G$ undergoes only edge deletions, each edge in $G$ is scanned at most once per tree level. Therefore, the total expected time spent updating these trees is bounded by:
    $$\onolog{|D_2| \cdot \tau \cdot m} = \onolog{\frac{n}{s_2} \cdot \tau \cdot n^2} = \onolog{\frac{\tau n^3}{s_2}} = \onolog{\tau n^{7/3}}$$

    For the second iteration, we maintain Monotone ES-Trees from each $w \in D_1$ up to depth $\tau+2$ within the subgraph $G_2 = (V, E_2)$. By definition, $E_2$ contains only edges incident to vertices whose degree falls below $s_2$, meaning the maximum degree in $G_2$ is strictly bounded by $s_2$, and hence the total number of edges in $E_2$ is bounded by $\onolog{n \cdot s_2}$. Since $E_2 \subseteq G$ and $G$ is unweighted, each edge has only one possible update: deletion.
    Consequently, the expected total update time for the second iteration is bounded by:
    
    $$\onolog{|D_1| \cdot \tau \cdot n s_2} = \onolog{\frac{n}{s_1} \cdot \tau \cdot n s_2} = \onolog{\tau n^{7/3}}$$

    Consider now the third iteration. From each $u \in D_0 = V$, a Monotone ES-Tree is maintained on $G_1(u)=(V, E_1 \cup E^* \cup (D_2 \times V) \cup (\singleton{u} \times D_1))$.
    
    Similarly to $E_2$, each vertex can contribute at most $s_1$ edges to $E_1$, and hence the total number of edges in $E_1$ is bounded by $\bigo{n\cdot s_1} = \bigo{n^{4/3}}$.
    
    For the size of $E^*$, it follows from \clmref{uw+2:doryetal-total-edges-estar} that the total number of edges ever present in $E^*$ is bounded by $\onolog{\frac{m}{s_1}+\frac{n\cdot s_1}{s_2}}=\onolog{n^{4/3}}$.
    
    Since the expected size $\expectation{\abs{D_2 \times V}} = \onolog{n^{\frac{4}{3}}}$ and $\expectation{\abs{D_1 \times \singleton{u}}} = \expectation{\abs{D_1}} = \onolog{n^{\frac{2}{3}}}$, and these sets are immutable once set,  the expected total number $\mathcal{E}\paren{G_1(u)}$  of edges ever present in $G_1(u)=\paren{V,E_1 \cup E\sstar \cup \paren{D_2 \times V} \cup \paren{D_1 \times \singleton{u}}}$ is bounded by $\onolog{n^{\frac{4}{3}}}$ for every $u \in V$.
    
    Since we have weighted edges in $G_1(u)$, we must consider $\mathcal{W}\paren{G_1(u)}$ - the total number of updates in $G_1(u)$. Observe that the only weighted edges in $G_1(u)$ are shortcuts, and each shortcut edge can have at most $\tau+2$ updates, as this is the number of times the distance between its endpoints can change. We conclude that $\mathcal{W}(G_1(u))\le (\tau+3) \cdot \mathcal{E}(G_1(u))$ for every $u \in V$.
    Therefore, the expected total update time spent on maintaining a Monotone ES-Tree on $G_1(u)$ for every $u \in V$ is $\onolog{\tau \cdot n \cdot n^{\frac{4}{3}}} = \onolog{\tau n^{\frac{7}{3}}}$.
\end{proof}

We now turn to analyze the approximation guarantee. The proof is by induction on $t$. We begin with $t=0$.

 \begin{lemma}
    \label{lem:uw+2:approx-base-case}
     At the beginning of the algorithm, it holds that $\dist{u,v} \le d[u,v] \le \dist{u,v} + 2$ for every $u,v \in V$ such that $\dist{u,v} \le \tau$.
 \end{lemma}
\begin{proof}
    We consider three cases:
    
    \ccase{1} There exists a shortest path between $u$ and $v$ that contains a vertex $w$ from $V_2$.
    
    Let $w' \in D_2$ such that $(w, w') \in E$.
    The edges $(u, w')$ and $(w', v)$ belong to $G_1(u)$, on which a Monotone ES-Tree is maintained from $u$.
    
    The weights of these edges are $\level{w'}{u}$ and $\level{w'}{v}$, the estimates found by the ES-Tree on $G$ from $w' \in D_2$.
    Moreover, it holds that $\dist{w',v} \le \dist{w,v} + 1$ and, similarly, $\dist{w',u} \le \dist{w,u} + 1$. Also, $\dist{w',v} \le \dist{w,v} + 1 \le \dist{u,v} + 1 \le \tau +1$, and identically $\dist{w',u} \le \tau + 1$.

    \figref{uw+2:base-case} (a) demonstrates this case.
    
    Since $w' \in D_2$, we maintain an ES-Tree from $w'$ on $G$ up to depth $\tau+2$, and hence these distances are explored and $\level{w'}{u}=\dist{w',u}$ and $\level{w'}{v}=\dist{w',v}$.
    
    Consequently, the graph $G_1(u)$, on which $u$ maintains a Monotone ES-Tree (as $u \in D_0 = V$), contains edges $(u,w')$ and $(w',v)$ with weights $\weight{u, w'}=\level{w'}{u}$ and $\weight{w',v}=\level{w'}{v}$, which form a path from $u$ to $v$ of weight $\weight{u,w'} + \weight{w',v} =\level{w'}{u} + \level{w'}{v}  \le \dist{u, w} + 1 + \dist{w, v} + 1 = \dist{u, v} + 2$. The weight of this path is at most $\tau+2$ and hence is considered within $u$'s tree.
    
    Since this is the beginning of the algorithm, it follows from \lemref{prelim:monotone-es-properties} that $\dist{u,v}{G_1(u)} \le \level{u}{v} \le \weight{u,w'} + \weight{w',v} \le \dist{u,v} + 2$.
    
    \ccase{2} There exists a shortest path between $u$ and $v$ that contains vertices from $V_1$ but not from $V_2$.
    
    Let $w$ be the last vertex on the path that belongs to $V_1$.
    All edges on the path from $w$ to $v$ touch vertices not in $V_1$ and therefore belong to $E_1 \subseteq G_1(u)$, on which a Monotone ES-Tree is run from $u$.
    
    Let $w' \in D_1$ such that $(w, w') \in E\sstar$.
    The graph $G_1(u)$ contains weighted edges connecting $u$ to all of the vertices in $D_1$, and in particular, a weighted edge $\paren{u, w'}$ with $\weight{u, w'} = \level{w'}{u}$.
    
    Since all edges on the path from $u$ to $w$, as well as $(w, w')$, belong to $E_2$, there exists a path $P_{w',u}^* = w' \rightarrow w \squiggly u$ of length at most $\dist{u, w} + 1\le \tau+1$, and hence this path is explored by the Monotone ES-Tree from $w' \in D_1$ on $G_2$. Since this is the beginning of the algorithm, $\level{w'}{u}=\dist{u,w'}{G_2} \le \dist{u, w}{G} + 1$.
    
    Then $G_1(u)$ contains a path $P_{u,v}^* = u \rightarrow w' \squiggly v$ of length at most
    $\level{w'}{u} + \dist{w', w} + \dist{w, v} \le \dist{u, w} + 1 + 1 + \dist{w,v} \le \dist{u,v} + 2$.
    
    This path is explored by the tree. Since this is the beginning of the algorithm, $\dist{u,v}{G_1(u)} \le \dist{u,v} + 2$.

    An example can be seen in \figref{uw+2:base-case} (b).
    
    \ccase{3} There exists a shortest path between $u$ and $v$ that does not contain any vertex from $V_1$.
    
    This shortest path is contained in $E_1$. Since this is the beginning of the algorithm, we have $\level{u}{v}=\dist{u,v}{G_1(u)}=\dist{u,v}$.
    
    \begin{figure}
    \centering
    \begin{tikzpicture}[
        vertex/.style={circle, draw, minimum size=18pt, inner sep=0pt, font=\small},
        shortcut/.style={draw, thick, color={rgb,255:red,74; green,144; blue,226}},
        invisibleEdge/.style={draw, thick, dashed}
    ]
        
        \begin{scope}
            \node[vertex] (ua) at (0,0) {$u$};
            \node[vertex] (wa) at (2.2,0) {$w$};
            \node[vertex] (va) at (4.4,0) {$v$};
            \node[vertex] (wpa) at (2.2,1.6) {$w'$};
            
            \draw[thick, dashed, decorate, decoration={snake, amplitude=1.5pt, segment length=6pt}] (ua) -- (wa);
            \draw[thick, dashed, decorate, decoration={snake, amplitude=1.5pt, segment length=6pt}] (wa) -- (va);
            \path[invisibleEdge] (wa) -- (wpa);
            \path[shortcut] (ua) -- (wpa);
            \path[shortcut] (wpa) -- (va);
            
            \node[font=\small] at (2.2,-0.6) {(a)};
        \end{scope}

        \begin{scope}[shift={(6.5,0)}]
            \node[vertex] (ub) at (0,0) {$u$};
            \node[vertex] (wb) at (2.2,0) {$w$};
            \node[vertex] (vb) at (4.4,0) {$v$};
            \node[vertex] (wpb) at (2.2,1.6) {$w'$};
            
            \draw[thick, dashed, decorate, decoration={snake, amplitude=1.5pt, segment length=6pt}] (ub) -- (wb);
            \draw[thick, decorate, decoration={snake, amplitude=1.5pt, segment length=6pt}] (wb) -- (vb);
            \path[draw, thick] (wb) -- (wpb);
            \path[shortcut] (ub) -- (wpb);
            
            \node[font=\small] at (2.2,-0.6) {(b)};
        \end{scope}

    \end{tikzpicture}
    \caption{Paths seen by $u$ and $w'$ in cases 1 and 2. Straight lines are edges, curly lines are paths. Blue edges are shortcut edges. Dotted lines are edges/paths not seen by $u$. All black edges are seen by $w'$.}
    \label{fig:uw+2:base-case}
\end{figure}
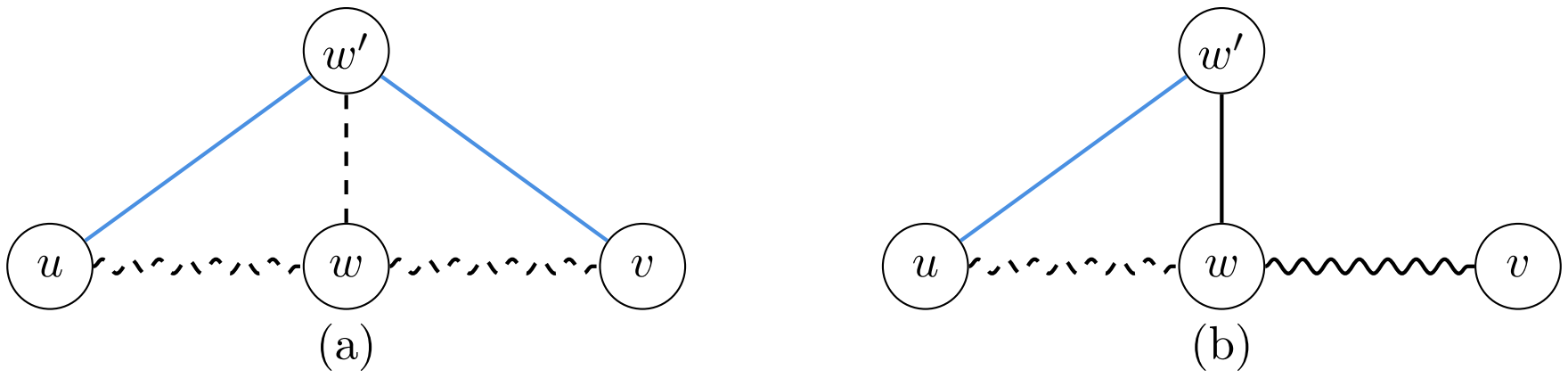

    Finally, by Monotone ES-Tree properties, $\level{u}{v} \ge \dist{u,v}$ holds in all cases.
\end{proof}

Using \lemref{uw+2:approx-base-case} as the base case, we now show that the claim holds for any time $t$.

\begin{lemma}
    \label{lem:uw+2:approx-guarantee}
    At any time $t$, for every $v,u \in V$ such that $\dist{u,v} \le \tau$, it holds that $\dist{u,v} \le d[u,v] \le \dist{u,v} + 2$.
\end{lemma}
\begin{proof}
    The base case was established in \lemref{uw+2:approx-base-case}.
    We now assume correctness for every time $t' < t$ and prove correctness at time $t$.
    
    Again, we consider the following three cases:
    
    \ccase{1} There exists a shortest path $P_{u,v}$ between $u$ and $v$ that contains a vertex $w$ from $V_2$.
    
    Let $w'' \in D_2$ such that $(w, w'') \in E\sstar$.
    The edges $(u, w'')$ and $(w'', v)$ belong to $G_1(u)$, on which a Monotone ES-Tree is run from $u$.
    Similarly to the base case, the weights of these edges are $\dist{u, w''}$ and $\dist{w'', v}$, the (exact) distances found by the ES-Tree on $G$ from $w'' \in D_2$, since $\dist{w'',u}\le\dist{w,u} + 1\le \tau+1$ and similarly for $\dist{w'',v}$.
    We note that even though the edge $\paren{w'', w}$ may not belong to $E^*$ at time $0$, the existence of some $w'' \in D_2$ with $\paren{w, w''} \in E^*$ is guaranteed at every time $t$, since $D_2$ dominates $V_2$ throughout the algorithm.
    
    It follows that the graph $G_1(u)$, on which $u$ maintains a Monotone ES-Tree, contains edges $(u,w'')$ and $(w'',v)$ with weights $\weight{u, w''}=\level{w''}{u}$ and $\weight{w'',v}=\level{w''}{v}$, which form a path $P_{u,v}^*$ from $u$ to $v$ of weight $\weight{u,w''} + \weight{w'',v} =\level{w''}{u} + \level{w''}{v}  \le \dist{u, w} + 1 + \dist{w, v} + 1 = \dist{u, v} + 2$. The weight of this path is at most $\tau+2$ and is considered by $u$'s tree.

    By definition, $P_{u,v}$ exists in $G$ from the beginning of the algorithm, and $w$ belongs to $V_2$ continuously from the beginning up to time $t$.
    Since $D_2$ dominates $V_2$ throughout the algorithm, at any time $t$ there exists such a vertex $w''$, and hence at any time $t$ there exists such a path $P_{u,v}^*$ between $u$ and $v$, and $\level{u}{v} \le \weight{P_{u,v}^*} \le \dist{u,v} + 2$ always holds.
    Under the case assumption, $\hops{P_{u,v}} = \dist{u,v}$ at time $t$, and hence $\level{u}{v} \le \dist{u,v} + 2$.
     
    \ccase{2} There exists a shortest path $P_{u,v}$ between $u$ and $v$ that contains vertices from $V_1$ but not from $V_2$.
    
    Let $w$ be the last vertex on $P_{u,v}$ that belongs to $V_1$.
    All edges on the path from $w$ to $v$ touch vertices not in $V_1$ and therefore belong to $E_1 \subseteq G_1(u)$, on which a Monotone ES-Tree is run from $u$. Denote this portion by $P_{w,v}$.
    
    Let $w'' \in D_1$ such that $(w, w'') \in E\sstar$ at time $t$.
    The graph $G_1(u)$ contains weighted edges connecting $u$ to all vertices in $D_1$ - in particular, a weighted edge $(u, w'')$.
    The weight of this edge is the level of $u$ in $w''$'s tree in $G_2$: $\level{w''}{u}$.
    We show by induction that $\level{w''}{u} \le \dist{u,w} + 1$.
    We begin with time $0$. Since $w \in V_1$ at time $t$, it must be in $V_1$ at time $0$ (otherwise it cannot belong to $V_1$ at time $t$). Thus at time $0$ we have $(w'',w) \in E_2$ (though possibly $(w'',w) \notin E^*$). The analysis of the base case shown earlier holds. An example can be seen in \figref{uw+2:induction-case-2}.

    \begin{figure}
    \centering
    \begin{tikzpicture}[
        vertex/.style={circle, draw, minimum size=18pt, inner sep=0pt, font=\small},
        shortcut/.style={draw, thick, color={rgb,255:red,74; green,144; blue,226}},
        invisibleEdge/.style={draw, thick, dashed}
    ]
        
        \begin{scope}
            \node[vertex] (ul) at (0,1.2) {$u$};
            \node[vertex] (wl) at (2.4,1.2) {$w$};
            \node[vertex] (vl) at (4.8,1.2) {$v$};
            \node[vertex] (wpl) at (2.4,2.7) {$w'$};
            \node[vertex] (wppl) at (2.4,-0.3) {$w''$};
            
            \draw[thick, dashed, decorate, decoration={snake, amplitude=1.5pt, segment length=6pt}] (ul) -- (wl);
            \draw[thick, decorate, decoration={snake, amplitude=1.5pt, segment length=6pt}] (wl) -- (vl);
            \draw[thick] (wl) -- (wpl);
            \draw[invisibleEdge] (wl) -- (wppl);
            \draw[shortcut] (ul) -- (wpl);
            \draw[shortcut] (ul) -- (wppl);
        \end{scope}

        \begin{scope}[shift={(6.5,0)}]
            \node[vertex] (ur) at (0,1.2) {$u$};
            \node[vertex] (wr) at (2.4,1.2) {$w$};
            \node[vertex] (vr) at (4.8,1.2) {$v$};
            \node[vertex] (wpr) at (2.4,2.7) {$w'$};
            \node[vertex] (wppr) at (2.4,-0.3) {$w''$};
            
            \draw[thick, dashed, decorate, decoration={snake, amplitude=1.5pt, segment length=6pt}] (ur) -- (wr);
            \draw[thick, decorate, decoration={snake, amplitude=1.5pt, segment length=6pt}] (wr) -- (vr);
            \draw[thick] (wr) -- (wppr);
            \draw[shortcut] (ur) -- (wpr);
            \draw[shortcut] (ur) -- (wppr);
        \end{scope}

    \end{tikzpicture}
    \caption{Dominating vertices change at time $t$ due to the deletion of $(w,w')$. The left side shows the graph at time $t-1$ and the right side shows the graph at time $t$. As before, straight lines are edges, curly lines are paths. Blue edges are shortcut edges. Dotted lines are edges/paths not seen by $u$. All black edges are seen by both $w',w''$.}
    \label{fig:uw+2:induction-case-2}
\end{figure}
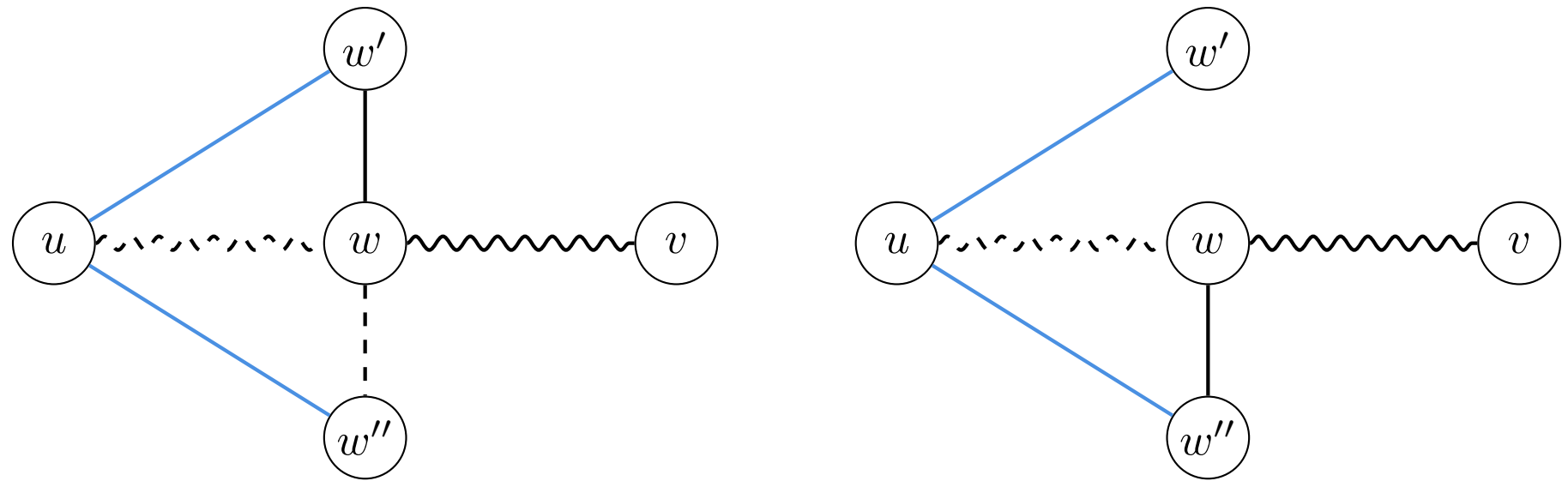

    We assume correctness for any time $t' < t$ and prove correctness for $t$.
    
    Denote by $P_{w,u}$ the portion of the path from $u$ to $w$.
    Since all edges on $P_{w,u}$, as well as $(w, w'')$, belong to $E_2$, we have $\dist{u, w''}{G_2} \leq \dist{u, w} + 1 \le \tau+1$, and hence these distances are explored by the Monotone ES-Tree from $w'' \in D_1$ on $G_2$.
    
    If none of the edges on $P_{w,u} \circ \paren{w, w''}$ are stretched in $w''$'s tree, then $\level{w''}{u}$ is set to the length of this path.

    Otherwise, let $\paren{x',x}$ be the last stretched edge on the path, traversing from $w''$ to $u$.
    
    By the induction hypothesis, the tree sets $\level[t]{w''}{x} = \level[t-1]{w''}{x} \le \dist{w,x} + 1$. Since the rest of the path is not stretched in $w''$'s tree, we conclude that $\level[t]{w''}{u} = \level[t-1]{w''}{x} + \dist{x,u} \le \dist{w,x} + 1 + \dist{x,u} = \dist{w,u} +1$.
    
    Now consider the portion of $P_{u,v}$ from $w$ to $v$ and denote it by $P_{w,v}$. As already discussed, all edges of $P_{w,v}$ are contained in $E_1 \subseteq G_1(u)$.
    Denote $P_{u,v}^* = \paren{u,w''} \circ \paren{w'',w} \odot P_{w,v}$.
    We conclude that $\weight{P_{u,v}^*} \le \level{w''}{u} + \dist{w'',w} + \dist{w,v} \le \dist{u,w} + 1 + 1 + \dist{w,v} = \dist{u,v} + 2$.
    $P_{u,v}^*$ is explored by $u$'s tree. If none of the edges on $P_{u,v}^*$ are stretched in $u$'s tree, $\level{u}{v} \le \weight{P_{u,v}^*}$.
    Otherwise, let $\paren{y',y}$ be the last stretched edge on $P_{u,v}^*$, traversing from $u$ to $v$.
    Note that we have already argued for $y = w''$, so we assume $y \neq w''$. Then $y$ lies on $P_{w,v}$.
    By the definition of a Monotone ES-Tree and the induction hypothesis, $u$'s tree sets $\level[t]{u}{y} = \level[t-1]{u}{y} \le \dist{u,y} + 2$. Since the rest of $P_{u,v}^*$ is not stretched in $u$'s tree, we conclude that $\level[t]{u}{v} = \level[t-1]{u}{y} + \dist{y,v}\le \dist{u,y} + \dist{y,v} + 2 = \dist{u,v} + 2$.
    
    \ccase{3} There exists a shortest path $P_{u,v}$ between $u$ and $v$ that does not contain any vertex from $V_1$.
    
    Then $P_{u,v} \subseteq E_1$. If none of the edges on $P_{u,v}$ are stretched in $u$'s tree, we have $\level{u}{v}=\dist{u,v}$.
    Otherwise, let $\paren{x',x}$ be the last stretched edge from $P_{u,v}$ in $u$'s tree, traversing from $u$ to $v$.
    By the induction hypothesis, the tree sets $\level[t]{u}{x} = \level[t-1]{u}{x} \le \dist{u,x} + 2$. Since the rest of the path is not stretched, we conclude that $\level[t]{u}{v} = \level[t-1]{u}{x} + \dist{x,v}\le \dist{u,x} + \dist{x,v} + 2 = \dist{u,v} + 2$.

    Finally, note that even though $u$ and $v$ may belong to different cases as the graph changes, all cases hold. By \lemref{prelim:monotone-es-properties}, it holds that $\level{u}{v} \ge \dist{u,v}$ at any time $t$.
\end{proof}

These are the main building blocks for \thmref{+2}. The proof is deferred to the next section.

\subsection{$(\onepluseps,2\cdot\paren{k-1})$-APASP}
\label{sec:uw+k}
In this section, we establish the following theorem:
\thmunweightedplusk*

\algref{uw+k:bounded} generalizes \algref{uw+2:bounded} to achieve an additive error of $+2(\lfloor k/3\rfloor + 1)$ for any $k \ge 3$ in a total update time of $\onolog{n^{2+\frac{1}{k}}}$.
While the overall structure is similar, the generalization introduces non-trivial complications in both the correctness analysis and the handling of degree changes under edge deletions, which we address below.

The algorithm of Dor, Halperin, and Zwick \cite{DorHalZwi2000} computes $k-1$ hitting sets $V = D_0 \supseteq D_1 \supseteq D_2 \supseteq \dots \supseteq D_{k-1} \supseteq D_k = \emptyset$ based on $k-1$ degree thresholds and sets $V = V_0 \supseteq V_1 \supseteq \dots V_{k-1} \supseteq V_{k} = \emptyset$ \footnote{The indexing scheme in this lemma has been adapted from the original presentation in \cite{DorHalZwi2000} to match our convention where $D_0 = V$ is the largest dominating set and indices decrease with set size, i.e., $D_0 \supseteq D_1 \supseteq \cdots \supseteq D_{k-1}$. The original work used the opposite convention where $D_1$ was the smallest set.} and its correctness proof is based on the following lemma:
\begin{lemma*}[\cite{DorHalZwi2000}]
Fix $0 \le i_1,i_2 \le k-1$. If $u \in D_{i_1}$ and $v \in D_{i_2}$ are connected by a shortest path $P$ in which the vertex of highest degree belongs to $V_j \setminus V_{j-1}$, then
\[
d[u,v] \leq \dist{u,v} + e_{i_1,j,i_2}
\]
\end{lemma*}

where $e_{i_1,j,i_2}$ is part of a recursive sequence defined as follows:
\[
e_{i_1,j,i_2} =
\begin{cases}
0 & \text{if } i_1 \ge j \\
2 & \text{if } i_1 + j + i_2 \ge k-1 \\
e_{i_1+1,j,i_2} + 2 & \text{otherwise}
\end{cases}
\]

However, as edges are deleted from $G$, the shortest path between any two vertices $u,v \in V$ may change arbitrarily. Since the approximation guarantee of~\cite{DorHalZwi2000} depends on properties of a shortest path (specifically, the highest-degree vertex on it), a deletion may simultaneously increase $\dist{u,v}$ and reduce the applicable error term $e_{i_1,j,i_2}$. Correcting for this would require decreasing levels in the Monotone ES-Tree, which may lead to undesired total update time, while ignoring it could yield a vacuous guarantee.

To circumvent this limitation, we strengthen the lemma to hold for \emph{any} path $P_{u,v}$, rather than exclusively for the shortest path. In the decremental model, any path present at time $t$ must have existed at time $0$, and its weight is monotonically non-decreasing. By evaluating arbitrary paths, we can safely leverage structural properties of $P_{u,v}$ established at earlier stages of the edge deletion sequence, decoupling our invariant from the volatile nature of the exact shortest path.

Albeit, analyzing arbitrary paths resolves only part of the difficulty. The lemma of \cite{DorHalZwi2000} also assumes that the vertex with the highest degree on the path belongs to $V_j$. Under edge deletions, however, every vertex may lose neighbors and drop from $V_j$ to $V_r \setminus V_{r+1}$ for some $r < j$. To handle this, we relax the inductive claim: rather than fixing the index $j$, we show that
$d[u,v] \leq \weight{P_{u,v}} + e_{i_1,j',i_2}$ for some $j \le j' \le k$.

\begin{algorithm}
\caption{$\overline{\texttt{dapasp}}_k$ up to depth $\tau$}
\label{alg:uw+k:bounded}
\textbf{Input:} An unweighted, undirected graph $G = (V, E)$ and a depth parameter $\tau$.

\textbf{Output:} A matrix $\{ d[u,v] \}_{u,v}$ of estimated distances.

\For {$i = 1$ to $k-1$}
{
    $s_i \gets n^{i/k}$
}
\For {$i = 1$ to $k-1$}
{
    $V_i \gets \{v \in V \mid \deg(v) \ge s_i\}$
}
\For {$i = 1$ to $k-1$}
{
    $E_i \gets \{(u,v) \in E \mid \deg(u) \le s_i \text{ or } \deg(v) \le s_i \}$
}

$D_0 \gets V$

\For {$i = 1$ to $k-1$}
{
    $(D_i,E_i^*) \gets$ dominating set that dominates $G$ at scale $s_i$
}

$E^* \gets  \bigcup_{i=1}^{k-1} E_i^*$

$E_k \gets E$, $E_0 \gets \emptyset$

For every $u \in D_{k-1}$, maintain an ES-Tree up to length $\tau+\frac{2}{3}(k+3)$ on $G$.

\For{$i = k-2$ to $0$}{
    For every $u \in D_i$, maintain Monotone ES-Tree up to length $\tau+\frac{2}{3}(k+3)$ on the graph $G_{i+1}(u)=(V, E_{i+1} \cup E^* \cup (\{u\}\times V) \cup (\cup_{i+j_1+j_2\ge k-1}D_{j_1}\times D_{j_2}))$.
}
\end{algorithm}

\textbf{Algorithm Description.} The structure of \algref{uw+k:bounded} 
is similar to that of \algref{uw+2:bounded}.
\algref{uw+k:bounded} employs $k-1$ degree thresholds
$\{s_i = n^{i/k}\}_{i=1}^{k-1}$ and $k-1$ sets of heavy vertices
$\{V_i\}_{i=1}^{k-1}$ and light edges
$\{E_i\}_{i=1}^{k-1}$ accordingly. 
\algref{uw+k:bounded} then constructs $k-1$ dominating sets
$\{D_i\}_{i=1}^{k-1}$ and edges $\{E_i\sstar\}_{i=1}^{k-1}$
such that $D_i$ dominates the vertices of $V_i$.
As before, we define $E\sstar = \bigcup_{i=1}^{k-1} E_i\sstar$.
We also set $D_0 = V$.

\algref{uw+k:bounded} maintains ES-Trees from every $u \in D_{k-1}$
on $G$ up to depth $\tau + \frac{2}{3}(k+3)$,
and maintains Monotone ES-Trees from every $u \in D_i$ for
$i = k-2, \ldots, 0$ up to depth $\tau + \frac{2}{3}(k+3)$
on $G_{i+1}(u)=(V, E_{i+1} \cup E\sstar \cup (\singleton{u} \times V) \cup (\cup_{i+j_1+j_2\ge k-1}D_{j_1}\times D_{j_2}))$.
The final distance estimate is $d[u,v] = \mmin{\level{u}{v},\level{v}{u}}$.

\textbf{Maintaining Distances Under Edge Deletions.}
We use methods similar to those introduced in \algref{uw+2:bounded}.
The sets $D_0, D_1, \ldots, D_{k-1}$ are created at the beginning of the algorithm and remain fixed. Maintaining the edge sets and the Monotone ES-Trees is more complicated.

Each time the distance estimate between $x \in D_i$ and a node $u$ increases, we update the weight
of the corresponding shortcut edge in every graph containing it. We can have at most $\tau + \frac{2}{3}(k+3)$ updates per shortcut. If the distance estimate exceeds $\tau + \frac{2}{3}(k+3)$, $u$ leaves $x$'s tree and we remove the shortcut edge from every graph containing it.

When an edge $(v, x) \in E_i\sstar$ between a node $v \in V_i$ and $x \in D_i$ is deleted from
the graph $G$, we distinguish between two cases. If $v$ still has a neighbor $w \in D_i$, we first add
a new edge $(v, w)$ to the set $E_i\sstar$ and then delete the edge $(v, x)$.
Otherwise, we add all edges incident to $v$ in $G$ to $E_{i+1}$, and then delete the edge $(v, x)$.

\begin{lemma}[Total Update Time]
    \label{lem:uw+k:total-update-time}
    \algref{uw+k:bounded} has  an expected total update time of $\onolog{\tau k^5 n^{2+\frac{1}{k}}}$
\end{lemma}

\begin{proof}
    Similarly to \algref{uw+2:bounded}, for every $1 \le i \le k-1$, $E[|D_i|] = \bigo{n/s_i}$ and $\mathcal{E}_i = \bigo{ns_i}$, where $\mathcal{E}_i$ denotes the total number of edges ever present in $E_i$.
    
    For the size of $E\sstar$, it follows from \clmref{uw+2:doryetal-total-edges-estar} that the total number of edges ever added to $E\sstar$ is at most $\bigo{\frac{m\log n}{s_1} + \sum_{i=1}^{k-1}\frac{n\log n \cdot s_{i+1}}{s_i}} = \bigo{mn^{-\frac{k-1}{k}}\log n + kn^{1+\frac{1}{k}}\log n}=\bigo{n^{1+\frac{1}{k}}\log n}$.
    
    The set $\paren{\singleton{u}\times V}$ contains at most $n$ edges.
    
    Putting all these together, the total number of edges that ever exist in $G_{i+1}(u)$ is bounded by \begin{align*}
        &\bigo{ns_{i+1} + n^{1+\frac{1}{k}}\log n + n + \frac{n^2}{s_{j_1} \cdot s_{j_2}}} =\\
        &\onolog{n^{1+\frac{i+1}{k}}+\sum_{i+j_1+j_2 \ge k-1}\frac{n^2}{n^{\frac{j_1+j_2}{k}}}}=\\
        &\onolog{n^{1+\frac{i+1}{k}}+\sum_{i+j_1+j_2 \ge k-1}n^{2-\frac{j_1+j_2}{k}}} =\\
        &\onolog{n^{1+\frac{i+1}{k}}+k^2 \cdot n^{2-\frac{k-1-i}{k}}} =\\
        &\onolog{k^2 \cdot n^{1+\frac{i+1}{k}}}
    \end{align*}
    
    The total update time of maintaining a single Monotone ES-Tree from $u \in D_i$ up to length $\tau + \frac{2}{3}(k+3)$ on $G_{i+1}(u)$ is $\onolog{(\tau+k)k^3n^{1+\frac{i+1}{k}}}=\onolog{\tau k^4n^{1+\frac{i+1}{k}}}$. This includes weight updates to shortcuts, as each shortcut can be updated $\bigo{\tau+k}$ times.
    
    Summing over all $u \in D_i$, the total update time of the $i$-th iteration is $\onolog{\frac{n}{s_i} \cdot \tau k^4n^{1+\frac{i+1}{k}}}=\onolog{n^{1-\frac{i}{k}} \cdot \tau k^4n^{1+\frac{i+1}{k}}}=\onolog{\tau k^4n^{2 + \frac{1}{k}}}$.
    
    Summing over all $k$ iterations adds another $k$ factor, yielding an expected total update time of $\onolog{\tau k^5n^{2 + \frac{1}{k}}}$.
\end{proof}

\begin{lemma}[Approximation Guarantee]
    \label{lem:uw+k:approx-guarantee}
    \algref{uw+k:bounded} maintains a distance estimate $d[\cdot,\cdot]$ such that for every $u,v \in V$, if $\dist{u,v} \le \tau$, then $\dist{u,v} \le d[u,v] \le \dist{u,v} + \dhztwothirdapprox$
\end{lemma}

We establish the base case of the induction (time $t = 0$) with the following lemma, which generalizes the base case structure of \secref{uw+2}.

\medskip
\begin{claim}
\label{clm:uw+k:approx-time-zero}
At the beginning of the algorithm, if $u \in D_{i_1}$ and $v \in D_{i_2}$ are connected by a path $P_{u,v}$ in which the vertex of highest degree belongs to $V_j \setminus V_{j-1}$ and $\hops{P_{u,v}} \le \tau$, then 
\[
\dist{u,v} \le \mmin{\level{u}{v},\level{v}{u}} \leq \hops{P_{u,v}} + e_{i_1,j,i_2}
\]
\end{claim}

\begin{proof}
Assume, without loss of generality, that $i_1 \le i_2$. We partition into cases.

\ccase{1} $i_1 \ge j$.

In this case, $P_{u,v} \subseteq E_{i_1+1} \subseteq G_{i_1+1}(u)$.
Since this is the beginning of the algorithm, it follows from \lemref{prelim:monotone-es-properties} that $\level{u}{v} = \hops{P_{u,v}}$ as required.

\ccase{2} $i_1 + j + i_2 \ge k - 1$ and $j > i_1$.

In this case, $D_j \times D_{i_2} \subseteq G_{i_1+1}(u)$.
Let $w$ be a vertex on $P_{u,v}$ that belongs to $V_j$.
Denote by $P_{u,w}$  the portion of $P_{u,v}$ from $u$ to $w$, and by $P_{w,v}$  the portion of $P_{u,v}$ from $w$ to $v$.
By path decomposition, $\hops{P_{u,v}} = \hops{P_{u,w}} + \hops{P_{w,v}}$. 

Let $w' \in D_j$ such that $\paren{w, w'} \in E\sstar$.
Then $P_{u,v} \subseteq G_{j+1}(w')$, and since this is the beginning of the algorithm, $\level{w'}{u} = \dist{w',u}{G_{j+1}\paren{w'}} \leq \hops{P_{u,w}} + 1$ and $\level{w'}{v} = \dist{w',v}{G_{j+1}\paren{w'}} \leq \hops{P_{w,v}} + 1$.

Since $j > i_1$, there exist edges $(u,w')$ and $(w',v)$ in $G_{i_1+1}(u)$ such that $\weight{u,w'} = \level{w'}{u}$ and $\weight{v,w'} = \level{w'}{v}$.
As this is the beginning of the algorithm, it follows from \lemref{prelim:monotone-es-properties} that $\level{u}{v} = \dist{u,v}{G_{i_1+1}\paren{u}} \leq \weight{u,w'} + \weight{v,w'} \le \hops{P_{u,w}} + 1 + \hops{P_{w,v}} + 1 = \hops{P_{u,v}} + 2$.

\ccase{3} $i_1 + j + i_2 < k - 1$ and $j > i_1$.

Since $j > i_1$, by definition $V_j \subseteq V_{i_1 + 1}$.
Let $w$ be the \textit{last} vertex on $P_{u,v}$ that belongs to $V_{i_1 + 1}$, traversing from $u$ to $v$.
Denote by $P_{u,w}$  the portion of $P_{u,v}$ from $u$ to $w$, and by $P_{w,v}$ the portion of $P_{u,v}$ from $w$ to $v$.
Let $w' \in D_{i_1 + 1}$ such that $\paren{w, w'} \in E\sstar$.

Now consider the path $P_{w',u} = \paren{w, w'} \circ P_{u,w} = \paren{w, w'} \circ P_{w,u}$. The path $P_{w',u}$ starts at $w' \in D_{i_1 + 1}$ and ends at $u \in D_{i_1}$. The vertex with the highest degree on it belongs to $V_j$, and $\hops{P_{w',u}} = \hops{P_{u,w}} + 1$.

By applying the claim inductively to $P_{w',u}$ (note that $w' \in D_{i_1+1}$ has a higher index than $u \in D_{i_1}$, so this is an inductive application on the dominating set hierarchy), $\level{w'}{u} \leq \hops{P_{u,w}} + 1 + e_{i_1 + 1,j, i_1}$.

Since $w$ is the last vertex from $V_{i_1 + 1}$ on $P_{u,v}$, it follows that $P_{w,v} \subseteq E_{i_1+1} \subseteq G_{i_1+1}(u)$. The graph $G_{i_1+1}(u)$ also contains the edge $\paren{w, w'} \in E\sstar$ and a weighted edge $\paren{u, w'}$ with $\weight{u,w'} \le \level{w'}{u}$.
We conclude that there exists a path $P_{u,v}\sstar = \paren{u, w'} \circ \paren{w, w'} \circ P_{w,v} \subseteq G_{i_1+1}(u)$ such that $\weight{P_{u,v}\sstar} \leq (\hops{P_{u,w}} + 1 + e_{i_1+1,j,i_1}) + (1 + \hops{P_{w,v}})= \hops{P_{u,v}} + (e_{i_1+1,j,i_1} + 2) = \hops{P_{u,v}} + e_{i_1,j,i_1} = \hops{P_{u,v}} + e_{i_1,j,i_2}$, where the last equality follows from the assumption that $i_1 \le i_2$.

We will argue later that these paths are indeed explored by the Monotone ES-Tree, as it explores paths up to bounded length $\tau + 2/3(k+3)$.

Finally, note that every graph $H$ used in the analysis contains edges that are either edges of $G$ or shortcut edges representing distances in $G$; thus for any $x,y \in V$, $\dist{x,y}{H} \ge \dist{x,y}{G}$.
Since we use Monotone ES-Trees, \lemref{prelim:monotone-es-properties} ensures that $\dist{u,v}{G} \le \dist{u,v}{H} \le \level{u}{v}$.
\end{proof}

We now show that a similar approximation holds throughout distance changes.
\begin{claim}
At any time $t$, for any $u \in D_{i_1}$ and $v \in D_{i_2}$ such that the highest vertex on $P_{uv}$ belongs to $V_j \setminus V_{j-1}$, there is some $j \le j' \le k$ such that
\[
\dist{u,v} \le \mmin{\level{u}{v},\level{v}{u}} \le  \hops{P_{u,v}} + e_{i_1,j',i_2}
\]
\end{claim}

\begin{proof}
The proof proceeds by double induction on time $t$ and on $\hops{P_{u,v}}$. The base case of $t=0$ was established in \clmref{uw+k:approx-time-zero}. We assume correctness for any time $t' < t$ and prove the claim for time $t$.

For paths of length $1$ at time $t$, $(u,v) \in G_i(u)$ for every $u$, and hence it is never stretched. We assume correctness for any path of length $\ell < \hops{P_{u,v}}$ and prove the claim for $\hops{P_{u,v}}$.

We again assume that all paths shown in this analysis are explored by the trees. We will discuss this later.

We assume, without loss of generality, that $i_1 \le i_2$ and partition into cases.

\ccase{1} $i_1 \ge j$.

The previous analysis shows that the path between $u$ and $v$ is in $G_{i_1+1}(u)$. If none of the edges on this path are stretched in $u$'s tree, then $\level{u}{v} =  \dist{u,v}$.

Otherwise, let $\paren{z',z}$ be the last stretched edge on $P_{u,v}$ in $u$'s tree, traversing from $u$ to $v$. Let $P_{u,z}$ denote the portion of $P_{u,v}$ from $u$ to $z$, and let $P_{z,v}$ denote the portion of $P_{u,v}$ from $z$ to $v$.

Let $t' < t$ be the last time when $\level{u}{z}$ was set. Then $\level{u}{z} = \level[t']{u}{z}$.
Since $u \in D_{i_1}$ throughout all edge updates, and $j$ can only increase as edges are deleted, we conclude that the case hypothesis applied at time $t'$. It follows from the analysis of the non-stretched case that $\level[t']{u}{z} \le \hops{P_{u,z}}$.

As none of the edges on $P_{z,v}$ are stretched in $u$'s tree, we conclude that $\level{u}{v} = \level[t-1]{u}{z} + \weight{P_{z,v}} \le \hops{P_{u,z}} + \hops{P_{z,v}} \le \hops{P_{u,v}}$.

\ccase{2} $i_1 + j + i_2 \ge k - 1$ and $j > i_1$.

In this case, $D_j \times D_{i_2} \subseteq G_{i_1+1}(u)$. Let $w$ be a vertex on $P_{u,v}$ that belongs to $V_j$.
Let $P_{u,w}$ denote the portion of $P_{u,v}$ from $u$ to $w$, and let $P_{w,v}$ denote the portion of $P_{u,v}$ from $w$ to $v$.

By path decomposition, $\hops{P_{u,v}} = \hops{P_{u,w}} + \hops{P_{w,v}}$.
Let $w'' \in D_j$ such that $(w, w'') \in E^*$.
Then $\dist{u,w''} \le \hops{P_{u,w}} + 1$ and $\dist{w'',v} \le \hops{P_{w,v}} + 1$.
We prove by induction on time $t$ and path length that $\level{w''}{u} \le \hops{P_{u,w}} + 1$ and $\level{w''}{v} \le \hops{P_{w,v}} + 1$.

For the base case of $t=0$, we note that each vertex's degree can only decrease. We assume correctness for any time $t' < t$ and prove for $t$.
First, since $w \in V_{j}$ at time $t$, at any time $t' < t$, $w \in V_{j'}$ for some $j' \le j$, and hence $w \in V_j$ and there always exists a vertex $w'' \in D_j$ dominating it. It may be that $w'' \neq w'$, the dominating vertex of $w$ from time $0$, but the edge $(w,w'')\in E_j$ is contained in $G_{j+1}(w'')$ from the beginning of the algorithm and hence the base case applies to $w''$ as well.

For the base case of a length-$1$ path, let $x$ be a vertex such that $\dist{w,x} = 1$. Then either $\level{w''}{x} = \dist{w'',x} \le \dist{w'',w} + \dist{w,x} = 1 + 1 = \dist{w,x} + 1$, or it is stretched and $\level{w''}{x}=\level[t-1]{w''}{x} \le \dist{w,x} + 1$ by the induction hypothesis on time.

Now assume correctness for path length $\ell < \dist{w,u}$ at time $t$, and let $x$ be the last neighbor before $u$ on the path. By induction, $\level{w''}{x} \le \dist{w,x} + 1$. Then either $\level{w''}{u} = \level{w''}{x} + 1 \le \dist{w,x} + 2 = \dist{w,u} + 1$, or the edge $(x,u)$ is stretched and $\level{w''}{u} = \level[t-1]{w''}{u} \le \dist{w,u} + 1$ by the induction hypothesis. Similarly for $v$.

Thus there exists a path $P_{u,v}\sstar = u \straight w'' \straight v \subseteq G_{i_1+1}(u)$ with $\weight{P_{u,v}\sstar} \le \dist{w,u} + 1 + \dist{w,v} + 1 \le \hops{P_{u,v}} + 2 = \hops{P_{u,v}} + e_{i_1,j,i_2}$.

If none of the edges on $P_{u,v}\sstar$ are stretched in $u$'s tree, then $\level{u}{v} \le \dist{u,v} + 2 \le \hops{P_{u,v}} + 2$.

Otherwise, since $w''$ cannot be stretched in $u$'s tree, we conclude that $\level{u}{v} = \level[t-1]{u}{v} \le  \hops{P_{u,v}} + e_{i_1,j',i_2}$ by the induction hypothesis.

\ccase{3} $i_1 + j + i_2 < k - 1$ and $j > i_1$.

Since $j > i_1$, by definition $V_j \subseteq V_{i_1 + 1}$. Let $w$ be the \textit{last} vertex on $P_{u,v}$, traversing from $u$ to $v$, that belongs to $V_{i_1 + 1}$.
Let $w'' \in D_{i_1 + 1}$ such that $(w, w'') \in E^*$.
Let $P_{u,w}$ and $P_{w,v}$ be, as before, the portions of $P_{u,v}$ from $u$ to $w$ and from $w$ to $v$, respectively.

Now consider the path $P_{w'',u}$ composed of the edge $(w, w'')$ and the portion of $P_{u,v}$ from $w$ to $u$. The path $P_{w'',u}$ starts at $w'' \in D_{i_1 + 1}$ and ends at $u \in D_{i_1}$. The vertex of highest degree on it belongs to $V_j$. The length of $P_{w'',u}$ is $\hops{P_{u,w}} + 1$.
By applying the claim inductively to $P_{w'',u}$ (note that $w'' \in D_{i_1+1}$ has a higher index than $u \in D_{i_1}$, so this is an inductive application on the dominating set hierarchy), we conclude that there exists some $j' \ge j$ such that $\level{w''}{u} \leq \hops{P_{u,w}} + 1 + e_{i_1 +1,j',i_1}$.

Since $w$ is the last vertex from $V_{i_1 + 1}$ on $P_{u,v}$, it holds that $P_{w,v} \subseteq E_{i_1+1} \subseteq G_{i_1+1}(u)$.
The graph $G_{i_1+1}(u)$ also contains the edge $\paren{w, w''} \in E\sstar$ and a weighted edge $\paren{u, w''}$ of weight $\level{w''}{u}$. Thus there exists a path $P\sstar_{u,v} \subseteq G_{i_1+1}(u)$ with $\weight{P_{u,v}\sstar} \leq (\hops{P_{u,w}} + 1 + e_{i_1 + 1,j',i_1}) + (1 + \hops{P_{w,v}})= \hops{P_{u,v}} + (e_{i_1+1,j',i_1} + 2) = \hops{P_{u,v}} + e_{i_1 ,j',i_1}$.

If none of the vertices on $P_{u,v}\sstar$ are stretched in $u$'s tree, then $\level{u}{v}=\weight{P_{u,v}\sstar}$.

Otherwise, let $\paren{z',z}$ be the last stretched vertex on $P_{u,v}\sstar$ in $u$'s tree. Note that $z \neq w''$, as levels are monotone. Thus $z \in P_{w,v}$, and hence $z \notin V_{i_1+1}$. Let $i'$ be maximal such that $\deg(z) \ge s_{i'}$. By the induction hypothesis, there exists some $j' \ge j$ such that $\level{u}{z} = \level[t-1]{u}{z} \le \hops{P_{u,z}} + e_{i',j',i_2} \le \hops{P_{u,z}} + e_{i_1,j',i_2}$, where the last inequality follows from the definition of $e_{i_1,j',i_2}$, as it is decreasing in $i_1$.

In all cases, $\dist{u,v} \le \dist_{G_{i_1+1}(u)}(u,v) \le \level{u}{v}$, since we use a Monotone ES-Tree.
\end{proof}

Now let $e_{i,j} = e_{i,j,i+1}$. Substituting $i_2 = i+1$ into the recursion gives:

\[
e_{i,j} =
\begin{cases}
0 & \text{if } i \ge j \\
2 & \text{if } 2i+j \ge k-2 \\
e_{i+1,j} + 2 & \text{otherwise}
\end{cases}
\]

Unwinding the recursion: starting at level $i_1 = i$, each step increments $i_1$ by $1$ and adds $2$. The first case (return $0$) triggers when $i_1 = j$, after $j - i$ steps. The second case (return $2$) triggers when $2i_1 + j \ge k - 2$, i.e.\ at $i_1 = \lceil(k-2-j)/2\rceil$, after $\lceil(k-2-j)/2\rceil - i$ steps. Hence:

\[
e_{i,j} =
\begin{cases}
0 & \text{if } i \ge j \\
2 & \text{if } 2i+j \ge k-2 \\
\min\!\left\{2(j-i),\; 2\!\left(\left\lceil\tfrac{k-j}{2}\right\rceil - i\right)\right\} & \text{otherwise}
\end{cases}
\]

Note that $e_{i_1,j,i_2}$ is non-increasing with respect to $i_1$ and $i_2$.

Finally:

\[
e_{0,j,0} =
\begin{cases}
0 & \text{if } j=0 \\
2 & \text{if } j=k-1 \\
2\mmin{j,\lceil\tfrac{k-j}{2}\rceil} & \text{otherwise}
\end{cases}
\]

Evaluating $e_{0,j,0}$ for each $0 \le j \le k-1$ yields $e_{0,j,0} \le \dhztwothirdapprox$.
This completes the proof of \lemref{uw+k:approx-guarantee}.

We can now prove \thmref{+k}.
    
\begin{proof}
    We maintain simultaneously our dynamic data structure to handle short distances of at most $\tau$
    and the data structure from \lemref{uw:hkn14-apsp} to handle distances larger than $\tau$, for a parameter $\tau$ that we specify next.

    We run the algorithm from \lemref{uw:hkn14-apsp} with parameters $\varepsilon' = \varepsilon / 2$ and $k'=\mmin{3k-4,\Theta\paren{\sqrt{\log n}}}$. This algorithm provides a $(\onepluseps', \beta)$-approximation for
    $\beta = 2(1 + \frac{2}{\varepsilon'})^{k' - 2}$.
    We define $\tau = \frac{\beta}{\varepsilon'}$.
    
    We use \algref{uw+k:bounded} up to depth $\tau$ with total update time $\onolog(\tau k'^5 n^{2+\frac{1}{k'}})$.

    To answer a distance query for a pair of nodes $u$ and $v$, we return the minimum of the values obtained from both algorithms.
    Since both algorithms have constant query time, we obtain constant query time.
    
    For the stretch analysis, if $\dist{u,v} \leq \tau$, then by \lemref{uw+k:approx-guarantee}, we obtain an additive stretch of at most $+2(\lfloor (\frac{3(k-2)+2}{3})\rfloor + 1) = +2(k-1)$.
    
    If $\dist{u,v} \geq d = \frac{\beta}{\varepsilon'}$, then the distance estimate from the $(\onepluseps', \beta)$-approximation algorithm is at most
    \[
    (\onepluseps')\dist{u,v} + \beta \leq (\onepluseps')\dist{u,v} + \varepsilon' \dist{u,v} = (1 + 2\varepsilon')\dist{u,v} = (\onepluseps)\dist{u,v}
    \]
    Hence, overall we obtain a $\paren{\onepluseps, 2(k-1)}$-approximation.
    
    We now analyze the time complexity.
    It follows from \lemref{uw+k:total-update-time} that the total update time of \algref{uw+k:bounded} is $\onolog{\tau k'^5 n^{2+\frac{1}{k'}}}=\onolog{\frac{\beta}{\varepsilon} k'^5 n^{2+\frac{1}{k'}}}=\onolog{\varepsilon^{-(k'-2)} k'^5 n^{2+\frac{1}{k'}}}$.
    Using \lemref{uw:hkn14-apsp}, the total update time of the near-additive approximation algorithm is
    \[
    \onolog{n^{2 + \frac{1}{k'}} \left(\tfrac{37}{\varepsilon}\right)^{k' - 1}} = \onolog{\varepsilon^{-(k'-1)}n^{2+\frac{1}{k'}}}
    \]

    Since $k' = \mmin{3k-4,\Theta\paren{\sqrt{\log n}}}$, we conclude that the total update time of our algorithm is $\onolog(n^{2+\frac{1}{3k-4}+o(1)})$ for any constant $\varepsilon$.
    
\end{proof}

The proof of \thmref{+2}
is identical, with the additive error reduced to $+2$ by \lemref{uw+2:approx-guarantee}, the total update time is $\onolog{\tau n^{\frac{7}{3}}}$ by \lemref{uw+2:total-update-time} and the depth parameter set to $\tau = \frac{\beta}{\varepsilon'}$ as above for $\beta = 2(1 + \frac{2}{\varepsilon'})$.

\section{Weighted APASP for Bounded Depth}
\label{sec:weighted:bounded}
Having established decremental $\paren{\onepluseps, 2}$ and $\paren{\onepluseps, 2(k-1)}$-APASP algorithms for unweighted graphs, we now consider undirected weighted graphs $G = \paren{V, E, \weight}$, where each edge $e \in E$ has a positive weight $\weight{e} > 0$.

\subsection{Decremental Hierarchy-Induced Edge Sets}~\label{sec:hierarchy}

In this section, we develop the decremental variant of hierarchy-induced edge sets up to a bounded depth $\tau$, which serves as the common structural foundation for all of our weighted decremental APASP algorithms.

In the static setting, a common approach for achieving additive approximations parameterized by heaviest edges $W_i$
efficiently is constructing a sparse subgraph that preserves light paths. 
Static algorithms generally employ one of two paradigms to construct this subgraph: 
a deterministic approach (e.g., Cohen and Zwick~\cite{CohZwi1997}) that selects a collection of the lightest incident edges 
at each vertex, 
or a randomized approach (e.g., Baswana and Kavitha~\cite{BK}) based on \emph{hierarchy-induced edge sets} $E_S$, consisting of 
all edges whose weight is strictly smaller than the distance from their endpoints to a sampled pivot set $S \subseteq V$:
\[
    E_S \;=\; \bigcup_{u \in V} \bracce{(u,v) \in E \midline \weight{u,v} < \dist{u,S}{G}}.
\]
The definition of $E_S$ creates a clear metric dichotomy: for any path $P$, either all of its edges belong to
 $E_S$ (so the path is preserved intact within the sparse subgraph), or $P$ contains an edge $e = (x,y) \notin E_S$. 
In the latter case, the definition of $E_S$ guarantees that an endpoint of $e$ is close to a sampled pivot in 
$S$ ($\dist{x,S}{G} \le \weight{e}$), which enables a detour through $S$ with overhead bounded by $2\weight{e} \le 2W_i(P)$. 
In static graphs, when each vertex is sampled into $S$ independently with probability $q$, the expected size of 
$E_S$ is bounded by $\bigo{\frac{n}{q}}$, ensuring the sparsity of the constructed subgraphs. The deterministic version would
instead pick the $\bigo{\frac{1}{q}}$ lightest edges incident on each vertex into $E_S$ and utilize an hitting set capturing any vertex $v$ with $\deg\paren{v} \ge \frac{1}{q}$. The additive error analysis is exactly the same.

In the decremental setting, however, the deterministic paradigm of maintaining lightest incident edges fails:
as edges are deleted from the graph, formerly heavier edges continually become the new lightest 
incident edges of a vertex. 
Consequently, an adversary that repeatedly deletes the current lightest incident edges forces the cumulative set of edges 
ever present in $E_S$ to become $\Omega(m)$, destroying cumulative sparsity. 
The randomized hierarchy-induced formulation $E_S$ is therefore essential to preserving cumulative sparsity under deletions.

Yet, maintaining even the randomized edge set $E_S$ dynamically introduces a similar algorithmic challenge. 
As edges are deleted from the host graph $G$, distances $\dist{u,S}{G}$ to the pivot set monotonically increase. 
Consequently, edges whose weights previously exceeded $\dist{u,S}{G}$ now fall below it, which forces new edges to be inserted into $E_S$, and the cumulative sparsity is not trivially achieved. 
Moreover, while the input graph $G$ undergoes only edge deletions, the subgraph $G_S = (V, E_S)$ experiences continuous edge insertions.

In this section, we construct the first decremental variant of this hierarchy-induced edge set. 
We start by restricting the set to vertices at distance up to a bounded depth $\tau$ from $S$, and denote the set by $E_S^\tau$. 
By capping the exploration depth at $\tau + 1$, the distance $\dist{u,S}{G}$ for any vertex $u$ can increase at most $\tau$ 
times before exceeding the threshold. 
We prove that this parameterization bounds the expected number of edges inserted into $E_S^\tau$ across all decremental updates by $\bigo{\frac{n\tau}{q}}$, thereby preserving 
cumulative sparsity in the decremental setting. 
Furthermore, we provide a maintenance procedure that maintains $E_S^\tau$ in 
$\bigo{\frac{n\tau \log n}{q}}$ expected total update time.

$E_S^\tau$ forms the basis for our algorithms. 
While $E_S^\tau$ operates directly on graphs where paths are restricted to distance at most $\tau$, 
we later show in \secref{tools:hierarchy} how to extend this construction to general, unbounded-depth graphs.

When generalized to a multi-level hierarchy of sampled vertex sets $V = S_0 \supseteq S_1 \supseteq \dots \supseteq S_k = \emptyset$, 
this construction induces a nested sequence of hierarchy-induced edge sets 
$\emptyset = E_{S_0}^\tau \subseteq E_{S_1}^\tau \subseteq \dots \subseteq E_{S_k}^\tau = E$. 
These bounded hierarchy-induced edge sets form the core building block for our bounded-depth APASP algorithms:
in \secref{w+2w2:bounded}, we deploy a single sampled set to construct a $+2W_2$-APASP algorithm;
in \secref{w+2w1+2w2}, we use two nested hitting levels to achieve a $+2W_1+2W_2$-APASP algorithm;
and in \secref{w+2kw1}, we generalize this structure to $k$ levels to obtain $+2(k-1)W_1$-APASP algorithms. 
Subsequently, in \secref{tools}, we generalize these bounded-depth constructions to unbounded graph distances. 

We begin by reviewing the structural and metric properties established by Baswana and Kavitha~\cite{BK} for the static setting. 
For a set $S\subseteq V$ of vertices sampled independently from $V$ with probability $q\in\paren{0,1}$, the following properties hold:

\begin{lemma}[\cite{BK}]
\label{lem:tools:bk-es-properties}
Given a graph $G = \paren{V,E}$ and $S \subseteq V$, the following assertions are
 true:
 \begin{itemize}
     \item[1.] For any two vertices $u,v \in V$, if $\dist{u,v} < \dist{u,\piv[S]{u}}$, then the subgraph $G_S =\paren{V,E_S}$ preserves the exact distance between $u$ and $v$.
     \item[2.] The distance between  $u$ and $\piv[S]{u}$ in the graph  $\paren{V,E_S \cup E(\piv[S]{u})}$ is $\dist{u,\piv[S]{u}}$    
 \end{itemize}
\end{lemma}

\begin{lemma}[\cite{BK}]
    \label{lem:tools:bk-es-size}
    If the set $S \subseteq V$ is formed by selecting each vertex independently
    with probability $q$, then the expected size of $E_S(v)$ for any $v \in V$ is $\bigo{\frac{1}{q}}$.
\end{lemma}

\lemref{tools:bk-es-properties} applies directly to the decremental setting, as its correctness relies solely on the definition of $E_S$. However, adapting \lemref{tools:bk-es-size} requires extending the original static analysis. By bounding the expected number of edges entering $E_S$ after each update, we establish a decremental analogue that bounds the expected  \textit{total} number of edges inserted into $E_S$ across the entire update sequence. 

\begin{lemma}
    \label{lem:tools:es-size}
    If the set $S \subseteq V$ is formed by selecting each vertex independently
    with probability $q$, and the maximal number of distance updates between a vertex $u \in V$ and $S$ is $\alpha$, then the total number of edges in $E_S(u)$ throughout all edge deletions is $\bigo{\frac{\alpha}{q}}$.
\end{lemma}

\begin{proof}
Fix $u \in V$. If $u \in S$, then $\dist{u,S} = 0$ throughout and
$E_S(u) = \emptyset$, so the claim holds trivially. Assume $u \notin S$. Since $G$ is decremental, no edge re-enters $E$ after deletion.
Consequently, an edge $\paren{u,v}$ is inserted into $E_S\paren{u}$ only when $\dist{u,S}$ increases to a value strictly greater than $\weight{u,v}$.

As $\dist{u,S}$ increases at most $\alpha$ times, we bound the expected number of edges entering $E_S\paren{u}$ per distance update. Consider a single distance update where $\dist{u,S}$ increases from $\delta_{\mathrm{old}}$ to $\delta_{\mathrm{new}} > \delta_{\mathrm{old}}$. The edges entering $E_S\paren{u}$ are precisely those incident to neighbors $v_{1}, v_{2}, \ldots, v_{k}$ satisfying $\weight{u,v_i} \in \left[ \delta_{\mathrm{old}},\delta_{\mathrm{new}} \right)$. 

Assume these are arranged in an increasing order:
$ \weight{u,v_1} \le \weight{u,v_2} \le \ldots \le \weight{u,v_k} $.  For an edge $\paren{u,v_i}$ to be added to $E_{S}$ it should hold that $\dist{u,S} > \weight{u,v_i}$, which enforces that $v_1, \ldots, v_i \notin S$, as they are arranged in an increasing order. Because $S$ is sampled independently, the probability is bounded by:
$$ \Pr\bracke{\paren{u,v_i} \textnormal{ enters } E_S\paren{u}} \le \Pr\bracke{v_1, \ldots, v_i \notin S} = \mprod{\Pr\bracke{v_j\notin S}}{j=1}{i}= \paren{1-q}^i $$

By linearity of expectation, the expected number of edges added to $E_S\paren{u}$ by this update is at most:

$$ \sum_{i=1}^{k} \paren{1-q}^i < \sum_{i=1}^{\infty} \paren{1-q}^i = \frac{1-q}{q} = \bigo{\frac{1}{q}} $$

Summing over all $\alpha$ distance updates, the expected total number of edges ever inserted into $E_S\paren{u}$  is bounded by $\bigo{\frac{\alpha}{q}}$.
\end{proof}

\begin{corollary}
\label{cor:tools:es-max-edges}
Let $S \subseteq V$ be formed by selecting each vertex independently
with probability $q$, and let $\alpha$ be the maximum number of times
$\dist{u,S}$ increases over all edge deletions, over all $u \in V$.
Then the expected total number of edge-vertex incidences ever inserted
into $E_S$ throughout all edge deletions is
\[
    \mathcal{E}_S = \bigo{\frac{\alpha n}{q}}.
\]
\end{corollary}

\begin{proof}
By \lemref{tools:es-size}, for each fixed $u \in V$ the expected
total number of edges ever inserted into $E_S(u)$ is $O(\alpha/q)$.
Applying linearity of expectation and summing over all
$n$ vertices gives
\[
    \mathbb{E}\!\left[\mathcal{E}_S\right]
    \;=\;
    \sum_{u \in V}
    \mathbb{E}\!\left[\,\left|\text{edges ever in } E_S(u)\right|\,\right]
    \;\leq\;
    n \cdot O\!\left(\frac{\alpha}{q}\right)
    \;=\;
    O\!\left(\frac{\alpha n}{q}\right).
    \qedhere
\]
\end{proof}

To bound the number of distance changes between $u$ and $\piv[S]{u}$, we
introduce a distance parameter $\tau \leq nW_{\max}$ and maintain distances from $S$ only
up to $\tau$ using a $\tau$-bounded ES-Tree. We define the truncated edge set
\[
    E_S^\tau(u) := \{ (u,v) \in E \mid \weight{u,v} < \min\{\dist{u,S},\, \tau + 1\} \},
\]
where the $\tau + 1$ cap ensures that edges of weight greater than $\tau$ are never
inserted. Once an edge enters $E_S^\tau$, it remains there until deleted from $G$,
so $E_S^\tau(u)$ grows monotonically. Since $\dist{u,S}$ can increase at most $\tau$
times before exceeding $\tau$, applying \corref{tools:es-max-edges}
with $\alpha = \tau$ gives $\mathcal{E}_S = \bigo{n\tau/q}$, where $\mathcal{E}_S$
denotes the expected total number of edge-vertex incidences ever inserted into
$E_S^\tau$ across all vertices and all edge deletions.

The proof of \lemref{tools:es-size} also provides an algorithm for maintaining
$E_S^\tau$ over time. Whenever $\dist{u,S}$ increases to a new value $d'$, we pop
edges from a min-heap over $E(u)$ in ascending order of weight, inserting each edge
$(u,v)$ into $E_S^\tau(u)$ as long as $\weight{u,v} < \min\{d',\, \tau+1\}$, and
stopping as soon as the condition fails. Each edge-vertex incidence is inserted at
most once, so the total number of heap operations is $\mathcal{E}_S$, giving a total
update time of $\bigo{\mathcal{E}_S \log n} = \bigo{\frac{n\tau \log n}{q}}$.

\lemref{tools:bk-es-properties} remains valid for $E_S^\tau$ as long as
$\dist{u,\piv[S]{u}} \leq \tau$, since in this range $E_S^\tau(u)$ agrees with
$E_S(u)$. For every edge $(u,v) \notin E_S^\tau(u)$, either
$\weight{u,v} \geq \dist{u,S}$ or $\weight{u,v} > \tau$.
\begin{corollary}
\label{cor:tools:modified-es-properties}
Let $E_S^\tau(u) := \{(u,v)\in E \mid \weight{u,v} < \min\{\dist{u,S},\tau+1\}\}$
and $E_S^\tau := \bigcup_{u \in V}E_S^\tau(u)$. Then:
\begin{itemize}
    \item[1.] As long as $\dist{u,\piv[S]{u}} \le \tau$, the subgraph
    $\paren{V, E_S^\tau \cup E(\piv[S]{u})}$ preserves the exact distance between
    $u$ and $\piv[S]{u}$.

    \item[2.] For every edge $(u,v) \notin E_S^\tau(u)$ with $\weight{u,v} \le \tau$, $\weight{u,v} \ge \dist{u,S}$.

    \item[3.] Let $\mathcal{E}_S^\tau$ 
    be the total number of edges present in $E_S^\tau$ over
    the entire update sequence. Then $\mathcal{E}_S^\tau = \bigo{\frac{n\tau}{q}}$.

    \item[4.] There is an algorithm for maintaining $E_S^\tau$ in
    $\bigo{\frac{n\tau\log n}{q}}$ total update time.

\end{itemize}
\end{corollary}

\begin{observation}
    For parameters $k \ge 1 ,\tau \in \mathbb{N}\cup\singleton{W}$ and $0 < q < 1$, Let $S_0,\dots,S_k = \mathcal{R}\paren{q,k}$.
    Then, $V = S_0 \supseteq S_1 \supseteq \dots \supseteq S_{k-1} \supseteq S_k = \emptyset$ and $\emptyset = E^{\tau}_{S_0} \subseteq E^{\tau}_{S_1} \subseteq \dots \subseteq E^{\tau}_{S_{k-1}} \subseteq E^{\tau}_{S_k} = E$
\end{observation}

The bounded-depth hierarchy-induced edge sets are one of the main building blocks of the bounded-depth APASP algorithms presented in this paper. In \secref{w+2w2:bounded} we construct our first bounded-depth algorithm. Then, in \secref{tools:hierarchy} we show how to generalize the hierarchy-induced edge sets to unbounded depths, which, along with several other tools, will enable the construction of unbounded-depth algorithms that are based on the bounded-depth algorithms.

\subsection{$+2W_{2}$-APASP}
\label{sec:w+2w2:bounded}

In this section, we present \algref{w+2w2:bounded}, which computes a $+2W_{2}$-APASP up to a bounded distance $\tau$ in
$\onolog{\tau^2 n^{\frac{3}{2}} m^{\frac{1}{2}}}$ expected total update time. Restricting the exploration to a bounded distance $\tau$ is essential,
as decremental Even-Shiloach trees and Monotone Even-Shiloach trees have total update time
proportional to their maximum exploration depth, which can be as large as $W_{\max} \cdot \paren{n-1}$.
\algref{w+2w2:bounded} serves as the subroutine for our general construction:
subsequently, in \secref{w+2w2:unbounded}, we leverage this bounded-depth algorithm
to achieve an unbounded $\paren{\onepluseps, 2W_2}$-APASP with an expected total update time
of $\onolog{n^{\frac{3}{2}}m^{\frac{1}{2}}}$ if $m = n^{1+\Omega\paren{1}}$ or $\onolog{n^{\frac{3}{2}+\subpol}m^{\frac{1}{2}}}$
otherwise, for any constant $\varepsilon > 0$.

The algorithmic foundation of our approach is the hierarchy-induced edge set $E_{S}^\tau$ introduced in \secref{hierarchy} which is well demonstrated in the approximation analysis.
Furthermore, we introduce here the concept of \emph{auxiliary pivots}. We elaborate on the matter in the algorithm description.

\begin{algorithm}[t]
\caption{$+2W_{2}$-APASP$\paren{G=\paren{V,E,\weight}}$, up to depth $\tau$}
\label{alg:w+2w2:bounded}

\BlankLine

 \textbf{Input:} A weighted undirected graph $G = \paren{V,E,\weight}$ and a depth parameter  $\tau$.

\textbf{Output:} A $+2W_{2}$-APASP for any pair of vertices whose distance $\leq \tau$.

\hrulefill

Initialize $d$ as the edges' weights and $\infty$ otherwise

Let $q\in\paren{0,1}$ to be fixed later

$S \gets \textit{sample}(V,q)$

Maintain an ES-Tree up to depth $2\tau$ for the distances from any $u\in V$ to its pivot $\piv{u}$ and update distances to $\tilde p \paren{u}$

Maintain the set $E_S  \gets E_{S}^{\tau}$

\ForEach{ $s \in S$}
        {
        Maintain an ES-Tree from $u$ up to depth $2\tau$ in $G$ and update $d$ accordingly.
    }

Let $\tilde S$ be the set of auxiliary pivots $\tilde p \paren{u}$ for all $u\in V$

$Q_{\tilde S} \gets \bracce{\paren{u,\auxpiv{u}}}$
    
\ForEach{$u \in V$}
        {
        $F (u)\gets(V,E_S \cup Q_{\tilde{S}} \cup (\singleton{u} \times \tilde{S}) \cup E(u))$
        
        Maintain a Monotone ES-Tree from $u$ up to depth $2\tau$ in $F(u)$ and update $d$ accordingly.
    }
    
\ForEach{ $u \in V$}
        { ${\hat F}(u)\gets (V,E_S \cup Q_{\tilde{S}} \cup (\singleton{u} \times \tilde{S}) \cup (\singleton{u} \times V))$
        
        Maintain a Monotone ES-Tree from $u$ up to depth $2\tau$ in ${\hat F}(u)$ and update $d$ accordingly.
   }

\end{algorithm}

\paragraph{Algorithm Description.} \algref{w+2w2:bounded} receives as input a weighted graph $G=\paren{V,E,\weight}$ and a  parameter $\tau$ that indicates the maximal distance that would be explored. It considers a parameter $q\in \paren{0,1}$, whose value will be determined as part of the runtime analysis.

Our algorithm works as follows: first, we initialize 
our distance estimation matrix $d$ to hold the weight of each edge or $\infty$ otherwise. 
We sample each vertex, independently, with probability $q$, to form the set $S\subseteq V$. We  then maintain an ES-Tree,
up to bounded depth $2\tau$, from an auxiliary vertex $\hat s$ over the edge set $E\cup \paren{\bracce{\hat s}\times S}$, where 
the weight of the auxiliary edges $\bracce{\hat s}\times S$ is $0$. This allows us to maintain the pivot $\piv{u}$ of each 
vertex $u\in V$, the distance $d\bracke{\piv{u},u} = \dist{\piv{u},u}$ and the edge set $E_S \gets E_S^\tau$ and the 
weights to auxiliary edges $d\bracke{\tilde p \paren{u},v} \leftarrow d\bracke{\piv{u},v} $. We observe that as we 
consider an ES-Tree with a bounded depth $2\tau$, some vertices may not have a pivot. We also introduce a set of auxiliary pivots
$\tilde S =\bracce{\tilde p \paren{u} \mid u\in V}$, maintaining the distances 
$d\bracke{\tilde p \paren{u},v}=d\bracke{ p \paren{u},v}$, alongside a corresponding auxiliary edge set 
$Q_{\tilde S} =\bracce{\paren{u,\tilde p \paren{u}} \mid u\in V}$.

We now maintain additional $2n+\abs{S}$ instances of ES-Trees throughout the execution of the algorithm, 
up to a depth of $2\tau$. Each of these instances can be seen as one of three \qoute{types}. The first is a standard 
ES-Tree for every $s\in S$, while both the second and the third are Monotone ES-Trees, for every $u\in V$. The difference 
between the second and the third lies in the set of edges adjacent to $u$: the second considers only 
the original graph edges $E\paren{u}$, while the third considers any auxiliary edge $\bracce{u}\times V$.

It is important to note that while these trees are maintained dynamically over the sequence of edge deletions, 
the loops are processed sequentially one after another,
with each loop fully completed \textbf{for all the vertices} before the next loop starts. 

The first type is an ES-Tree for each $s\in S$ over all $E$, up to depth $2\tau$. The second type is 
a Monotone ES-Tree, for each $u\in V$, that utilizes the distances computed by the first type of tree. 
It considers the weighted graph $F\paren{u} = \paren{V,E_{S} \cup Q_{\tilde S}\cup \paren{\bracce{u}\times \tilde S}\cup E\paren{u}}$,
where:

\begin{itemize}
    \item $E_{S}$ is the hierarchy-induced edge set, defined as the union over \textbf{all} $v\in V$ of edges $\paren{v,w}$ such that $\weight{v,w} < \dist{v,\piv{v}}$,
    \item $Q_{\tilde S}$ are edges from each vertex $v\in V$ to its auxiliary pivot $\auxpiv{v}$,
    \item $\bracce{u} \times \tilde S$ are all of the auxiliary edges from the current vertex $u$ to any auxiliary pivot $\auxpiv{x} \in \tilde S$, representing distances previously computed,
    \item $E\paren{u}$ are the edges from  $u$ to its neighbors from $V$.
\end{itemize} 

It is important to note that while these trees are maintained dynamically over the sequence of edge deletions, 
they are processed in three sequential phases:
upon each update, the pivot ES-Trees are fully updated first, 
followed by the first-phase Monotone ES-Trees \textbf{for all vertices}, 
and finally the second-phase Monotone ES-Trees \textbf{for all vertices}.

While the third type of a tree is also a Monotone ES-Tree, it considers a slightly different edge set. Each $u\in V$  considers the weighted graph ${\hat F}\paren{u}= \paren{V,E_{S} \cup Q_{\tilde S}\cup \paren{\bracce{u}\times \tilde S}\cup \paren{\bracce{u}\times V}}$. This graph ${\hat F}\paren{u}$ can now utilize distance estimations of the form $\bracce{u}\times V$ that were previously computed. 

Finally, we set $d\bracke{u,v} \leftarrow \mmin{\levelh{u}{v} , \levelh{v}{u}} $, where $\levelh{u}{\cdot}$ is the level in $\hat{F}\paren{u}$. We then return the distance estimate matrix $d: V\times V \rightarrow \mathbb{N}$.

Before we proceed to the formal correctness proofs, 
we elaborate on the key challenges of \algref{w+2w2:bounded} and explain, at a high level, 
why they are necessary in the decremental setting and how they support the desired approximation guarantee.

\tikzstyle{pathnode} = [circle, draw=black, fill=white, inner sep=2.5pt, minimum size=5pt]
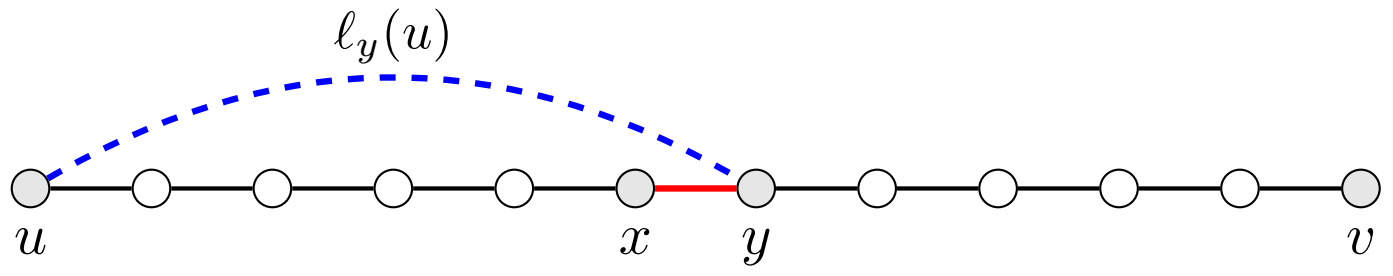
\begin{figure}[htbp]
  \centering
  \begin{tikzpicture}[scale=0.8, >=stealth]

    \def\dx{1} 

    \node[pathnode, fill=gray!20, label=below:$u$] (u) at (0*\dx, 0) {};
    \node[pathnode, fill=gray!20, label=below:$x$] (x) at (5*\dx, 0) {};
    \node[pathnode, fill=gray!20, label=below:$y$] (y) at (6*\dx, 0) {};
    \node[pathnode, fill=gray!20, label=below:$v$] (v) at (11*\dx, 0) {};

    \foreach \i in {1, 2, 3, 4} {
      \node[pathnode] (a\i) at (\i*\dx, 0) {};
    }
    \foreach \k in {1, 2, 3, 4} {
      \pgfmathsetmacro{\pos}{\k+6}
      \node[pathnode] (b\k) at (\pos*\dx, 0) {};
    }

    \draw[thick, black] (u) -- (a1);
    \draw[thick, black] (a1) -- (a2);
    \draw[thick, black] (a2) -- (a3);
    \draw[thick, black] (a3) -- (a4);
    \draw[thick, black] (a4) -- (x);

    \draw[thick, black] (y) -- (b1);
    \draw[thick, black] (b1) -- (b2);
    \draw[thick, black] (b2) -- (b3);
    \draw[thick, black] (b3) -- (b4);
    \draw[thick, black] (b4) -- (v);

    \draw[very thick, red] (x) -- (y);

    \draw[blue, dashed, very thick, bend left=30] (u) to node[above, inner sep=2pt, font=\small, black] {$\ell_y(u)$} (y);

  \end{tikzpicture}
  \caption{The path $P_{u,v}$ between $u$ and $v$. The edge $\paren{x,y}$ is the only edge not in $E_S$, yet it is observed by $y$ in $F\paren{y}$. The blue dashed edge is an auxiliary edge between $u$ and $y$ in ${\hat F}\paren{u}$, whose weight holds the updated estimate $\level{y}{u}$.}
  \label{fig:two-monotone-trees-example}
\end{figure}

First, we address the necessity of maintaining two separate Monotone ES-Trees, $F\paren{u}$ and $\hat{F}\paren{u}$, for each vertex $u\in V$.
In the static setting, the classical approach of Cohen and Zwick~\cite{CohZwi1997} relies on a two round SSSP computation.
Intuitively, this approach is useful when the shortest path between $u$ and $v$ has the structure
$u\squiggly v = u\squiggly x \straight y \squiggly v$ where $u\squiggly x, y\squiggly v\subseteq E_{S}$, 
as illustrated in \figref{two-monotone-trees-example}. In the first round, 
the exact distance $d\bracke{y,u} =\dist{y,u}$ is computed. In the second round, the auxiliary edge $\paren{u,y}$ 
is taken into consideration as well. Its weight is the previously computed $d\bracke{u,y} =\dist{u,y}$. 
As the rest of the path $y\squiggly v\subseteq E_{S}$, the algorithm would compute the exact distance $d\bracke{u,v} = \dist{u,v}$. 

Extending this multi-round approach to the decremental setting introduces two distinct difficulties that require maintaining two separate trees:
\begin{itemize}
    \item \textbf{Stale auxiliary edges:} 
    Assume $F\paren{u}$ had the same edges as $\hat{F}\paren{u}$. 
    Since the adversary may modify only original edges, an auxiliary edge $\paren{u,y}$ remains in $F\paren{u}$, 
    from iterations at earlier times, with the weight assigned to it at time zero, i.e., 
    at the beginning of the execution. This weight may represent the weight of a path whose edges were already deleted, 
    and thus no longer represent a valid distance. 
    We overcome this issue by distinguishing between $F\paren{u}$ and $\hat{F}\paren{u}$: in $F\paren{u}$, 
    we only consider edges that were originally in the graph, and compute the current weight for the auxiliary edges. 
    In $\hat{F}\paren{u}$, we utilize these auxiliary edges (e.g. $\paren{u,y}$ from \figref{two-monotone-trees-example}) 
    with their updated weight.

    \item \textbf{Inter-round distance monotonicity:} 
    The second difficulty arises due to the possible values of distance estimates across two rounds (at the same point in time). 
    In the static setting, distance estimates obtained after the second round may be smaller than those obtained after 
    the first round. However, the structure of a Monotone ES-Tree does not permit distance values to decrease, which may occur, 
    for example, for an auxiliary edge such as $\paren{u,y}$. By maintaining two separate Monotone ES-Trees,
    $F\paren{u}$ and $\hat{F}\paren{u}$, we avoid this issue as well: each tree is monotone in itself.
\end{itemize}

Second, unlike the static setting, pivot assignments may change dynamically. Even if the distance $\dist{u,S}$ remains constant, 
edge deletions may sever the path to the current pivot, forcing $u$ to switch its assignment to another equidistant vertex 
in $S$. This has a major effect on the total update time. The total update time of a Monotone ES-Tree depends on the 
total number of edges inserted to it. If the distance to the pivot does not change, yet the pivot assignment does, 
our algorithm would have expensive total update time costs. In order to avoid these costs, we introduce the 
notion of \emph{auxiliary pivots}. Namely, this is a set $\tilde{S} = \bracce{\auxpiv{u}  \mid u\in V}$, together with a 
set of auxiliary edges $Q_{\tilde{S} }=\bracce{\paren{u,\auxpiv{u}} \mid {u \in V}}$. The weights of these 
auxiliary edges are initially provided by the ES-Tree we invoke from each $s \in S$. However, when the pivot of 
a vertex $u$ changes from $s \in S$ to $s' \in S$, we only update the weights of the auxiliary edges $\paren{\auxpiv{u}, x}$ 
if the distance increases. Otherwise, we leave the previous weight. 

Third, we address a matter unique to weighted additive algorithms in the decremental setting: the non-monotonicity of local heaviest edges along shortest paths.
As discussed earlier, the upper bound $\dist{u,v} + 2W_{2}\paren{u\squiggly v}$ may, counterintuitively,
decrease over time as edge deletions occur and shortest paths shift.
Because decremental shortest-path data structures are designed to track non-decreasing distances,
they cannot efficiently maintain an upper bound that shrinks over time.
Restricting the analysis to shortest paths is therefore insufficient.
Instead, we establish that our approximation guarantee holds for
\textit{any} path $P_{u,v}$ between $u$ and $v$.
Because the graph is decremental, a path in this context is defined as a 
\emph{constant set of edges, present from the beginning and until the deletion of its first edge};
thus, any path $P_{u,v}$ that is available at time $t$ was inherently present at all earlier times $r \leq t$, 
satisfying $W_{i,r}\paren{P_{u,v}} \leq W_{i,t}\paren{P_{u,v}}$ for every $1 \le i \le \hops{P_{u,v}}$.
To establish correctness at any time $t$, we prove that every such path $P_{u,v}$ in $G$ admits an alternative $P^{*}_{u,v}$ in $\hat{F}\paren{u}$ satisfying    
$d\bracke{u,v} \leq \weight{P^{*}_{u,v}} \leq \weight{P_{u,v}} + 2W_{2}\paren{P_{u,v}}$.

It should be observed that we establish a stronger invariant than necessary:
by proving the upper bound holds for any available path $P_{u,v}$ at time $t$,
the standard $+2W_{2}$-APASP guarantee follows immediately as a special case when $P_{u,v}$ is a shortest path at time $t$.

We now turn to prove the correctness of \algref{w+2w2:bounded} by induction on the time $t$. Our correctness proofs throughout this paper
follow a common two-part induction scheme: we first establish the base case at time $t=0$ by identifying a short alternative 
path $P^*_{u,v}$ in the maintained graph and arguing it is explored by the Monotone ES-Tree. For the inductive step at time 
$t>0$, we show that the Monotone ES-Tree either computes the alternative path directly, or inherits a valid estimate from 
time $t-1$ via a stretched edge. This paradigm is used consistently in all subsequent algorithm analyses. 

For the analysis of \algref{w+2w2:bounded}, at each time $t$, 
we consider, for any path $P_{u,v}$ between $u,v\in V$, the following three cases that cover all possibilities:

\begin{itemize}
    \item \case{1}: $\sizeindex{P_{u,v}}{ E_{S}} = 0 $,
    \item \case{2}: $\sizeindex{P_{u,v}}{ E_{S}} = 1 $,
    \item \case{3}: $\sizeindex{P_{u,v}}{ E_{S}} > 1 $.
\end{itemize}

We begin by proving for time $t=0$:

\begin{lemma}\label{lem:w+2w2:bounded-approx-t0}
    Let $P_{u,v}$ be a path between $u$ and $v$  in $G$ at time $t=0$ with weight $\weight{P_{u,v}} \leq \tau$. Then there exists a path $P^{*}_{u,v}$ between $u$ and $v$ in $\hat{F} \paren{u}$ such that $d\bracke{u,v} \leq w\paren{P^{*}_{u,v}} \leq w\paren{P_{u,v}} +2W_{2}\paren{P_{u,v}}$.
\end{lemma}
\begin{proof}

Let $\level{u}{v}$ be the level of $v$ in $u$'s first Monotone ES-Tree and  let $\levelh{u}{v}$ be the level of $v$ in $u$'s second Monotone ES-Tree. We consider the three cases:

\ccase{1} $\sizeindex{P_{u,v}}{ E_{S}} = 0$.

Hence, $P_{u,v}\subseteq E_{S}$. As $F\paren{u}$ contains all the edges of $E_S$, we may  set $P^{*}_{u,v} = P_{u,v}$. As $  \weight{P_{u,v}} \leq \tau$, it follows that  $d\bracke{u,v} \leq \levelh{u}{v} \leq  \weight{P^{*}_{u,v}} = \weight{P_{u,v}} $.

\ccase{2} $\sizeindex{P_{u,v}}{ E_{S}} = 1 $.
Let $\paren{x,y} = \rightabovearrow{P_{u,v}}{E_{S}} = \leftabovearrow{P_{u,v}}{E_{S}}$ and denote $P_{u,v} = P_{u,y} \circ P_{y,v}$. 
 We observe that $P_{u,y}\subseteq F\paren{y}$ and $\weight{P_{u,y}} \leq \weight{P_{u,v}} \leq \tau$. Hence, $\level{y}{u}  \leq \weight{P_{u,y}}$. During the invocation of the second Monotone ES-Tree from $u$, the edge $\paren{u,y}$ will be included in ${\hat F}\paren{u}$ with the weight $\level{y}{u} \leq \weight{P_{u,y}}$. As $P_{y,v}\subseteq E_S\subseteq {\hat F}\paren{u}$, we may set $P^{*}_{u,v} = \bracce{\paren{u,y}} \circ P_{y,v}$. Hence: $d\bracke{u,v} \leq \levelh{u}{v} \leq  \weight{u,y} + \weight{P_{y,v}} = \weight{P^*_{u,v}}\leq \weight{P_{u,y}}+\weight{P_{y,v}} = \weight{P_{u,v}}$.

\ccase{3} $\sizeindex{P_{u,v}}{ E_{S}} > 1 $. Let $\paren{x_1,y_1} = \rightabovearrow{P_{u,v}}{E_{S}}$ and $\paren{x_2,y_2} = \leftabovearrow{P_{u,v}}{E_{S}}$ (See: \figref{w+2w2:approx-case-3}). Since $\paren{x_1,y_1} \neq \paren{x_2,y_2}$, assume w.l.o.g. that $\weight{x_2,y_2} \leq \weight{x_1,y_1}$. Let $P_{u,v} = P_{u,y_{2}} \circ P_{y_{2},v}$. Because $\paren{x_2,y_2} \notin E_S$, we have $\dist{y_2, \piv{y_2}} \leq \weight{x_2,y_2}$. As $\piv{y_2} \in S$, an ES-Tree is maintained from
$\piv{y_2}$ over all $G$, hence
$\level{\piv{y_2}}{u} = \dist{u,\piv{y_2}} \leq \dist{u,y_2} + \weight{x_2,y_2}$.

During $u$'s first Monotone ES-Tree over $F\paren{u}$, the auxiliary edges $\paren{u,\auxpiv{y_2}}$ and $\paren{\auxpiv{y_2}, y_2}$ hold weights bounded by $\weight{P_{u,y_{2}}}+\weight{x_2,y_2}$ and $\weight{x_2,y_2}$, respectively. Since $P_{y_2,v} \subseteq E_{S} \subseteq F\paren{u}$, we may set $P^{*}_{u,v} = \bracce{\paren{u,\auxpiv{y_{2}}}} \circ \bracce{\paren{\auxpiv{y_{2}},y_{2}}} \circ P_{y_{2},v}$, hence:
$$\level{u}{v} \leq \weight{P_{u,y_{2}}} + 2 \weight{x_2,y_2} + \weight{P_{y2,v}}  = \weight{P^{*}_{u,v}} \leq \weight{P_{u,v}} + 2W_{2}\paren{P_{u,v}}$$

Where the last inequality holds due to our w.l.o.g. assumption. 

\tikzstyle{mainnode} = [circle, draw=black, fill=gray!20, inner sep=2pt, minimum size=7pt, font=\small]
\tikzstyle{pathnode} = [circle, draw=black, fill=white, inner sep=2.5pt, minimum size=5pt] 
\tikzstyle{specialnode} = [circle, draw=black, fill=blue!10, inner sep=2pt, minimum size=7pt, font=\small]
\tikzstyle{dotsep} = [scale=2] 

\begin{figure}[htbp]
  \centering
  \begin{tikzpicture}[scale=0.8, >=stealth]

    \def\xsep{1.2} 

    \node[mainnode, label=above:$u$] (u) at (0*\xsep, 0) {};
    \node[mainnode, label=above:$x_2$] (x2) at (4*\xsep, 0) {};
    \node[mainnode, label=above:$y_2$] (y2) at (5*\xsep, 0) {};
    \node[mainnode, label=above:$v$] (v) at (9*\xsep, 0) {};

    \foreach \i in {1, 3} {
      \node[pathnode] (a\i) at (\i*\xsep, 0) {};
    }

    \node[dotsep] (dots1) at (2*\xsep, 0) {$\cdots$};

    \foreach \k in {1, 2, 3} {
      \pgfmathsetmacro{\pos}{\k+5} 
      \node[pathnode] (b\k) at (\pos*\xsep, 0) {};
    }

    \node[specialnode, label=below:$\auxpiv{y_{2}}$] (py2) at (5*\xsep, -1.8) {};

    \draw[thick, black] (u) -- (a1);
    \draw[thick, black] (a3) -- (x2);
    
    \draw[thick, black] (y2) -- (b1);
    \draw[thick, black] (b1) -- (b2);
    \draw[thick, black] (b2) -- (b3);
    \draw[thick, black] (b3) -- (v);

    \draw[very thick, red] (x2) -- (y2);

    \draw[blue, dashed, thick] (py2) -- node[midway, black, inner sep=2pt, font=\small, anchor=south west] {$\le \weight(x_2,y_2)$} (y2);
    
    \draw[blue, dashed, very thick, bend left=50] (py2) to (u);

  \end{tikzpicture}
  \caption{\label{fig:w+2w2:approx-case-3}The edge $(x_2,y_2)$ is the last edge not in $E_S$. The graph $\hat{F}(u)$ contains the black edges and the blue dashed edges. As $(x_2,y_2) \notin E_S$ it holds that $\dist(y_2,p(y_2)) \le \weight(x_2,y_2)$}
\end{figure}
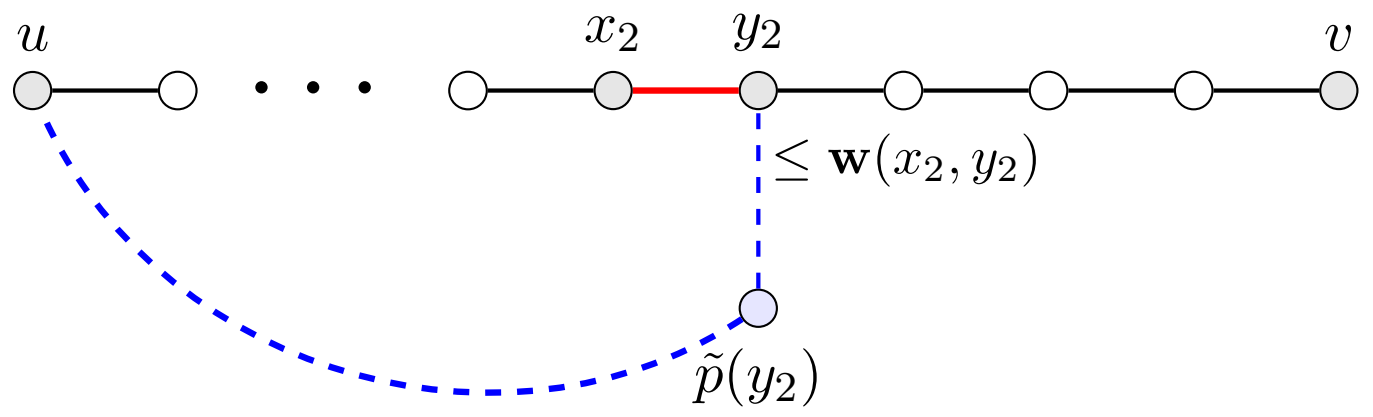

\end{proof}

Having established the base case for time $t=0$, 
we proceed to extend \lemref{w+2w2:bounded-approx-t0} by induction over the time $t$. 
As discussed in our technical overview, we establish the upper bound for any available path $P_{u,v}$ at time $t$, 
from which the standard $+2W_{2}$-APASP guarantee follows directly when $P_{u,v}$ is a shortest path at time $t$.

\begin{lemma}\label{lem:w+2w2:bounded-approx}
     Let $P_{u,v}$ be a path between $u$ and $v$  in $G$ at time $t\geq 0$ with weight $\weight{P_{u,v}} \leq \tau$. Then $d\bracke{u,v} \leq w\paren{P_{u,v}} +2W_{2}\paren{P_{u,v}}$.
\end{lemma}
\begin{proof}
The proof proceeds by induction on the time $t$. The base case for $t=0$ holds by \lemref{w+2w2:bounded-approx-t0}. Assume this claim holds for any time $r < t$. We now prove it for time $t$. We consider the same three cases:

\ccase{1} $\sizeindex{P_{u,v}}{ E_{S}} = 0$.

Hence, $P_{u,v}\subseteq E_{S}$. If $P_{u,v}$ contains no stretched edges in $u$'s tree, the analysis of \lemref{w+2w2:bounded-approx-t0} applies identically. Otherwise, let $\paren{w,z}$ be the last stretched edge on $P_{u,v}$ when traversing from $u$ to $v$, and denote $P_{u,v} = P_{u,z} \circ P_{z,v}$. Because $\paren{w,z}$ is stretched in $u$'s tree, $\level[t]{u}{z} = \level[t-1]{u}{z}$. By the inductive hypothesis: $\level[t-1]{u}{z} \leq \weight{P_{u,z}} + 2W_{2} \paren{P_{u,z}}$. Furthermore, since $P_{z,v}$ contains no stretched edges and $P_{z,v} \subseteq P_{u,v} \subseteq E_{S}$, it follows that: $\level[t]{u}{v}  \leq \level[t]{u}{z} +\weight{P_{z,v}} \leq \weight{P_{u,z}} + 2W_{2}\paren{P_{u,z}} +\weight{P_{z,v}} \leq \weight{P_{u,v}} + 2W_2\paren{P_{u,v}}$, where the last inequality holds as $W_{2}\paren{P_{u,z}} \leq W_{2}\paren{P_{u,v}}$.

\ccase{2} $\sizeindex{P_{u,v}}{ E_{S}} = 1 $.
Let $\paren{x,y} = \rightabovearrow{P_{u,v}}{E_{S}}=\leftabovearrow{P_{u,v}}{E_{S}}$ and denote $P_{u,v} = P_{u,y} \circ P_{y,v}$. Let $P^{*}_{u,v} = \paren{u,y}\circ P_{y,v} \subseteq \hat{F}\paren{u}$ and recall that $\weight{u,y} \le \level{y}{u}$.

Let us first consider the case where $P^{*}_{u,v}$ has no stretched edges in $u$'s tree. 
If additionally $P_{u,y}$ has no stretched edges in $y$'s tree, the analysis of \case{2} of \lemref{w+2w2:bounded-approx-t0} applies identically.
Otherwise, assume that $P_{u,y}$ has stretched edges in $y$'s tree. For a clearer reading, we reverse the order of the edges and write 
$P_{y,u}$ instead of $P_{u,y}$. Let $\paren{h,f}\in P_{y,u}$ be the last stretched edge from $P_{y,u}$ in $y$'s tree, when traversing 
from $y$ to $u$ (See: \figref{w+2w2:approx-case-2-stretched}(a)). Denote: $P_{y,u} = P_{y,f}\circ P_{f,u}$. As $\paren{h,f}$ is 
stretched in $y$'s tree, it follows that $\level[t]{y}{f} = \level[t-1]{y}{f}$. By the induction hypothesis: 
$\level[t-1]{y}{f} \leq \weight{P_{y,f}}{t-1} +2W_{2,t-1}\paren{P_{y,f}}$. 
As $P_{f,u}\subseteq E_{S}$ and has no stretched edges, it follows that 
$\level[t]{y}{u}  \leq \level[t]{y}{f} + \weight{P_{f,u}}{t} \leq \weight{P_{y,f}}{t-1} + 
2W_{2,t-1}\paren{P_{y,f}} + \weight{P_{f,u}}{t} \leq \weight{P_{y,u}}{t} + 2W_{2,t}\paren{P_{y,u}} = 
\weight{P_{u,y}}{t} + 2W_{2,t}\paren{P_{u,y}}$,
where the last inequality holds as $W_{2,t-1}\paren{P_{y,f}} \leq W_{2,t}\paren{P_{y,f}} \leq W_{2,t}\paren{P_{u,y}}$. 
As $\weight{u,y} \leq \level{y}{u}$ and $P^{*}_{u,v}$ has no stretched edges in $u$'s tree, it follows that 
$d\bracke{u,v} \leq \levelh[t]{u}{v}  \leq \weight{u,y}{t}+\weight{P_{y,v}}{t} \leq \weight{P_{u,y}}{t} + 2W_{2,t}\paren{P_{u,y}} + 
\weight{P_{y,v}}{t} \leq \weight{P_{u,v}}{t} + 2W_{2,t}\paren{P_{u,v}}$.

We are left with the case where $P^{*}_{u,v}$ contains a stretched edge in $u$'s tree. Let $\paren{w,z} \in P^{*}_{u,v}$ be the last 
stretched edge in $u$'s tree when traversing the path from $u$ to $v$ (See: \figref{w+2w2:approx-case-2-stretched}(b)). Hence, 
$\levelh[t]{u}{z} = \levelh[t-1]{u}{z}$. We observe that $\paren{w,z}=\paren{u,y}$ or 
$\paren{z,w}\in P_{y,v}$. Denote: $P_{u,v}=P_{u,z}\circ P_{z,v}$, where $P_{u,z}=P_{u,y}\circ P_{y,z}$, by our observation.

By the induction hypothesis: $\levelh[t]{u}{z} = \levelh[t-1]{u}{z} \leq \weight{P_{u,z}}{t-1}+2W_{2,t-1}\paren{P_{u,z}}
\leq \weight{P_{u,z}}{t}+2W_{2,t}\paren{P_{u,z}}$.  
Regardless of whether $\paren{w,z}=\paren{u,y}$ or $\paren{z,w}\in P_{y,v}$, 
it holds that $P_{z,v}\subseteq E_{S}$ and contains no stretched edges in $u$'s tree. Thus:
$\levelh[t]{u}{v}  \leq \levelh[t]{u}{z} + \weight{P_{z,v}}{t} \leq \weight{P_{u,z}}{t} + 2W_{2,t}\paren{P_{u,z}} + 
\weight{P_{z,v}}{t} \leq \weight{P_{u,v}}{t} + 2W_{2,t}\paren{P_{u,v}} \leq \weight{P_{u,v}}{t} + 2W_{2,t}\paren{P_{u,v}}$.

\tikzstyle{pathnode} = [circle, draw=black, fill=white, inner sep=2.5pt, minimum size=5pt]
\begin{figure}[htbp]
  \centering
  \begin{tikzpicture}[scale=0.8, >=stealth]

    \def\dx{1} 

    \begin{scope}[xshift=-5cm,local bounding box=scope1]
        \node[pathnode, fill=gray!20, label=below:$u$] (u) at (0*\dx, 0) {};
        \node[pathnode, fill=gray!20, label=below:$x$] (x) at (5*\dx, 0) {};
        \node[pathnode, fill=gray!20, label=below:$y$] (y) at (6*\dx, 0) {};
        \node[pathnode, fill=gray!20, label=below:$f$] (w) at (2*\dx, 0) {};
        \node[pathnode, fill=gray!20, label=below:$h$] (f) at (3*\dx, 0) {};
        \node[] (dots) at (7*\dx, 0) {$\dots$};
        \node[pathnode, fill=gray!20, label=below:$v$] (v) at (8*\dx, 0) {};
    
        \foreach \i in {1, 4} {
          \node[] (a\i) at (\i*\dx, 0) {$\dots$};
        }
    
        \draw[thick, black] (u) -- (a1);
        \draw[thick, black] (a1) -- (w);
        \draw[thick, black] (f) -- (a4);
        \draw[thick, black] (a4) -- (x);
    
        \draw[thick, black] (y) -- (dots);
        \draw[thick, black] (dots) -- (v);
    
        \draw[very thick, red] (x) -- (y);
        \draw[very thick, brown] (w) -- (f);
    
        \draw[blue, dashed, very thick, bend left=30] (u) to node[black, above=5pt] 
        {\scriptsize $\level{y}{u}$} (y);
        \draw[yellow, dashed, very thick, bend left=55] (y) to node[black, below=5pt] 
        {\scriptsize $\level{y}{f} \le \weight\paren{P_{y,f}} + 2W_{2}\paren{P_{y,f}}$} (w);
        \draw [decorate, decoration={brace, amplitude=10pt,mirror}] (y) -- node [midway,below=12pt] {$\subseteq E_S$} (v);
    \end{scope}
    \node[below=12pt] at (scope1.south) {(a)};

    \begin{scope}[xshift=5cm,local bounding box=scope2]
        \node[pathnode, fill=gray!20, label=below:$u$] (u) at (0*\dx, 0) {};
        \node[pathnode, fill=gray!20, label=below:$x$] (x) at (2*\dx, 0) {};
        \node[pathnode, fill=gray!20, label=below:$y$] (y) at (3*\dx, 0) {};
        \node[] (dots) at (1*\dx, 0) {$\dots$};
        \node[pathnode, fill=gray!20, label=below:$w$] (h) at (5*\dx, 0) {};
        \node[pathnode, fill=gray!20, label=below:$z$] (z) at (6*\dx, 0) {};
        \node[pathnode, fill=gray!20, label=below:$v$] (v) at (8*\dx, 0) {};
    
        \foreach \k in {1, 4} {
          \pgfmathsetmacro{\pos}{\k+3}
          \node[] (b\k) at (\pos*\dx, 0) {$\dots$};
        }
    
        \draw[thick, black] (u) -- (dots);
        \draw[thick, black] (dots) -- (x);
    
        \draw[thick, black] (y) -- (b1);
        \draw[thick, black] (b1) -- (h);
        \draw[thick, black] (z) -- (b4);
        \draw[thick, black] (b4) -- (v);
    
        \draw[very thick, red] (x) -- (y);
        \draw[very thick, brown] (h) -- (z);
    
        \draw[blue, dashed, very thick, bend left=30] (u) to node[black, above=5pt, midway] {\scriptsize $\level{y}{u}$} (y);
        \draw[yellow, dashed, very thick, bend right=45] (u) to node[black, below=5pt, midway] {\scriptsize $\level{u}{z} \le \weight\paren{P_{u,z}} + 2W_2\paren{P_{u,z}}$} (z);
        \draw [decorate, decoration={brace, amplitude=10pt,mirror}] (z) -- node [midway,below=12pt] {$\subseteq E_S$} (v);
    \end{scope}
    \node[below=5pt] at (scope2.south) {(b)};

  \end{tikzpicture}
  \caption{Visualization of \case{2}, 
  $\protect\paren{x,y}=\protect\rightabovearrow{P_{u,v}}{E_{S}}=\protect\leftabovearrow{P_{u,v}}{E_{S}}$,
  demonstrating the two subcases with stretched edges:
  (a) $P^{*}_{u,v}$ contains no stretched edges in $u$'s tree, while $\protect\paren{h,f}$ is the last stretched edge on 
  $P_{y,u}$ in $y$'s tree;
  (b) $\protect\paren{w,z}$ is the last stretched edge on $P^{*}_{u,v}$ in $u$'s tree, shown here for the case.}
  \label{fig:w+2w2:approx-case-2-stretched}
\end{figure}
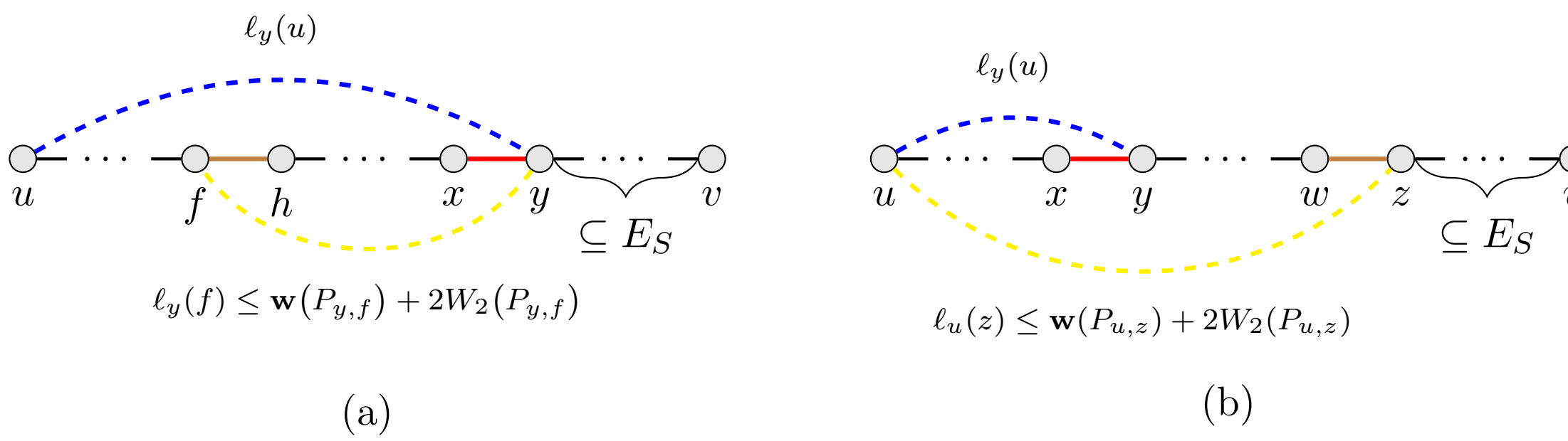

\ccase{3} $\sizeindex{P_{u,v}}{ E_{S}} > 1 $.
Let $\paren{x_{1},y_{1}} = \rightabovearrow{P_{u,v}}{E_S}$ and $\paren{x_{2},y_{2}} = \leftabovearrow{P_{u,v}}{E_S}$, 
where $\paren{x_{1},y_{1}}\neq \paren{x_{2},y_{2}}$. Assume, w.l.o.g., 
that $\weight{x_{2},y_{2}} \leq \weight{x_{1},y_{1}}$, therefore: 
$\weight{x_{2},y_{2}} \leq W_2\paren{P_{u,v}}$. As $\paren{x_{2},y_{2}} \notin E_{S}$ and $ \weight{P_{u,v}} \leq \tau$,
it follows that $\dist{y_2,\piv{y_2}} \leq \weight{x_2,y_2}$. Denote $P_{u,v} = P_{u,y_{2}} \circ P_{y_{2},v}$. We now consider the path  $P^{*}_{u,v} = \paren{u,\auxpiv{y_{2}}}\circ \paren{\auxpiv{y_2},y_2}\circ P_{y_{2},v} \subseteq F\paren{u}$. At first, assume no edges are stretched on $P^{*}_{u,v}$ in $u$'s tree. Therefore: $\weight{u,\auxpiv{y_{2}}} \leq \level[t]{\piv{y_2}}{u}$ and  $\weight{\auxpiv{y_{2}},y_{2}} \leq \level[t]{\piv{y_{2}}}{y_{2}}$. 
Recall that $\piv{y_2} \in S$. Hence, we maintain an ES-Tree from $\piv{y_2}$ on the entire graph $G$ and 
$\level[t]{\piv{y_2}}{u} = \dist{u,\piv{y_2}}{t} \leq \dist{u,y_2}{t} + \dist{y_2,\piv{y_2}}{t} \leq 
\weight{P_{u,y_2}}{t} + \weight{x_2,y_2}{t} \leq \weight{P_{u,y_{2}}}{t} +W_{2,t}\paren{P_{u,v}}$. 
Likewise, $\level[t]{\piv{y_2}}{y_{2}} \leq W_{2,t}\paren{P_{u,v}}$. 
Thus: $d\bracke{u,v} \leq \weight{P^{*}_{u,v}}{t} =  \weight{u,\auxpiv{y_{2}}}{t} + 
\weight{\auxpiv{y_{2}},y_{2}}{t} +\weight{P_{y_{2},v}}{t} \leq \weight{P_{u,y_{2}}}{t}+\weight{P_{y_{2},v}}{t} 
+2W_{2,t}\paren{P_{u,v}} = \weight{P_{u,v}}{t}+2W_{2,t}\paren{P_{u,v}}$.

Now, assume $P^{*}_{u,v}$ has a stretched edge in $u$'s tree. 
Let $\paren{w,z}$ be the last stretched edge on $P^{*}_{u,v}$. We consider two possibilities: 
First, when $\paren{w,z} \in \bracce{\paren{\auxpiv{y_{2}},y_{2}}} \circ P_{y_{2},v}$. 
We observe that $z$ is a vertex on $P_{y_{2},v}\subseteq P_{u,v}$, hence $P_{z,y_{2}}\subseteq E_{S}$. 
Denote: $P_{u,v} = P_{u,z} \circ P_{z,v}$. As the edge $\paren{w,z}$  is stretched in $u$'s tree it follows that 
$\levelh[t]{u}{z} = \levelh[r]{u}{z}$ for a previous point in time $r<t$. By the induction hypothesis on $r<t$, 
it follows that $\levelh[r]{u}{z} \leq \weight{P_{u,z}}{r} +2W_{2,r}\paren{P_{u,z}} \leq \weight{P_{u,z}}{t} +2W_{2,t}\paren{P_{u,z}}$. 
Since $\paren{w,z}$ is the last stretched edge and $P_{z,v}\subseteq E_{S}$, it follows that 
$d\bracke{u,v} \leq \levelh[t]{u}{v} \leq \levelh[t]{u}{z} + \weight{P_{z,v}}{t} 
\leq \weight{P_{u,z}}{t} + 2W_{2,t}\paren{P_{u,z}} + \weight{P_{z,v}}{t} \leq \weight{P_{u,v}}{t} +2W_{2,t}\paren{P_{u,v}}$. 

We are therefore left with the second option for the last stretched edge from $P^{*}_{u,v}$ in $u$'s tree: 
$\paren{w,z} = \paren{u,\auxpiv{y_{2}}}$. Let $r<t$ be the point in time where 
$\levelh[r]{u}{\auxpiv{y_{2}}}$ was last changed. Let $p^{\paren{r}}\paren{y_{2}}$ and $p^{\paren{t}}\paren{y_{2}}$ 
be the pivots of $y_{2}$ at times $r$ and $t$, respectively. As $\levelh[r]{u}{\auxpiv{y_{2}}}$ was changed at time $r$, 
it follows that $\levelh[r]{u}{\auxpiv{y_{2}}} \leq \weight{u,\auxpiv{y_{2}}}{r} \leq \levelh[r]{p^{\paren{r}}\paren{y_{2}}}{u}$. 
Recall that  $p^{\paren{r}}\paren{y_{2}} \in S$, and therefore maintains an ES-Tree. Therefore, 
$\levelh[r]{p^{\paren{r}}\paren{y_{2}}}{u} = \dist{u,p^{\paren{r}}\paren{y_{2}}}{G}{r} \leq 
\weight{P_{u,y_{2}}}{r} +\weight{x_{2},y_{2}}{r} \leq \weight{P_{u,y_{2}}}{t} + \weight{x_{2},y_{2}}{t} \leq 
\weight{P_{u,y_{2}}}{t} +W_{2,t}\paren{P_{u,v}}$. Thus, 
$\levelh[t]{u}{\auxpiv{y_{2}}} = \levelh[r]{u}{\auxpiv{y_{2}}} 
\leq \weight{P_{u,y_{2}}}{t} +W_{2,t}\paren{P_{u,v}}$. 
Furthermore, by definition it holds that $\weight{\auxpiv{y_{2}},y_{2}} = \delta_{G}^{t} \paren{y_{2},p^{\paren{t}}\paren{y_{2}}} \leq \weight{x_{2},y_{2}}\leq W_{2}\paren{P_{u,v}}$.
As all the edges on $\bracce{\paren{\auxpiv{y_{2}},y_{2}}}\circ P_{y_{2},v}$
are not stretched and are in $\hat{F}\paren{u}$ it follows that 
$d\bracke{u,v}\leq \levelh[t]{u}{v} \leq \levelh[t]{u}{\auxpiv{y_{2}}}+ 
\weight{\auxpiv{y_{2}},y_{2}}{t} + \weight{P_{y_{2},v}}{t} \leq \weight{P_{u,y_{2}}}{t} +W_{2,t}\paren{P_{u,v}} + 
W_{2,t}\paren{P_{u,v}} + \weight{P_{y_{2},v}}{t} = \weight{P_{u,v}}{t} +2W_{2,t}\paren{P_{u,v}}$.

\end{proof}

Next, we discuss the total update time of \algref{w+2w2:bounded}.

\begin{lemma}\label{lem:w+2w2:bounded-total-update-time}
The expected total update time of \algref{w+2w2:bounded} is $\onolog{\tau^2 n^{\frac{3}{2}}m^{\frac{1}{2}}}$.
\end{lemma}
\begin{proof}
    Observe that $\mathbb{E}\bracke{\abs{S}} = \onolog{nq}$. By \corref{tools:modified-es-properties}, it follows that, the expected total number of edges to ever be present in $E_{S}$ is  $\onolog{\tau \cdot \frac{n}{q}}$. It is worth mentioning that defining $E_{S}$ as the $\frac{1}{q}$ lightest edges\footnote{This was the approach in the static setting, by Cohen and Zwick \cite{CohZwi1997}.} of each vertex fails, as this would eventually lead to any edge being taken into consideration within $E_S$.

    We observe that both $\abs{F\paren{u}}$ and $\abs{{\hat F}\paren{u}}$ contain $E_S$ and additional $\onolog{n}$ edges. Therefore, the total update time of each Monotone ES-Trees depends on the total number of edges ever present in $E_S$. Each edge may have at most $\tau$ weight updates, as we assume the weights are integers and an edge is either an original edge or an auxiliary edge, in which case it will consider only a path of weight at most $\tau$. 

    By \lemref{prelim:monotone-es-update-time} we conclude that
    the total update time of each Monotone ES-Tree is $\onolog{\tau ^2 \frac{n}{q}}$ and for each ES-Tree is $\onolog{\tau m}$. As we maintain a Monotone ES-Tree from every $u\in V$ and an ES-Tree from every $s\in S$, these total update times become $\onolog{\tau^2 \frac{n^2}{q}}$ and $\onolog{\tau m\cdot nq }$, respectively.

    Additional runtimes include maintaining the set $E_{S}$ and the auxiliary pivots of each vertex, both require only $\onolog{m}$ time.

    Hence, our expected total update time is $\onolog{\tau^2\frac{n^2}{q} +\tau m \cdot n q}$. We set $q= \paren{\frac{n}{m}}^{\frac{1}{2}} $. The total update time becomes $\onolog{\tau^2 n^{\frac{3}{2}} m^{\frac{1}{2}}}$.
\end{proof}

By \lemref{w+2w2:bounded-approx} and \lemref{w+2w2:bounded-total-update-time}, we conclude:

\begin{conclusion} \label{cnc:w+2w2:bounded}

    For every pair of vertices $u,v \in V$ satisfying $\dist{u,v} \leq \tau$, \algref{w+2w2:bounded} maintains a  distance estimate $d\bracke{u,v}$ such that:
    $\dist{u,v} \leq d\bracke{u,v} \leq \dist{u,v} + 2W_2\paren{u \squiggly v}$. Moreover, the expected total update time of the algorithm is $\onolog{\tau^2 n^{\frac{3}{2}} m ^{\frac{1}{2}}}$. 
\end{conclusion}

As one can observe, the value of $\tau$ held an important role in the runtime analysis, as it is an upper bound on the number of distance updates. 

\begin{remark}\label{rmk:w+2w2:distance-updates}
    Let $SSSP\paren{G,s,D}$ be any decremental SSSP algorithm from a source  vertex $s$ up to a depth parameter $D$ in a graph $G$, in which the number of distance updates between the source vertex and each other vertex is $\alpha$, with a total update time of $T_{SSSP}\paren{n,m,D}$.
    If we employ $SSSP(G,s,D)$ as our algorithm to maintain distances from pivots, \algref{w+2w2:bounded} has a total update time of $\onolog{\tau \cdot \alpha \cdot n^{\frac{3}{2}}m^{\frac{1}{2}} + T_{SSSP}\paren{n,m,D}}$.
\end{remark}

\section{Tools for Weighted Graphs for Unbounded Depth}~\label{sec:tools}

In \secref{hierarchy} and \secref{w+2w2:bounded},
our algorithms and structural components operated under an explicit distance bound $\tau$.
In general weighted graphs,
distances can be as large as $\Omega(n W_{\max})$,
making depth-bounded Even-Shiloach trees rather expensive when applied directly.
To generalize our framework to arbitrary, unbounded distances,
we require algorithmic tools that eliminate the polynomial dependence on $W_{\max}$.

In this section,
we develop the core technical toolkit utilized by all of our unbounded decremental APASP algorithms:
\begin{itemize}
    \item \textbf{Edge Weight Rounding (\secref{tools:scaling}):}
    We review the geometric scaling and rounding technique of Bernstein~\cite{BernsteinRounding},
    which converts path weights in $G$ to hop-lengths.
    \item \textbf{Unbounded Hierarchy-Induced Edge Sets (\secref{tools:hierarchy}):}
    We construct an unbounded-depth counterpart $E_S$ of the hierarchy-induced edge sets,
    which serves as the core component of our unbounded-depth APASP algorithms.
    \item \textbf{Hopset Construction for Hierarchy-Induced Subgraphs (\secref{tools:hopset}):}
    We design a hopset tailored to our hierarchy-induced subgraphs,
    enabling to reduce the number of hops of each path present in $E_S$.
\end{itemize}
Together,
these tools form the algorithmic foundation
for the unbounded APASP algorithms presented in \secref{w+2w2:unbounded},
\secref{w+2w1+2w2},
and \secref{w+2kw1}.

\subsection{Edge Weight Rounding}\label{sec:tools:scaling}

A critical obstacle in designing decremental shortest path algorithms for weighted graphs
is the dependence of the update time on the distance estimates.
The Even-Shiloach tree and the Monotone Even-Shiloach tree maintain distances
from a fixed source up to a distance bound $\tau$ with total update time $O(\tau m)$.
In weighted graphs,
the maximal distance between two vertices may be as large as $\Omega(n W_{\max})$,
making maintenance of distances without bounding the depth rather expensive.

To overcome this bottleneck,
we employ the edge weight rounding technique introduced by Bernstein~\cite{BernsteinRounding}.
This method trades a $(1+\varepsilon)$ multiplicative approximation and $\bigo{\log\paren{n W_{\max}}}$
factor in the total update time for a substantial reduction in the required tree depth,
shifting the focus from weighted distances to the \emph{hop-length} of paths.
In our framework, this rounding technique serves two purposes:
In \secref{tools:hierarchy}, it is used in order to construct our unbounded hierarchy-induced edge sets $E_S$.
Additionally, in all of our unbounded decremental APASP algorithms
(\secref{w+2w2:unbounded} to \secref{w+2kw1}),
it enables the maintenance of Monotone ES-Trees on low-hop graphs,
ensuring that update times depend only on the graph hopbound
rather than $W_{\max}$ itself.

While the weighted distance between a pair of vertices is a fixed property of the graph,
the hop-length of a shortest path can be made small artificially
by adding shortcut edges (a hop set).
This reduction of hop-length is what makes the approach efficient for weighted graphs.
The use of shortcuts introduces new challenges: the shortcuts must be maintained throughout all edge deletions,
kept sparse, and shown not to distort distances. We address these in \secref{tools:hopset}.

We now describe the rounding technique of Bernstein~\cite{BernsteinRounding}.
For each $0 \le i \le \log(nW_{\max})$,
let $\alpha_i = \frac{\epsilon \cdot 2^i}{h}$ for a hop-length parameter $h$.
The scaled graph $G_i$ is constructed as follows:
\begin{enumerate}
    \item Include only edges of weight $\le 2^{i+1}$, discarding all heavier edges.
    \item Round all remaining edge weights up to the nearest integer multiple of
    $\alpha_i$.
    \item Divide all edge weights by $\alpha_i$.
\end{enumerate}
After this transformation, every edge in $G_i$ has a positive integer weight, and for any path $p$ of at most $h$ hops with $2^i \le \weight{p} \le 2^{i+1}$ in $G$,
the weight of $p$ in $G_i$ satisfies $\weight{p}{G_i} \le \ceil{\frac{4h}{\varepsilon}}$. This means an ES-tree on $G_i$ up to depth 
$\ceil{\frac{4h}{\varepsilon}}$ runs in $O(m \cdot \frac{h}{\varepsilon})$ time per scale, independent of $W_{\max}$.

Furthermore, Bernstein~\cite{BernsteinRounding} shows that the distance estimate obtained in $G_i$, when scaled back by $\alpha_i$,
gives a $(\onepluseps)$-approximation of the true distance in $G$ for every pair $u,v$ with $\dist{u,v} \in [2^i, 2^{i+1}]$.  
Running one instance per scale and taking the minimum over all scales gives a $(\onepluseps)$-approximate SSSP algorithm
with total update time $\onolog{m \cdot \log(nW_{\max})\cdot \frac{h}{\varepsilon}}$.

For completeness, the proof is provided in \appref{bernstein-rounding-process}.

\subsection{Unbounded Hierarchy-Induced Edge Sets}~\label{sec:tools:hierarchy}
The bounded hierarchy-induced edge set $E_S^\tau$ presented in \secref{hierarchy} is one of the core components of our bounded algorithms.
We now construct an unbounded-depth version denoted by $E_S$, enabling us to generalize our bounded algorithms to unbounded depth.

To remove the dependence on $\tau$, we replace the ES-Tree used to construct $E_S^{\tau}$ with a near-linear $(\onepluseps)$-SSSP algorithm.
Specifically, we maintain a $(\onepluseps)$-SSSP from a dummy source vertex connected to all vertices $s \in S$ via zero-weight edges.
While there are many variations of near-linear $(\onepluseps)$-SSSP~\cite{HKNLinSSSP,ChechikADO,LackiNazari22} that can replace the ES-Tree, in order to align with our APASP algorithms and offer a size-update time tradeoff we use the decremental hopset guarantee of \L\k{a}cki and Nazari~\cite{LackiNazari22} and maintain an SSSP algorithm on top of it.

\begin{theorem}[Theorem 1 of \cite{LackiNazari22}]
\label{thm:tools:ln-hopset}
Given an undirected graph $G = (V, E)$ with polynomial weights, subject to edge deletions, one can maintain a $(\beta, 1 + \varepsilon)$-hopset of size $\onolog{n^{1 + \frac{1}{2^k-1}}}$ in total expected update time $\onolog{\frac{\beta}{\varepsilon} \cdot m n^\rho}$, where
\[
    \beta \;=\; \paren{\bigo{\frac{\log n}{\varepsilon} \cdot \paren{k + \frac{1}{\rho}}}}^{k+\frac{1}{\rho}+1},
\]
for any integer $k \ge 1$, $0 < \varepsilon < 1$, and $\frac{2}{2^k-1} < \rho < 1$.
\end{theorem}

We use \thmref{tools:ln-hopset} on the input $G$ with parameter $2 \le k = O\paren{\sqrt{\log n}}$ and $\rho = \frac{1}{k}$.
Under this choice of parameters, the hopbound satisfies:
\[
    \beta \;=\; \paren{\bigo{\frac{\log n}{\varepsilon}\paren{k + \frac{1}{\rho}}}}^{k + \frac{1}{\rho} + 1}
    \;=\; \paren{\bigo{\frac{k \log n}{\varepsilon}}}^{2k+1}
\]

Using this hopset, we apply Bernstein's~\cite{BernsteinRounding} edge weight rounding algorithm and maintain $(\onepluseps)$-approximate single-source distances on $\ceil{\log(nW_{\max})}$ scaled graphs via Monotone ES-Trees up to depth $D = \ceil{\frac{2(2\beta+1)}{\varepsilon}} = \bigo{\frac{\beta}{\varepsilon}}$\footnote{This is the approach in \cite{LackiNazari22} to obtain $\onepluseps$ SSSP}.
Let $\tilde{\delta}(u, S)$ denote the approximate distance from $u$ to $S$ produced by this data structure, satisfying
\[
    \dist{u,S} \;\le\; \tilde{\delta}(u, S) \;\le\; (\onepluseps)\,\dist{u,S}.
\]
We use $\tilde{\delta}$ as our distance metric throughout this section in place of the exact distance $\dist$.

In each scaled graph, every edge has a strictly positive integer weight, and Monotone ES-Tree distance labels increase monotonically from $0$ up to at most $D$. Thus, for each scaled graph, the level of each vertex can increase at most $D$ times.
Across all $\ceil{\log(nW_{\max})}$ scales, the approximate distance $\tilde{\delta}(u, S)$ is taken as the minimum of the unscaled level estimates, so the maximum number of distance updates to $\tilde{\delta}(u, S)$ per vertex is bounded by $\bigo{D \log(nW_{\max})} = \onolog{\frac{\beta}{\varepsilon}}$.

The total update time required to maintain the decremental hopset of \cite{LackiNazari22} with $\rho = \frac{1}{k}$ and the Monotone ES-Trees from the dummy source is $\onolog{\frac{\beta}{\varepsilon}(m + n^{1+\frac{1}{2^k-1}})n^\rho} = \onolog{\frac{\beta}{\varepsilon} m n^{\frac{1}{k}}}$.

Since $\tilde{\dist}$ is an approximation of the true distance, the pivots induced by
$\tilde{\dist}$ are \emph{approximate pivots}: for each $u \in V$, the approximate pivot
$\piv[S]{u}$ is defined as $\arg\min_{s \in S}\tilde{\dist}(u, s)$, i.e., the vertex in $S$
closest to $u$ with respect to the approximate distance metric $\tilde{\dist}$.

We now redefine $E_S$ to use the approximate distance. Formally, let $E_S(u) \;:=\; \{(u,v) \in E \mid \weight{u,v} < \tilde{\dist}(u, S)\}$ and $E_S \;:=\; \bigcup_{u \in V} E_S(u)$.
Since the approximate pivots are \emph{closest with respect to $\tilde{\dist}$}, the
arguments underlying the proofs of \secref{hierarchy} carry over unchanged: the key property
used in the proofs is that $\piv[S]{u}$ is the nearest vertex of $S$ under the operative metric,
which holds here by definition of the approximate pivots. Consequently, all results of
\secref{hierarchy} remain valid when $\dist$ is replaced by $\tilde{\dist}$ throughout.

We summarize these properties in the following corollary.
\begin{corollary}
\label{cor:tools:modified-es-unbounded}
Let $\tilde{\dist}(u,S)$ be the approximate distance from $u$ to $S$ produced by \thmref{tools:ln-hopset} with parameters $2 \le k = \bigo{\sqrt{\log n}}$ and $\rho = \frac{1}{k}$,
and let $\piv[S]{u} := \arg\min_{s \in S}\tilde{\dist}(u, s)$ be the resulting approximate pivot.
Define
\[
    E_S(u) \;:=\; \{(u,v)\in E \mid \weight{u,v} < \tilde{\dist}(u,S)\},
    \qquad
    E_S \;:=\; \bigcup_{u \in V}E_S(u).
\] and let $\beta = \paren{\bigo{\frac{k \log n}{\varepsilon}}}^{2k+1}$.
Then:
\begin{itemize}
    \item[1.] The subgraph $\paren{V, E_S \cup E(\piv[S]{u})}$ preserves the \textbf{exact} distance between
    $u$ and the approximate pivot $\piv[S]{u}$ in $G$.

    \item[2.] For every edge $\paren{u,v} \notin E_S(u)$, $\weight{u,v} \ge \tilde{\dist}(u,S)$.

    \item[3.] Let $\mathcal{E}_S$ denote the expected total number of edges that ever exist in $E_S$ throughout
    all decremental updates. Then $\mathcal{E}_S = \onolog{\frac{\beta}{\varepsilon}\frac{n}{q}}$.

    \item[4.] There is an algorithm for maintaining $E_S$ in
    $\onolog{\frac{\beta}{\varepsilon}\paren{m n^{\frac{1}{k}} + \frac{n}{q}}}$ expected total update time.
\end{itemize}
\end{corollary}

\begin{remark}
\label{rem:tools:es-tradeoff-linear-vs-size}
For applications prioritizing nearly linear preprocessing and update time over the edge set size, the approximate distances $\tilde{\delta}(u, S)$ can alternatively be maintained in $\onolog{m^{1+\subpol}}$ total update time, using existing near-linear $\paren{\onepluseps}$-SSSP algorithms \cite{HKNLinSSSP,ChechikADO,LackiNazari22}. In those algorithms, the number of distance updates per vertex is $\beta = 2^{\bigo{\sqrt{\log n}}} = n^{\subpol}$, resulting in $\mathcal{E}_S = \onolog{\frac{n^{1+\subpol}}{q}}$. The total maintenance time would be $\onolog{m^{1+\subpol} + \frac{n^{1+\subpol}}{q}}$.
\end{remark}

For simplicity, we retain the notation $\piv[S]{u}$ and refer to approximate pivots simply as pivots.

\subsection{Hopset Construction for Hierarchy-Induced Subgraphs}\label{sec:tools:hopset}

\subsubsection{Overview}\label{sec:tools:hopset-overview}
In \secref{w+2w2:bounded}, \secref{w+2w1+2w2} and \secref{w+2kw1}, our APASP algorithms share a common structure: for each $s \in S_i$ a Monotone ES-Tree is maintained over a set of edges that includes $E_{i+1}^{\tau}$, up to a bounded depth $\tau$. Consequently, their total update time has a quadratic dependence on the depth $\tau$, which is at most $n \cdot W_{\max}$, making the algorithm highly inefficient.

To address this, we construct a hopset tailored to our hierarchy-induced subgraphs.
This construction differs from standard decremental hopsets (e.g., \cite{HKNLinSSSP, ChechikADO, LackiNazari22}) in two main aspects:

\begin{enumerate}
    \item  Existing constructions consider a single decremental graph. We consider a hopset for a series of subgraphs $G_{i} = \paren{V, E_{i}}$ that are not decremental: edges may be added to $E_{i}$ when the distance to a pivot  increases.
    \item We extend the hopset guarantee to cover not only shortest paths but rather \textbf{any arbitrary} path $P_{u,v} \subseteq E_{i}$, for $i\in\bracke{k}$. This    is essential for our techniques to obtain additive APASP bounds for weighted graph in the decremental setting.
\end{enumerate}

Furthermore, our APASP algorithms also require the construction of a hitting set hierarchy. Rather than maintaining an independent hierarchy for the hopset, we utilize the exact same structure $V = S_0 \supseteq S_1 \supseteq \ldots \supseteq S_k = \emptyset$ for both the APASP and the hopset frameworks. This guarantees that the  size of the hopset remains  bounded by $\bigo{\abs{E_{1}}}$. Consequently, the hopset remains sufficiently sparse to be efficiently processed by a Monotone ES-Tree originating from any vertex $u \in V$. To properly balance the size of the hopset against the total update time of the Monotone ES-Tree invocations, we set the hopbound to $\beta \approx \paren{k \log n}^{k}$.

Finally, we observe that establishing a hopbound $\beta$ alone does not suffice for weighted graphs. Because a single edge can have a weight up to $W_{\max}$, a path of $\beta$ hops may correspond to depth $\tau = \beta \cdot W_{\max}$. To overcome this we combine the hopset construction with Bernstein's scaling framework~\cite{BernsteinRounding} to bound the depth. Instead of a single Monotone ES-Tree invocation, we perform $\log\paren{nW_{\max}}$ invocations for different weight scales (See: \secref{tools:scaling}), where the maximal weight of an edge is bounded by a small value\footnote{More precisely, the weight of each edge in any given scaled graph is bounded by $O\paren{\frac{1}{\varepsilon}}$ and $\beta = O\paren{\paren{\frac{k\cdot \log n}{\varepsilon}}^k}$, for an arbitrary $\varepsilon>0$.}. This allows us to replace the dependency in $\beta\cdot W_{\max}$ with  $\beta \cdot \log\paren{nW_{\max}}$.

\subsubsection{Extending the \L\k{a}cki-Nazari Hopset for Arbitrary Paths }\label{sec:tools:hopset-original}
We review the hopset framework of \L\k{a}cki and Nazari~\cite{LackiNazari22} under a broader viewpoint. Specifically, while outlining their construction, we demonstrate that their proof is not restricted to shortest paths, but rather applies to any arbitrary path. 

Let $k \in \mathbb{N}$, $\rho \in \paren{\frac{2}{2^k-1}, \frac{1}{2}}$, and $\varepsilon, \gamma > 0$ be parameters balancing the hopset size $\abs{F}$, the hopbound $\beta$, and the total update time. We set $\tilde{k} = k + \ceil{\frac{1}{\rho}} + 1$ and construct the hierarchy $V = S_0 \supseteq S_1 \supseteq \ldots \supseteq S_{\tilde{k}-1} \supseteq S_{\tilde{k}} = \emptyset$, where each member of $S_{i-1}$ is independently sampled into $S_i$ with probability $q_i \in \paren{0,1}$, where $q_i = \mmax{n^{-\frac{2^i}{2^k-1}}, n^{-\rho}}$ and $i \in \bracke{\tilde{k}-1}$. 
We construct the hopset $F$ as the union of two \qoute{types} of shortcut edges added across all levels of the hierarchy:
\begin{equation*}
    F = \bracce{ \paren{u,v} \mid \exists i \in \bracke{\tilde{k}-1} \textnormal{ s.t. } u,v \in S_i  \wedge u \in \bunch{v}{i}} \cup \bracce{ \paren{u,v} \mid \exists i \in \bracke{\tilde{k}-1} \textnormal{ s.t. } u = \piv[i]{v}}
\end{equation*}

The first set introduces a shortcut edge for every pair $x,y \in S_{i}$ such that $x \in \bunch{y}{i}$ (equivalently, $y \in \cluster{x}{i}$). The second set ensures that for every vertex $u \in V$, a pivot edge $\paren{u, \piv[i]{u}}$ is maintained. To guarantee the stretch bound, the construction maintains the property that every inserted edge $\paren{x,y} \in F$ is assigned a weight representing an $\paren{1+\varepsilon}$-approximation of the current distance between its endpoints in the graph:  $\weight{x,y} \in \bracke{\dist{x,y}, \paren{\onepluseps} \cdot \dist{x,y}}$.

To maintain $F$ in the decremental setting, the algorithm maintains a Monotone ES-tree per vertex $w \in S_i$  at each hierarchy level $i \in \bracke{\tilde{k}-1}$. Any such tree is rooted at $w$ and 
takes into consideration only vertices within $\cluster{w}{i}$, in the following way: an edge $\paren{x,y}$ is relaxed in the tree of $w$ only if
$$\level{w}{x} + \weight{x,y} < \frac{\approxdist\bracke{x, S_{i+1}}}{1+\varepsilon},$$
where $\approxdist\bracke{x, S_{i+1}}$ is the approximate distance between  $x$ and its next  level pivot $S_{i+1}$. Enforcing this condition ensures the tree of $w$ extends to $y$ only as long as the distance between $y$ and $w$ is smaller than the distance between $y$ and $S_{i+1}$, which is exactly the condition required by the cluster $\cluster{w}{i}$. Consequently, a vertex $v$ joins $\cluster{w}{i}$ only when $\approxdist\bracke{v, S_{i+1}}$ increases.

To avoid an inefficient runtime, the trees are maintained up to a bounded depth $\tau = \bigo{\frac{\beta}{\varepsilon}}$, using the path doubling and scaling ideas introduced in \cite{LackiNazari22}. 
The algorithm therefore maintains $\ceil{\log\paren{nW_{\max}}}$ 
weight scales and maintains a sequence of separate hopsets   $H_1, \ldots, H_{\ceil{\log\paren{nW_{\max}}}}$. Each hopset $H_j$, for $j \in \bracke{\ceil{\log\paren{nW_{\max}}}}$, is a hopset whose edges belong to $F$ \textbf{and} restricted exclusively to the $j^{\textnormal{th}}$ weight  scale. This ensures that:

$$ \mcup{H_{j}}{j=1}{\ceil{\log\paren{nW_{\max}}}} = F $$

The stretch analysis relies on threefold  induction over the distance scale $j \in \bracke{\log\paren{nW_{\max}}}$, the hierarchy level $i \in \bracke{\tilde{k}}$ of both endpoints, and the time step $t > 0$. The induction nesting is as follows:

\begin{itemize}
    \item \textbf{Outer Induction (Scale $j$):} Assumes the restricted hopsets $H_{1}, \ldots, H_{j-1}$ are correctly maintained with a multiplicative stretch of $\paren{1+\varepsilon}$ up to distance $2^{j}$. We use the distances from these lower scales to bound the edge weights in $H_{j}$, extending the stretch guarantee to the $2^{j+1}$ scale.
    \item \textbf{Middle Induction (Level $i$):} Within the fixed scale $j$ we consider the level of the vertex within the hitting set hierarchy $S_{0} \supseteq S_{1} \supseteq \ldots \supseteq S_{\tilde{k}}$. By assuming hop bounded path for vertices in $S_{i}$, we construct paths for vertices in $S_{i+1}$.
    \item \textbf{Inner Induction (Time $t$):} Dynamic maintenance for a fixed scale $j$ and level $i+1$ over the time $t$. Because distances strictly increase in the decremental setting, the algorithm enforces monotonicity within the ES-tree. When an edge becomes ``stretched'' upon a deletion at time $t$, we bypass recalculation by substituting its valid $t-1$ distance estimate.
\end{itemize}

Originally, \L\k{a}cki and Nazari \cite{LackiNazari22} considered a shortest path $u \squiggly v$ and proved the existence of an alternative path $\paren{u \squiggly v}^{*} \subseteq E \cup \mcup{H_{r}}{r=1}{j}$  with $\hops{\paren{u \squiggly v}^{*}} \leq \beta$.  However, a finer inspection of their analysis reveals that their proof does not actually require the path to be shortest. 

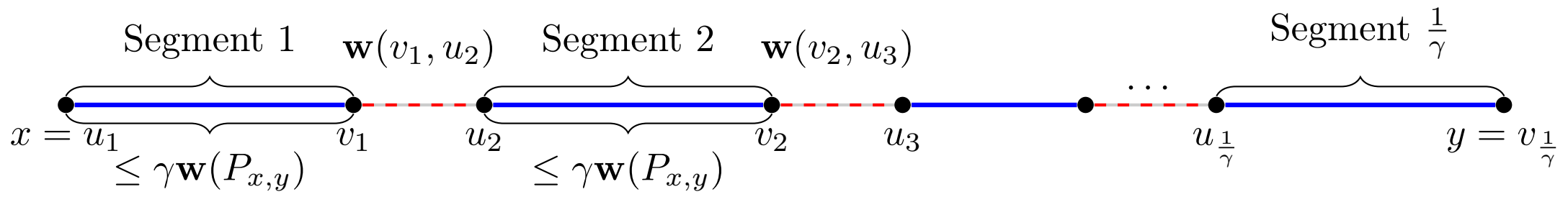
\begin{figure}[ht]
    \centering
    \begin{tikzpicture}[
        scale=1.2,
        every node/.style={font=\small},
        basepath/.style={thick, gray, opacity=0.4},
        segment/.style={very thick, blue},
        gapedge/.style={thick, red, dashed},
        dot/.style={circle, fill=black, inner sep=1.5pt}
    ]
    
    \draw[basepath] (0,0) -- (11,0);
    
    \node[dot, label=below:{$x = u_1$}] (x) at (0,0) {};
    \node[dot, label=below:{$v_1$}] (v1) at (2.2,0) {};
    
    \node[dot, label=below:{$u_2$}] (u2) at (3.2,0) {};
    \node[dot, label=below:{$v_2$}] (v2) at (5.4,0) {};
    
    \node[dot, label=below:{$u_3$}] (u3) at (6.4,0) {};
    \node[dot] (v3) at (7.8,0) {};
    
    \node[dot, label=below:{$u_{\frac{1}{\gamma}}$}] (uk) at (8.8,0) {};
    \node[dot, label=below:{$y = v_{\frac{1}{\gamma}}$}] (y) at (11,0) {};
    
    \draw[segment] (x) -- (v1);
    \draw[segment] (u2) -- (v2);
    \draw[segment] (u3) -- (v3);
    \draw[segment] (uk) -- (y);
    
    \draw[gapedge] (v1) -- (u2) node[midway, above, text=black, yshift=2mm] {$\weight{v_1,u_2}$};
    \draw[gapedge] (v2) -- (u3) node[midway, above, text=black, yshift=2mm] {$\weight{v_2,u_3}$};
    \draw[gapedge] (v3) -- (uk) node[midway, above, text=black] {$\dots$};
    
    \draw[decorate, decoration={brace, amplitude=5pt}] 
        (x.north) -- (v1.north) node[midway, above=6pt] {Segment 1};
    \draw[decorate, decoration={brace, amplitude=5pt, mirror}] 
        (x.south) -- (v1.south) node[midway, below=6pt] {$\leq \gamma \weight{P_{x,y}}$};
        
    \draw[decorate, decoration={brace, amplitude=5pt}] 
        (u2.north) -- (v2.north) node[midway, above=6pt] {Segment 2};
    \draw[decorate, decoration={brace, amplitude=5pt, mirror}] 
        (u2.south) -- (v2.south) node[midway, below=6pt] {$\leq \gamma \weight{P_{x,y}}$};
        
    \draw[decorate, decoration={brace, amplitude=5pt}] 
        (uk.north) -- (y.north) node[midway, above=6pt] {Segment $\frac{1}{\gamma}$};

    \end{tikzpicture}
    \caption{Segmentation of  $P_{x,y}$ into $\frac{1}{\gamma}$ bounded sub-paths. This decomposition isolates the bounded-hop approximations evaluated in the inductive step. $u_a$ is the first node on $P_{x,y}$ at distance $\ge (a-1)\gamma \weight{P_{x,y}}$ from $x$. $v_a$ is the furthest node on $P_{x,y}$ at distance $\le a\gamma \weight{P_{x,y}}$ from $x$. The dashed red lines represent exactly one edge of $P_{x,y}$ connecting the bounded segments.}
    \label{fig:ln-hopset:path-segmentation}
\end{figure}

We begin the analysis by fixing the distance scale $j$. Let $P_{u,v}$ be an arbitrary path between $u$ and $v$ with weight $\weight{P_{u,v}} \in \bracke{2^{j}, 2^{j+1}}$. Under the outer induction on $j$, we assume that the lower scale hopsets $H_{1}, \ldots, H_{j-1}$ are correctly maintained up to distance $2^{j}$. This provides the valid distance estimates required to bound path segments of lower scale using edges from these hopsets.

\begin{figure}[ht]
    \centering
    \begin{tikzpicture}[
        scale=1.1,
        every node/.style={font=\small},
        basepath/.style={very thick, gray, opacity=0.5},
        hopset/.style={very thick, blue, -{Stealth[length=3mm]}},
        subpath/.style={thick, dashed, red!80!black},
        dot/.style={circle, fill=black, inner sep=1.5pt}
    ]
    
    \draw[basepath] (0,0) -- (8,0);
    \node[dot, label=below:$u_l$] (ul) at (1,0) {};
    \node[dot, label=below:$v_l$] (vl) at (3,0) {};
    
    \node[dot, fill=blue, label=below right:{$z_l \in S_{i+1}$}] (zl) at (2.5, 1.5) {};
    \node[dot, fill=blue, label=below:{$z_r \in S_{i+1}$}] (zr) at (7, 1.5) {};
    
    \node[dot, fill=orange, label=above:{$p(z_l) \in S_{i+2}$}] (pzl) at (4.5, 3.5) {};
    
    \draw[subpath] (ul) to[out=60, in=180] node[midway, left] {$\leq \paren{\frac{3}{\gamma}}^{i-1}$ hops $\,\,$} (zl);
    
    \draw[hopset, orange] (zl) -- (pzl) node[midway, left, text=black] {Hopset edge $\paren{z_l, \piv[i+2]{z_l}}$};

    \end{tikzpicture}
    \caption{Alternative bridging utilizing the pivot $\piv[i+2]{z_l}$ in the subsequent hierarchy level $S_{i+2}$ when $z_r$ is absent from the bunch of $z_l$. If $z_r \notin B(z_l)$, we cannot add $(z_l, z_r)$. Instead, we bridge upward to $p(z_l)$.}
    \label{fig:ln-hopset:pivot-alternative}
\end{figure}
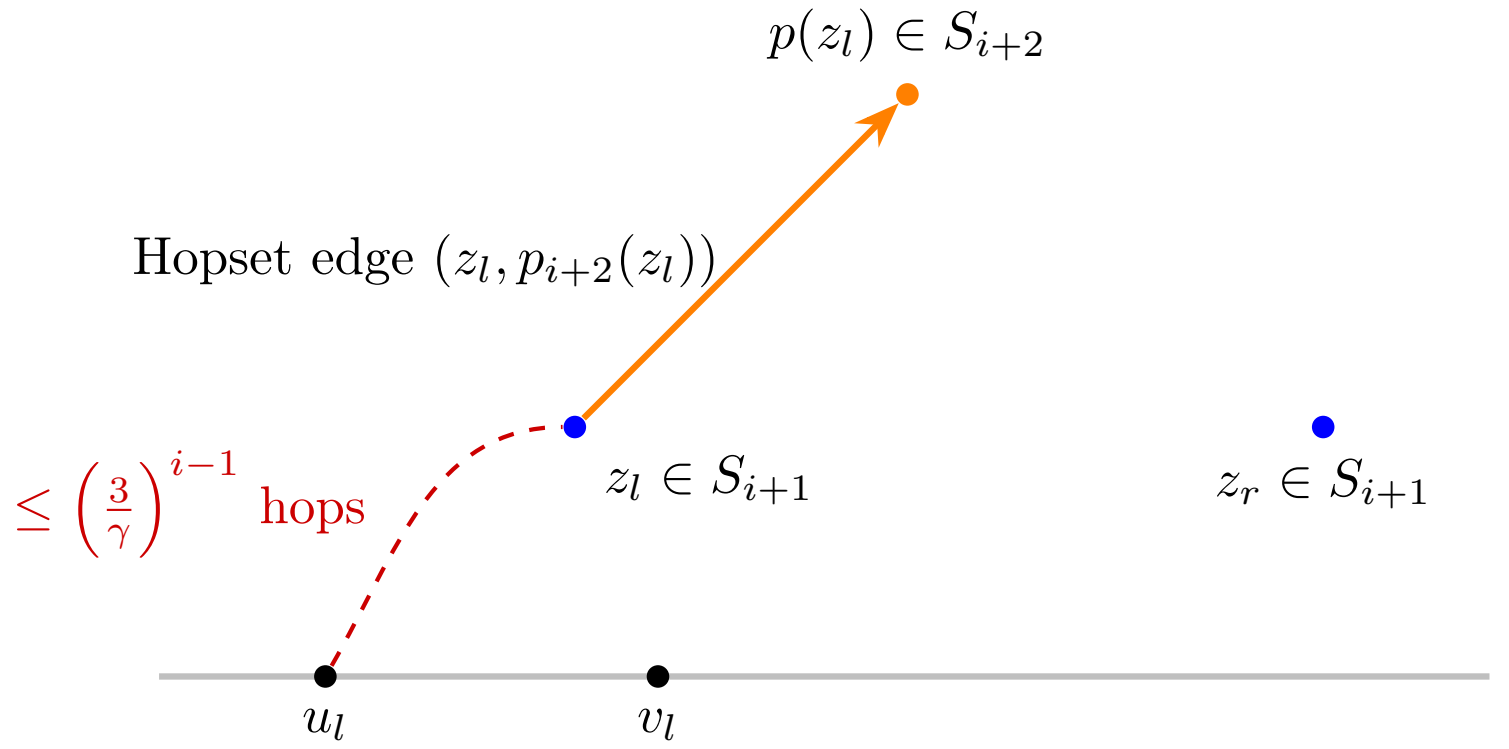

To construct the alternative path $P^{*}_{u,v} \subseteq E \cup \mcup{H_{r}}{r=1}{j}$, we partition $P_{u,v}$ into $\floor{\frac{1}{\gamma}}$ segments for $\gamma = \frac{1}{8\tilde{k}}$, ensuring each segment has weight at most $\gamma \cdot \weight{P_{u,v}}$ (See: \figref{ln-hopset:path-segmentation}). Within this fixed scale $j$, we apply the middle induction over the hierarchy level $i$. For each segment $\bracke{u_{a}, v_{a}}$, we assume inductively that it either admits an alternative path using lower scale shortcuts or can be bridged by the hitting set hierarchy. The inductive step on  $i$ proceeds as follows:

\begin{enumerate}
    \item If the starting endpoint $u_{a} \in S_{i}$ lies within the $i^{\textnormal{th}}$ level bunch of the terminating endpoint $v_{a}$, i.e., $u_{a} \in \bunch{v_{a}}{i}$, we bridge the segment using edges from $\mcup{H_{r}}{r=1}{j-1}$. This yields an alternative path of at most $\paren{\frac{3}{\gamma}}^{i-1} \cdot \frac{1}{\gamma} = \paren{\frac{3}{\gamma}}^{i}$ hops.
    
    \item  If $u_{a} \notin \bunch{v_{a}}{i}$, we locate vertices $z_{l}, z_{r} \in S_{i+1}$ such that the distances from the segment endpoints to these vertices are bounded.
    \begin{itemize}
        \item If $z_{l} \in \bunch{z_r}{i+1}$, a single shortcut edge $\paren{z_{l}, z_{r}} \in H_{j}$ connects them  (See: \figref{ln-hopset:bridging}).
        \item If $z_{l} \notin \bunch{z_{r}}{i+1}$, the distance from $z_{l}$ to  $\piv[i+2]{z_{l}} \in S_{i+2}$ is bounded by the distance between $z_{l}$ and $z_{r}$. We connect $z_{l}$ to $\piv[i+2]{z_{l}}$ using a single shortcut in $H_{j}$, and $\piv[i+2]{z_{l}}$ serves as the starting vertex for the next inductive step at the $\paren{i+2}^{\textnormal{th}}$ level (See: \figref{ln-hopset:pivot-alternative}). 
    \end{itemize}
\end{enumerate}

\begin{figure}[ht]
    \centering
    \begin{tikzpicture}[
        scale=1.1,
        every node/.style={font=\small},
        basepath/.style={very thick, gray, opacity=0.5},
        hopset/.style={very thick, blue, -{Stealth[length=3mm]}},
        subpath/.style={thick, dashed, red!80!black},
        successpath/.style={thick, dashed, green!60!black},
        dot/.style={circle, fill=black, inner sep=1.5pt}
    ]
    
    \draw[basepath] (0,0) -- (12,0);
    
    \node[dot, label=below:$x$] (x) at (0,0) {};
    \node[dot, label=below:$y$] (y) at (12,0) {};
    
    \node[dot, label=below:$u_l$] (ul) at (2.5,0) {};
    \node[dot, label=below:$v_l$] (vl) at (4,0) {};
    \node[dot, label=below:$u_r$] (ur) at (8,0) {};
    \node[dot, label=below:$v_r$] (vr) at (9.5,0) {};
    
    \draw[decorate, decoration={brace, amplitude=5pt, mirror}] 
        (ul.south) -- (vl.south) node[midway, below=6pt] {$\le \gamma \weight{P_{x,y}}$};
        
    \node[dot, fill=blue, label=above:{$z_l \in S_{i+1}$}] (zl) at (3.5, 2.5) {};
    \node[dot, fill=blue, label=above:{$z_r \in S_{i+1}$}] (zr) at (8.5, 2.5) {};
    
    \draw[successpath] (x) to[out=45, in=135] node[midway, above] {$\le \paren{\frac{3}{\gamma}}^{i-1}$ hops} (ul);
    \draw[successpath] (vr) to[out=45, in=135] node[midway, above] {$\le \paren{\frac{3}{\gamma}}^{i-1}$ hops} (y);
    
    \draw[subpath] (ul) to[out=70, in=180] node[midway, left] {$\le \paren{\frac{3}{\gamma}}^{i-1}$ hops $\,$} (zl);
    \draw[subpath] (zr) to[out=0, in=110] node[midway, right] {$\le \paren{\frac{3}{\gamma}}^{i-1}$ hops} (vr);
    
    \draw[hopset] (zl) -- (zr) node[midway, above, text=black] {\tiny Single Hopset Edge in $\bigcup_{j' < j}H_j$};
    
    \end{tikzpicture}
    \caption{Inductive construction of the bounded-hop path $P^*_{x,y}$ in $G \cup \bigcup_{j' < j}H_{j'}$. The original path $P_{x,y}$ is divided into segments. When the first segment $[u_l, v_l]$ and the last segment $[u_r, v_r]$ fail the stretch condition, they are bypassed using higher-level cluster centers $z_l, z_r \in S_{i+1}$ and a single hopset edge.}
    \label{fig:ln-hopset:bridging}
\end{figure}
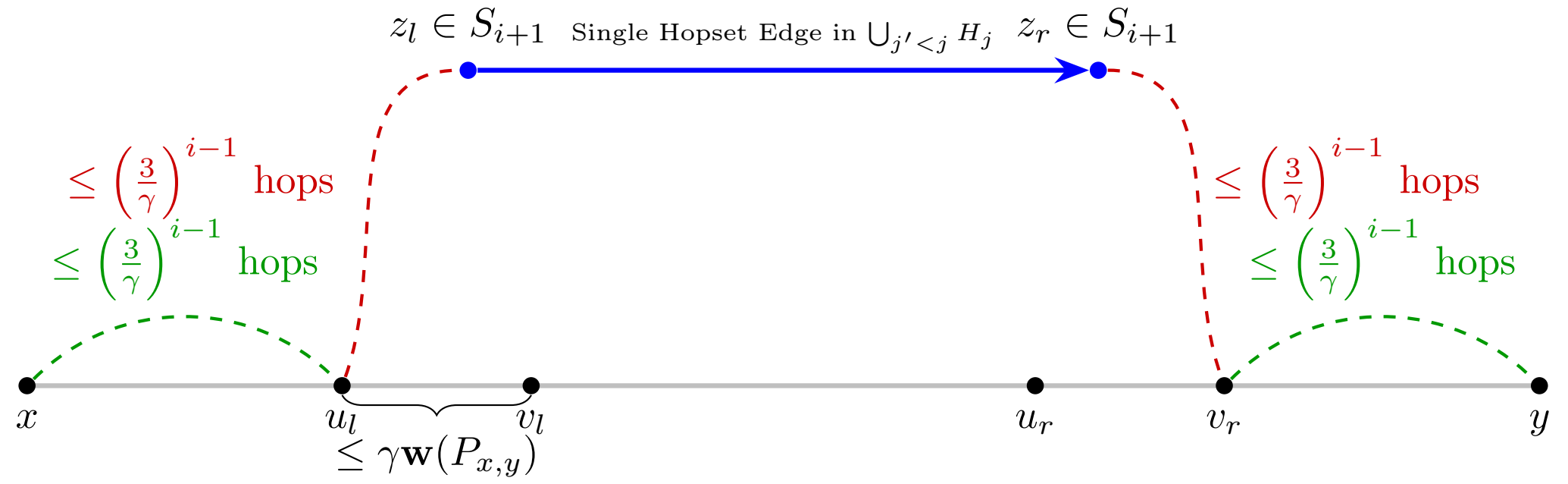

Eventually, the middle induction on $i$ halts since $z_{l} \in \bunch{z_{r}}{\tilde{k}}$ holds for every $z_{l}, z_{r} \in S_{\tilde{k}-1}$, ensuring the path is fully bridged.

Finally, for a fixed scale $j$ and hierarchy level $i$, the distance estimation is preserved by the inner induction over time $t$. Because distances are non-decreasing in the decremental setting, when an edge insertion causes an edge to become ``stretched'' in a Montone ES-Tree, the algorithm retains the distance estimate from time $t-1$, satisfying the required upper stretch bound by the inductive hypothesis. As for the lower bound, we observe that if an edge becomes ``stretched,'' the distance is actually smaller than it was before, therefore the same estimate still holds. Lastly, if a subsequent deletion increases the distance, the Montone ES-Tree recalculates and increases the estimate accordingly.

This bridging process bounds the stretch of $P^{*}_{u,v}$ purely as a function of the input weight $\weight{P_{u,v}}$. Consequently, while \L\k{a}cki and Nazari \cite{LackiNazari22} restrict their proof to shortest paths, our analysis demonstrates that the same structural bound holds for an arbitrary path $P_{u,v}$. We formalize this in the following theorem. Here, $\varepsilon_{j-1}$ denotes the stretch accumulated up to scale $j$, and $\tilde{\varepsilon}$ is the approximation error introduced by the rounding procedure specifically at the current $j^{\textnormal{th}}$ scale.

\begin{theorem}[\L\k{a}cki and Nazari \cite{LackiNazari22}, adapted for arbitrary paths]
\label{thm:tools:ln-hopbound}
Given a graph $G=\paren{V,E}$, assume we have correctly maintained a $\paren{2^{j},\beta, 1+\varepsilon_{j-1}}$-restricted hopset $\mcup{H_{r}}{r=1}{j-1}$. Let $H_{j}$ be the hopset obtained from $\tilde{G} = G \cup \mcup{H_{r}}{r=1}{j-1}$ with a given rounding parameter $0 \leq \tilde{\varepsilon} < 1$. Fix $0 < \gamma \leq \frac{1}{8\tilde{k}}$ and define the hopbound limit $\beta_{i} = \paren{\frac{3}{\gamma}}^{i}$. Consider any pair $x, y \in V$ connected by an arbitrary path $P_{x,y} \subseteq G$ such that $\weight{P_{x,y}} \in \bracke{2^{j},2^{j+1}}$. Then, for any level $0 \leq i \leq \tilde{k}$, at least one of the following conditions holds:
\begin{enumerate}
    \item $\hopdist{x,y}{\beta_{i}}{\tilde{G} \cup H_{j}} \leq \paren{1+8\gamma i}\paren{1+\varepsilon_{j-1}}\paren{1+\tilde{\varepsilon}}\cdot\weight{P_{x,y}}$, or
    \item There exists a center $z \in S_{i+1}$ such that $\hopdist{x,z}{\tilde{\beta}}{\tilde{G} \cup H_{j}} \leq 2\paren{1+\varepsilon_{j-1}}\paren{1+\tilde{\varepsilon}}\cdot\weight{P_{x,y}}$.
\end{enumerate}
Furthermore, by maintaining a monotone ES-tree on $G^{j}$ up to depth $\ceil{\frac{2\paren{2\beta+1}}{\tilde{\varepsilon}}}$, and applying the rounding procedure, we maintain a $\paren{1+\varepsilon_{j}}$-approximate single-source distances up to distance $2^{j+2}$ from a fixed source $s$ on $G$, where $1+\varepsilon_{j}=\paren{1+\varepsilon_{j-1}}\cdot\paren{1+\tilde{\varepsilon}}^{2}\cdot\paren{1+\varepsilon}$ and $\beta=\paren{\frac{3}{\gamma}}^{\tilde{k}}$.
\end{theorem}

As each vertex belongs to an expected $\bigo{\frac{1}{q_i}}$ clusters at level $i$, its incident edges are scanned $\bigo{\frac{1}{q_i}}$ times by vertices of $S_{i}$.  The total number of edges in $F$, $\mathcal{E}\paren{F}$ is bounded by \lemref{tools:ln-cluster-overlap-bound}:

\begin{lemma}[Lemma 25 of \cite{LackiNazari22}]
\label{lem:tools:ln-cluster-overlap-bound}
For every vertex $v \in V$ and hierarchy level $i\in \bracke{ \tilde{k}-1}$, the expected total number of times the edges incident on $v$ are scanned by the Monotone ES-Trees rooted at all centers $w \in S_{i}$ is bounded by $\bigo{\frac{\tau}{q_{i}}}$, where $q_{i}$ is the sampling probability and $\tau$ is the depth bound of the trees.
\end{lemma}

By summing over all time updates $t$, weight scales $j \in \bracke{\ceil{\log\paren{nW_{\max}}}}$ and levels
$i \in \bracke{\tilde{k}-1}$, the expected total update time becomes 
$\tilde{O}\paren{\beta \cdot m n^{\rho} \cdot \frac{1}{\varepsilon}}$.

\subsubsection{Integrating the Hopset for Hierarchy-Induced Subgraphs}\label{sec:tools:hopset-our}
We construct a single  hopset $F$ for the sequence of subgraphs $\bracce{G_i}_{i=1}^{k}$ utilized by our APASP algorithms in \secref{w+2w2:unbounded} through \secref{w+2kw1}. To maintain the same clusters, we must align the hopset with our algorithmic structure and consider the same hitting set hierarchy in both. We then set the sampling probability of the hierarchy to be  $q_i = q$ for all $i \in \bracke{k}$, where $S_0 = V$, $S_k = \emptyset$ and $\tilde{k} = k$. For an arbitrary $\varepsilon > 0$, the  definition of the hopset $F$ remains identical as in \secref{tools:hopset-original}.

While the definition of $F$ remains unchanged, it must satisfy a different condition:
\L\k{a}cki and Nazari~\cite{LackiNazari22} assumed the input graph is a single decremental graph. In contrast, we consider a series of subgraphs $\bracce{G_i}_{i=1}^{k-1}$, which are not decremental. When a pivot for a vertex changes, the distance to that pivot does not decrease. As a result, more edges may be added to a set $E_{i+1}$ for $i\in\bracke{k-1}$. 
To maintain the hopset within this framework, we restrict the Monotone ES-trees to operate exclusively over the hierarchy-induced edge 
sets $E_{i+1}$ rather than the global set $E$. We prove in \lemref{tools:es-preserves-clusters} that considering the clusters in 
$G_{i+1}\paren{u}=\paren{V,E_{i+1}\cup E\paren{u}}$ for $u\in S_{i}$ is  equivalent to considering the clusters in $G$.

In \thmref{tools:ln-hopbound} we established that  the construction of \L\k{a}cki and Nazari \cite{LackiNazari22} can be extended to arbitrary paths. That is, for any path $P_{u,v}$ there exists an alternative path $P^{*}_{u,v} \subseteq P_{u,v} \cup F$ with $\hops{P^*_{u,v}} \leq \beta$. We now show that this property holds despite restricting the Monotone ES-Trees to the subgraph $E_{i+1} \cup E\paren{u}$. Thus, for an arbitrary path $P_{u,v} \subseteq E_{i+1}\cup E\paren{u}$, there exists an alternative path $P^{*}_{u,v} \subseteq E_{i+1} \cup E\paren{u} \cup F$ with $\hops{P^*_{u,v}} \leq \beta$. This property is essential, as our APASP algorithms  in \secref{w+2w2:unbounded} through \secref{w+2kw1} achieve their additive stretch bounds by evaluating arbitrary paths rather than exact shortest paths.

Because  $P^{*}_{u,v}$ utilizes edges from $F$ that represent distances in the graph $G$, these edges are not necessarily present in $P_{u,v}$ or $E_{i+1}\cup E\paren{u}$. Consequently, the weight of $P^{*}_{u,v}$ might be  smaller than the weight of the evaluated path $P_{u,v}$. However, since every shortcut inherently approximates a path in $G$, up to a multiplicative factor of $1+\varepsilon$, the weight is bounded  below by  $\dist{u,v}{G}$. In \lemref{tools:hopset-hopbound-applies-for-ei}, we show that:

$$ \dist{u,v}{G} \leq \weight{P^*_{u,v}} \leq \paren{\onepluseps} \cdot \weight{P_{u,v}} $$

Although this relaxed lower bound deviates from the standard definition of a hopset, it suffices because our APASP algorithms approximate distances with respect to $G$. Therefore, a lower bound of $\dist{u,v}{G}$ suffices. Additionally, 
we preserve the  $\paren{\onepluseps} \cdot \weight{P_{u,v}}$ upper bound, 
within $E_{i+1}\cup E\paren{u} \cup F$, which is  sparser than $E \cup F$. Lastly, \lemref{tools:hopset-shortcut-set-size} bounds the size of $F$.

We now formally prove the correctness of our implementation. We first demonstrate that operating over the restricted edge set $G_{i+1}\paren{u}=\paren{V,E_{i+1}\cup E\paren{u}}$  preserves the required cluster formations.

\begin{lemma}
\label{lem:tools:es-preserves-clusters}
For any $u \in S_{i}$ and $v \in \cluster{u}{G}$, the subgraph $G_{i+1}\paren{u} = \paren{V, E_{i+1} \cup E\paren{u}}$ satisfies: 
\begin{enumerate}
    \item $\dist{u,v}{G} = \dist{u,v}{G_{i+1}\paren{u}}$. 
    \item The clusters of $u$ with respect to $G$ and $G_{i+1}\paren{u}$ are identical: $\cluster{u}{G_{i+1}\paren{u}} = \cluster{u}{G}$.
\end{enumerate}
\end{lemma}

\begin{proof}
We begin by establishing the first property. Let $v \in \cluster{u}{G}$. By the definition of a cluster, $\dist{u,v}{G} < \dist{u,S_{i+1}}{G} = \dist{u,\piv[i+1]{u}}{G}$. This inequality satisfies the condition of \lemref{tools:bk-es-properties} applied with  $S = S_{i+1}$ and the source vertex $u$. Because $E\paren{u}$ contains the edges incident to $u$ and $E_{i+1}$ contains the subsequently scanned edges, the subgraph $G_{i+1}\paren{u}$ fully preserves the shortest path to $v$, yielding $\dist{u,v}{G} = \dist{u,v}{G_{i+1}\paren{u}}$.

To establish the second property, we prove mutual inclusion. We first show that $\cluster{u}{G} \subseteq \cluster{u}{G_{i+1}\paren{u}}$. Consider a vertex $v \in \cluster{u}{G}$, which implies $\dist{u,v}{G} < \dist{v, S_{i+1}}{G}$. By the first property of this lemma, $\dist{u,v}{G_{i+1}\paren{u}} = \dist{u,v}{G}$. Since $G_{i+1}\paren{u} \subseteq G$, any path in $G_{i+1}\paren{u}$ is a valid path in $G$, ensuring that $\dist{v, S_{i+1}}{G_{i+1}\paren{u}} \geq \dist{v, S_{i+1}}{G}$. Combining these bounds yields:
\begin{equation*}
    \dist{u,v}{G_{i+1}\paren{u}} = \dist{u,v}{G} < \dist{v, S_{i+1}}{G} \leq \dist{v, S_{i+1}}{G_{i+1}\paren{u}}
\end{equation*}
Therefore, $v \in \cluster{u}{G_{i+1}\paren{u}}$. 

We now show that $\cluster{u}{G_{i+1}\paren{u}} \subseteq \cluster{u}{G}$. Let $v \in \cluster{u}{G_{i+1}\paren{u}}$, hence $\dist{u,v}{G_{i+1}\paren{u}} < \dist{v, S_{i+1}}{G_{i+1}\paren{u}}$. By 
\lemref{tools:bk-es-properties} applied for $S_{i+1}$ we get  that $\dist{v,\piv[i+1]{v}}{G_{i+1}\paren{u}} = \dist{v,\piv[i+1]{v}}{G} $, hence $\dist{v, S_{i+1}}{G_{i+1}\paren{u}} = \dist{v, S_{i+1}}{G}$.  Because $G_{i+1}\paren{u} \subseteq G$, it holds that $\dist{u,v}{G} \leq \dist{u,v}{G_{i+1}\paren{u}}$. We obtain:
\begin{equation*}
    \dist{u,v}{G} \leq \dist{u,v}{G_{i+1}\paren{u}} < \dist{v, S_{i+1}}{G_{i+1}\paren{u}} = \dist{v, S_{i+1}}{G}
\end{equation*}
Therefore, $v \in \cluster{u}{G}$, concluding the proof.

\end{proof}

The construction of \L\k{a}cki and Nazari~\cite{LackiNazari22} maintains $\paren{1+\varepsilon}$-approximate clusters, defined as $\cluster{u} = \bracce{ v \in V \mid \dist{u,v}{G} < \frac{\dist{v, S_{i+1}}{G}}{\onepluseps} }$, rather than exact clusters. Nonetheless, \lemref{tools:es-preserves-clusters} holds under approximate clusters as well. For clarity of presentation, we omit the $\paren{\onepluseps}$ factor from the clusters at each of the following steps.
We now establish  the  hop bound and stretch guarantee for arbitrary paths in $G_{i+1}$.

\begin{lemma}
    \label{lem:tools:hopset-hopbound-applies-for-ei}

Let $u,v \in V$ and let $P_{u,v}$ be a path between $u$ and $v$. For any $i \in \bracke{k-1}$ such that $P_{u,v} \subseteq E_{i+1}$:
$$ \dist{u,v}{G} \leq \hopdist{u,v}{\beta}{G_{i+1} \cup F} \leq \paren{\onepluseps} \cdot \weight{P_{u,v}} $$
\end{lemma}

\begin{proof}
Let $H \subseteq G$ be a decremental subgraph such that $P_{u,v} \in H$.
By \thmref{tools:ln-hopbound}, there exists a $P^{*}_{u,v} \subseteq P_{u,v} \cup F$ satisfying $\hops{P^{*}_{u,v}} \leq \beta$ and $\dist{u,v}{G} \leq \weight{P^{*}_{u,v}} \leq \paren{\onepluseps}\weight{P_{u,v}}$. As $ \hopdist{u,v}{\beta}{H \cup F} \leq \weight{P^{*}_{u,v}} $, it follows that:
$$ \dist{u,v}{G} \leq \hopdist{u,v}{\beta}{H \cup F} \leq \paren{\onepluseps}\weight{P_{u,v}} $$

Now, let $i \in \bracke{k-1}$ and assume $P_{u,v} \subseteq E_{i+1}$. By \lemref{tools:es-preserves-clusters}, $\cluster{x}{G_{i+1}\paren{x}} = \cluster{x}{G}$ for all $x \in S_i$. Because the construction of $F$ depends exclusively on these clusters, the weights assigned to the edges of $F$ are identical whether the clusters are in $G$ or $G_{i+1}\paren{x}$ for $x\in S_{i}$. . Thus, the guarantee of \thmref{tools:ln-hopbound} applies  to $G_{i+1}$ as well.
\end{proof}

This concludes the correctness proof for the hopset construction over the sequence of subgraphs $\bracce{G_{i+1}}_{i=0}^{k-1}$. We now bound the total update time required for the maintenance of the hopset and analyze its size.

\begin{lemma}
\label{lem:tools:hopset-bounded-cluster-overlap}
Let $i \in \bracke{0, k-2}$. Throughout all edge updates, every vertex $v \in V$ belongs to at most $\onolog{\frac{1}{q}}$ clusters rooted at vertices in $S_{i}$. Furthermore, there exists an algorithm that maintains these clusters up to a bounded depth $\tau$ for every root $w \in S_{i} $, such that, for every $v \in V$, the expected  number of times  edges incident to $v$ are scanned over all such clusters is at most  $\onolog{\frac{\tau}{q}}$.
\end{lemma}

\begin{proof}
The proof directly follows the structural analysis preceding \lemref{tools:ln-cluster-overlap-bound} (Lemma 25 of \cite{LackiNazari22}), substituting their sampling probability $q_{i}$ with our uniform parameter $q$.
\end{proof}

We proceed to bound the total number of edges ever included in the hopset $F$ across all updates.

\begin{lemma}
\label{lem:tools:hopset-shortcut-set-size}
Let $\mathcal{E}\paren{F}$ denote the total number of edges ever included in  $F$. Then: $$\expectation{\mathcal{E}\paren{F}} = \onolog{\paren{k-1} \frac{n}{q} + n^2 q^{2\paren{k-1}}} $$
\end{lemma}

\begin{proof}
Fix $i \in \bracke{k-2}$ and let $u,v\in S_{i}$ such that $v \in \bunch{u}{i}$. Equivalently,   $u \in \cluster{v}$. Consequently, the number of shortcut edges $\bracce{ \paren{u,v} \mid v \in S_{i}, v \in \bunch{u}{i} }$ is bounded by the total number of clusters containing $u$. By \lemref{tools:hopset-bounded-cluster-overlap}, the number of clusters that $u$ joins is bounded by $\onolog{\frac{1}{q}}$ per hierarchy level. Summing over all $n$ vertices and $k-1$ levels yields an  expected total of $\onolog{\paren{k-1}\frac{n}{q}}$ edges throughout all updates.

Next, consider the set of pivot edges $\bracce{ \paren{u,v} \mid \exists i \text{ such that } u = \piv[i]{v} }$. Note that $\piv[k-1]{u} \in \bunch{u}{k}$, therefore $u\in \cluster{\piv[i]{u}}$ for all $i \in \bracke{k-2}$. Hence, these edges are already countered for within the $\onolog{\paren{k-1}\frac{n}{q}}$ edges of vertices within clusters. 

Lastly, we consider $S_{k-1} \times S_{k-1}$, which can be seen as edges from members in $S_{k-1}$ to vertices in bunches from the $k^{\textnormal{th}}$ level.
As the distance to the $k^{\textnormal{th}}$ level is \qoute{$\infty$}, these are exactly $S_{k-1} \times S_{k-1}$.
The expected number of these edges is $\onolog{n^2 q^{2\cdot\paren{k-1}}}$.
Summing these expressions concludes the proof.
\end{proof}

The underlying hopset construction of \L\k{a}cki and Nazari~\cite{LackiNazari22} maintains clusters at each scale $j$ using $\tau$-bounded Monotone ES-Trees, where $\tau = \bigo{\frac{\beta}{\varepsilon}}$. We proceed to analyze the expected total update time required to maintain these hopsets.

\begin{lemma}
\label{lem:tools:hopset-total-update-time-analysis}
Let $\Delta$ denote the maximum number of changes in $\dist{u,S_{i}}$ for every $u \in V$. The expected total update time for maintaining $\tau$-bounded clusters from every root $w \in S_{i}$ is $ \onolog{\tau \paren{\Delta n q^{-\paren{i+2}} + n^2 q^{2k-3}}} $.
\end{lemma}

\begin{proof}
From each root $w \in S_{i}$, the algorithm maintains a $\tau$-bounded Monotone ES-Tree over the edge set $E_{i+1} \cup F$.
By \corref{tools:modified-es-properties}, the expected number of edges to ever be inserted  into $E_{i+1}$ is $\expectation{\mathcal{E}_{i+1}} = \onolog{\Delta n q^{-\paren{i+1}}}$. Furthermore, by \lemref{tools:hopset-shortcut-set-size}, the expected size of edges to ever be present within the hopset is $\expectation{\mathcal{E}\paren{F}} = \onolog{\paren{k-1} \frac{n}{q} + n^2 q^{2\cdot\paren{k-1}}}$.
Lastly, 
by \lemref{tools:hopset-bounded-cluster-overlap}, every vertex $v \in V$ belongs to at most $\onolog{\frac{1}{q}}$ clusters. The expected total update time across all such trees is therefore bounded by:
\begin{align*}
    \onolog{\sum_{v \in V} \frac{1}{q} \cdot \tau \cdot \tilde{\deg}\paren{v}} &= \onolog{\frac{\tau}{q} \sum_{v \in V} \tilde{\deg}\paren{v}} \\
    &= \onolog{\frac{\tau}{q} \paren{\expectation{\mathcal{E}_{i+1}} + \expectation{\mathcal{E}\paren{F}}}} \\
    &= \onolog{\tau \paren{\Delta n q^{-\paren{i+2}} + \frac{n}{q} + n^2 q^{2k-3}}}
\end{align*}

The term $\frac{n}{q}$ is  dominated by the rest of the terms, resulting in the desired bound.

\end{proof}

Combining the stretch bound, the expected hopset size, and the total update time yields the full guarantees of the dense hopset.

\begin{theorem}[Dense Hopset]
\label{thm:tools:hopset-on-es-dense}
Let $G$ be a weighted, undirected graph undergoing edge deletions and edge weight increases. For a parameter $2 \leq k = \bigo{\sqrt{\log n}}$, let the hitting set hierarchy be $S_{0}, \dots, S_{k} = \mathcal{R}\paren{n^{-\frac{1}{k}}, k}$, and let $0 < \varepsilon < 1$, and let $\beta = \bigo{\paren{\frac{k \log n}{\varepsilon}}^k}$. There exists an algorithm that maintains a set of shortcut edges $F$ such that:
\begin{enumerate}
    \item The expected total update time of the algorithm is bounded by:
    \[
        \onolog{k \cdot \frac{\beta}{\varepsilon} \cdot n^{2+\frac{1}{k}}}.
    \]
    \item For any pair of vertices $u,v \in V$ and any path $P_{u,v} \subseteq E_{i+1}$, 
    \[
        \dist{u,v}{G} \leq \hopdist{u,v}{\beta}{E_{i+1} \cup F} \leq \paren{\onepluseps} \cdot \weight{P_{u,v}},
    \]
    \item The expected total number of edges ever present in $F$ is bounded by $\onolog{k\cdot n^{1+\frac{1}{k}}}$. 
\end{enumerate}
\end{theorem}

\begin{proof}
We first establish the bound on the total update time (Property 1). As $q=n^{-\frac{1}{k}}$, it follows by \lemref{tools:hopset-total-update-time-analysis} that the total update time required to maintain the $\tau$-bounded Monotone ES-Trees from every root $w \in S_{i}$ is at most $\onolog{\tau \cdot\paren{\Delta n^{1+\frac{i+2}{k}} + n^{\frac{3}{k}}}} = \onolog{\tau\Delta n^{1+\frac{i+2}{k}}}$, where $\Delta = \onolog{\frac{\beta}{\varepsilon}}$ denotes the maximum number of changes in $\dist{u,S_{i}}$ for every $u \in V$ by \corref{tools:modified-es-unbounded}.

We recall that the cluster construction of \cite{LackiNazari22} fixes the depth parameter $\tau = \bigo{\frac{\beta}{\varepsilon}}$ for the $\tau$-bounded Monotone ES-Trees.
Summing across all hierarchy levels $i \in \bracke{0, k-2}$, the overall total update time for maintaining these Monotone ES-trees is at most:
\[
    \onolog{k \tau \Delta n^{1+\frac{k-2+2}{k}}} \;=\; \onolog{k \paren{\frac{\beta}{\varepsilon}}^2 \cdot n^{2}} \;=\; \onolog{k \frac{\beta}{\varepsilon} \cdot n^{2}}.
\]
Furthermore, the Monotone ES-trees maintained from all members of $S_{k-1}$ require an expected total update time of $\onolog{\beta m n q^{k-1}} \leq \onolog{\beta n^{2+\frac{1}{k}}}$.
Combining these contributions yields the total expected update time of $\onolog{k \cdot \frac{\beta}{\varepsilon} \cdot n^{2+\frac{1}{k}}}$.

Finally, the stretch guarantee (Property 2) follows directly from \lemref{tools:hopset-hopbound-applies-for-ei}, and the size bound (Property 3) follows from substituting $q=n^{-\frac{1}{k}}$ into \lemref{tools:hopset-shortcut-set-size}.
\end{proof}

\subsubsection{Sparsifying the Hopset}\label{sec:tools:hopset-sparse}
The APASP algorithms presented in \secref{w+2w2:unbounded} through \secref{w+2kw1} utilize one of two sampling probabilities: $q = n^{-\frac{1}{k}}$ or $q = \paren{\frac{m}{n}}^{-\frac{1}{k}}$. By \lemref{tools:hopset-shortcut-set-size}, the dense hopset construction guarantees that the expected total number of edges ever present in $F$ is bounded by $\onolog{k\frac{n}{q} + n^{2}q^{2k-2}}$.
When considering the sampling probability $q = n^{-\frac{1}{k}}$, this evaluates  to $\onolog{n^{1+\frac{1}{k}} + n^{\frac{2}{k}}}$. Because $n^{\frac{2}{k}} \in o\paren{n^{1+\frac{1}{k}}}$, the expected size of $F$ is  bounded by $\onolog{\frac{n}{q}} = \onolog{n^{1+\frac{1}{k}}}$. This, in turn, yields an efficient total update time in the APASP algorithms where $q=n^{-\frac{1}{k}}$. That is, we use \thmref{tools:hopset-on-es-dense} for \algref{+2w1+2w2:unbounded} in \secref{w+2w1+2w2} and \algref{w+2kw1-dense:unbounded} in \secref{w+2kw1}.

However, when considering the sampling probability $q=\paren{\frac{m}{n}}^{-\frac{1}{k}}$, the total update time evaluates to
$\onolog{n^{1-\frac{1}{k}} m^{\frac{1}{k}} + n^{4-\frac{2}{k}} m^{\frac{2}{k}-2}}$.
The term $n^{4-\frac{2}{k}} m^{\frac{2}{k}-2}$ arises from the Cartesian product $S_{k-1} \times S_{k-1}$.
For sparse\footnote{For this purpose, we refer to a graph as \textit{sparse} if $m\in \onolog{n^{\frac{3}{2}}}$. } graphs,
the update time requires $\onolog{\frac{n}{q}} = \onolog{n^{1-\frac{1}{k}}m^{\frac{1}{k}}}$ to be the dominant term.

To resolve this, we sparsify the final level of the hierarchy by constructing an auxiliary hierarchy extending beyond $S_{k-1}$. We distinguish two regimes based on graph density: moderately dense graphs where $m = n^{1+\Omega(1)}$, and extremely sparse graphs where $m = n^{1+o(1)}$.

Assume first that $m = n^{1+\Omega\paren{1}}$.
To reduce the size of the final set, we construct an auxiliary hierarchy by sampling $k'$ additional levels $S_k, S_{k+1}, \dots, S_{k+k'-1}$ beyond $S_{k-1}$, for a parameter $k'$ that is specified next.
The first $k$ levels $V = S_0 \supseteq S_1 \supseteq \dots \supseteq S_{k-1}$ are maintained exactly as before.
For each auxiliary level $i \in \{k, \dots, k+k'-1\}$, each vertex of $S_{i-1}$ is sampled into $S_i$ independently with probability $q' = \paren{\frac{m}{n}}^{-\frac{1}{k}} = q$.
In each auxiliary round $i \in \{k, \dots, k+k'-1\}$, each vertex $v \in V$ belongs to an expected $\onolog{\frac{1}{q'}}$ clusters by \lemref{tools:hopset-bounded-cluster-overlap}. Summing across all $n$ vertices, each auxiliary round adds
$\onolog{\frac{n}{q'}} = \onolog{\frac{n}{q}} = \onolog{n^{1-\frac{1}{k}} m^{\frac{1}{k}}}$ shortcut edges.

We now bound the size of the final hitting set $S_{k+k'-1}$. Across the entire hierarchy, $S_{k+k'-1}$ is obtained from $V = S_0$ through $k + k' - 1$ independent sampling steps, each with probability $q = \paren{\frac{m}{n}}^{-\frac{1}{k}}$. By linearity of expectation, the expected size of $S_{k+k'-1}$ is:
\[
    \expectation{\abs{S_{k+k'-1}}} \;=\; n \cdot q^{k+k'-1} \;=\; n \cdot \paren{\frac{m}{n}}^{-\frac{k+k'-1}{k}}.
\]
To ensure that $\expectation{\abs{S_{k+k'-1} \times S_{k+k'-1}}} = \onolog{\frac{n}{q}}$, we require:
\[
    n^2 \cdot \paren{\frac{m}{n}}^{-\frac{2(k+k'-1)}{k}} \;\le\; \frac{n}{q}
    \quad\iff\quad
    \paren{\frac{m}{n}}^{\frac{2k+2k'-1}{k}} \;\ge\; n
    \quad\iff\quad
    2k + 2k' - 1 \;\ge\; k \cdot \frac{\log n}{\log(m/n)}.
\]
We therefore set:
\[
    k' \;=\; \left\lceil \frac{k}{2} \cdot \frac{\log n}{\log(m/n)} - k + \frac{1}{2} \right\rceil \;=\; \Theta\paren{k \cdot \frac{\log n}{\log(m/n)}}.
\]
The Cartesian product $S_{k+k'-1} \times S_{k+k'-1}$ over the final level then has expected size of $\onolog{\frac{n}{q}}$.
The total number of shortcut edges ever added to $F$ across all $k + k' - 1$ levels is therefore bounded by:
\[
    \expectation{\mathcal{E}(F)} \;=\; (k + k' - 1) \cdot \onolog{\frac{n}{q}} + \expectation{\abs{S_{k+k'-1} \times S_{k+k'-1}}} \;=\; \bigo{k' \cdot n^{1-\frac{1}{k}} m^{\frac{1}{k}}} + \onolog{\frac{n}{q}} \;=\; \onolog{k n^{1-\frac{1}{k}} m^{\frac{1}{k}}}.
\]
Furthermore, the hopbound across the $k + k'$ levels satisfies:
\[
    \beta \;=\; \bigo{\paren{\frac{(k+k') \log n}{\varepsilon}}^{k+k'}} \;=\; \bigo{\paren{\frac{\frac{k \log^2 n}{\log(m/n)}}{\varepsilon}}^{\frac{k \log n}{\log(m/n)}}} \;=\; \bigo{\paren{\frac{k \log n}{\varepsilon}}^{\bigo{k}}}
\]
where the last equality follows from the assumption that $m = n^{1+\Omega(1)}$, and hence $\frac{\log n}{\log(m/n)} = \bigo{1}$.
The total update time required to maintain the clusters across all $k+k'$ levels up to depth $\tau = \bigo{\frac{\beta}{\varepsilon}}$ is:
\[
    \onolog{\frac{\beta}{\varepsilon} \paren{n q^{-(k+1)} + \frac{m}{q}}} \;=\; \onolog{\frac{\beta}{\varepsilon} n^{2-\frac{1}{k}} m^{\frac{1}{k}} + m^{1+\frac{1}{k}} n^{-\frac{1}{k}}} \;=\; \onolog{\frac{\beta}{\varepsilon} n^{2-\frac{1}{k}} m^{\frac{1}{k}}}.
\]

We now address the regime of very sparse graphs where $m = n^{1+o(1)}$.
Note that on such graphs $\log(m/n) = \subpol$ and hence $\beta$ becomes rather large. 
To maintain efficiency on such graphs with a sub-polynomial hopbound, we choose a slightly different strategy:
again, the first $k-1$ hierarchy levels are maintained exactly the same.
We then maintain an auxiliary hierarchy following the framework of \L\k{a}cki and Nazari~\cite{LackiNazari22}. For parameters $k' \ge 1,\rho < 1$ which we specifiy later, for each $k \le i < \tilde k = k + k' + \ceil{\frac{1}{\rho}} + 1$, we sample each vertex from $S_{i-1}$ into $S_i$ with probability:
\[
    q_i \;=\; \mmax{n^{-\frac{2^i}{2^{k+k'}-1}}, n^{-\rho}}.
\]
Setting $k' = \Theta\paren{\sqrt{\log n} - k}$ and $\rho = \Theta\paren{\frac{1}{2^{k+k'} - 1}}$ yields $n^\rho = n^{\subpol}$ and a subpolynomial hopbound $\beta = 2^{\bigo{\sqrt{\log n}}}$ for any constant $0 < \varepsilon < 1$.
The expected total number of edges ever present in $F$ is then $\onolog{\tilde k\paren{n^{1-\frac{1}{k}} m^{\frac{1}{k}} + n^{1+\subpol}}}$, and maintaining these auxiliary levels requires an expected total update time of $\onolog{m n^{\subpol}}$.

We summarize the sparse hopset guarantees in the following theorem:

\begin{theorem}[Sparse Hopset]
\label{thm:tools:hopset-on-es-sparse}
Let $G$ be a weighted, undirected graph with $m = n^{1+\Omega(1)}$ undergoing edge deletions and edge weight increases. For a parameter $2 \leq k = \bigo{\sqrt{\log n}}$, let the hitting set hierarchy be $S_{0}, \dots, S_{k-1}$ sampled with probability $q = \paren{\frac{m}{n}}^{-\frac{1}{k}}$, let $0 < \varepsilon < 1$, and let $\beta = \bigo{\paren{\frac{k \log n}{\varepsilon}}^{\bigo{k}}}$. There exists an algorithm that maintains a set of shortcut edges $F$ such that:
\begin{enumerate}
    \item The expected total update time of the algorithm is bounded by:
    \[
        \onolog{\frac{\beta}{\varepsilon} \cdot n^{2-\frac{1}{k}} m^{\frac{1}{k}}}.
    \]
    \item For any pair of vertices $u,v \in V$ and any path $P_{u,v} \subseteq E_{i+1}$,
    \[
        \dist{u,v}{G} \leq \hopdist{u,v}{\beta}{E_{i+1} \cup F} \leq \paren{\onepluseps} \cdot \weight{P_{u,v}}.
    \]
    \item The expected total number of edges ever present in $F$ is bounded by $\onolog{k \cdot n^{1-\frac{1}{k}} m^{\frac{1}{k}}}$.
\end{enumerate}
For graphs with $m = n^{1+o(1)}$, it is possible to maintain $F$ with size $\onolog{n^{1-\frac{1}{k}} m^{\frac{1}{k}} + n^{1+\subpol}}$ in expected total update time $\onolog{n^{2-\frac{1}{k}+\subpol} m^{\frac{1}{k}} + m n^{\subpol}}$ and $\beta = 2^{\bigo{\sqrt{\log n}}}$.
\end{theorem}

Consequently, we invoke the sparse hopset guarantees of \thmref{tools:hopset-on-es-sparse} for \algref{w+2w2:unbounded} in \secref{w+2w2:unbounded} and \algref{w+2kw1:unbounded} in \secref{w+2kw1-sparse:unbounded}.

\section{$\paren{1+\varepsilon,2W_{2}}$-APASP for Unbounded Depth}\label{sec:w+2w2:unbounded}
We are now ready to prove \thmref{+2W2}:
\thmsecedge*

\begin{algorithm}[t]
\caption{$\paren{\onepluseps,2W_{2}}$-APASP$\paren{G=\paren{V,E,\weight}}$}
\label{alg:w+2w2:unbounded}

\BlankLine

 \textbf{Input:} A weighted undirected graph $G = \paren{V,E,\weight}$.

\textbf{Output:} A $\paren{\onepluseps,2W_{2}}$-APASP for any pair of vertices $u,v \in V$.

\hrulefill

Initialize $d$ as the edges' weights and $\infty$ otherwise

Let $q\in\paren{0,1}$ to be fixed later

$S \gets \textit{sample}(V,q)$

Maintain the set $E_S$ and pivot distances using \corref{tools:modified-es-unbounded} with parameter $k=2$, and update distances to $\auxpiv{u}$

Construct a $(\beta,\varepsilon)$-hopset $H_S = E_S \cup F$ using \thmref{tools:hopset-on-es-sparse} with parameter $k=2$.

\For{ $s \in S$}
        {
        Maintain a $(\onepluseps)$-SSSP algorithm from $s$ on $G$ and update $d$ accordingly.
    }

Let $\tilde S$ be the set of auxiliary pivots $\auxpiv{u}$ for all $u\in V$

$Q_{\tilde S} \gets \bracce{\paren{u,\auxpiv{u}}}$

$R \gets \paren{V,H_S \cup Q_{\tilde{S}}}$
$\{R^j\}_{j=1}^{\ceil{\log nW_{\max}} + 1} \gets \textit{SCALE}\paren{R,\onepluseps,\beta + 2}$

\For{$u \in V$}{
    $T(u) \gets (\singleton{u} \times \tilde{S}) \cup E(u)$.
    
    $\hat T(u) \gets (\singleton{u} \times \tilde{S}) \cup (\singleton{u} \times V)$.
    
    $\{T^j(u)\}_{j=1}^{\ceil{\log nW_{\max}} + 1} \gets \textit{SCALE}\paren{T(u),\onepluseps,\beta + 2}$.
    
    $\{\hat T^j(u)\}_{j=1}^{\ceil{\log nW_{\max}} + 1} \gets \textit{SCALE}\paren{\hat T(u),\onepluseps,\beta + 2}$.

    \For{$i = 1, \dots, \ceil{\log nW_{\max}} + 1$}
    {
        $R^i(u)\gets R^i \cup T^i(u)$.
        
        $\hat R^i(u)\gets R^i \cup \hat T^i(u)$.
        
        Maintain a Monotone ES-Tree from $u$ up to depth $\ceil{\frac{4\paren{\beta + 2}}{\varepsilon}}$ in $R^i(u)$ with level estimate $\level[i]{u}{\cdot}$.

        Maintain a Monotone ES-Tree from $u$ up to depth $\ceil{\frac{4\paren{\beta + 2}}{\varepsilon}}$ in $\hat R^i(u)$ with level estimate $\level[i]{u}{\cdot}$.
        
    }
}

\end{algorithm}

\paragraph{Algorithm Description.}
\algref{w+2w2:unbounded} extends the $\tau$-depth restricted \algref{w+2w2:bounded} from \secref{w+2w2:bounded} to unbounded depth. In \algref{w+2w2:bounded}, the Monotone ES-trees explore paths up to a total depth $\tau$, resulting in an expected total update time of $\onolog{\tau^2 n^{\frac{3}{2}} m^{\frac{1}{2}}}$, which is highly inefficient when $\tau$ is large. To overcome this, we integrate the three tools introduced in \secref{tools}, each taking care of a different aspect of the $\tau$-dependence.

\textbf{Step 1: Replacing $E_{S}^\tau$ with $E_{S}$.}
The bounded hierarchy-induced edge set $E_{S}^\tau$ maintains edges of weight at most $\tau$, which limits \algref{w+2w2:bounded} to paths whose weight is at most $\tau$. \algref{w+2w2:unbounded} replaces $E_{S}^\tau$ with the unbounded hierarchy-induced edge set $E_{S}$, maintained as in \corref{tools:modified-es-unbounded} with parameter $k=2$. 
By \corref{tools:modified-es-unbounded}, the expected total number of edges ever inserted into $E_{S}$ is bounded by $\mathcal{E}_S = \onolog{\frac{\beta}{\varepsilon}\frac{n}{q}}$, regardless of the depth $\tau$. For a constant $\varepsilon > 0$ and $k=2$, $\beta = \bigo{(\log n)^{\bigo{1}}} = \onolog{1}$, so $\mathcal{E}_S = \onolog{\frac{n}{q}}$.

\textbf{Step 2: Bounding the hop-length with our hopset.}
Even with an unbounded $E_{S}$, the Monotone ES-trees must explore paths of arbitrary depth. To overcome this, \algref{w+2w2:unbounded} constructs a $\paren{\beta, \varepsilon}$-hopset $F$ by applying \thmref{tools:hopset-on-es-sparse} with parameter $k=2$, where $\beta = \bigo{\paren{\frac{k \log n}{\varepsilon}}^{\bigo{k}}} = \onolog{1}$ for constant $\varepsilon$. Set 
$H_{S} = E_{S} \cup F$. 
By \thmref{tools:hopset-on-es-sparse}, for any path $P_{u,v} \subseteq E_{S}$, there exists an alternative path $P_{u,v}^{\prime} \subseteq H_{S}$ satisfying $\hops{P_{u,v}^{\prime}} \leq \beta$ and $\dist{u,v}{G} \leq \weight{P_{u,v}^{\prime}} \leq \paren{\onepluseps} \cdot \weight{P_{u,v}}$.

In \lemref{w+2w2:bounded-approx} of \algref{w+2w2:bounded}, every evaluated path consists of a sub-path $P_{y,v} \subseteq E_{S}$ and at most two auxiliary edges linking $u$ to some $y\in P_{u,v}$. Integrating with \thmref{tools:hopset-on-es-sparse} guarantees that every relevant path can be traversed within $\beta + 2$ hops in $H_{S}$, regardless of the depth.

\textbf{Step 3: Scaling to bound the tree depth.}
A hopbound alone does not bound the total weight of a path, as individual edges can be \qoute{heavy}. \algref{w+2w2:unbounded} applies the edge rounding procedure of Bernstein \cite{BernsteinRounding} with scale factor $\alpha = \frac{2\varepsilon}{\beta + 2}$ to 3 graphs: $R = \paren{V, H_{S} \cup Q_{\tilde{S}}}$, $T(u) = (\singleton{u} \times \tilde{S}) \cup E(u)$
and $\hat T(u) = (\singleton{u} \times \tilde{S}) \cup (\singleton{u} \times V)$. The scaling procedure produces $\ceil{\log nW_{\max}} + 1$ scaled copies of each graph which we denote by $\{R^j\}_{j=1}^{\ceil{\log nW_{\max}} + 1},\{T^j(u)\}_{j=1}^{\ceil{\log nW_{\max}} + 1},\{\hat T^j(u)\}_{j=1}^{\ceil{\log nW_{\max}} + 1}$. For each scale $i$, every path of weight at most $2^{i+1}$ and at most $\beta + 2$ hops has a scaled weight of at most $\ceil{\frac{4\paren{\beta+2}}{\varepsilon}}$. Therefore, the Monotone ES-trees can be bounded to a maximum depth of $\ceil{\frac{4\paren{\beta+2}}{\varepsilon}}$, removing the dependence on $\tau,n$ or $W_{\max}$.

\textbf{Two tree invocation.}
For each scale $i$ and each vertex $u \in V$, \algref{w+2w2:unbounded} maintains two Monotone ES-trees up to depth $\ceil{\frac{4\paren{\beta+2}}{\varepsilon}}$, repeating the two tree invocation of \algref{w+2w2:bounded}:
\begin{itemize}
    \item The tree on $R^{i}\paren{u} = R^{i} \cup T^i(u)$ produces the estimates $\level[i]{u}{\cdot}$.
    \item The tree on ${\hat R}^{i}\paren{u} = R^{i} \cup \hat T^i(u)$ produces the estimates $\levelh[i]{u}{\cdot}$.
\end{itemize}
The final distance estimate is given by $d\bracke{u,v} = \mmin{\levelh[i]{u}{v},\levelh[i]{v}{u}}{i=1}{\ceil{\log nW_{\max}} + 1}$.

We briefly pause the presentation of \algref{w+2w2:unbounded} to remark that all of our weighted algorithms follow the same structure: first, we present a bounded depth version and then extend it to unbounded depth by using these exact 3 steps, adjusted per algorithm.

We now return to \algref{w+2w2:unbounded}. We begin with the stretch analysis. As the same case analysis presented in 
\lemref{w+2w2:bounded-approx} applies here, we mainly highlight the modifications introduced by the use of the unbounded hierarchy-induced edge sets, hopset and scaling.

\begin{lemma}[Approximation Analysis]
    \label{lem:w+2w2:unbounded-approx}
    Fix $u,v \in V$ and let $P_{u,v}$ be any simple path between $u$ and $v$. Then $\dist{u,v} \le \mmin{\levelh[i]{u}{v},\levelh[i]{v}{u}}{i=1}{\ceil{\log nW_{\max}} + 1} \le (\onepluseps)\weight{P_{u,v}} + 2W_2(P_{u,v})$
\end{lemma}
\begin{proof}
    We begin with time $t=0$.
    We consider the same three cases of \lemref{w+2w2:bounded-approx}:

    \ccase{1} $\sizeindex{P_{u,v}}{E_S} = 0$.
    By \thmref{tools:hopset-on-es-sparse}, there exists a path $P_{u,v}' \subseteq H_S$ with $\hops{P_{u,v}'} \le \beta$ and $\dist{u,v} \le \weight{P_{u,v}'} \le (\onepluseps)\weight{P_{u,v}}$. Let $j$ such that $2^j \le \weight{P_{u,v}'} \le 2^{j+1}$. The scaling procedure places $P_{u,v}'$ in the $j$-th scaled graph $R^j$, where it has scaled weight $\weight{P_{u,v}'}{R^j} \le \ceil{\frac{4\beta}{\varepsilon}}$. Unscaling by the factor $\alpha_j = \frac{\varepsilon 2^j}{\beta+2}$ gives $\dist{u,v} \le \alpha_j \cdot \weight{P_{u,v}'}{R^j} \le (\onepluseps)\weight{P_{u,v}'} \le (\onepluseps)^2\weight{P_{u,v}}{G_S}$.
    The rest follows the proof of \lemref{w+2w2:bounded-approx}.

    \ccase{2} $\sizeindex{P_{u,v}}{ E_{S}} = 1 $.
    Let $\paren{x,y}=\rightabovearrow{P_{u,v}}{E_{S}}=\leftabovearrow{P_{u,v}}{E_{S}}$ and write $P_{u,v} = P_{u,y} \circ P_{y,v}$. Since $P_{u,y} \subseteq E_S$, by \thmref{tools:hopset-on-es-sparse} there exists $P_{u,y}' \subseteq H_S$ with $\hops{P_{u,y}'} \le \beta$ and $\weight{P_{u,y}'} \le (\onepluseps)\weight{P_{u,y}}$. Let $j_1$ such that $2^{j_1} \le \weight{P_{u,y}'} \le 2^{j_1+1}$. The Monotone ES-Tree rooted at $y$ on $R^{j_1}(y)$ satisfies $\level[j_1]{y}{u} \le (\onepluseps)\weight{P_{u,y}}$. Then, for every $j$ the graph $\hat R^j\paren{u}$ contains the edge $\paren{u,y}$ with $\weight{u,y} \le \level[j_1]{y}{u} \le (\onepluseps)\weight{P_{u,y}}$.
    Applying \thmref{tools:hopset-on-es-sparse} to $P_{y,v} \subseteq E_S$ likewise gives $P_{y,v}' \subseteq H_S$ with $\hops{P_{y,v}'} \le \beta$ and $\weight{P_{y,v}'} \le (\onepluseps)\weight{P_{y,v}}$. Let $P^*_{u,v} = \paren{u,y} \circ P_{y,v}'$. The path $P^*_{u,v}$ satisfies $\hops{P^*_{u,v}} \le \beta + 1$ and $\weight{P^*_{u,v}} \le (\onepluseps)\weight{P_{u,y}} + (\onepluseps)\weight{P_{y,v}} \le (\onepluseps)\weight{P_{u,v}}$.Let $j_2$ such that $2^{j_2} \le \weight{P^*_{u,v}} \le 2^{j_2 + 1}$. $\hat R^{j_2}\paren{u}$ then contains the path $\weight{P^*_{u,v}}$ with weight at most $\ceil{\frac{4(\beta+1)}{\varepsilon}}$ and hence it is explored by $u$'s Monotone ES-Tree on $\hat R^{j_2}\paren{u}$.
    Finally, unscaling yields a path of weight at most $(\onepluseps)^2\weight{P_{u,v}}$.

    \ccase{3} $\sizeindex{P_{u,v}}{E_{S}} > 1 $.
    Let $\paren{x_{1},y_{1}}=\rightabovearrow{P_{u,v}}{E_{S}} \neq \leftabovearrow{P_{u,v}}{E_{S}}=\paren{x_{2},y_{2}}$.
    As in \lemref{w+2w2:bounded-approx}, the path is routed through two auxiliary edges via $\auxpiv{y_2}$ followed by $P_{y_2,v} \subseteq E_S$. By \thmref{tools:hopset-on-es-sparse}, there exists $P_{y_2,v}' \subseteq H_S$ with $\hops{P_{y_2,v}'} \le \beta$ and $\weight{P_{y_2,v}'} \le (\onepluseps)\weight{P_{y_2,v}}$. Let $P_{u,v}''$ be the concatenation of $P_{u,v}'$ with the two auxiliary edges. The path $P_{u,v}''$ has at most $\beta + 2$ hops and satisfies $\weight{P_{u,v}''} \le (\onepluseps)\weight{P_{u,v}} + 2(\onepluseps)W_2(P_{u,v})$, where the $(1+\varepsilon)$ factor on the additive term reflects the use of a $(\onepluseps)$-SSSP algorithm for pivot distances in place of exact ES-Trees. Again, let $j$ be such that $2^j \le \weight{P_{u,v}''} \le 2^{j+1}$. Then, there is an alternative path $P_{u,v}^* \subseteq R^J\paren{u}$ with $\weight_{R^j}(P_{u,v}^*) \le \ceil{\frac{4(\beta+2)}{\varepsilon}}$ and hence it is explored by the Monotone ES-Tree.
    Scaling then yields $\dist{u,v} \le \alpha_j \cdot \weight_{R^j}(P_{u,v}^*) \le (\onepluseps)^2\weight{P_{u,v}} + 2(\onepluseps)^2 W_2(P_{u,v})$.

    The analysis for any time $t$ and the stretched-edge sub-cases within each case carry over from \lemref{w+2w2:bounded-approx}, with one technicality: we must also address the case where $P_{u,v}$ transitions from scale $j$ to a different scale $\ell > j$ to justify the use of previous estimates (notice that a fixed path can only move to a higher scale). In the edge rounding process of \cite{BernsteinRounding} a path with total weight $\le 2^j$ exists also in the $\ell$'th scale for every $\ell > j$ and it is still guaranteed to be of total weight $\ceil{\frac{4\paren{\beta + 2}}{\varepsilon}}$ in $G_\ell$. 
    However, the error after unscaling each edge is guaranteed to be $\onepluseps$ only for the $j$'th scale. 
    This does not change the correctness of our claim as we maintain the trees on the rounded graphs, and hence maintaining the trees on each scale separately suffices.

    Finally, combining the $(\onepluseps)^2$ multiplicative overhead from the three cases, we set $\varepsilon' = \varepsilon/12$ and apply the bound $W_2(P_{u,v}) \le \frac{1}{2}\weight{P_{u,v}}$ to obtain the stated guarantee.
\end{proof}

\begin{remark}
    It may happen that $2^j \le \weight{P_{u,v}} \le 2^{j+1}$ while $2^i \le \weight{P_{u,v}'} \le 2^{i+1}$ for $i \neq j$ (and specifically for $i < j$) as the path $P_{u,v}'$ (or $P_{y_2,v}'$) returned by the hopset is guaranteed to have a weight of $\dist{u,v}{G} \le \weight{P_{u,v}'} \le (\onepluseps)\weight{P_{u,v}}$ and in fact may be better than $\weight{P_{u,v}}$. It does not change the correctness of our algorithm.
\end{remark}

Finally, we analyze the total update time of \algref{w+2w2:unbounded}.

\begin{lemma}[Total Update Time Analysis]\label{lem:w+2w2:unbounded-total-update-time}
    For a constant $0 < \varepsilon < 1$, the expected total update time of \algref{w+2w2:unbounded} is $\onolog{n^{\frac{3}{2}} m^{\frac{1}{2}}}$ if $m = n^{1+\Omega(1)}$ or $\onolog{n^{\frac{3}{2}+\subpol} m^{\frac{1}{2}}}$ otherwise.
\end{lemma}

\begin{proof}
    First assume that $m=n^{1+\Omega(1)}$.
    We set the sampling probability to $q = \paren{\frac{m}{n}}^{-\frac{1}{2}}$ and the number of levels within the hierarchy to $k = 2$.
    For a constant $\varepsilon > 0$, the hopbound is $\beta = \bigo{\paren{\frac{k \log n}{\varepsilon}}^{\bigo{k}}} = \onolog{1}$.
    By \corref{tools:modified-es-unbounded}, maintaining $E_S$ takes expected total update time $\onolog{m n^{1/2} + \frac{n}{q}} = \onolog{m n^{1/2}}$, and the expected total number of edges ever present in $E_S$ is $\mathcal{E}_S = \onolog{\frac{n}{q}} = \onolog{n^{3/2} m^{-1/2}}$.
    By \thmref{tools:hopset-on-es-sparse}, constructing the hopset $H_S = E_S \cup F$ takes $\onolog{\beta n^{3/2}m^{1/2}} = \onolog{n^{3/2}m^{1/2}}$ update time, and the number of edges in $F$ is bounded by $\onolog{\beta \frac{n}{q}} = \onolog{\frac{n}{q}}$.
    The $(\onepluseps)$-SSSP algorithm from each of the $|S| = \onolog{nq} = \onolog{n^{3/2} m^{-1/2}}$ sources on $G$ takes $|S| \cdot \onolog{m} = \onolog{n^{3/2} m^{1/2}}$ total update time.

    For each scale, the Monotone ES-Trees are maintained up to depth $\tau = \ceil{\frac{4(\beta+2)}{\varepsilon}} = \onolog{1}$. 
    The total number of edges in $R^i(u)$ and $\hat{R}^i(u)$ across all $u \in V$ is 
    $\sum_{u \in V} (|H_S| + |Q_{\tilde{S}}| + 2n + \deg(u)) = \onolog{n \cdot \frac{n}{q} + n^2 + m} = \onolog{n^{3/2} m^{1/2} + m} = \onolog{n^{3/2} m^{1/2}}$. 
    Maintaining the Monotone ES-Trees across all $u \in V$ and all $\bigo{\log(nW_{\max})}$ scales thus takes 
    $\onolog{n^{3/2} m^{1/2}}$ total update time.
    Summing all components yields an expected total update time of $\onolog{n^{\frac{3}{2}} m^{\frac{1}{2}}}$.
    If $m=n^{1+\subpol}$ we use $\beta=n^{\subpol}$ instead and the total update time becomes $\onolog{n^{\frac{3}{2}+\subpol} m^{\frac{1}{2}}}$ by the same reasoning.

\end{proof}

Combining \lemref{w+2w2:unbounded-approx} and \lemref{w+2w2:unbounded-total-update-time} concludes the proof of \thmref{+2W2}.

\section{$\paren{\onepluseps,2W_1+2W_2}$-APASP}
\label{sec:w+2w1+2w2}
In the static setting, Cohen and Zwick~\cite{CohZwi1997} introduced the \textit{$STRETCH_{7/3}$} algorithm, which computes a $\paren{1,\mmin{2W_1,4W_2}}$-APASP with time complexity $\onolog{n^{\seventhirds}}$, achieving a multiplicative stretch of $\seventhirds$.

While quite similar, the algorithm is suitable for dense graphs as it removes the update time depenence on $m$, and utilizes a hierarchy of 2 hitting sets instead of 1, along with additional sets of shortcut edges $(S_2 \times V)$ and $(S_1 \times S_1)$. \algref{w+2w1+2w2:bounded} adapts a similar framework to the decremental setting.

In this section, we present a $\paren{\onepluseps,2W_{1} + 2W_{2}}$ APSP algorithm for decremental graphs, and establish the following theorem:
\thmtwoheaviestedges*

Expanding upon the framework developed in \algref{w+2w2:bounded}, our analysis evaluates arbitrary paths $P_{u,v}$ rather than strictly shortest paths and proves that for any path $P_{u,v}$, we have $\dist{u,v} \le d[u,v] \le (\onepluseps)\weight{P_{u,v}} + 2W_{1}\paren{P_{u,v}} + 2W_{2}\paren{P_{u,v}} \le (3 + \varepsilon) \weight{P_{u,v}}$.

The reliance on shortcut edges in the decremental model introduces analytical challenges, ultimately yielding the $+2W_1+2W_2$ error term instead of the classic static bound. We elaborate on these challenges following the algorithm description.

Crucially, the scheme achieves an overall multiplicative stretch of $3+\varepsilon$. To the best of our knowledge, this is the fastest decremental algorithm achieving a $3+\varepsilon$-stretch for dense graphs where $m = \Omega\paren{n^{\frac{11}{6}}}$.

We first establish a $\tau$-bounded algorithm with total update time $\onolog{\tau^2 n^{\seventhirds}}$, which serves as the core framework for handling unbounded distances afterwards.

\subsection{Bounded Depth $\tau$}

\begin{algorithm}[h]
\caption{$+2W_1+2W_2$ up to depth $\tau$}
\label{alg:w+2w1+2w2:bounded}
\BlankLine
\textbf{Input:} A weighted, undirected graph $G = \paren{V, E}$, a depth parameter $\tau$ and a sampling probability $q$.

\textbf{Output:} A matrix $\{ d[u,v] \}_{u,v \in V}$ of estimated distances.

$S_0 \gets V, S_3 = \emptyset$

$S_1 \gets \operatorname{sample}(S_0,q)$

$S_2 \gets \operatorname{sample}(S_1,q)$

$E_3 \gets E, E_0 \gets \emptyset$

$E_2 \gets E_{S_2}^{\tau}$

$E_1 \gets E_{S_1}^{\tau}$

$\tilde{S}_1 \gets \{\auxpiv[1]{v}\}_{v \in V}$

$\tilde{S}_2 \gets \{\auxpiv[2]{v}\}_{v \in V}$

$Q_{\tilde{S}_1} \gets \{(v,\auxpiv[1]{v})\}_{v \in V}$

$Q_{\tilde{S}_2} \gets \{(v,\auxpiv[2]{v})\}_{v \in V}$

\ForEach{$u \in S_2$}
        {
        Maintain an ES-Tree on $G$ up to depth $3 \tau$
    }

\ForEach{$i \in [1,0]$ in a descending order and $u \in S_i$}
{
    Maintain a Monotone ES-Tree from $u$ up to depth $3 \tau$ in $F_{i+1}(u)=(V,E_{i+1} \cup (S_2 \times V)  \cup (\singleton{u} \times \tilde{S}_{i+1}) \cup  Q_{\tilde{S}_{i+1}} \cup E(u))$ with distance estimates $\level{u}{\cdot}$.
}

\ForEach{$i \in [1,0]$ in a descending order and $u \in S_i$}
{
    Maintain another Monotone ES-Tree from $u$ up to depth $3 \tau$ in $\hat{F}_{i+1}(u)=(V,E_{i+1}' \cup (S_2 \times V) \cup (\singleton{u} \times \tilde{S}_{i+1})  \cup  Q_{\tilde{S}_{i+1}} \cup (\singleton{u} \times V))$ with distance estimates $\levelh{u}{\cdot}$.
}
\end{algorithm}

\paragraph{Algorithm Description.} 

Adapting the framework of \algref{w+2w2:bounded} to a three-level hierarchy, \algref{w+2w1+2w2:bounded} receives as input a weighted graph $G=\paren{V,E,\weight}$ and a parameter $\tau$ indicating the maximal distance to be explored. It considers a parameter $q\in \paren{0,1}$, whose value will be determined as part of the total update time analysis.

\algref{w+2w1+2w2:bounded} first initializes the distance estimation matrix $d$ with the corresponding edge weights and $\infty$ otherwise.
It then constructs the hierarchy by sampling a set $S_1$ from $S_0 = V$ with probability $q$ and subsequently sampling a set $S_2$ from $S_1$ with probability $q$.

For each $i \in \{1,2\}$, \algref{w+2w1+2w2:bounded} maintains $\piv[i]{u}$ for every $u \in V$ and the hierarchy-induced edge sets $E_1 = E_{S_1}^\tau$ and $E_2 = E_{S_2}^\tau$, using the framework developed in \secref{hierarchy}.

We also define $S_3 = \emptyset$, $E_0 = \emptyset$, and $E_3 = E$, establishing the relations $\emptyset = S_3 \subseteq S_2 \subseteq S_1 \subseteq S_0 = V$ and $E = E_3 \supseteq E_2 \supseteq E_1 \supseteq E_0 = \emptyset$.

For each vertex $u \in V$, we create corresponding auxiliary vertices $\auxpiv[1]{u}$ and $\auxpiv[2]{u}$. We define the sets $\tilde{S}_1=\{\auxpiv[1]{v}\}_{v \in V}$, $\tilde{S}_2=\{\auxpiv[2]{v}\}_{v \in V}$, $Q_{\tilde{S}_1}=\{(v,\auxpiv[1]{v}\}_{v \in V}$, and $Q_{\tilde{S}_2}=\{(v,\auxpiv[2]{v}\}_{v \in V}$. Note that there are $2n$ such auxiliary vertices and edges.

For each $u \in S_2$, \algref{w+2w1+2w2:bounded} maintains a standard ES-Tree on $G$ up to depth $3 \tau$.

For $i \in \{1,0\}$ in descending order, \algref{w+2w1+2w2:bounded} maintains two Monotone ES-Trees up to depth $3 \tau$ from each $u \in S_i$. The first tree is maintained on the graph $F_{i+1}(u)=(V,E_{i+1} \cup (S_2 \times V)  \cup (\singleton{u} \times \tilde{S}_{i+1}) \cup  Q_{\tilde{S}_{i+1}} \cup E(u))$ with level estimates $\level{u}{v}$. The second tree is maintained on $\hat{F}_{i+1}(u)$, which replaces the adjacency edges $E(u)$ with shortcuts to all vertices, $\paren{\singleton{u} \times V}$, utilizing the levels computed in the previous step as edge weights. The level estimates produced by this tree are denoted by $\levelh{u}{v}$.

Similarly to \algref{w+2w2:bounded}, the purpose of the two trees is to mimic the behavior of the two-rounded Dijkstra scheme of Cohen and Zwick~\cite{CohZwi1997} while adhering to the monotonic constraints of the decremental setting. This dual-tree structure is, in fact, what enables the refined $+2W_1+2W_2$ additive error analysis.

To ensure consistency, edge deletions must be processed in strict cascading order, similarly to \algref{w+2w2:bounded}: updates are first processed across all trees of $S_1$ on $F_2$, followed by all trees of $S_0$ on $F_1$, then all trees of $S_1$ on $\hat{F}_2$, and finally all trees of $S_0$ on $\hat{F}_1$.

The global distance estimate is continuously updated as $d[u,v] = \mmin{\levelh{u}{v}, \levelh{v}{u}}$.

\paragraph{Differences from the static algorithm of Cohen and Zwick \cite{CohZwi1997}.}
Our algorithm diverges from the static $+2W_1+2W_2$ framework of Cohen and Zwick~\cite{CohZwi1997}
in several key algorithmic aspects.

The primary difference lies in the distances computed from the hitting set $S_1$.
In the static algorithm, distances from $S_1$ are computed exactly on $G$.
However, maintaining ES-Trees on $G$ from every $s \in S_1$ would require $\Omega(\tau m |S_1|)$ total update time, exceeding our update time budget.
Consequently, vertices in $S_1$ maintain trees restricted to the sparse subgraphs $F_2(u)$ and $\hat{F}_2(u)$,
and as those graphs may have edges inserted to them, we use a Monotone ES-Tree.
However, in a Monotone ES-Tree distance estimates can become stretched.
In particular, consider the case where a path $P_{u,v}$ was not initially contained in $E_1$. Then a vertex $s \in S_1$ did not see
the entire path within its tree, and must use an alternative path that approximates $P_{u,v}$ via $S_2$ to reach either $u$ or $v$. 
If edges remain stretched from that time period, having $P_{u,v} \subseteq E_1$ at some time $t$ in the future is not enough to
achieve the required approximation error for distances computed from $S_1$, as the distance estimates would carry the error from earlier times. We refer to this as the \emph{Stretch Propagation Challenge}.

Furthermore, although we demonstrate that the error bound for the vertices of $S_1$ is strictly superior to the general case, we have only succeeded in establishing an additive error approximation of $+2W_1$, falling short of the $+2W_2$ bound achieved in the static algorithm.
This limitation is a consequence of the structural requirement that, to successfully preclude the scenario of a single missing edge $(x,y)$ within $E_2$,
the vertex $y$ must possess full visibility of the entire path connecting $x$ and the source vertex $s$. 
In our decremental framework, we were unable to guarantee this condition.

We begin with approximation analysis.

\begin{lemma}
    \label{lem:+2w1+2w2:bounded:approximation-guarantee-base-case}
    At the beginning of the algorithm, for every $u,v \in V$ and for any path $P_{u,v}$ between $u$ and $v$ satisfying $\weight{P_{u,v}} \le \tau$, $\dist{u,v} \le d[u,v] \le \weight{P_{u,v}} + 2W_1\paren{P_{u,v}}$
\end{lemma}
\begin{proof}
    Fix $u,v \in V$ and a path $P_{u,v}$ between $u$ and $v$ with $\weight{P_{u,v}} \le \tau$.
    
    We consider four cases:
    
    \ccase{1} $\sizeindex{P_{u,v}}{E_{2}} \geq 1$.
    
    Let $\paren{x,y}\in P_{u,v}$ be any edge that does not belong to $E_{2}$. As $\paren{x,y}\notin E_{2}$ and $\weight{x,y} \leq \weight{P_{u,v}} \leq \tau$
    , it follows from \corref{tools:es-max-edges} that there exists $w = \piv[2]{y} \in S_2$ such that $\dist{y,w} < \weight{x,y}$.
    In the ES-Tree of $w$ over $E$, it holds that $\level{w}{u} = \dist{w,u} \leq \dist{w,y} + \dist{y,u} \leq \weight{P_{u,y}} + \weight{x,y}$ and similarly $\level{w}{v} = \dist{w,v} \le \dist{w,y} + \dist{y,v} \le \weight{P_{y,v}} + \weight{x,y}$.
    
    Since $w \in S_{2}$, the graph $F_{1}\paren{u}$ on which we maintain a Monotone ES-Tree from $u$ contains the edges $\paren{u,w} \in \singleton{u} \times S_{2}$ with $\weight{u,w} = \level{w}{u} \leq \weight{P_{u,y}} + \weight{x,y}$ and $\paren{w,v} \in S_{2} \times V$ with weight $\weight{w,v} = \level{w}{v} \leq \weight{P_{y,v}} + \weight{x,y}$.
    
    Concatenated, these two edges form a path $P_{u,v}\sstar$ with $\weight{P_{u,v}\sstar} \leq \weight{P_{u,y}} + \weight{x,y} + \weight{P_{y,v}} + \weight{x,y} = \weight{P_{u,v}} + 2\weight{x,y} \leq \weight{P_{u,v}} + 2W_{1}\paren{P_{u,v}}$.
    As this is the start of the algorithm, $\level{u}{v} \leq \weight{P_{u,v}\sstar}$.
    
    \ccase{2} $\sizeindex{P_{u,v}}{E_{2}} = 0$ and $\sizeindex{P_{u,v}}{E_{1}} > 1$.
    
    Let $\paren{x_{1},y_{1}}=\rightabovearrow{P_{u,v}}{E_{1}} \neq \leftabovearrow{P_{u,v}}{E_{1}} = \paren{x_{2},y_{2}}$ and assume, without loss of generality, that $\weight{x_{2}, y_{2}} \leq \weight{x_{1}, y_{1}}$. Let $P_{u,y_2}$ denote the portion of $P_{u,v}$ between $u$ and $y_2$, and let $P_{y_2,v}$ denote the portion of $P_{u,v}$ between $y_2$ and $v$.
    Since $\paren{x_2,y_2} \notin E_1$ and $\weight{x_2,y_2} \leq \weight{P_{u,v}} \leq \tau$, it follows that $\dist{y_2,\piv[1]{y_2}} \leq \weight{x_2,y_2}$.

    Let $w = \piv[1]{y_2}$, and recall that $w \in S_1$. Hence, we maintain a Monotone ES-Tree from $w$ on $F_2\paren{w}$.
    Note that $P_{u,v} \subseteq E_2 \subseteq F_2\paren{w}$, and hence $\dist{w,u}{F_2\paren{w}} \le \weight{P_{u,v}} + \dist{y_2,w} \le \weight{P_{u,v}} + \weight{x_2,y_2}$.
    
    Consider the first Monotone ES-Tree from $u$. There exists an auxiliary edge $\paren{u,\auxpiv[1]{y_2}}$ of weight $\level{w}{u}$ and an auxiliary edge $\paren{\auxpiv[1]{y_2},y_2}$ of weight $\dist{\piv{y_2},y_2}$.
    Since $P_{y_2,v}\subseteq E_1$, there exists a path $P^*_{u,v} \subseteq F_1\paren{u}$ with
    $\weight{P^*_{u,v}} \leq \level{w}{u} + \dist{\piv{y_2}}{y_2} + \weight{P_{y_2,v}} \leq \level{w}{u} + \weight{x_2,y_2} + \weight{P_{y_2,v}}$.

    Let $P' = P_{u,y_2} \circ y_2 \squiggly w$. Since this is the beginning of the algorithm, $\level{w}{u} \le \weight{P'}\le \weight{P_{u,y_2}} + \dist{y_2,w}$.

    Let $P^*_{u,v} = \paren{u,\auxpiv[1]{y_2}} \circ \paren{\auxpiv[1]{y_2},y_2} \circ P_{y_2,v}$. From the previous discussion, $\weight{P^*_{u,v}} \le \weight{P_{u,v}} + 2\weight{x_{2},y_{2}}$.

    Since this is the beginning of the algorithm, $\level{u}{v} \le \weight{P^*_{u,v}}$.
    
    \ccase{3} $\sizeindex{P_{u,v}}{E_{2}} = 0$ and $\sizeindex{P_{u,v}}{E_{1}} = 1$.
    
    Let $ \paren{x, y} = \rightabovearrow{P_{u,v}}{E_{1}} = \leftabovearrow{P_{u,v}}{E_{1}}$.
    This case is similar to Case 2 of \lemref{w+2w2:bounded-approx}: $\paren{x,y}$ and all the edges of $P_{u,y}$ belong to the graph $F_{1}\paren{y}$, on which we maintain a Monotone ES-Tree from $y$ in the first loop. As this is the beginning of the algorithm, we conclude that $\level{y}{u} \le \weight{P_{u,y}}$.
    
    The graph $\hat{F}_{1}\paren{u}$, on which we maintain a Monotone ES-Tree from $u$ in the second loop, therefore includes an edge $(u,y)$ of weight $\level{y}{u} \le \weight{P_{u,y}}$ and all the edges of $P_{y,v}$.
 Consequently, there exists a path $P_{u,v}\sstar \subseteq \hat{F}_{1}\paren{u}$ with $\weight{P_{u,v}\sstar} \le \weight{P_{u,y}}+ \weight{P_{y,v}} \le \weight{P_{u,v}}$.
    Since this is the beginning of the algorithm, $\level{u}{v} \le \weight{P_{u,v}\sstar}$.
    
    \ccase{4} $\sizeindex{P_{u,v}}{E_{1}} = 0$.
    
    The Monotone ES-Tree from $u \in S_0$ on $F_{1}\paren{u}$ contains $E_1$ and hence the path $P_{u,v}$. As this is the beginning of the algorithm, $\level{u}{v} \le \weight{P_{u,v}}$. By \lemref{prelim:monotone-es-properties}, it holds that $\level{u}{v} \ge \dist_{\hat{F_1}(u)}(u,v) \ge \dist{u,v}$, ensuring the lower bound holds across all cases.
\end{proof}

Having established the base case, we state a key observation that will be used in the inductive step of \lemref{+2w1+2w2:bounded:approximation-guarantee}.

\begin{observation}
    \label{obv:+2w1+2w2:bounded:approximation-from-vertices-of-s1}
    At any time $t$, for any $w\in S_1$, $u \in V$ and any path $P_{w,u}$ between $w$ and $u$ satisfying $\weight{P_{w,u}} \le \tau$, it holds that $\level{w}{u} \le \weight{P_{w,u}} + 2W_{1}\paren{P_{w,u}}$.
\end{observation}
\begin{proof}
    The observation follows from \algref{w+2w2:bounded} and \lemref{w+2w2:bounded-approx}, as the graphs $F_{2}\paren{w}$ and $\hat{F}_{2}\paren{w}$ of \algref{w+2w1+2w2:bounded} contain the corresponding graphs from \algref{w+2w2:bounded}.
 There are two cases to consider: 
$\sizeindex{P_{w,u}}{E_{2}} = 0$ and $\sizeindex{P_{w,u}}{E_{2}} \geq 1$. The proofs are identical to those of Cases 1 and 3 of \lemref{w+2w2:bounded-approx}.
\end{proof}

\begin{note}
    The proof of \lemref{w+2w2:bounded-approx} also considers a case where there is only one edge not in $E_S$, crucial for achieving a $+2W_2$ approximation as opposed to $+2W_1$. The proof relies on the presence of $E_S$ in $F(y)$, where $y \in V$ is one of the endpoints of the missing edge.
    We cannot guarantee this for the case of $E_S = E_2$.
\end{note}

We can now complete the inductive argument:

\begin{lemma} [Approximation Guarantee]
    \label{lem:+2w1+2w2:bounded:approximation-guarantee}
    At any time $t$, for every $u,v \in V$ and any path $P_{u,v}$ between $u$ and $v$ satisfying $\weight{P_{u,v}} \le \tau$, $\dist{u,v} \le d[u,v] \le \weight{P_{u,v}} + 2W_{1}\paren{P_{u,v}} + 2W_{2}\paren{P_{u,v}}$
\end{lemma}
\begin{proof}
    Fix $u,v \in V$ and a path $P_{u,v}$ between $u$ and $v$ with $\weight{P_{u,v}} \le \tau$.
    The base case of $t=0$ was established in \lemref{+2w1+2w2:bounded:approximation-guarantee-base-case}. We therefore assume the claim holds for all times $t' < t$ and prove it for time $t$.
    
    Following the structure of \lemref{+2w1+2w2:bounded:approximation-guarantee-base-case}, we consider four cases:
    
    \ccase{1} $\sizeindex{P_{u,v}}{E_{2}} \geq 1$.
    
    This case structurally repeats Case 1 of \lemref{+2w1+2w2:bounded:approximation-guarantee-base-case}. The underlying graph properties guaranteeing the existence of a higher-level pivot $w = \piv[2]{y} \in S_{2}$ and the formation of an alternative path $P_{u,v}^{*} \subseteq F_{1}(u)$ with bounded weight $\weight{P_{u,v}^{*}} \le \weight{P_{u,v}} + 2W_{1}(P_{u,v})$ remain valid at any time $t$.

    Importantly, because the weights of the specific auxiliary edges comprising this alternative path can never decrease,
they are never stretched in the Monotone ES-Tree.
Consequently, the tree computes the path with exact distance, yielding $\level{u}{v} \le \weight{P_{u,v}^{*}}$.
    
    \ccase{2} $\sizeindex{P_{u,v}}{E_{2}} = 0$ and $\sizeindex{P_{u,v}}{E_{1}} > 1$.
    
    Let $\paren{x_{1},y_{1}}=\rightabovearrow{P_{u,v}}{E_{1}} \neq \leftabovearrow{P_{u,v}}{E_{1}} = \paren{x_{2},y_{2}}$ and assume, without loss of generality, that $\weight{x_{2},y_{2}} \leq \weight{x_{1},y_{1}}$, hence  $\weight{x_2,y_2} \leq W_{2}(P_{u,v})$. The pivot $w = \piv[1]{y_{2}} \in S_{1}$ induces an alternative path $P_{u,v}^{*} = (u,\auxpiv[1]{y_{2}}) \circ (\auxpiv[1]{y_{2}},y_{2}) \circ P_{y_{2},v} \subseteq F_{1}(u)$, where $\weight{\auxpiv[1]{y_{2}},y_{2}} = \dist{y_2,\piv[1]{y_2}} \le \weight{x_2,y_2}$ and $\weight{u,\auxpiv[1]{y_{2}}} \le \level{w}{u}$.

    Let $P' = P_{u,y_2} \circ y_2 \squiggly w$. It follows from \obvref{+2w1+2w2:bounded:approximation-from-vertices-of-s1} that $\level{w}{u} \le \weight{P'} + 2W_{1}\paren{P'} \le \weight{P_{u,y_2}} + \dist{y_2,w} + 2W_{1}\paren{P'}$.
    Consider the heaviest edge on $P'$. The edge can belong to either $P_{u,y_2}$ or $y_2 \squiggly w$. Since $\dist{w,y_2} = \weight{y_2 \squiggly w} \le \weight{x_{2},y_{2}} \le W_{2}\paren{P_{u,y_2}}$, we conclude that $W_{1}\paren{P'} \le W_{1}\paren{P_{u,y_2}}$ and consequently $\level[t]{w}{u} \le \weight{P_{u,y_2}} + \dist{y_2,w} + 2W_{1}\paren{P_{u,y_2}}$.

    Let $P^*_{u,v} = \bracce{\paren{u,\auxpiv[1]{y_2}}} \circ \bracce{\paren{\auxpiv[1]{y_2},y_2}} \circ P_{y_2,v}$. From the previous discussion, $\weight{P_{u,v}} \le \weight{P^*_{u,v}} \le \weight{P_{u,v}} + 2W_{1}\paren{P_{u,v}} + 2\weight{x_{2},y_{2}} \le \weight{P_{u,v}} + 2W_{1}\paren{P_{u,v}} + 2W_{2}\paren{P_{u,v}}$.
    A demonstration of the path is shown in \figref{2w12w2:approximation-analysis-case-2}.
    \usetikzlibrary{positioning, arrows.meta, calc, shapes.geometric, decorations.pathreplacing}

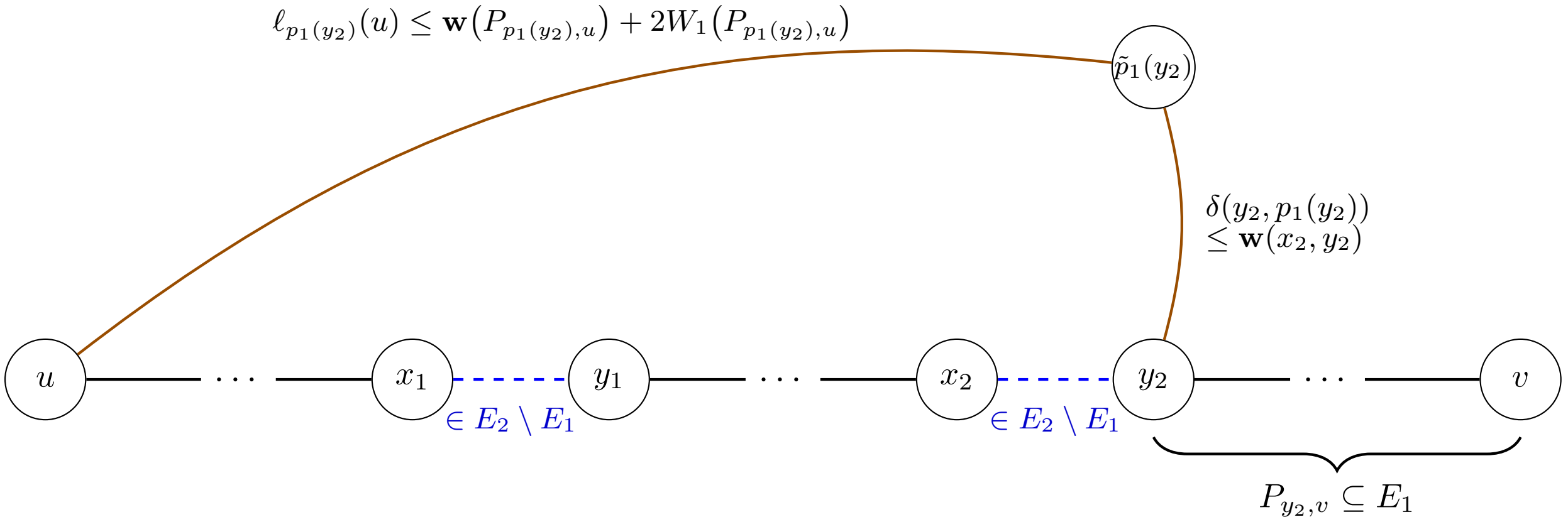
\begin{figure}[htbp]
    \centering
    \resizebox{\textwidth}{!}{%
    \begin{tikzpicture}[
        node distance=1.5cm,
        vertex/.style={circle, draw=black, fill=white, inner sep=0pt, minimum size=24pt, font=\small},
        path edge/.style={thick},
        dashed edge/.style={thick, dashed},
        aux edge/.style={thick, orange!60!black},
        label font/.style={font=\footnotesize}
    ]

        \node[vertex] (u) at (0,0) {$u$};
        
        \node[right=1.2cm of u] (dots1) {$\dots$};
        \draw[path edge] (u) -- (dots1);
        
        \node[vertex, right=1.0cm of dots1] (x1) {$x_1$};
        \node[vertex, right=1.2cm of x1] (y1) {$y_1$};
        \draw[path edge] (dots1) -- (x1);
        \draw[dashed edge, blue] (x1) -- (y1);
        \node[below=4pt of $(x1)!0.5!(y1)$, font=\footnotesize, text=blue!80!black] {$\in E_2 \setminus E_1$};
        
        \node[right=1.0cm of y1] (dots2) {$\dots$};
        \draw[path edge] (y1) -- (dots2);
        
        \node[vertex, right=1.0cm of dots2] (x2) {$x_2$};
        \node[vertex, right=1.2cm of x2] (y2) {$y_2$};
        \draw[path edge] (dots2) -- (x2);
        \draw[dashed edge, blue] (x2) -- (y2);
        \node[below=4pt of $(x2)!0.5!(y2)$, font=\footnotesize, text=blue!80!black, align=center] {$\in E_2 \setminus E_1$};
        
        \node[right=1.0cm of y2] (dots3) {$\dots$};
        \node[vertex, right=1.2cm of dots3] (v) {$v$};
        \draw[path edge] (y2) -- (dots3);
        \draw[path edge] (dots3) -- (v);

        \node[vertex, above=2.4cm of y2] (aux_y2) {\scriptsize $\auxpiv[1]{y_2}$};
        
        \draw[aux edge, bend left=15] (aux_y2) to node[right=3pt, black, font=\footnotesize, align=left] {$\dist{y_2,\piv[1]{y_2}}$ \\[-2pt] $\le \weight{x_2,y_2}$} (y2);
        
        \draw[aux edge, bend left=22] (u) to node[above=12pt, black, font=\footnotesize] {$\level{\piv[1]{y_2}}{u} \le \weight{P_{\piv[1]{y_2},u}} 
        + 2W_1\paren{P_{\piv[1]{y_2},u}}$} (aux_y2);

        \draw[decorate, decoration={brace, amplitude=10pt, mirror, raise=5pt}, thick]
            (y2.south) -- (v.south) 
            node[midway, below=15pt, font=\small] {$P_{y_2,v} \subseteq E_1$};

    \end{tikzpicture}
    }
    \caption{Graph illustration of Case~2, where $P_{u,v} \subseteq E_2$ 
    and it contains at least two edges not in $E_1$. 
    The edges $\paren{x_1,y_1}$ and $\paren{x_2,y_2}$ 
    belong to $E_2 \setminus E_1$, with $\weight{x_2,y_2} \le \weight{x_1,y_1}$. 
    Since $\paren{x_2,y_2}$ is the last edge not in $E_1$, the suffix satisfies 
    $P_{y_2,v} \subseteq E_1$. In $F_1(u)$, the algorithm uses the 
    alternative path $P^*_{u,v} = \paren{u,\auxpiv[1]{y_2}} \circ \paren{\auxpiv[1]{y_2},y_2} \circ P_{y_2,v}$.
    (orange auxiliary edges followed by the black edges of $P_{y_2,v}$), 
    with total weight bounded by 
    $\weight{P_{u,v}} + 2W_1\paren{P_{u,v}} + 2W_2\paren{P_{u,v}}$.}
    \label{fig:2w12w2:approximation-analysis-case-2}
\end{figure}

    If none of the edges of $P^*_{u,v}$ are stretched in $u$'s tree, we are done.
    Otherwise, let $\paren{z',z}$ be the last stretched edge on $P^*_{u,v}$, and consider two cases:

    \ccase{2.1} $z = \auxpiv[1]{y_2}$.
    
    We first prove by induction that $\level{u}{\auxpiv[1]{y_2}} \le \weight{P_{u,y_2}} + W_{1}\paren{P_{u,y_2}} + \weight{x_{2},y_{2}}$.
    The base case was established in \lemref{+2w1+2w2:bounded:approximation-guarantee-base-case}.
    Fix some time $t$. If $\paren{u,\auxpiv[1]{y_2}}$ is not stretched in $u$'s tree at time $t$, the claim follows from the previous discussion.
    
    Assume then that $\paren{u,\auxpiv[1]{y_2}}$ is stretched in $u$'s tree at time $t$.
    
    Note that $\weight{u,\auxpiv[1]{y_2}} = \level{\piv[1]{y_2}}{u}$ and it exists at any given time. Hence, the value of $\weight{u,\auxpiv[1]{y_2}}$ can decrease only when $\piv[1]{y_2}$ changes from $w$ to $w'$ such that $\level{w}{u} > \level{w'}{u}$.
    We conclude that $\level{u}{\auxpiv[1]{y_2}} = \level[t-1]{u}{\auxpiv[1]{y_2}} \le \level[t-1]{w}{u} \le \weight{P_{u,y_2}} + W_{1}\paren{P_{u,y_2}} + \weight{x_{2},y_{2}}$ by the induction hypothesis.

    Since $\paren{u,\auxpiv[1]{y_2}}$ is the last stretched edge in $u$'s tree, it follows from the previous discussion that $\level{u}{v} \le \level{u}{\auxpiv[1]{y_2}} + \weight{x_{2},y_{2}} + \weight{P_{y_2,v}} \le \weight{P_{u,v}} + 2W_{1}\paren{P_{u,v}} + 2W_{2}\paren{P_{u,v}}$.
    
    \ccase{2.2} $z=y_2$ or $\paren{z',z} \in P_{y_2,v}$.
    
    By the properties of a Monotone ES-Tree and the induction hypothesis, $\level{u}{z} = \level[t-1]{u}{z} \le \weight{P_{u,z}} + 2W_{1}\paren{P_{u,z}} + 2W_{2}\paren{P_{u,z}}$.

    Since $\paren{z',z}$ is the last stretched edge on $P^*_{u,v}$ in $u$'s tree and $P_{z,v} \subseteq P_{y_2,v} \subseteq E_1$, we conclude that $\level{u}{v} \le \level{u}{z} + \weight{P_{z,v}} \le \weight{P_{u,z}} + 2W_{1}\paren{P_{u,z}} + 2W_{2}\paren{P_{u,z}} + \weight{P_{z,v}} \le \weight{P_{u,v}} + 2W_{1}\paren{P_{u,v}} + 2W_{2}\paren{P_{u,v}}$.
    
    \ccase{3} $\sizeindex{P_{u,v}}{E_{2}} = 0$ and $\sizeindex{P_{u,v}}{E_{1}} = 1$.
    
    Let $ \paren{x, y} = \rightabovearrow{P_{u,v}}{E_{1}} = \leftabovearrow{P_{u,v}}{E_{1}}$.
     The sub-path $P_{u,y}\subseteq P_{u,v}$ between $u$ and $y$ belongs to $F_{1}(y)$. If none of its edges are stretched in $y$'s tree, $\level{y}{u} \le \weight{P_{u,y}}$. Otherwise, by the induction hypothesis on time $t$, $\level{y}{u} \le \weight{P_{u,y}} + 2W_{1}(P_{u,y}) + 2W_{2}(P_{u,y})$.

    As $P_{y,v} \subseteq E_1$, the graph $\hat{F}_{1}(u)$ on which we maintain the second Monotone ES-Tree contains $P_{u,v}^* = \paren{u,y} \circ P_{y,v}$ with $\weight{P_{u,v}^*} = \weight{u,y} + \weight{P_{y,v}} \le \level{y}{u} + \weight{P_{y,v}} \le \weight{P_{u,v}} + 2W_{1}(P_{u,y}) + 2W_{2}(P_{u,y})$. If no edges on $P_{u,v}^*$ are stretched in $u$'s tree, the bound holds directly.
    
    Otherwise, let $\paren{w',w}$ be the last stretched edge on $P_{u,v}^*$ in $u$'s tree, traversing from $u$ to $v$. Then, by the induction hypothesis we have $\level{u}{w} \le \weight{P_{u,w}} + 2W_{1}\paren{P_{u,w}} + 2W_{2}\paren{P_{u,w}}$. As $P_{w,v} \subseteq P_{y,v} \subseteq E_1 \subseteq \hat{F}_{1}\paren{u}$ and none of the edges on $P_{w,v}$ are stretched in $u$'s tree, we get that $\level{u}{v} \le \weight{P_{u,v}} + 2W_{1}\paren{P_{u,w}} + 2W_{2}\paren{P_{u,w}}$ as required.
    
    \ccase{4} $\sizeindex{P_{u,v}}{E_{1}} = 0$.
    
    The Monotone ES-Tree rooted at $u \in S_0$ on $F_{1}\paren{u}$ contains $E_1$ and hence the path $P_{u,v}$. If none of the vertices on $P_{u,v}$ are stretched in $u$'s tree, we are done.
    
    Otherwise, let $\paren{w',w}$ be the last stretched edge on $P$ in $u$'s tree, traversing from $u$ to $v$. Then, by the induction hypothesis we have $\level{u}{w} \le \weight{P_{u,w}} + 2W_{1}\paren{P_{u,w}} + 2W_{2}\paren{P_{u,w}}$. As $P_{w,v} \subseteq E_1 \subseteq \hat{F}_{1}\paren{u}$ and none of the vertices on $P_{w,v}$ are stretched, we get that $\level{u}{v} \le \weight{P_{u,v}} + 2W_{1}\paren{P_{u,w}} + 2W_{2}\paren{P_{u,w}}$ as required.

    Using \lemref{prelim:monotone-es-properties}, we also have $\level{u}{v} \geq \dist{u,v}{\hat{F}_{1}\paren{u}}\geq \dist{u,v}$.
\end{proof}

We are now ready to argue for the expected total update time of the algorithm.

\begin{lemma}[Total Update Time Analysis]
    \label{lem:+2w1+2w2:bounded:total-update-time-analysis}
    The expected total update time of \algref{w+2w1+2w2:bounded} is $\onolog{\tau^2n^{\frac{7}{3}}}$.
\end{lemma}
\begin{proof}
    Observe that $\expectation{\abs{S_1}} = \onolog{nq}$ and $\expectation{\abs{S_2}} = \onolog{nq^2}$.
    Applying \corref{tools:es-max-edges}, the expected total number of edges ever inserted into the sets $E_1$ and $E_2$ is bounded by $\mathcal{E}_1 = \onolog{\frac{\tau n}{q}}$ and $\mathcal{E}_2 = \onolog{\frac{\tau n}{q^{2}}}$, respectively. Also, $\abs{S_0} = n$ and $\abs{E_3} = m = \bigo{n^2}$.
    
    From every $s \in S_2$ we maintain an ES-Tree up to depth $3 \tau$ on $G$. As each tree requires $\onolog{\tau m}$ total update time, summing across all $s \in S_2$ yields an expected total update time of $\onolog{\tau n^3q^2}$ for this step.
    
    To analyze the cost of maintaining Monotone ES-Trees, we must account for both the total number of edges and the total number of updates in the augmented graphs.
    
    The sets $Q_{\tilde{S_1}},Q_{\tilde{S_2}},\singleton{u} \times \tilde{S_1},\singleton{u} \times \tilde{S_2}$ have a constant size of $n$ edges.
Since we assume integer edge weights, and each edge - whether original from $E_1, E_2$ or auxiliary - represents a path of weight at most $\tau$, each edge undergoes at most $\tau$ weight updates.
Summing these costs, the expected total update time is bounded by:
    
    \[\onolog{ \frac{\tau^{2} n^{2}}{q} + \tau n^3q^2 + m} = \onolog{\frac{\tau^{2}n^{2}}{q} + \tau n^3q^2}\]

    This yields the desired total update time by setting $q=n^{-\frac{1}{3}}$.
\end{proof}

Utilizing \lemref{+2w1+2w2:bounded:approximation-guarantee} and \lemref{+2w1+2w2:bounded:total-update-time-analysis}, we conclude:

\begin{conclusion}
    \label{con:+2w1+2w2:bounded-theorem}
    \algref{w+2w1+2w2:bounded} maintains a distance estimate $d[\cdot,\cdot]$ such that for any pair of vertices $u,v \in V$ with $\dist{u,v} \le \tau$,  $\dist{u,v} \le d[u,v] \le \dist{u,v} + 2W_{1}\paren{\paren{u\squiggly v}_{t}} + 2W_{2}\paren{\paren{u\squiggly v}_{t}}$ at any time. The expected total update time of \algref{w+2w1+2w2:bounded} is $\onolog{\tau^2 n^{\frac{7}{3}}}$.
\end{conclusion}

\begin{note}
        We can also set $q= \paren{\frac{\tau}{n}}^{\frac{1}{3}}$ to achieve a slightly better total update time.
        We intentionally avoid this to fit the next algorithms.
\end{note}

\begin{observation}
    \label{obv:+2w1+2w2:number-of-distance-changes-different-than-tau}
    Similarly to \algref{w+2w2:bounded}, given a SSSP algorithm that ensures $\alpha$ distance estimate changes, the total update time can be expressed as $\onolog{\tau \alpha n^{\frac{7}{3}}}$.
\end{observation}

\subsection{Unbounded Depth}
To overcome the dependency on the depth parameter $\tau$, which renders the Monotone ES-Trees prohibitively expensive for large, arbitrary distances, we transition to an unbounded framework.
\begin{algorithm}[H]
\caption{$\paren{\onepluseps, 2W_1 + 2W_2}$-APASP$\paren{G = \paren{V, E, \weight}}$}
\label{alg:+2w1+2w2:unbounded}
\BlankLine
\textbf{Input:} A weighted, undirected graph $G = \paren{V, E, \weight}$

\textbf{Output:} A matrix $\{ d[u,v] \}_{u,v \in V}$ of estimated distances.

$S_0 \gets V, S_3 = \emptyset$

$S_1 \gets \operatorname{sample}(S_0,q)$

$S_2 \gets \operatorname{sample}(S_1,q)$

$H_3 \gets E, H_0 \gets \emptyset$

$E_2 \gets E_{S_2}$

$E_1 \gets E_{S_1}$

$F,\beta \gets \textit{Make-Hopset}\paren{S_0,S_1,S_2,S_3}$

$H_2 \gets E_2 \cup F$

$H_1 \gets E_1 \cup F$

\ForEach{$u \in S_2$}
{
    Maintain a $(\onepluseps)$-SSSP tree on $G$.
}

\ForEach{$i \in [1,2]$}
{
     $F_{i} \gets (V,H_{i} \cup (S_2 \times V) \cup  Q_{\tilde{S}_{i}})$
     
     $\{F_i^j\}_{j=1}^{\ceil{\log nW_{\max}} + 1} \gets \textit{SCALE}\paren{F_i,\onepluseps,\beta + 2}$
}

\ForEach{$i \in [1,0]$, $u \in S_i$}
{
    $R_{i+1}(u) = (\singleton{u} \times \tilde{S}_{i+1}) \cup E(u)$
    
    $\{R_{i+1}^j(u)\}_{j=1}^{\ceil{\log nW_{\max}} + 1} \gets \textit{SCALE}\paren{R_{i+1}(u),\onepluseps,\beta + 2}$
    
    \ForEach{$1 \le j \le \ceil{\log nW_{\max}} + 1$}
    {
    Maintain a Monotone ES-Tree from $u$ up to depth $\ceil{\frac{4\paren{\beta + 2}}{\varepsilon}}$ in $F_{i+1}^j(u)= F_{i+1}^j \cup R_{i+1}^j(u)$ with distance estimates $\level[j]{u}{\cdot}$.
    }
}

\ForEach{$i \in [1,0]$, $u \in S_i$}
{
    $\hat{R}_{i+1}(u) = (\singleton{u} \times \tilde{S}_{i+1}) \cup (\singleton{u} \times V)$
    
    $\{\hat{R}_{i+1}^j(u)\}_{j=1}^{\ceil{\log nW_{\max}} +1} \gets \textit{SCALE}\paren{\hat{R}_{i+1}(u),\onepluseps,\beta + 2}$
    
    \ForEach{$1 \le j \le \ceil{\log nW_{\max}} + 1$}
    {
    Maintain a Monotone ES-Tree from $u$ up to depth $\ceil{\frac{4\paren{\beta + 2}}{\varepsilon}}$ in $\hat{F}_{i+1}^j(u)= F_{i+1}^j \cup \hat{R}_{i+1}^j(u)$ with distance estimates $\levelh[j]{u}{\cdot}$.
    }
}
\end{algorithm}

\paragraph{Algorithm Description.}
\algref{+2w1+2w2:unbounded} generalizes \algref{w+2w1+2w2:bounded} to handle unbounded distances,
following the same three-stage strategy as \algref{w+2w2:unbounded} of \secref{w+2w2:unbounded}.
We recall that the $\tau$-dependence in the total update time of \algref{w+2w1+2w2:bounded} arises
because the Monotone ES-Trees must explore paths up to weight $\tau$;
when $\tau$ is large, this becomes prohibitively expensive.
We address this by employing the tools of \secref{tools} within the three-level hierarchy employed in \algref{w+2w1+2w2:bounded}.

Concretely, \algref{+2w1+2w2:unbounded} constructs an hierarchy with $k=3$ levels, setting $S_0 = V$
and sampling $S_i$ from $S_{i-1}$ with probability $q=n^{-\frac{1}{3}}$ for $i \in \bracke{2}$.
\algref{+2w1+2w2:unbounded} maintains the edge sets $E_1 = E_{S_1}$ and $E_2 = E_{S_2}$ via \corref{tools:modified-es-unbounded}
with parameter $k=3$, and then invokes \thmref{tools:hopset-on-es-dense} with parameter $k=3$ to produce a shortcut set $F$
of size $\onolog{k \cdot n^{1+\frac{1}{k}}} = \onolog{n^{\frac{4}{3}}}$ such that $H_1 = E_1 \cup F$ and $H_2 = E_2 \cup F$
are $\paren{\beta=\bigo{\paren{\frac{\log n}{\varepsilon}}^3},\onepluseps}$-hopsets for $E_1$ and $E_2$, respectively.
From each $u \in S_2$, \algref{+2w1+2w2:unbounded} maintains a near-linear $(\onepluseps)$-SSSP tree on $G$
rather than an exact ES-Tree.

Next, for each $i \in \{1,2\}$, \algref{+2w1+2w2:unbounded} forms the augmented graph 
$F_{i} \gets (V,H_{i} \cup (S_2 \times V) \cup Q_{\tilde{S}_{i}})$ and applies the edge-rounding algorithm of 
Bernstein~\cite{BernsteinRounding} with hop-length target parameter $\beta + 2$ to obtain $\bigo{\log nW_{\max}}$ 
scaled copies $\{F_i^j\}$. Similarly, for each $i \in \{1,0\}$ and $u \in S_i$, \algref{+2w1+2w2:unbounded} forms the 
per-source edge sets $R_{i+1}(u) = (\singleton{u} \times \tilde{S}_{i+1}) \cup E(u)$ and 
$\hat{R}_{i+1}(u) = (\singleton{u} \times \tilde{S}_{i+1}) \cup (\singleton{u} \times V)$, and scales them accordingly 
to obtain $\{R_{i+1}^j(u)\}$ and $\{\hat{R}_{i+1}^j(u)\}$. For each $i \in \{1,0\}$ and $u \in S_i$, \algref{+2w1+2w2:unbounded} 
then maintains, for every scale $j$, a $\tau$-bounded Monotone ES-Tree with $\tau=\ceil{\frac{4(\beta + 2)}{\varepsilon}}$ rooted at 
$u$ on $F_{i+1}^j(u)= F_{i+1}^j \cup R_{i+1}^j(u)$ with level estimates $\level[j]{u}{\cdot}$, and a second Monotone ES-Tree on
 $\hat{F}_{i+1}^j(u)= F_{i+1}^j \cup \hat{R}_{i+1}^j(u)$ with level estimates $\levelh[j]{u}{\cdot}$. Finally, we set
  $d[u,v] = \mmin{\levelh[j]{u}{v},\levelh[j]{v}{u}}{j=1}{\ceil{\log nW_{\max}} + 1}$.

These modifications introduce only minor changes to the approximation and total update time analyses.

\begin{lemma}[Approximation Analysis]
    \label{lem:w+2w1+2w2:unbounded-approximation-analysis}
    Fix $u,v \in V$ and let $P_{u,v}$ be any path between $u$ and $v$. Then, at any time $t$,
    $$\dist{u,v} \le \mmin{\hat{\level[j]{u}{v}},\hat{\level[j]{v}{u}}}{j=1}{\ceil{\log nW_{\max}} + 1} \le (\onepluseps)\weight{P_{u,v}} + 2W_{1}\paren{P_{u,v}} + 2W_{2}\paren{P_{u,v}}$$
\end{lemma}
\begin{proof}
    The case analysis is similar to that of \lemref{+2w1+2w2:bounded:approximation-guarantee}.
    
    Note that in all examined paths, we use at most two auxiliary edges and then finally a path $P_{w,z} \subseteq E_i$ for some $w,z \in V$ and $i \in \{1,2\}$.
    Consequently, every relevant path has at most $\beta + 2$ hops in $H_i$. This step introduces a $(\onepluseps)$ error.
    An additional $(\onepluseps)$ error occurs by the trees of $S_2$, since we use a $(\onepluseps)$-SSSP algorithm rather than an exact ES-Tree.
    We denote the resulting path so far by $P'$.

    Subsequently, since we scale the graphs $F_i(u)$ and $\hat{F}_i(u)$, it suffices to maintain trees up to depth $\ceil{\frac{4\paren{\beta + 2}}{\varepsilon}}$. The rounding process introduces another $(\onepluseps)$ error in every path approximation.
    We denote the scaled path by $P''$.
    
    The case analysis remains structurally identical to that of \lemref{+2w1+2w2:bounded:approximation-guarantee} after these modifications, and the inductive claim applies for each scale separately, as done in \lemref{w+2w2:unbounded-approx}.
    Similarly to \lemref{w+2w2:unbounded-approx}, it should be stated that Bernstein's edge rounding process~\cite{BernsteinRounding}, a path with total weight $\le 2^j$ is guaranteed to exist also in the $\ell$-th scale for every $\ell > j$, and it is still guaranteed to be of total weight at most $\ceil{\frac{4\paren{\beta + 2}}{\varepsilon}}$ in $G_{\ell}$.
    However, the error after unscaling each edge is guaranteed to be $\onepluseps$ only for the $j$-th scale.
    This does not change the correctness of our claim, as we maintain the trees on the rounded graphs.

    Finally, since the resulting error is at most $(\onepluseps)^2$, we scale the error parameter to $\varepsilon' = \frac{\varepsilon}{16}$ and use the fact that for every path $P \subseteq G$ $2W_1\paren{P} + 2W_2\paren{P} \le 2\weight{P}$ to obtain the desired result.
\end{proof}

\begin{lemma}[Total Update Time Analysis]
    \label{lem:w+2w1+2w2:unbounded-total-update-time}
    For any $0 < \varepsilon < 1$, the expected total update time of \algref{+2w1+2w2:unbounded} is $\onolog{\onediveps[9] n^{\seventhirds}}$.
\end{lemma}
\begin{proof}
    We set $q = n^{-\frac{1}{3}}$ and $k = 3$.
    By \thmref{tools:hopset-on-es-dense}, the total number of edges ever present in $F$ is $\onolog{k \cdot n^{1+\frac{1}{k}}} = \onolog{n^{\frac{4}{3}}}$, with hopbound $\beta = \bigo{\paren{\frac{k\log n}{\varepsilon}}^k} = \onolog{\onediveps[4]}$, and maintaining $F$ requires expected total update time $\onolog{k \cdot \frac{\beta}{\varepsilon} \cdot n^{2+\frac{1}{k}}} = \onolog{\onediveps[4] n^{\seventhirds}}$.

    By \corref{tools:modified-es-unbounded} with $k=3$, the total number of edges ever inserted into $E_1$ and $E_2$ are bounded by $\mathcal{E}_{S_1} = \onolog{\frac{\beta}{\varepsilon}\frac{n}{q}} = \onolog{\onediveps[4] n^{\frac{4}{3}}}$ and $\mathcal{E}_{S_2} = \onolog{\frac{\beta}{\varepsilon}\frac{n}{q^2}} = \onolog{\onediveps[4] n^{\frac{5}{3}}}$, respectively, and their maintenance requires $\onolog{\frac{\beta}{\varepsilon} (m n^{\frac{1}{3}} + \frac{n}{q^2})} = \onolog{\onediveps[4] n^{\seventhirds}}$ expected total update time.
    From each $u \in S_2$, the $(\onepluseps)$-SSSP tree on $G$ requires $\onolog{m} \le \onolog{n^2}$ total update time, contributing $\onolog{\expectation{|S_2|} \cdot  n^2} = \onolog{n^{\frac{1}{3}} \cdot n^2} = \onolog{n^{\seventhirds}}$.

    For the Monotone ES-Trees, each tree is bounded to depth $\tau = \ceil{\frac{4(\beta+2)}{\varepsilon}} = \onolog{\onediveps[5]}$.
    
    For the $|S_0| = n$ trees at the base level, the graph has size $\mathcal{E}_{S_1} + |F| + \onolog{n^{\frac{4}{3}}} = \onolog{\onediveps[4] n^{\frac{4}{3}}}$, so the total update time of these trees is bounded by $\onolog{n \cdot \tau \cdot \mathcal{E}_{S_1}} = \onolog{n \cdot \onediveps[5] \cdot \onediveps[4] n^{\frac{4}{3}}} = \onolog{\onediveps[9] n^{\seventhirds}}$. 
    
    For the trees rooted at vertices of $S_1$, the graph has an expected size of $\onolog{\mathcal{E}_{S_2} + |F| + n + n \cdot \expectation{\abs{S_2}}} = \onolog{\onediveps[4] n^{\frac{5}{3}}}$, so the total update time of these trees is bounded by $\onolog{|S_1| \cdot \tau \cdot \mathcal{E}_{S_2}} = \onolog{n^{\frac{2}{3}} \cdot \onediveps[5] \cdot \onediveps[4] n^{\frac{5}{3}}} = \onolog{\onediveps[9] n^{\seventhirds}}$.
    Summing across all $\bigo{\log(nW_{\max})}$ scales, the total update time is bounded by $\onolog{\onediveps[9] n^{\seventhirds}}$.
\end{proof}

Together, \lemref{w+2w1+2w2:unbounded-approximation-analysis} and \lemref{w+2w1+2w2:unbounded-total-update-time} constitute \thmref{+2w1+2w2}.

\section{$\paren{\onepluseps,2\cdot\paren{k-1} W_{1}}$-APASP}
\label{sec:w+2kw1}
In the static setting, Cohen and Zwick~\cite{CohZwi1997} establish a $+2\sum_{i=1}^{k-1} W_i$ additive approximation APASP algorithm for weighted, undirected graphs, operating with a time complexity of $\onolog{n^{2-\frac{1}{k}}m^{\frac{1}{k}}}$. The algorithm is a generalization of their $+2W_2$-APASP algorithm.

In this section, we present an $\paren{\onepluseps,2(k-1)W_1}$-APASP algorithm for decremental graphs with an expected total update time of $\onolog{\paren{\frac{k \log n}{\varepsilon}}^{\bigo{k}} n^{2-\frac{1}{k}}m^{\frac{1}{k}}}$ if $m=n^{1+\Omega(1)}$ or $\onolog{n^{2-\frac{1}{k}+\subpol}m^{\frac{1}{k}}}$ otherwise, which is a decremental variant of their framework, and establish the following theorem:
\thmtwokheaviest*
Note that the total update time is always bounded by $n^{2-\frac{1}{k}+\subpol}m^{\frac{1}{k}}$.

Similarly to Cohen and Zwick, \algref{w+2kw1:unbounded} presented in this section is a generalization of our $+2W_2$ decremental algorithm presented in \algref{w+2w2:unbounded}.

There is a notable difference in the additive approximation guarantees of our algorithm and its static counterpart: $2\cdot\paren{k-1}W_{1}$ as opposed to $+2\msum{W_{i}}{i=1}{k-1}$. The reason for that is discussed  following the description of \algref{w+2kw1:bounded} .

\subsection{Sparse Graphs}
\label{sec:w+2kw1-sparse}

Following the same structure of the previous algorithms, we begin with a $\tau$-bounded depth algorithm and generalize it to an unbounded depth algorithm.
\subsubsection{Bounded Depth}
\label{sec:w+2kw1-sparse:bounded}

\begin{algorithm}[t]
\caption{$+2\paren{k-1}W_{1}$-APASP up to depth $\tau$}
\label{alg:w+2kw1:bounded}

\BlankLine

 \textbf{Input:} A weighted undirected graph $G = \paren{V,E,\weight}$, a parameter $k$ and a depth parameter  $\tau$.

\textbf{Output:} A $+2\paren{k-1}W_{1}$-APASP for any pair of vertices whose distance $\leq \tau$.

\hrulefill

Initialize $d$ as the edge weights and $\infty$ otherwise

Let $q\in\paren{0,1}$ be fixed later

$V = S_0,S_1,\dots,S_{k-1} \gets \mathcal{R}(q,k)$

\ForEach{$i \in \{0,\dots,k-1\}$ in a descending order}
{
    Maintain an ES-Tree up to depth $(2\paren{k-i-1})\tau$ for the distances from any $u\in V$ to its pivot $\piv[i]{u}$ and update distances to $\auxpiv[i]{u}$
    
    Maintain the set $E_i  \gets E_{S_i}^{(2\paren{k-i-1})\tau}$
}

\ForEach{$s \in S_{k-1}$}
        {
        Maintain an ES-Tree from $u$ up to depth $\tau$ in $G$ and update $d$ accordingly.
    }

\ForEach{$i \in \{0,\dots,k-1\}$ in a descending order}
{
    Let $\tilde S_i$ be the set of auxiliary pivots $\auxpiv[i]{u}$ for all $u\in V$

    $\tilde S \gets \cup_{i=1}^{k-1} \tilde S_i$

    $\tilde Q_i \gets \bracce{\paren{u,\auxpiv[i]{u}}}$

    $\tilde Q = \cup_{i=1}^{k-1} \tilde Q_i$
    
    \ForEach{$u \in S_i$}
            {
            $F_{i+1}(u)\gets \paren{V,E_{i+1} \cup \tilde Q \cup (\singleton{u} \times \tilde{S}) \cup E(u)}$
            
            Maintain a Monotone ES-Tree from $u$ up to depth $(2\paren{k-i-1})\tau$ in $F_{i+1}(u)$ and update $d$ accordingly.
        }
}

\end{algorithm}

\paragraph{Algorithm Description.}
\algref{w+2kw1:bounded} extends the hierarchical approach established in \algref{w+2w2:bounded} to achieve an additive error bounded by $+2\paren{k-1}W_{1}$ for paths of weight at most $\tau$.

The algorithm receives a weighted undirected graph $G=(V,E,\weight)$, an integer approximation parameter $k$, and a depth parameter $\tau$. It begins by initializing a $k$-level hierarchy of vertex sets $V = S_0 \supseteq S_1 \supseteq \dots \supseteq S_{k-1}$, where each successive set $S_i$ is formed by sampling vertices from $S_{i-1}$ independently with probability $q$.

The algorithm also maintains, for every vertex $u$ and every $1 \le j \le k-1$, its $j$-th pivot $\piv[j]{u}$ and maintains the hierarchy-induced edge sets $E_i = E_{S_i}^{\paren{2\paren{k-i-1}}\tau}$ as described in \secref{hierarchy}.

For the top-level vertices $s \in S_{k-1}$, exact ES-Trees are maintained on the original graph $G$ up to depth $2\tau$ to provide accurate baseline distance estimates.

To compute the final approximated distances, the algorithm constructs a Monotone ES-Tree from each vertex $u \in S_i$, iterating sequentially downward through the hierarchy levels. These Monotone ES-Trees maintain distances up to depth $\paren{2\paren{k-i-1}}\tau$ over the graph $F_{i+1}\paren{u} = \paren{V, E_{i+1} \cup \tilde{Q} \cup (\{u\} \times \tilde{S}) \cup E(u)}$, where $\tilde S$ is the collection of all auxiliary pivots and $\tilde Q$ contains auxiliary edges that link every vertex to the auxiliary representations of its $k-1$ pivots.

This construction explicitly generalizes the single-level pivot mechanism introduced in \algref{w+2w2:bounded}. 
Similarly to \algref{w+2w2:bounded}, the $k$-level hitting set hierarchy allows us to maintain $k-1$ hierarchy-induced edge sets 
$E_1,\dots,E_{k-1}$ with sizes proportional to $S_1,\dots,S_{k-1}$, and the sets $\tilde S$ and $\tilde Q$ are used 
in $F_{i+1}\paren{\cdot}$ to skip edges absent from $E_{i+1}$. This mechanism guarantees that the algorithm accumulates an 
additive error of at most $+2W_{1}$ per level transition, ultimately yielding the cumulative $+2\paren{k-1}W_{1}$ stretch bound.

\paragraph{Comparison with the Static Algorithm of Cohen and Zwick~\cite{CohZwi1997}}
Our decremental algorithm achieves a $+2(k-1)W_1$ additive error, 
which is weaker than the $+2\sum_{i=1}^{k-1} W_{i}$ bound of the static algorithm of
Cohen and Zwick. We now explain why the static analysis does not carry over directly
 and how the decremental setting leads to a weaker error bound.

\subparagraph{Static analysis.}
Fix $u \in S_{i}$ and a path $P_{u,v}$ with $\sizeindex{P_{u,v}}{E_{i+1}} > 0$. 
Let $e' = (x, y) = \leftabovearrow{P_{u,v}}{E_{i+1}}$ be the last edge on $P_{u,v}$ not in $E_{i+1}$, and let $j > i$ be the unique index with $e' \in E_{j+1} \setminus E_j$.%
Since $e' \notin E_j$, the pivot $w = \piv[j]{y}$ satisfies $\dist{y, w} \le \weight{e'}$. The static algorithm routes \footnote{The hierarchy in \cite{CohZwi1997} is deterministic, the edge sets $E_i$ are formed by the lightest $q^{-i}$ edges and the vertex order is reversed. We have adapted the presentation to match our conventions. Also, we use an implied any-path evaluation instead of shortest paths.} the estimate from $u$ to $v$ via the path
  $$P_{u,v}^* = u \squiggly w \circ w \squiggly y \circ P_{y,v}.$$
The sub-path $P_{y,v}$ is entirely contained in $E_{j+1}$. Let $P_{w,u} = w \squiggly y \circ P_{u,v}$. By induction, $w$'s distance estimate for $u$ contributes an additive error of $+2\sum_{e \in \varphi_{k-j-1}(P_{w,u})}\weight{e}$, where $\varphi_{k-j-1}\paren{P_{w,u}}$ is a set of $k-j-1$ edges not in $E_{j+1}$.
Since every edge on $y \squiggly w$ belongs to $E_{j+1}$, $\varphi_{k-j-1}\paren{P_{w,u}} \cap y \squiggly w = \emptyset$
and hence $\varphi_{k-j-1}\paren{P_{w,u}} \subseteq P_{u,v}$.

As $\dist{y,\piv[j]{y}} \le \weight{e'}$, and as $e' \in E_{j+1}$, it follows that $e' \notin \varphi_{k-j-1}\paren{P_{u,v}}$ and the total additive error incurred at this level is $+2\weight{e'} +2\sum_{e \in \varphi_{k-j-1}(P_{w,u})}\weight{e}$.
Notably, because $e' \notin \varphi_{k-j-1}(P_{w,u})$, the edge $e'$ is distinct from all edges charged for error in $P_{w,u}$,
completing the inductive argument.
Hence, the static analysis counts each path edge at most once, giving a final bound of $+2\sum_{i=1}^{k-1} W_i(P_{u,v})$.

\subparagraph{Decremental complication.}
We generalize the same route $P_{u,v}^*$ to the decremental setting via the auxiliary edges
of $\tilde{Q}$ and $\tilde{S}$ in $F_{i+1}(u)$. However, as edges are deleted from $G$,
the sets $E_i$ are monotonically expanding: an edge may enter $E_j$ at some time $t > 0$, \emph{after}
initialization. This creates the following difficulty:

Let $t$ be the current time and suppose that $u$ is stretched in $w=\piv[j]{y}$'s tree. Consequently, $\level{u}{\auxpiv[j]{y}}$ reflects
the level $\level[t']{w}{u}$, computed via the path $P_{w,u} = P_{u,y} \circ (y \squiggly w)$ at time $t'$. 
The detour $y \squiggly w$ was a shortest path \emph{at time $t'$} but, if $w$ was not $y$'s $j$'th pivot at time $t'$,
we cannot use \lemref{tools:bk-es-properties} to guarantee that $y \squiggly w$ was indeed contained in $E_{j+1}$ at time $t'$.
As a result, the edges of $P_{w,u}$ that contribute to the error are \emph{not}
necessarily edges of $P_{u,v}$, so the static counting argument breaks down.

To recover a valid bound, we use the pivot guarantee that holds at \emph{current} time $t$:
since $e' = (x,y) \notin E_j$ at time $t$, the current pivot $w = \piv[j]{y}$
satisfies $\dist{y, w} \leq \weight{e'} \leq W_1(P_{u,v})$. We therefore bound all
error-inducing edges - regardless of whether they lie on $P_{u,v}$ or on a detour outside it
- by $W_1(P_{u,v})$. This yields a per-level additive error of $+2W_1(P_{u,v})$,
and summing over all $k-i-1$ levels gives the final bound of $+2(k-i-1)W_1(P_{u,v})$.

\tikzstyle{pnode}   = [circle, draw=black, fill=white,   inner sep=2pt, minimum size=6pt]
\tikzstyle{pivnode} = [circle, draw=black, fill=gray!30, inner sep=2pt, minimum size=6pt]

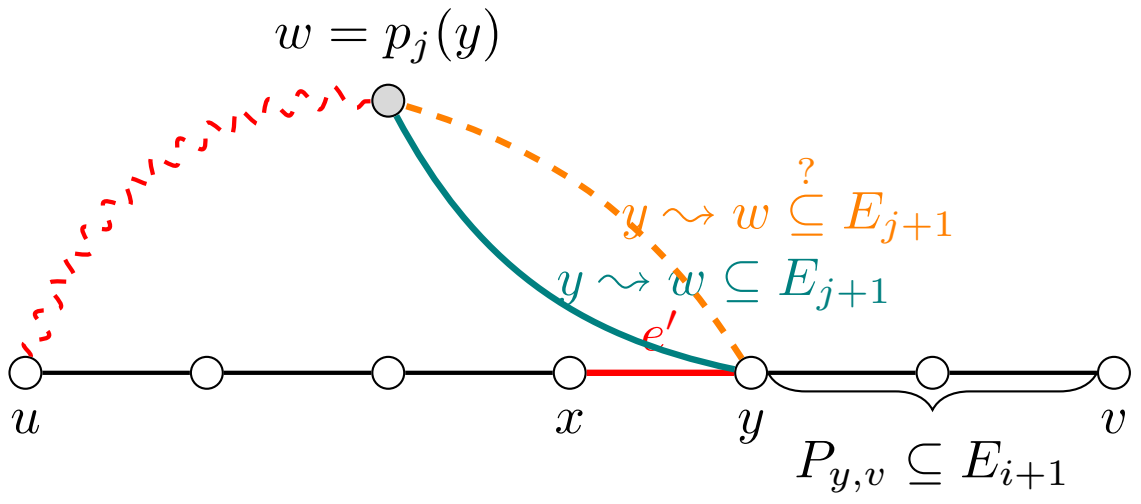
\begin{figure}[htbp]
\centering
\begin{tikzpicture}[>=stealth, scale=1.0,
  thinpath/.style  = {thick, black},
  pathcurve/.style = {thick, red, dashed,
                      decorate,
                      decoration={snake, amplitude=2pt, segment length=6pt}},
  static/.style    = {very thick, teal},
  stale/.style     = {very thick, orange, dashed},
  annot/.style     = {font=\small},
]

\begin{scope}

  \node[pnode, label=below:{\small$u$}] (u) at (0,   0) {};
  \node[pnode]                          (a1) at (1.2, 0) {};
  \node[pnode]                          (a2) at (2.4, 0) {};
  \node[pnode, label=below:{\small$x$}] (x)  at (3.6, 0) {};
  \node[pnode, label=below:{\small$y$}] (y)  at (4.8, 0) {};
  \node[pnode]                          (b1) at (6.0, 0) {};
  \node[pnode, label=below:{\small$v$}] (v)  at (7.2, 0) {};

  \node[pivnode, label=above:{\small$w = p_j(y)$}] (w) at (2.4, 1.8) {};

  \draw[thinpath] (u)  -- (a1);
  \draw[thinpath] (a1) -- (a2);
  \draw[thinpath] (a2) -- (x);

  \draw[very thick, red] (x) -- node[above, annot]{$e'$} (y);

  \draw[thinpath] (y)  -- (b1);
  \draw[thinpath] (b1) -- (v);

  \draw[pathcurve] (u) to[bend left=35] (w);

  \draw[static] (y) to[bend left=25]
    node[right, annot, text=teal]{$y \leadsto w \subseteq E_{j+1}$} (w);

  \draw[stale] (y) to[bend right=20]
    node[right, annot, text=orange]{$y \leadsto w \overset{?}{\subseteq} E_{j+1}$} (w);

  \draw[decorate, decoration={brace, amplitude=7pt, mirror}]
    (y) -- node[below=9pt, annot]{$P_{y,v} \subseteq E_{i+1}$} (v);

\end{scope}

\end{tikzpicture}
\caption{%
  The path $P_{u,v}$ is drawn horizontally with $u \in S_i$.
  The edge $e'$ is the last edge of
  $P_{u,v}$ absent from $E_{i+1}$; $j > i$ is the unique index with
  $e' \in E_{j+1} \setminus E_j$.
  The pivot $w = \piv[j]{y}$ satisfies $\dist{y,w} \le \weight{e'}$.
  The red arc is the path $u \leadsto w$ of weight at most
  $\weight{P_{u,y}} + \dist{y,w}$; the teal and orange arcs
  represent how the time dimension affects $y \leadsto w$.
  The time dimension has an effect on whether $y \leadsto w$ is contained
  within $E_{j+1}$ or not when $\level{w}{u}$ is determined, whereas
  in the static setting it is known with certainty that
  $y \leadsto w \subseteq E_{j+1}$, guaranteeing that the additive error
  is derived from distinct edges of $P_{u,v} \setminus \singleton{e'}$.
}
\label{fig:w+2kw1:static-vs-decremental}
\end{figure}

We first establish the approximation guarantee at initialization.

\begin{lemma}
    \label{lem:w+2kw1:approximation-guarantee-t0}
    Fix $0 \le i \le k-1$, $u \in S_i$, $v \in V$, and a path $P_{u,v}$ between $u$ and $v$ with $\weight{P_{u,v}} \le \tau$. Then, at the beginning of the algorithm, $\level{u}{v} \le \dist{u,v} + 2(k-i-1) \cdot W_1\paren{P_{u,v}}$.
\end{lemma}

\begin{proof}
    We prove the claim by induction on $k-i-1$.

    For the base case of $i=k-1$, \algref{w+2kw1:bounded} maintains exact ES-Trees on the original graph $G$ from the vertices of $S_{k-1}$. Consequently, for every $u\in S_{k-1}$ and $v\in V$, $\level{u}{v}=\dist{u,v}$, establishing the claim for $i=k-1$.
    
    We assume the claim holds for every $j$ such that $i<j\le k-1$ and prove it for $i$. Let $P_{u,v}$ be a path connecting $u\in S_{i}$ and $v\in V$ such that $\weight{P_{u,v}}\le\tau$. We consider two cases:
    
    \ccase{1} $\sizeindex{P_{u,v}}{ E_{i+1}} = 0 $.
    
As $P_{u,v}\subseteq E_{i+1}$ 
it follows that   the Monotone ES-Tree rooted at $u$ on $F_{i+1}\paren{u}$ contains $E_{i+1}$ and hence the path $P_{u,v}$. Therefore, $\dist{u,v}{F_{i+1}\paren{u}}\le \weight{P_{u,v}}$. As this is the beginning of the algorithm, $\level{u}{v}=\dist{u,v}{F_{i+1}\paren{u}} \le \weight{P_{u,v}}$, and the claim holds.
    
    \ccase{2}  $\sizeindex{P_{u,v}}{ E_{i+1}} > 0 $.
    
     Let $e'=\paren{u',v'}=\leftabovearrow{P_{u,v}}{E_{i+1}}$ and let $i<j\le k-1$ be the distinct index for which $e'\in E_{j+1}\setminus E_{j}$.
    As $e' \notin E_{j}$, there exists a vertex $w= \piv[j]{v'} \in S_{j}$ such that $\dist{v',w} \le \weight{e'}$.
    Correspondingly, $F_{i+1}\paren{u}$ contains the following auxiliary edges:
    \begin{itemize}
        \item $\paren{\auxpiv[j]{v'},v'}$ with weight $\weight{\auxpiv[j]{v'},v'} \le \dist{w,v'}$
        \item $\paren{u,\auxpiv[j]{v'}}$ with weight $\weight{u,\auxpiv[j]{v'}} = \level{w}{u}$
    \end{itemize}
    
    Let $P_{w,u}=P_{u,v'}\circ v' \squiggly w \subseteq G$. Since $w\in S_{j}$, it follows from the induction hypothesis that $\level{w}{u}\le \weight{P_{w,u}}+2\paren{k-j-1}W_{1}\paren{P_{w,u}}$.
    
    We first show that $W_{1}\paren{P_{w,u}} \le W_{1}\paren{P_{u,v}}$. Consider the heaviest edge $e^*$ on $P_{w,u}$:
    
    If $e^*\in P_{u,v'}$, then $W_{1}\paren{P_{w,u}}=W_{1}\paren{P_{u,v'}}\le W_{1}\paren{P_{u,v}}$. Conversely, if $e^*\in v'\squiggly w$, then $\weight{e^*}\le\dist{v',w}\le \weight{e'}\le W_{1}\paren{P_{u,v}}$.
    
    In either case, $\level{w}{u}\le \weight{P_{w,u}}+2\paren{k-j-1}W_{1}\paren{P_{u,v}}$.
    
    Finally, consider the path $P_{u,v}^{*}=\paren{u,\auxpiv[j]{v'}} \circ \paren{\auxpiv[j]{v'},v'} \circ P_{v',v} \subseteq F_{i+1}\paren{u}$. The total weight of $P_{u,v}^{*}$ can be bounded by:
    
    \begin{align*}
    &\weight{P_{u,v}^{*}} \le \level{w}{u}+\dist{w,v'}+\weight{P_{v',v}} \\
    &\le \weight{P_{w,u}}+2\paren{k-j-1}W_{1}\paren{P_{u,v}}+\dist{w,v'}+\weight{P_{v',v}} \\
    &\le \weight{P_{u,v'}}+\dist{w,v'}+2\paren{k-j-1}W_{1}\paren{P_{u,v}}+\dist{w,v'}+\weight{P_{v',v}} \\
    &\le \weight{P_{u,v}}+2\dist{w,v'}+2\paren{k-j-1}W_{1}\paren{P_{u,v}} \\
    &\le \weight{P_{u,v}}+2\weight{e'}+2\paren{k-j-1}W_{1}\paren{P_{u,v}} \\
    &\le \weight{P_{u,v}}+2\paren{k-j}W_{1}\paren{P_{u,v}} \\
    &\le \weight{P_{u,v}}+2\paren{k-i-1}W_{1}\paren{P_{u,v}}
    \end{align*}
    
    Since this is the beginning of the algorithm, $\level{u}{v}\le \weight{P_{u,v}^{*}}$.
\end{proof}

We now show that the guarantee to hold at every time step. Using double induction on time $t$ and hierarchy level $i$, the following lemma establishes that the Monotone ES-Tree estimates remain within the $+2(k-i-1)W_1$ additive bound throughout all updates.

\begin{lemma}[Approximation Guarantee]
    \label{lem:w+2kw1:approximation-guarantee}
    Fix $0 \le i \le k-1$, $u \in S_i$, $v \in V$, and a path $P_{u,v}$ between $u$ and $v$ with $\weight{P_{u,v}} \le \tau$. Then, at any time $t$, $\level{u}{v} \le \dist{u,v} + 2(k-i-1) \cdot W_1\paren{P_{u,v}}$.
\end{lemma}

\begin{proof}
    We prove the claim by double induction on $t$ and $k-i-1$.

    The base case of $t=0$ was established in \lemref{w+2kw1:approximation-guarantee-t0}. We assume the claim holds for any time $t' < t$ and prove it for time $t$.

    For the base case of $i=k-1$, \algref{w+2kw1:bounded} maintains exact ES-Trees on the original graph $G$ from the vertices of $S_{k-1}$. Consequently, for every $u\in S_{k-1}$ and $v\in V$, $\level{u}{v}=\dist{u,v}$ at any time, so the claim holds for $i=k-1$.
    
    We assume the claim holds for every $j$ such that $i<j\le k-1$ and prove it for $i$. Let $P_{u,v}$ be a path connecting $u\in S_{i}$ and $v\in V$ such that $\weight{P_{u,v}}\le\tau$. We consider two cases:
    
    \ccase{1} $\sizeindex{P_{u,v}}{ E_{i+1}} = 0 $.
    
    As $P_{u,v}\subseteq E_{i+1}$
    the Monotone ES-Tree from $u$ on $F_{i+1}\paren{u}$ contains $E_{i+1}$ and hence the path $P_{u,v}$. Therefore, $\dist{u,v}{F_{i+1}\paren{u}}\le \weight{P_{u,v}}$.

    If none of the edges of $P_{u,v}$ are stretched in $u$'s tree, then $\level{u}{v} \le \weight{P_{u,v}}$.

    Otherwise, let $\paren{z',z}$ be the last stretched edge on $P_{u,v}$ in $u$'s tree, traversing from $u$ to $v$. Since $\paren{z',z}$ is stretched, $\level{u}{z} = \level[t-1]{u}{z}$, and by the induction hypothesis, $\level[t-1]{u}{z} \le \weight{P_{u,z}} + 2(k-i-1)W_1\paren{P_{u,z}}$.

    Since none of the edges on $P_{z,v}$ are stretched in $u$'s tree, it holds that $\level{u}{v} \le \level{u}{z} + \weight{P_{z,v}} \le \weight{P_{u,z}} + 2(k-i-1)W_1\paren{P_{u,z}} + \weight{P_{z,v}} \le \weight{P_{u,v}} + 2(k-i-1)W_1\paren{P_{u,v}}$.

    \ccase{2} $\sizeindex{P_{u,v}}{ E_{i+1}} > 0 $.
    
    Let $e'=\paren{u',v'}=\leftabovearrow{P_{u,v}}{E_{i+1}}$ and let $i<j\le k-1$ be the distinct index for which $e'\in E_{j+1}\setminus E_{j}$.
    Since $e' \notin E_{j}$, there exists a vertex $w= \piv[j]{v'} \in S_{j}$ such that $\dist{v',w} \le \weight{e'}$.
    Correspondingly, $F_{i+1}\paren{u}$ contains the following auxiliary edges:
    \begin{itemize}
        \item $\paren{\auxpiv[j]{v'},v'}$ with weight $\weight{\auxpiv[j]{v'},v'} \le \dist{w,v'}$
        \item $\paren{u,\auxpiv[j]{v'}}$ with weight $\weight{u,\auxpiv[j]{v'}} = \level{w}{u}$
    \end{itemize}

    These are structural graph properties that hold at any time $t$. Let $P_{w,u}=P_{u,v'}\circ v' \squiggly w \subseteq G$. Since $w\in S_{j}$, it follows from the induction hypothesis that $\level{w}{u}\le \weight{P_{w,u}}+2\paren{k-j-1}W_{1}\paren{P_{w,u}}$.
    
    Following the same arguments as in \lemref{w+2kw1:approximation-guarantee-t0}, it holds that $W_{1}\paren{P_{w,u}} \le W_{1}\paren{P_{u,v}}$.
    
    Consider the path $P_{u,v}^{*}=\paren{u,\auxpiv[j]{v'}} \circ \paren{\auxpiv[j]{v'},v'} \circ P_{v',v} \subseteq F_{i+1}\paren{u}$. The total weight of $P_{u,v}^{*}$ can be bounded by:
    
    \begin{align*}
    &\weight{P_{u,v}^{*}} \le \level{w}{u}+\dist{w,v'}+\weight{P_{v',v}} \\
    &\le \weight{P_{w,u}}+2\paren{k-j-1}W_{1}\paren{P_{u,v}}+\dist{w,v'}+\weight{P_{v',v}} \\
    &\le \weight{P_{u,v'}}+\dist{w,v'}+2\paren{k-j-1}W_{1}\paren{P_{u,v}}+\dist{w,v'}+\weight{P_{v',v}} \\
    &\le \weight{P_{u,v}}+2\dist{w,v'}+2\paren{k-j-1}W_{1}\paren{P_{u,v}} \\
    &\le \weight{P_{u,v}}+2\weight{e'}+2\paren{k-j-1}W_{1}\paren{P_{u,v}} \\
    &\le \weight{P_{u,v}}+2\paren{k-j}W_{1}\paren{P_{u,v}} \\
    &\le \weight{P_{u,v}}+2\paren{k-i-1}W_{1}\paren{P_{u,v}}
    \end{align*}
    
    If none of the edges on $P_{u,v}^*$ are stretched in $u$'s tree, then $\level{u}{v} \le \weight{P_{u,v}^*} \le \weight{P_{u,v}}+2\paren{k-i-1}W_{1}\paren{P_{u,v}}$.

    Otherwise, let $\paren{z',z}$ be the last stretched edge of $P_{u,v}^*$ in $u$'s tree. We examine two cases:

        \ccase{2.1} $z = \auxpiv[j]{v'}$.
        
        The edge $\paren{u,\auxpiv[j]{v'}}$ can be stretched only when $\piv[j]{v'}$ changes from $w \in S_j$ to some $w' \in S_j$ at time $t' \le t$ and $\level{w'}{u} < \level{w}{u}$, in which case $\level{u}{\auxpiv[j]{v'}} = \level[t']{w}{u}$.
        
        By construction, edges cannot leave $E_j$, and hence it must be that $e' \notin E_j$ at time $t'$. Following the same reasoning, $\level{u}{\auxpiv[j]{v'}} = \level{w}{u} \le \weight{P_{u,v'}} + \weight{e'} +2\paren{k-j-1}W_{1}\paren{P_{u,v}}$.

        Since $\paren{u,\auxpiv[j]{v'}}$ is the last stretched edge on $P_{u,v}^* = \paren{u,\auxpiv[j]{v'}} \circ \paren{\auxpiv[j]{v'},v'} \circ P_{v',v}$ in $u$'s tree, the remaining sub-path is computed exactly:
        \begin{align*}
            &\level{u}{v} \le \level{u}{\auxpiv[j]{v'}} + \dist{v',\piv[j]{v'}}{t} + \weight{P_{v',v}} \\
            &\le \weight{P_{u,v'}} + \weight{e'} +2\paren{k-j-1}W_{1}\paren{P_{u,v}} + \dist{v',\piv[j]{v'}}{t} + \weight{P_{v',v}}\\
            &\le \weight{P_{u,v}} + 2\weight{e'} +2\paren{k-j-1}W_{1}\paren{P_{u,v}}\\
             &\le \weight{P_{u,v}} + 2\paren{k-j}W_{1}\paren{P_{u,v}}\\
             &\le \weight{P_{u,v}} + 2\paren{k-i-1}W_{1}\paren{P_{u,v}}
        \end{align*}
        
        \ccase{2.2} $z = v'$ or $\paren{z',z} \in P_{v',v}$.
        
        Note that in both sub-cases, $P_{z,v} \subseteq P_{v',v}$.
        By the Monotone ES-Tree properties and the induction hypothesis on time, $\level{u}{z} = \level[t-1]{u}{z} \le \weight{P_{u,z}} +2\paren{k-i-1}W_{1}\paren{P_{u,z}}$, where $P_{u,z}$ is the portion of $P_{u,v}$ between $u$ and $z$.

        Since $\paren{z',z}$ is the last stretched edge in $u$'s tree, the tree accurately computes the remaining portion of the path, guaranteeing $\level{u}{v} \le \weight{P_{u,z}} +2\paren{k-i-1}W_{1}\paren{P_{u,z}} + \weight{P_{z,v}} \le \weight{P_{u,v}} +2\paren{k-i-1}W_{1}\paren{P_{u,v}}$.
\end{proof}

\lemref{w+2kw1:approximation-guarantee} establishes the correctness of \algref{w+2kw1:bounded}. We now address the total update time of the algorithm.

\begin{lemma}[Total Update Time Analysis]
\label{lem:w+2kw1:total-update-time}
The expected total update time of \algref{w+2kw1:bounded} is $\onolog{k^2 \tau^2 n^{2-\frac{1}{k}} m^{\frac{1}{k}}}$.
\end{lemma}
\begin{proof}
    Fix $0 \le i \le k-1$.
    
    Observe that $\mathbb{E}\bracke{\abs{S_i}} = \onolog{nq^i}$.
    It follows from \corref{tools:es-max-edges} that the expected total number of edges in $E_{i}$ throughout all edge updates is bounded by $\onolog{(2\paren{k-i-1}) \tau \cdot \frac{n}{q^i}}$.

    Consider the graph $F_{i+1}\paren{u}$ utilized by the Monotone ES-Tree rooted at $u$. Apart from $E_{i+1}$, there are only $\onolog{kn}$ additional auxiliary edges.

    Since we assume integer edge weights, each edge - whether original or auxiliary - represents a path of weight at most $\paren{2\paren{k-i-1}}\tau$ and thus undergoes at most $\paren{2\paren{k-i-1}}\tau$ weight updates.

    By \lemref{prelim:monotone-es-update-time}, the total update time of each Monotone ES-Tree rooted at $u \in S_i$ is $\onolog{k \tau ^2 \frac{n}{q^{i+1}}}$, and for each ES-Tree rooted at $s \in S_{k-1}$ is $\onolog{\tau m}$. Since there are $\onolog{nq^{i}}$ vertices in $S_{i}$ in expectation, maintaining the Monotone ES-Trees across all $u \in S_{i}$ requires an expected total update time of $\onolog{k\tau^{2}\frac{n^{2}}{q}}$.

    Similarly, for each $s \in S_{k-1}$, we maintain a standard ES-Tree on the original graph $G$, requiring an update time of $\onolog{\tau m}$ per tree. Summing over all $s \in S_{k-1}$ yields an expected total cost of $\onolog{nq^{k-1}\tau m}$.

    Maintaining the nested sets $E_{i}$ and computing the auxiliary pivots for all vertices requires an additional $\onolog{km}$ update time.

    Summing across all of the $k$ levels of the hierarchy, the overall expected total update time becomes $\onolog{k^2 \tau^2\frac{n^2}{q} +\tau m \cdot n q^{k-1} + km}$.
    By setting the sampling probability to $q= \paren{\frac{n}{m}}^{\frac{1}{k}}$, the total update time resolves to $\onolog{k^2 \tau^2 n^{2-\frac{1}{k}} m^{\frac{1}{k}}}$.
\end{proof}

\subsubsection{Unbounded Depth}
\label{sec:w+2kw1-sparse:unbounded}
Following the structure of the previous algorithms, we now generalize \algref{w+2kw1:bounded} to explore distances up to unbounded length.

\begin{algorithm}[h]
\caption{$\paren{\onepluseps, 2\paren{k-1}W_1}$-APASP$\paren{G=(V,E,\weight)}$}
\label{alg:w+2kw1:unbounded}
\textbf{Input:} A weighted, undirected graph $G=(V,E,\weight)$, an approximation parameter $k$, and an error parameter $\varepsilon > 0$.

\textbf{Output:} A matrix $\{d[u,v]\}_{u,v \in V}$ of estimated distances.

Initialize $d$ as the edge weights and $\infty$ otherwise

Let $q = \paren{\frac{n}{m}}^{\frac{1}{k}}$.

$V = S_0, S_1, \dots, S_{k-1} \gets \mathcal{R}(q,k)$

\For{$i = k-1$ down to $0$}
{
    Maintain the set $E_i \gets E_{S_i}$ and pivot distances using \corref{tools:modified-es-unbounded} with parameter $k$, and update distances to $\auxpiv[i]{u}$
}

Construct a $(\beta, \varepsilon)$-hopset $F$ using \thmref{tools:hopset-on-es-sparse} with parameter $k$

\ForEach{$i \in \bracke{k}$}
{
    $H_i \gets E_i \cup F$
}

\ForEach{$s \in S_{k-1}$}
{
    Maintain a $\paren{\onepluseps}$-SSSP algorithm from $s$ on $G$ and update $d$ accordingly
}

\For{$i = k-1$ down to $0$}
{
    Let $\tilde{S}_i$ be the set of auxiliary pivots $\auxpiv[i]{u}$ for all $u \in V$
    
    $\tilde{S} \gets \bigcup_{j=1}^{k-1} \tilde{S}_j$
    
    $\tilde{Q}_i \gets \{(u, \auxpiv[i]{u}) \mid u \in V\}$
    
    $\tilde{Q} \gets \bigcup_{j=1}^{k-1} \tilde{Q}_j$
    
    $F_{i+1} \gets (V, H_{i+1} \cup \tilde{Q})$
    
    $\{F_{i+1}^j\}_{j=1}^{\lceil \log nW_{\max} \rceil + 1} \gets \textit{SCALE}(F_{i+1}, \onepluseps, \beta+2)$
    
    \ForEach{$u \in S_i$}
    {
        $R_{i+1}(u) \gets (\{u\} \times \tilde{S}) \cup E(u)$
        
        $\{R_{i+1}^j(u)\}_{j=1}^{\lceil \log nW_{\max} \rceil + 1} \gets SCALE(R_{i+1}(u), \onepluseps, \beta+2)$
        
        \For{$j = 1$ to $\lceil \log nW_{\max} \rceil + 1$}
        {
            Maintain a Monotone ES-Tree from $u$ up to depth $\ceil{\frac{4\paren{\beta+2}}{\varepsilon}}$ in $F_{i+1}^j \cup R_{i+1}^j(u)$ with distance estimates $\level[j]{u}{\cdot}$ and update $d$ accordingly
        }
    }
}
\end{algorithm}

\paragraph{Algorithm Description.}
\algref{w+2kw1:unbounded} generalizes \algref{w+2kw1:bounded} to handle unbounded distances using the same strategy as the previous 
unbounded algorithms. The key difference from \algref{+2w1+2w2:unbounded} is that we operate in the sparse-graph regime, 
so \algref{w+2kw1:unbounded} invokes \thmref{tools:hopset-on-es-sparse} with parameter $k$ rather than \thmref{tools:hopset-on-es-dense} 
to produce a $(\beta, \onepluseps)$-hopset $F$ with hopbound $\beta = \bigo{\paren{\frac{k \log n}{\varepsilon}}^{\bigo{k}}}$, and 
sets $H_i = E_i \cup F$ for each level $i$. Due to the choice of $k = \bigo{\sqrt{\log n}}$ and $\varepsilon$, $\beta$ 
is always bounded by $n^{\subpol}$, and is polylogarithmic whenever $k$ and $\varepsilon$ are constants. 
Each hierarchy-induced edge set $E_i$ and its pivot distances
are maintained via \corref{tools:modified-es-unbounded} with parameter $k$. For each $i$, \algref{w+2kw1:unbounded} forms 
$F_{i+1} \gets (V, H_{i+1} \cup \tilde{Q})$, scales it to $\{F_{i+1}^j\}$ with hop-length target $\beta + 2$, and scales the per-source
edge set $R_{i+1}(u) = (\{u\} \times \tilde{S}) \cup E(u)$ for each $u \in S_i$. For each scale $j$, \algref{w+2kw1:unbounded} 
maintains a $\tau$-bounded Monotone ES-Tree with $\tau = \ceil{\frac{4(\beta+2)}{\varepsilon}}$ from every $u \in S_i$ on 
$F_{i+1}^j \cup R_{i+1}^j(u)$, and sets $d[u,v]$ to the minimum level estimate across all scales.

\begin{lemma}
\label{lem:w+2kw1:approximation-guarantee-unbounded}
Fix $u, v \in V$, and let $P_{u,v}$ be any path between $u$ and $v$. Then, at any time $t$, the distance estimate $d[u,v]$ satisfies 
$\dist{u,v} \le d[u,v] \le \paren{\onepluseps} \weight{P_{u,v}} + 2\paren{k-1}W_{1}(P_{u,v})$.
\end{lemma}

\begin{proof}
The case analysis parallels that of \lemref{w+2kw1:approximation-guarantee}. By \thmref{tools:hopset-on-es-sparse}, the hopset 
$F$ guarantees the existence of a corresponding path $P'_{u,v} \subseteq H_{i}$ with $\hops{P'_{u,v}} \le \beta$ and 
$\weight{P'_{u,v}} \le \paren{\onepluseps}\weight{P_{u,v}}$. It should be noted that $H_i$ preserves the property that for every 
edge $\paren{x,y} \notin H_i$,
$\dist{x,\piv[i]{x}} \le \weight{x,y}$, ensuring the same additive error bound is maintained.

When executing the bounded algorithm on the scaled graphs, the alternative paths constructed via auxiliary edges utilize at most 
$2$ additional edges. Thus, every relevant path has at most $\beta+2$ hops. By setting the depth parameter to 
$\tau = \ceil{\frac{4(\beta+2)}{\varepsilon}}$, the scaled paths are fully explored by the Monotone ES-Trees.

The scaling procedure introduces a multiplicative error of $\paren{\onepluseps}$, and the use of the near-linear SSSP algorithm for 
pivots introduces another $\paren{\onepluseps}$ factor. Incorporating these into the bounded analysis from 
\lemref{w+2kw1:approximation-guarantee} yields:

\begin{align*}
&d[u,v] \le \paren{\onepluseps}^2 \weight{P_{u,v}} + 2\paren{k-1}\paren{\onepluseps}^2 W_{1}(P_{u,v}).
\end{align*}

By appropriately scaling $\varepsilon' = \frac{\varepsilon}{8k}$ and maintaining \algref{w+2kw1:bounded} with parameter $\varepsilon'$, we absorb these accumulated multiplicative errors, concluding that $\dist{u,v} \le d[u,v] \le \paren{\onepluseps}\weight{P_{u,v}} + 2\paren{k-1}W_{1}(P_{u,v})$.
\end{proof}

\begin{lemma}
    \label{lem:w+2kw1:total-update-time-unbounded}
For any integer $2 \le k = \bigo{\sqrt{\log n}}$ and $0 < \varepsilon < 1$, the expected total update time of \algref{w+2kw1:unbounded} is $\onolog{\paren{\frac{k^2 \log n}{\varepsilon}}^{\bigo{k}} n^{2-\frac{1}{k}} m^{\frac{1}{k}}}$ if $m = n^{1+\Omega(1)}$ or $\onolog{n^{2-\frac{1}{k}+\subpol} m^{\frac{1}{k}}}$ otherwise.
\end{lemma}

\begin{proof}
    First assume that $m=n^{1+\Omega(1)}$.
    Let $\varepsilon' = \frac{\varepsilon}{8k}$.
    We set the sampling probability to $q = \paren{\frac{m}{n}}^{-\frac{1}{k}}$ and the number of levels within the hierarchy to $k$.
    The hopbound of the sparse hopset is $\beta = \bigo{\paren{\frac{k \log n}{\varepsilon'}}^{\bigo{k}}}$.
    By \corref{tools:modified-es-unbounded}, maintaining the hierarchy-induced edge sets $E_i$ and the pivot distances takes expected total 
    update time $\onolog{\beta n^{2-\frac{1}{k}} m^{\frac{1}{k}}}$, and the expected total number of edges ever present 
    in $E_i$ is $\mathcal{E}_{S_i} = \onolog{\frac{\beta}{\varepsilon'}\frac{n}{q^{k-1-i}}}$.
    By \thmref{tools:hopset-on-es-sparse}, constructing the hopset $F$ takes expected total update time 
    $\onolog{\beta n^{2-\frac{1}{k}} m^{\frac{1}{k}}}$, and the number of edges in $F$ is bounded by $\onolog{\beta \frac{n}{q}}$.
    The $(\onepluseps)$-SSSP algorithm from each of the 
    $|S_{k-1}| = \onolog{n q^{k-1}} = \onolog{n^{2-\frac{1}{k}} m^{-\paren{1-\frac{1}{k}}}}$ sources on $G$ takes 
    $|S_{k-1}| \cdot \onolog{m} = \onolog{n^{2-\frac{1}{k}} m^{\frac{1}{k}}}$ total update time.

    For each scale, the Monotone ES-Trees are maintained up to depth $\tau = \ceil{\frac{4(\beta+2)}{\varepsilon'}}$. At each level $i \in \{0, \dots, k-1\}$, trees are rooted at each $u \in S_i$, where $\expectation{\abs{S_i}} = n q^i$, on graphs with $\onolog{\mathcal{E}_{S_{i+1}} + |F| + n} = \onolog{\frac{\beta}{\varepsilon'}\frac{n}{q^{i+1}}}$ edges. Maintaining these trees across all $u \in S_i$ requires $\onolog{\tau \cdot \expectation{\abs{S_i}} \cdot \frac{\beta}{\varepsilon'}\frac{n}{q^{i+1}}} = \onolog{\paren{\frac{\beta}{\varepsilon'}}^2 \frac{n^2}{q}} = \onolog{\paren{\frac{\beta}{\varepsilon'}}^2 n^{2-\frac{1}{k}} m^{\frac{1}{k}}}$ update time.
    Since $\beta = \bigo{\paren{\frac{k \log n}{\varepsilon'}}^{\bigo{k}}}$, we have $\paren{\frac{\beta}{\varepsilon}}^2 = \bigo{\paren{\frac{k \log n}{\varepsilon'}}^{\bigo{k}}}$.
    Summing across all $k$ levels and all $\bigo{\log nW_{\max}}$ scales contributes 
    $\onolog{k \cdot \paren{\frac{\beta}{\varepsilon'}}^2 \frac{n^2}{q}} = \onolog{\paren{\frac{k \log n}{\varepsilon'}}^{\bigo{k}} n^{2-\frac{1}{k}} m^{\frac{1}{k}}}$.
    Summing all components and yields an expected total update time of $\onolog{\paren{\frac{k \log n}{\varepsilon'}}^{\bigo{k}} n^{2-\frac{1}{k}} m^{\frac{1}{k}}} = \onolog{\paren{\frac{k^2 \log n}{\varepsilon}}^{\bigo{k}} n^{2-\frac{1}{k}} m^{\frac{1}{k}}}$.

    If $m=n^{1+\subpol}$, we use $\beta=2^{\bigo{\sqrt{\log n}}} = n^{\subpol}$ instead and the expected total update time becomes $\onolog{n^{2-\frac{1}{k}+\subpol} m^{\frac{1}{k}}}$ by the same reasoning.
\end{proof}

Combining \lemref{w+2kw1:approximation-guarantee-unbounded} and \lemref{w+2kw1:total-update-time-unbounded} concludes the proof of \thmref{w+2kw1}.

\subsection{Dense Graphs}
\label{sec:w+2kw1-dense}

In this section, we present a decremental bounded-depth APASP algorithm for
dense weighted graphs that adapts the static framework of Roditty and
Sapir~\cite{Ariel+kDense} to the decremental setting, and generalizes \algref{w+2w1+2w2:bounded} to a $k$-level hierarchy.
The static algorithm achieves an additive error of
$+2\msum{W_i}{i=1}{k-1}$ in $\onolog{n^{2 + \frac{1}{3k-2}}}$ time using a hierarchy of $3k+2$ hitting sets 
and employing shortcut sets in addition to the hierarchy-induced edges. 

We show that by maintaining shortcut sets $\mathcal{F}_i$ across a
$(3k+2)$-level hierarchy in Monotone ES-Trees and accounting for
errors propagated along stretched edges, our decremental algorithm
achieves an additive error of $2\paren{k-1} W_1$
with an expected total update time of $\onolog{n^{2 +\frac{2}{3k - 2}+\subpol}}$.
Similarly to the previous algorithms, we begin with a bounded version for all vertex pairs at distance at most $\tau$.

\subsubsection{Bounded Depth}
\label{sec:w+2kw1-dense:bounded}

\begin{algorithm}[htbp]
\caption{$+\paren{4k+2}W_{1}$-APASP up to depth $\tau$}
\label{alg:w+2kw1-dense:bounded}

\BlankLine

\textbf{Input:} A weighted undirected graph $G = \paren{V, E, \weight}$, an
integer parameter $k \ge 1$, and a depth parameter $\tau$.

\textbf{Output:} An estimate matrix $d[u,v]$ satisfying $\dist{u,v} \le d[u,v]
\le \dist{u,v} + \xi_0 \cdot W_1\paren{u \squiggly v}$ for any pair $u, v \in
V$ with $\dist{u,v} \le \tau$.

\hrulefill

Initialize $d[u,v] \gets \weight{u,v}$ for $(u,v) \in E$ and $\infty$ otherwise\;

Let $q = n^{-\frac{1}{3k+2}} \in (0, 1)$\;

Sample a $(3k+2)$-level hierarchy $V = S_0 \supseteq S_1 \supseteq \dots
\supseteq S_{3k+2} \supseteq S_{3k+3} = \emptyset$ with probability $q$;

\ForEach{$i \in \{0, \dots, 3k+2\}$ in descending order}
{
    Maintain an ES-Tree up to depth $(4k+2)\tau$ for distances from any $u
    \in V$ to its pivot $\piv[i]{u} \in S_i$\;
    
    $E_i \gets E_{S_i}^{(4k+2)\tau}$\;
}
$E_0 \gets \emptyset, E_{3k+3} \gets E$\;

\ForEach{$s \in S_{3k+2}$}
{
    Maintain an exact ES-Tree from $s$ up to depth $(4k+2)\tau$ in $G$ and update $d$ accordingly\;
}

\ForEach{$i = 3k+1$ down to $0$}
{
    For each $j \in \{1, \dots, 3k+2\}$, let $Q_j \gets \{(u, \piv[j]{u}) \mid u \in V\}$ with weight $\dist{u, \piv[j]{u}}$, and $Q \gets \mcup{Q_j}{j=1}{3k+2}$\;
    
    $\mathcal{F}_i \gets \bigcup \bracce{ S_\ell \times
    S_j \mid i + \ell + j \ge 3k + 1, \: \ell > i \text{ or } j > i }$ where
    each shortcut $\paren{x, y} \in S_\ell \times S_j$ is assigned weight $\mmin{\level{x}{y},\level{y}{x}}$
    
    \ForEach{$s \in S_i$}
    {
        $F_{i+1}(s) \gets \paren{V, E_{i+1} \cup Q \cup (\singleton{s} \times V) \cup E(s) \cup \mathcal{F}_i}$\;
        
        Maintain a Monotone ES-Tree from $s$ up to depth $(4k+2)\tau$ in
        $F_{i+1}(s)$ and update $d$ accordingly\;
    }
}

\ForEach{$u, v \in V$}
{
    $d[u,v] \gets \mmin{\level{u}{v}, \level{v}{u}}$\;
}

\end{algorithm}

\paragraph{Algorithm Description.}
\algref{w+2kw1-dense:bounded} operates on a weighted undirected
graph $G = (V, E, \weight)$ undergoing edge deletions and edge weight
increases. It receives an integer approximation trade-off parameter $k \ge 1$
and a depth threshold $\tau$.

The algorithm constructs a $(3k+2)$-level hierarchy with probability $q=n^{-\frac{1}{3k+2}}$,
and constructs the hierarchy-induced edge sets $E_1 \subseteq E_2 \subseteq \dots \subseteq E_{3k+1} \subseteq E_{3k+2} = E$.
In addition to these sets, the algorithm constructs the sets $\{ \mathcal{F}_i \}_{i=1}^{3k+2} = \bracce{\bracce{ S_\ell \times S_j \mid i + \ell + j \ge 3k + 1, \: \ell > i \text{ or } j > i } }_{i=1}^{3k+2}$, $\{ Q_j \}_{j=1}^{3k+2} = \bracce{\{(u, \piv[j]{u}) \mid u \in V\}}_{j=1}^{3k+2}$ and $Q = \bigcup_{j=1}^{3k+2} Q_j$. 
Every vertex $s \in S_i$ maintains a Monotone ES-Tree on $F_{i+1}(s) = \paren{V, E_{i+1} \cup Q \cup (\singleton{s} \times V) \cup E(s) \cup \mathcal{F}_i}$.

Note that unlike all previous algorithms, the edges in $Q$ link vertices to their actual pivots in place of auxiliary pivots.
In previous algorithms (such as in \secref{w+2kw1-sparse}), auxiliary pivots $\auxpiv[j]{u}$ and auxiliary edge sets $\tilde{Q}$, $\singleton{u} \times \tilde{S}$ were used to skip edges absent from $E_{i+1}$.
Here, however, auxiliary pivots cannot trivially concatenate with the shortcut edges in $\mathcal{F}_i \subseteq S_\ell \times S_j$ because an auxiliary pivot $\auxpiv[j]{u}$ is distinct from the actual pivot $\piv[j]{u} \in S_j$.
Consequently, $F_{i+1}(s)$ directly incorporates the true pivot edge sets $Q_j$ alongside $(\singleton{s} \times V)$ (where each edge $(s, v) \in \singleton{s} \times V$ is assigned weight $\level{v}{s}$ maintained by higher levels).
\lemref{w+2kw1-dense:bounded:size-of-Q} accounts for the maintenance of $Q$ and shows it fits our target total update time.

Each shortcut edge $(x, y) \in \mathcal{F}_i$ connects a pair of vertices $x \in S_\ell$ and $y \in S_j$ whose indices satisfy the threshold condition $i + \ell + j \ge 3k+1$. The threshold $3k+1$ is calibrated so that tree updates over shortcut edges require $\onolog{\expectation{|S_i|} \cdot \expectation{|S_\ell \times S_j|}} = \onolog{n^{3 - \frac{i + \ell + j}{3k+2}}} = \onolog{n^{3 - \frac{3k+1}{3k+2}}} = \onolog{n^{2 + \frac{1}{3k+2}}}$, matching the total update time of maintaining trees on the hierarchy-induced edge sets - $\onolog{\expectation{|S_i|} \cdot \mathcal{E}_{i+1}} = \onolog{n^{2 + \frac{1}{3k+2}}}$.
The weight of $(x, y)$ is set to $\mmin{\level{x}{y},\level{y}{x}}$, which has been computed in earlier iterations since $\mmax{\ell, j} > i$.

\paragraph{The Recursive Error Relation.}
\label{sec:w+2kw1-dense:bounded:recurrence}
We define $\xi_i$ as a variable denoting the maximum cumulative additive error coefficient (in units of $W_1$) carried by the Monotone ES-Tree rooted at any vertex $u \in S_i$. That is, for any $u \in S_i$, any $v \in V$, and any path
$P_{u,v}$ of weight at most $\tau$ at time $t$:
\[ \level[t]{u}{v} \le \weight{P_{u,v}} + \xi_i \cdot W_1\paren{P_{u,v}}. \]
We prove that the error coefficients $\xi_i$ satisfy the recurrence relation:
\begin{equation}
\label{eq:w+2kw1-dense:bounded:xi-recurrence}
\xi_i = \mmax{\min \left\{ 2(\lambda - i), \quad 2\paren{ \left\lceil \frac{3k - \lambda}{2} \right\rceil - i + 1 } + 2\xi_\lambda \right\}}{\lambda \in \{i+1, \dots, 3k+2\}} ,
\end{equation}
with the base case $\xi_{3k+2} = 0$.

\paragraph{Approximation Analysis.}
\label{sec:w+2kw1-dense:bounded:approximation}
We establish the approximation guarantee of the algorithm in two stages:
first at initialization ($t = 0$), and then for any time $t > 0$.

\begin{lemma}
\label{lem:w+2kw1-dense:bounded:approx-t0}
For any $0 \le j \le i \le 3k+2$, $u \in S_i$, $v \in S_j$, and a path $P_{u,v}$ between $u$
and $v$ with $\weight{P_{u,v}} \le \tau$. At intialization (time $t=0$), it holds that:
\[
\level{u}{v} \le \weight{P_{u,v}} + \xi_i \cdot W_{1}\paren{P_{u,v}},
\]
where $\xi_i$ is defined by recurrence \eqref{eq:w+2kw1-dense:bounded:xi-recurrence}.
\end{lemma}

\begin{proof}
At time $t = 0$, by the properties of Monotone ES-Trees upon initialization,
$\level{u}{v} = \dist{u, v}{F_{i+1}(u)}$. We prove the claim by induction on
$3k + 2 - i$.

For the base case $i = 3k + 2$, \algref{w+2kw1-dense:bounded} maintains an
exact ES-Tree on $G$ up to depth $2\tau$. Thus $\level{u}{v} = \dist{u,v}{G} \le
\weight{P_{u,v}}$, establishing $\xi_{3k+2} = 0$.

For the inductive step ($i < 3k + 2$), assume the claim holds for all levels
$r > i$. Let $u \in S_i$ and $v \in S_j$, and let $\lambda$ be the sole index such that $P_{u,v} \subseteq E_{\lambda+1} \setminus E_\lambda$. We
consider three cases:

\ccase{1} $\lambda \le i$.

Then $P_{u,v} \subseteq E_{\lambda+1} \subseteq E_{i+1}$. Since $E_{i+1} \subseteq F_{i+1}(u)$, the
path $P_{u,v}$ exists in $F_{i+1}(u)$. Therefore, $\level{u}{v} =
\dist{u,v}{F_{i+1}(u)} \le \weight{P_{u,v}}$.

\ccase{2} $\lambda > i$ and $i + \lambda + j \ge 3k + 1$.

Let $e' = (w, z)$ be any edge
along $P_{u,v}$ that is not in $E_\lambda$.
The pivot $\piv[\lambda]{w} \in S_\lambda$ satisfies $\dist{w, \piv[\lambda]{w}} \le
\weight{e'} \le W_{1}(P_{u,v})$.
Because $i + \lambda + j \ge 3k + 1$, the shortcut edge $\paren{\piv[\lambda]{w}, v} \in
S_\lambda \times S_j$ belongs to $\mathcal{F}_i \subseteq F_{i+1}(u)$ with $\weight{\piv[\lambda]{w}, v} \le \level{\piv[\lambda]{w}}{v}$. $F_{i+1}(u)$ also contains the shortcut edge $\paren{u,\piv[\lambda]{w}} \in \singleton{u} \times V$ with weight $\weight{u,\piv[\lambda]{w}} \le \level{\piv[\lambda]{w}}{u}$.
By definition of $\lambda$, the paths $\piv[\lambda]{w} \squiggly w \circ P_{u,w}$ and $\piv[\lambda]{w} \squiggly w \circ P_{w,v}$ belong to $E_{\lambda+1} \subseteq F_{\lambda+1}(\piv[\lambda]{w})$. 
Since this is initialization, $\level{\piv[\lambda]{w}}{u} = \dist{\piv[\lambda]{w}, u} \le \weight{e'} + \weight{P_{u,w}}$ and $\level{\piv[\lambda]{w}}{v} = \dist{\piv[\lambda]{w}, v} \le \weight{e'} + \weight{P_{w,v}}$.
The path $P^*_{u,v} =  (u,\piv[\lambda]{w}) \circ (\piv[\lambda]{w}, v) \subseteq F_{i+1}(u)$
has weight:
\[
\weight{P^*_{u,v}} \le \weight{P_{u,w}} + \weight{e'} + \weight{e'} + \weight{P_{w,v}} = \weight{P_{u,v}} + 2\weight{e'} \le \weight{P_{u,v}} + 2W_{1}\paren{P_{u,v}}.
\]
Since $2 \le \xi_i$, the claim holds.

\ccase{3} $\lambda > i$ and $i + \lambda + j < 3k + 1$.

Let $e' = (x, y) = \leftabovearrow{P_{u,v}}{E_{i+1}}$. Then $P_{y,v} \subseteq E_{i+1} \subseteq F_{i+1}(u)$.
Let $\psi$ such that $e' \in E_{\psi +1} \setminus E_{\psi}$.

The path $P_{u,v}^* = (u, \piv[\psi]{y}) \circ (\piv[\psi]{y}, y)
\circ P_{y,v} \subseteq F_{i+1}(u)$ has weight bounded by:
\[
\weight{P_{u,v}^*} \le \level{\piv[\psi]{y}}{u} + \dist{\piv[\psi]{y}, y} + \weight{P_{y,v}}.
\]
Applying the induction hypothesis to $\piv[\psi]{y}$, $\level{\piv[\psi]{y}}{u} \le \weight{P_{u,y} \circ y \squiggly \piv[\psi]{y}} + \xi_\psi W_1\paren{P_{u,y} \circ y \squiggly \piv[\psi]{y}}$.
Observe that $W_1\paren{P_{u,y} \circ y \squiggly \piv[\psi]{y}} \le W_1\paren{P_{u,v}}$ as discussed in the previous algorithm.

As $P_{y,v} \subseteq E_{i+1} \subseteq F_{i+1}(u)$, we conclude that $\weight{P_{u,v}^*} \le \level{\piv[\psi]{y}}{u} + \dist{\piv[\psi]{y}, y} + \weight{P_{y,v}} \le \weight{P_{u,v}} + \paren{2 + \xi_\psi} W_1\paren{P_{u,v}}$.

Since $\psi \ge i + 1$, recurrence \eqref{eq:w+2kw1-dense:bounded:xi-recurrence} ensures $2 + \xi_\psi \le \xi_i$. Therefore, $\level{u}{v} = \dist{u,v}{F_{i+1}(u)} \le \weight{P_{u,v}^*} \le \weight{P_{u,v}} + \xi_i W_1(P_{u,v})$.
\end{proof}

\begin{lemma}
\label{lem:w+2kw1-dense:bounded:approx-time-t}
For any $0 \le j \le i \le 3k+2$, $u \in S_i$, $v \in S_j$, time $t$ and a path $P_{u,v}$ between $u$
and $v$ with $\weight{P_{u,v}} \le \tau$ at time $t$, it holds that:
\[
\level[t]{u}{v} \le \weight{P_{u,v}} + \xi_i \cdot W_{1}\paren{P_{u,v}},
\]
where $\xi_i$ is defined by recurrence \eqref{eq:w+2kw1-dense:bounded:xi-recurrence}.
\end{lemma}

\begin{proof}
We prove the claim by double induction on time $t$ and the hierarchy level $3k
+ 2 - i$.
The base case for time $t = 0$ was established in \lemref{w+2kw1-dense:bounded:approx-t0}.
We assume the claim holds for all times $t' < t$, for all hierarchy levels, and
prove it for time $t$.

For the base case $i = 3k + 2$, \algref{w+2kw1-dense:bounded} maintains an
exact ES-Tree on $G$ up to depth $2\tau$. By the properties of exact ES-Trees,
distance estimates are exact at all times: $\level[t]{u}{v} = \dist{u,
v}{G}{t} \le \weight{P_{u,v}}$, so the claim holds with $\xi_{3k+2} = 0$.

For the inductive step ($i < 3k + 2$), assume the claim holds at time $t$ for
all higher hierarchy levels $r > i$.
Let $P_{u,v}$ be a path between $u \in S_i$ and $v \in S_j$ with $\weight{P_{u,v}}
\\le \tau$ at time $t$.
Let $\lambda$ be the sole index such that $P_{u,v} \subseteq E_{\lambda+1} \setminus E_\lambda$. 
We again consider three cases:

\ccase{1} $\lambda \le i$.

All edges of $P_{u,v}$ belong to $E_{\lambda+1} \subseteq E_{i+1} \subseteq F_{i+1}(u)$ at time $t$. If none of the edges on $P_{u,v}$ are stretched in $u$'s tree at time $t$
the Monotone ES-Tree accurately maintains the path distance and $\level{u}{v} \le \weight{P_{u,v}}$.

Assume then that at least one edge on $P_{u,v}$ is stretched in $u$'s tree at time $t$.
Let $(z', z)$ be the \emph{last} such edge when traversing from $u$ to $v$ and write $P_{u,v} = P_{u,z} \circ P_{z,v}$. Because $(z', z)$ is stretched in $u$'s tree at time $t$, $\level{u}{z} = \level[t-1]{u}{z}$.
By the induction hypothesis on time $t - 1$ for vertex $u \in S_i$, $\level[t-1]{u}{z} \le \weight{P_{u,z}}{t-1} + \xi_i \cdot W_{1,t-1}\paren{P_{u,z}}$.
Since edge weights in $G$ are non-decreasing under edge deletions and weight
increases, and $P_{u,v}$ exists at time $t$, $\weight{P_{u,z}}{t-1} \le \weight{P_{u,z}}{t}$ and $W_{1,t-1}\paren{P_{u,z}} \le W_{1,t}\paren{P_{u,z}} \le W_{1,t}\paren{P_{u,v}}$.
Therefore, $\level[t]{u}{z} \le \weight{P_{u,z}} + \xi_i \cdot W_{1}\paren{P_{u,v}}$.
Now consider the remaining suffix $P_{z,v}$. Because $(z', z)$ was the last
stretched edge on $P_{u,v}$, none of the edges on $P_{z,v}$ are stretched in
$u$'s tree at time $t$. Furthermore, $P_{z,v} \subseteq P_{u,v} \subseteq
E_{i+1} \subseteq F_{i+1}(u)$. we conclude that
$\level{u}{v} \le \level{u}{z} + \weight{P_{z,v}} \le \weight{P_{u,z}} + \xi_i W_{1}\paren{P_{u,v}} + \weight{P_{z,v}} = \weight{P_{u,v}} + \xi_i \cdot W_{1}\paren{P_{u,v}}$.

\ccase{2} $\lambda > i$ and $i + \lambda + j \ge 3k + 1$.

Let $e' = (w, z)$ be any edge along $P_{u,v}$ that is not in $E_\lambda$ at time $t$.
The pivot $\piv[\lambda]{w} \in S_\lambda$ satisfies $\dist{w, \piv[\lambda]{w}} \le
\weight{e'} \le W_{1}(P_{u,v})$.
Because $i + \lambda + j \ge 3k + 1$, the shortcut edges $(u, \piv[\lambda]{w}) \in S_i
\times S_\lambda$ and $(\piv[\lambda]{w}, v) \in S_\lambda \times S_j$ belong to
$\mathcal{F}_i \subseteq F_{i+1}(u)$.
Consider the path $P^*_{u,v} = (u, \piv[\lambda]{w}) \circ (\piv[\lambda]{w}, v) \subseteq F_{i+1}(u)$ with 
$\weight{P^*_{u,v}} \le \level{\piv[\lambda]{w}}{u} + \level{\piv[\lambda]{w}}{v}$.
By the induction hypothesis, we have $\level{\piv[\lambda]{w}}{u} \le \weight{P_{u,w} \circ w \squiggly \piv[\lambda]{w}} + \xi_\lambda W_1\paren{{P_{u,w} \circ w \squiggly \piv[\lambda]{w}}} \le \weight{P_{u,w}} + \weight{e'} + \xi_\lambda W_1\paren{P_{u,v}}$ and similarly $\level{\piv[\lambda]{w}}{v} \le \weight{P_{w,v}} + \weight{e'} + \xi_\lambda W_1\paren{P_{u,v}}$.

If none of the edges of $P^*_{u,v}$ are stretched in $u$'s tree, we have $\level{u}{v} \le \weight{P_{u,v}} + 2\weight{e'} +2\xi_\lambda W_1\paren{P_{u,v}} \le \weight{P_{u,v}} + 2(\xi_\lambda+1)W_1\paren{P_{u,v}} \le \weight{P_{u,v}} + \xi_{i}W_1\paren{P_{u,v}}$.

Note that the edge $\paren{u,\piv[\lambda]{w}}$ may be stretched only if $\piv[\lambda]{w}$ switches from $w' \in S_\lambda$ to some $w'' \in S_\lambda$. However, as $w'' \neq w' \in S_\lambda$ and $\paren{u,w''} \in \singleton{u} \times V$ is present throughout the entire update sequence, we have $\level{u}{w''} \le \level{w''}{u} \le \weight{P_{u,w} \circ w \squiggly w''} + \xi_\lambda W_1\paren{P_{u,w} \circ w''} \le \weight{P_{w,v}} + \weight{e'} + \xi_\lambda W_1\paren{P_{u,v}}$ and hence $\level{u}{v} \le \weight{P_{u,v}} + 2\weight{e'} +2\xi_\lambda W_1\paren{P_{u,v}} \le \weight{P_{u,v}} + 2(\xi_\lambda+1)W_1\paren{P_{u,v}} \le \weight{P_{u,v}} + \xi_{i}W_1\paren{P_{u,v}}$.

So it is only left to consider the case where $\paren{\piv[\lambda]{w},v}$ is stretched in $u$'s tree. But then under the induction hypothesis on time we have $\level{u}{v} = \level[t-1]{u}{v} \le \weight{P_{u,v}} + \xi_i W_1\paren{P_{u,v}}$.

\ccase{3} $\lambda > i$ and $i + \lambda + j < 3k + 1$.

In this case, the shortcut edge $(\piv[\lambda]{w}, v)$ is not available in $\mathcal{F}_i$.
Let $e' = (x, y) = \leftabovearrow{P_{u,v}}{E_{i+1}}$ be the last edge along
$P_{u,v}$ from $u$'s side that does not belong to $E_{i+1}$. Hence, $P_{y,v} \subseteq
E_{i+1} \subseteq F_{i+1}(u)$. Let $\psi \ge i + 1$ be the sole index such that $e'
\in E_{\psi+1} \setminus E_\psi$. The pivot $\piv[\psi]{y} \in S_\psi$
satisfies $\dist{\piv[\psi]{y}, y} \le \weight{e'} \le W_1(P_{u,v})$.

Consider the alternative path $P_{u,v}^* = (u, \piv[\psi]{y}) \circ (\piv[\psi]{y}, y) \circ P_{y,v} \subseteq F_{i+1}(u)$.
We first bound the weight of $P_{u,v}^*$ in $F_{i+1}(u)$:
$(u, \piv[\psi]{y}) \in \singleton{u} \times V$ has weight $\weight{u, \piv[\psi]{y}} = \level{\piv[\psi]{y}}{u}$ computed
in the tree of $\piv[\psi]{y} \in S_\psi$. In the tree of $\piv[\psi]{y}$, the path
$P_{\piv[\psi]{y}, u} = (\piv[\psi]{y} \squiggly y) \circ P_{y, u}$ connects $\piv[\psi]{y}$ to $u$ with
$\weight{P_{\piv[\psi]{y}, u}} \le \dist{\piv[\psi]{y}, y} + \weight{P_{u, y}} \le \weight{P_{u, y}} + \weight{e'}$.
Because $\psi \ge i + 1 > i$, $\piv[\psi]{y} \in S_\psi$ has strictly higher priority in the hierarchy than $u \in S_i$. By the induction
hypothesis on the hierarchy level we have
$\level{\piv[\psi]{y}}{u} \le \weight{P_{\piv[\psi]{y}, u}} + \xi_\psi \cdot W_1(P_{\piv[\psi]{y}, u}) \le \weight{P_{u, y}} + \weight{e'} + \xi_\psi \cdot W_1(P_{u,v})$.

The edge $(\piv[\psi]{y}, y) \in Q_\psi$ has weight $\dist{\piv[\psi]{y}, y} \le \weight{e'}$, and all edges of the suffix $P_{y, v}$ belong to $E_{i+1} \subseteq F_{i+1}(u)$.
Combining everything, the total weight of $P_{u,v}^*$ is bounded by:
\begin{align*}
&\weight{P_{u,v}^*} = \weight{u, \piv[\psi]{y}} + \weight{\piv[\psi]{y}, y} + \weight{P_{y,v}} \\
&\le \level{\piv[\psi]{y}}{u} + \dist{\piv[\psi]{y}, y} + \weight{P_{y,v}} \\
&\le \weight{P_{u, y}} + \weight{e'} + \xi_\psi W_1\paren{P_{u,v}} + \weight{e'} + \weight{P_{y,v}} \\
&= \weight{P_{u, y}} + \weight{P_{y,v}} + 2\weight{e'} + \xi_\psi W_1\paren{P_{u,v}} \\
&= \weight{P_{u,v}} + 2\weight{e'} + \xi_\psi W_1\paren{P_{u,v}} \\
&\le \weight{P_{u,v}} + \paren{2 + \xi_\psi} W_1\paren{P_{u,v}} \\
&\le \weight{P_{u,v}} + \xi_i W_1\paren{P_{u,v}}.
\end{align*}

We evaluate the possible cases for the last stretched edge on $P_{u,v}^*$ in $u$'s
tree at time $t$:

\ccase{3.1} No edge on $P_{u,v}^*$ is stretched in $u$'s tree at time $t$.

Since all edges of $P_{u,v}^*$ belong to $F_{i+1}(u)$ and none are stretched,
$
\level[t]{u}{v} \le \weight{P_{u,v}^*} \le \weight{P_{u,v}} + \paren{2 + \xi_\psi} W_1\paren{P_{u,v}} \le \weight{P_{u,v}} + \xi_i \cdot W_1\paren{P_{u,v}}.
$

\ccase{3.2} $(u, \piv[\psi]{y})$ is the last stretched edge on $P_{u,v}^*$.

As discussed in case~2, we have $\level{u}{\piv[\psi]{y}} \le \weight{P_{u,y}} + \weight{e'} + \xi_{\psi}W_1\paren{P_{u,v}}$.
If $(u, \piv[\psi]{y})$ is the last stretched edge on $P_{u,v}^*$, the
subsequent subpath $(\piv[\psi]{y}, y) \circ P_{y, v}$ is unstretched
in $u$'s tree at time $t$. Then:
\[
\level[t]{u}{v} \le \level[t]{u}{\piv[\psi]{y}} + \dist{\piv[\psi]{y}, y} + \weight{P_{y,v}} \le \weight{P_{u,v}^*} \le \weight{P_{u,v}} + \paren{2 + \xi_\psi} W_1\paren{P_{u,v}} \le \weight{P_{u,v}} + \xi_i \cdot W_1\paren{P_{u,v}}.
\]

\ccase{3.3} An edge $(z', z) \in (\piv[\psi]{y},y) \circ P_{y, v}$ is the last stretched edge on $P_{u,v}^*$.

In this case, $z \in V\paren{P_{y,v}}$. Because $(z', z)$ is stretched in $u$'s tree at time
$t$, $u$'s estimate to $z$ was set at an earlier time $t' < t$, so
$\level[t]{u}{z} = \level[t']{u}{z}$.
By the induction hypothesis on time $t' < t$ for the pair $(u, z)$ along
the subpath $P_{u, z} \subseteq P_{u, v}$:
\[
\level[t]{u}{z} = \level[t']{u}{z} \le \weight{P_{u, z}} + \xi_i \cdot W_1\paren{P_{u, z}} \le \weight{P_{u, z}} + \xi_i \cdot W_1\paren{P_{u, v}}.
\]
Since none of the subsequent edges on the suffix $P_{z, v} \subseteq P_{y,
v} \subseteq E_{i+1}$ are stretched in $u$'s tree at time $t$, it holds that
$\level[t]{u}{v} \le \level[t]{u}{z} + \weight{P_{z, v}} \le \paren{\weight{P_{u, z}} + \xi_i W_1\paren{P_{u, v}}} + \weight{P_{z, v}} = \weight{P_{u, v}} + \xi_i \cdot W_1\paren{P_{u, v}}$.

This completes the verification of all cases at time $t$.

\end{proof}

\paragraph{Derivation of the Recursive Error Relation.}
\label{sec:w+2kw1-dense:bounded:recurrence-derivation}
We formalize recurrence relation~\eqref{eq:w+2kw1-dense:bounded:xi-recurrence} by analyzing
the detour chain of Case~3.

Consider a path $P_{u,v} \subseteq E_{\lambda+1} \setminus E_\lambda$.
If a pivot $z = \piv[\psi]{y}$ along the detour chain carries a distance label $\level[t]{z}{u}$ that was frozen at an earlier timestamp $t' < t$ when it held that $P_{u,v} \subseteq E_{\lambda'+1} \setminus E_{\lambda'}$ for $\lambda' \ge \lambda$, the historical stretch carried by that frozen label
telescopes to $2(\lambda' - i)W_1$.
Because recurrence~\eqref{eq:w+2kw1-dense:bounded:xi-recurrence} explicitly maximizes over all
possible levels $\lambda \in \{i+1, \dots, 3k+2\}$, the historical containment level
$\lambda'$ is unconditionally absorbed by the worst-case bound $\xi_i W_1\paren{P_{u,v}}$.
Consequently, bounding the detour chain under any fixed index $\lambda$ guarantees that
the distance estimates remain valid across all edge deletion sequences.

We now unwind the detour chain of Case~3 for $\lambda$ such that $P_{u,v} \subseteq E_{\lambda+1} \setminus E_\lambda$.
Starting with endpoints at levels $(i_0, j_0) = (i, j)$, each recursive step $r \ge 1$ of Case~3 incurs an additional error of $2W_1\paren{P_{u,v}}$, preserves the index $\lambda$, increases the higher endpoint index ($i_r \ge i_0 + r$), and advances the sum of endpoint levels ($i_r + j_r \ge 2i_0 + 2r - 1$).
The chain of detours terminates as soon as either Case~1 or Case~2 is reached:
\begin{enumerate}
    \item \textbf{Branch A (Halting via Case~1):}
    Occurs when the higher endpoint reaches $i_r \ge \lambda$.
    In each detour step, detouring from level $i$ to a pivot at level $\psi \ge i + 1$
    incurs an error of $2(\psi - i)W_1\paren{P_{u,v}}$ while delegating the remaining path
    to the pivot's tree, which carries recursive error $\xi_\psi W_1\paren{P_{u,v}}$.
    Along Branch~A, the pivot's error is itself inductively bounded by
    $\text{Cost}_A(\lambda, \psi) = 2(\lambda - \psi)W_1\paren{P_{u,v}}$.
    Summing the detour overhead with the pivot's error yields:
    \[
    2(\psi - i) + \xi_\psi \le 2(\psi - i) + 2(\lambda - \psi) = 2(\lambda - i).
    \]
    The intermediate pivot level $\psi$ cancels out completely. Unwinding this relation
    across a chain of intermediate pivots $i = i_0 < i_1 < \dots < i_r$ produces a
    telescoping sum:
    \[
    \sum_{k=1}^r 2(i_k - i_{k-1}) \cdot W_1\paren{P_{u,v}} + \xi_{i_r} \cdot W_1\paren{P_{u,v}} = 2(i_r - i) \cdot W_1\paren{P_{u,v}} + \xi_{i_r} \cdot W_1\paren{P_{u,v}}.
    \]
    When the pivot reaches $i_r \ge \lambda$, all edges of the remaining subpath belong
    to $E_{\lambda+1} \subseteq E_{i_r+1}$, meaning the subpath is fully contained in $E_{i_r+1}$
    (Case~1). In the unstretched candidate path, these physical edges are followed directly
    in $E_{i_r+1}$ without requiring further pivots or artificial shortcut edges from $\mathcal{F}$,
    contributing a base error of $\xi_{i_r}^{\text{contained}} = 0$.
    Thus, the recursive errors of all intermediate pivots telescope cleanly into:
    \begin{equation}
    \label{eq:w+2kw1-dense:bounded:cost-A}
    \text{Cost}_A(\lambda, i) = 2(\lambda - i) \cdot W_1\paren{P_{u,v}}.
    \end{equation}

    \item \textbf{Branch B (Halting via Case~2):}
    Occurs when the sum of endpoint levels reaches $i_r + j_r \ge 3k + 1 - \lambda$.
    In each recursive detour step of Case~3, detours alternate between endpoints,
    advancing the sum of endpoint levels by $2$ every two detour steps (specifically,
    $i_r + j_r \ge 2i_0 + 2r - 1$).
    To reach the threshold $(3k + 1 - \lambda)$, the detour steps must cover the
    level deficit $(3k - \lambda) - 2i$.
    Dividing this deficit by $2$ yields at most $r_B \le \lceil \frac{3k - \lambda}{2} \rceil - i$
    detour steps.
    Case~2 then executes a final shortcut step via artificial edges $(u,\piv[\lambda]{w}) \in \singleton{u} \times V$ and $(\piv[\lambda]{w}, v) \in \mathcal{F}_i$.
    The weights of the edges are the precomputed distance estimate $\level[t]{\piv[\lambda]{w}}{u}$ and $\level[t]{\piv[\lambda]{w}}{v}$ maintained
    by the independent Monotone ES-Tree of $\piv[\lambda]{w} \in S_\lambda$.
    By the hierarchy induction hypothesis on level $\lambda$, the distance labels are bounded
    by $\dist{\piv[\lambda]{w}, u} + \xi_\lambda \cdot W_1\paren{P_{u,v}}$ and $\dist{\piv[\lambda]{w}, v} + \xi_\lambda \cdot W_1\paren{P_{u,v}}$ correspondingly, importing the full worst-case error
    $\xi_\lambda \cdot W_1\paren{P_{u,v}}$ of the pivot's tree.
    Accounting for the $r_B + 1$ total detour and shortcut steps (contributing $2(r_B + 1)W_1\paren{P_{u,v}}$) alongside the
    two imported errors $2\xi_\lambda \cdot W_1\paren{P_{u,v}}$ from pivot $\piv[\lambda]{w}$, Branch B incurs an additive error of:
    \begin{equation}
    \label{eq:w+2kw1-dense:bounded:cost-B}
    \text{Cost}_B(\lambda, i) = \bracke{ 2\paren{ \left\lceil \frac{3k - \lambda}{2} \right\rceil - i + 1 } + 2\xi_\lambda } \cdot W_1\paren{P_{u,v}}.
    \end{equation}
\end{enumerate}

Because the Monotone ES-Tree explores all valid paths in the augmented graph,
the distance estimate is bounded by the minimum of the two halting branches for
any given $\lambda$:
\[
\Delta(\lambda, i) \le \min\Big( \text{Cost}_A(\lambda, i), \; \text{Cost}_B(\lambda, i) \Big).
\]
Since $\lambda \in \{i+1, \dots, 3k+2\}$ is chosen adversarially by the edge
deletion sequence, the invariant error $\xi_i$ must bound the worst-case choice
of $\lambda$:
\[
\xi_i = \max_{\lambda \in \{i+1, \dots, 3k+2\}} \min \left\{ 2(\lambda - i), \quad 2\paren{ \left\lceil \frac{3k - \lambda}{2} \right\rceil - i + 1 } + 2\xi_\lambda \right\},
\]
establishing recurrence \eqref{eq:w+2kw1-dense:bounded:xi-recurrence}.

\paragraph{Mathematical Derivation of the All-Pairs Bound.}
For every $v \in V \space (i=0)$, recurrence \eqref{eq:w+2kw1-dense:bounded:xi-recurrence} specializes to:
\begin{equation}
\label{eq:w+2kw1-dense:bounded:xi-0-recurrence}
\xi_0 = \max_{\lambda \in \{1, \dots, 3k+2\}} \min \left\{ 2\lambda, \quad 2\paren{ \left\lceil \frac{3k - \lambda}{2} \right\rceil + 1 } + 2\xi_\lambda \right\}.
\end{equation}
Here, $\text{Cost}_A(\lambda, 0) = 2\lambda$ is strictly increasing in $\lambda$.
Conversely, because hierarchy error coefficients are non-increasing with respect to the
level index ($\xi_0 \ge \xi_1 \ge \dots \ge \xi_{3k+2} = 0$), the error term $\xi_\lambda$
is non-increasing in $\lambda$; combined with the ceiling term, $\text{Cost}_B(\lambda, 0)$
is non-increasing in $\lambda$.
The adversarial choice of $\lambda$ maximizes the minimum of these two opposing branches,
which peaks at the crossover threshold $\lambda^*$ where $\text{Cost}_A(\lambda^*, 0) \approx \text{Cost}_B(\lambda^*, 0)$.

To determine the balance point $\lambda^*$, we equate the continuous approximations of the two branches:
\[
2\lambda^* \approx 2\paren{ \left\lceil \frac{3k - \lambda^*}{2} \right\rceil + 1 } + 2\xi_{\lambda^*}.
\]
Simplifying:
\[
2\lambda^* \approx (3k - \lambda^*) + 2 + 2\xi_{\lambda^*} \implies 3\lambda^* \approx 3k + 2 + 2\xi_{\lambda^*},
\]
which yields the balance point:
\[
\lambda^* \approx k + \frac{2\xi_{\lambda^*} + 2}{3}.
\]
Evaluating the increasing branch $\text{Cost}_A$ at $\lambda^*$:
\[
\xi_0 \le 2\lambda^* \le 2k + \frac{4\xi_{\lambda^*} + 4}{3}.
\]
This expression demonstrates that bounding $\xi_0$ reduces to bounding the error $\xi_{\lambda^*}$ carried by the pivot level.

We now verify that the optimal crossover point necessarily satisfies $\lambda^* \ge k$.
For any $\lambda \le k$, the increasing branch is bounded by:
\[
\text{Cost}_A(\lambda, 0) = 2\lambda \le 2k.
\]
For Branch~B, since $\lambda \le k$, we have $3k - \lambda \ge 2k$, which implies
$\lceil \frac{3k - \lambda}{2} \rceil \ge k$.
Because error coefficients are non-negative ($\xi_\lambda \ge 0$), we obtain:
\[
\text{Cost}_B(\lambda, 0) \ge 2(k + 1) + 2\xi_\lambda \ge 2k + 2.
\]
Thus, for every $\lambda \le k$, we have:
\[
\text{Cost}_A(\lambda, 0) \le 2k < 2k + 2 \le \text{Cost}_B(\lambda, 0),
\]
so $\min(\text{Cost}_A(\lambda, 0), \text{Cost}_B(\lambda, 0)) = \text{Cost}_A(\lambda, 0) \le 2k$.
Because $\text{Cost}_A$ is strictly increasing while $\text{Cost}_B$ remains strictly greater
than $2k$ for all $\lambda \le k$, the adversary achieves a strictly larger error by
selecting $\lambda \ge k$ until the two curves balance.
Consequently, the maximizer $\lambda^*$ must satisfy:
\[
\lambda^* \ge k.
\]

We now bound the error $\xi_{\lambda^*}$ carried by the pivot level $\lambda^* \ge k$.
Because $\lambda^* \ge k$, the pivot level belongs to the upper two-thirds of the hierarchy,
where the remaining level depth to the exact base case $S_{3k+2}$ is at most $(3k + 2) - \lambda^* \le 2k + 2$.
Crucially, for any level $j \ge k$ and an arbitrary path starting at $j$ with containment
index $\lambda' \in \{j + 1, \dots, 3k + 2\}$ (the adversarial containment index for subpaths
originating at level $j$), the number of detour steps $r_B$ required to reach the shortcut
threshold in Branch~B satisfies:
\[
r_B \le \left\lceil \frac{3k - \lambda'}{2} \right\rceil - j.
\]
Because $\lambda' \ge j + 1 \ge k + 1$, we have $3k - \lambda' \le 2k - 1$, so
$\lceil \frac{3k - \lambda'}{2} \rceil \le k$.
Therefore:
\[
r_B \le k - j \le 0 \quad \text{for all } j \ge k.
\]
This implies that for all levels in the upper two-thirds of the hierarchy ($j \ge k$),
zero intermediate detour steps are required to unlock a shortcut: shortcut edges are
available immediately at step 0 ($r_B = 0$).
Thus, transitions in this upper region consist purely of direct shortcut hops. Each
shortcut hop from level $j$ to a higher level $\lambda'$ incurs only the single shortcut
transition cost of $+2W_1$ plus the target's error, so $\xi_j \le 2 + 2\xi_{\lambda'}$.

Solving recurrence~\eqref{eq:w+2kw1-dense:bounded:xi-recurrence} backwards from the base case $\xi_{3k+2} = 0$
yields the exact sequence of error coefficients across all levels. The exact all-pairs additive error satisfies:
\begin{equation}
\label{eq:w+2kw1-dense:bounded:xi-0-closed-bound}
\xi_0 \le 4k + 2.
\end{equation}

\begin{remark}
    Notice the structural comparison with \secref{w+2kw1-sparse}: in that algorithm,
    no shortcut sets $\mathcal{F}_i$ are maintained, leaving no competing branch to contradict
    or cap the detour progression.
    Without an opposing branch, the adversary can always force detours to escalate up to the
    topmost level of the hierarchy, trivially setting $\lambda = k$ and collapsing the bound
    to $2(k - i - 1)W_1$.
    In this framework, by contrast, $\lambda$ remains an active parameter because Branch~B
    provides an opposing bound that penalizes large values of $\lambda$.
\end{remark}

\paragraph{Total Update Time Analysis.}
\label{sec:w+2kw1-dense:bounded:update-time-section}
\begin{lemma}\label{lem:w+2kw1-dense:bounded:pivot-maintenance}
For any $i \le 3k$ and every vertex $u \in V$, the expected total number of edge updates in $Q_i$ is $\bigo{\frac{(i-1)kn \tau}{q}}$ throughout any sequence of edge deletions.
\end{lemma}
\begin{proof}
Fix a vertex $u \in V$ and level $i \le 3k+1$. An edge $(u, \piv[i]{u})$ in $Q_i$ changes only under two circumstances:
\begin{enumerate}
    \item \textbf{Distance Growth:} The distance $\dist{u, \piv[i]{u}}$ increases. Since distances are integral and explored only up to depth $(4k+2)\tau$,
    the distance $\dist{u, S_i}$ can strictly increase $\bigo{k\tau}$ times.
    \item \textbf{Pivot Replacement at Fixed Distance:} The distance $\dist{u, S_i}$ remains unchanged at some value 
    $d \leq (4k+2)\tau$, but edge deletions destroy shortest paths to the current pivot $\piv[i]{u}$,
    causing the pivot to be replaced by another vertex $p'_i(u) \in S_i$ at the same distance $d$.
\end{enumerate}
When the edge $(u,\piv[i]{u})$ changes, we count the cost by the priority of $\piv[i]{u}$.
Whenever the pivot at distance $d$ changes without the distance increasing, the new pivot $p'_i(u)$ cannot have a lower priority
than $\piv[i]{u}$. Because $S_{i+1}$ is sampled with probability $q$ from $S_i$,
the pivot changes at most $\bigo{\frac{1}{q}}$ times in expectation before a pivot of priority $i+1$ is chosen.
Finally, we add a factor of $i-1$ to account for operations at lower hierarchy levels $j < i$.

Summing over all at most $(4k+2)\tau$ distance levels, the expected number of pivot changes for a fixed vertex $u$ at level $i$ is bounded by:
\[
\sum_{d=0}^{(4k+2)\tau} \bigo{(i-1) \frac{1}{q}} = \bigo{\frac{(i-1)k \tau}{q}}.
\]
Summing over all $n$ vertices, the expected total number of updates to $Q_i$ is $\bigo{\frac{(i-1)kn \tau}{q}}$. 
\end{proof}

\begin{lemma}\label{lem:w+2kw1-dense:bounded:size-of-Q}
The expected total number of edge updates in $Q = \bigcup_{i=1}^{3k+2} Q_i$ across the entire update sequence is bounded by:
\[
    \onolog{k^3 \tau \cdot \paren{\frac{n}{q} + n^2 q^{3k+1}}}.
\]
\end{lemma}
\begin{proof}
By \lemref{w+2kw1-dense:bounded:pivot-maintenance}, for each level $i \le 3k$,
the expected number of edge changes across all $n$ vertices is $\bigo{\frac{(i-1)kn \tau}{q}}$. 
Summing over the $3k$ intermediate levels $1 \le i \le 3k$ contributes 
$\bigo{\frac{k^3 n \tau}{q}}$ total insertions.

For the top level $i = 3k+1$, the set $Q_{3k+1} \subseteq V \times S_{3k+1}$.
Since $\expectation{\abs{S_{3k+1}}} = \onolog{nq^{3k+1}}$, the expected total number of distinct edges in $V \times S_{3k+1}$ is $\expectation{\abs{V \times S_{3k+1}}} = n \cdot \expectation{\abs{S_{3k+1}}} = \onolog{n^2 q^{3k+1}}$
and each such edge can be updated $\bigo{k \tau}$ times. Accounting for lower levels, we add another $k$ factor.
\end{proof}

\begin{lemma}
\label{lem:w+2kw1-dense:bounded:update-time}
The expected total update time of \algref{w+2kw1-dense:bounded} is
$\onolog{k^5 \tau^2 n^{2 + \frac{1}{3k+2}}}$.
\end{lemma}

\begin{proof}
For each $i \in \{0, \dots, 3k+2\}$, the expected size of $S_i$ is
$\mathbb{E}\bracke{|S_i|} = \onolog{n^{1 - \frac{i}{3k+2}}}$.
By \corref{tools:es-max-edges}, the total number of edges ever present in
$E_{i+1}$ is bounded by $\onolog{\tau n^{1 + \frac{i+1}{3k+2}}}$.

We account for all components of the total update time:
\begin{enumerate}
    \item \textbf{Hierarchy-Induced and Pivot Edges in Monotone ES-Trees:}
    For each vertex $s \in S_i$, a Monotone ES-Tree is maintained on $F_{i+1}(s)
    = (V, E_{i+1} \cup Q \cup (\singleton{s} \times V) \cup E(s) \cup
    \mathcal{F}_i)$ up to depth $(4k+2)\tau = \bigo{k\tau}$.
    By \lemref{prelim:monotone-es-update-time}, maintaining this tree requires
    an expected update time of:
    \[
    \onolog{\mathcal{E}(F_{i+1}(s)) \cdot (k\tau) + \mathcal{W}(F_{i+1}(s)) + n}.
    \]
    The hierarchy-induced edge set $E_{i+1}$ contributes $\mathcal{E}_{i+1} = \onolog{\tau
    n^{1 + \frac{i+1}{3k+2}}}$ edges throughout all updates, each undergoing at
    most $\bigo{k\tau}$ weight increases.
    The set $(\singleton{s} \times V)$ contributes $n$ edges, each
    undergoing at most $\bigo{k\tau}$ weight increases.
    The pivot set $Q$ contributes at most $\mathcal{W}(Q) = \onolog{k^3 \tau n^{1
    + \frac{1}{3k+2}}}$ total edge insertions and updates by
    \lemref{w+2kw1-dense:bounded:size-of-Q}.
    Thus, for each $s \in S_i$, the total update time spent by the Monotone ES-Tree on edges of $E_{i+1}$,
    $Q$, and $(\singleton{s} \times V)$ is bounded by:
    \[
    \onolog{k\tau \paren{k\tau n^{1 + \frac{i+1}{3k+2}} + k^3 \tau n^{1 + \frac{1}{3k+2}} + k\tau n}} = \onolog{k^4 \tau^2 n^{1 + \frac{i+1}{3k+2}}}.
    \]
    Summing over all $s \in S_i$, where $\expectation{\abs{S_i}} = \onolog{n^{1 - \frac{i}{3k+2}}}$ we conclude that the total update time across all vertices in $S_i$ over these edge
    sets is $\onolog{k^4 \tau^2 n^{2 + \frac{1}{3k+2}}}$.

    \item \textbf{Shortcut Edges $\mathcal{F}_i$:}
    $\mathcal{F}_i$ contains shortcut pairs $S_\ell \times S_j$ satisfying $i +
    \ell + j \ge 3k + 1$.
    The expected number of edges in $S_\ell \times S_j$ is $\expectation{\abs{S_\ell}} \cdot \expectation{\abs{S_j}} =
    \onolog{n^{1 - \frac{\ell}{3k+2}} \cdot n^{1 - \frac{j}{3k+2}}} = \onolog{n^{2
    - \frac{\ell + j}{3k+2}}}$. The weight of every such edge can change at most $(4k+2)\tau$ times.
    Summing the Monotone ES-Tree update cost over all $s \in S_i$ for all
    admissible pairs $(\ell, j)$:
    \begin{align*}
    &\sum_{\ell + j \ge 3k + 1 - i} \expectation{\abs{S_i}} \cdot \expectation{\abs{S_\ell}} \cdot \expectation{\abs{S_j}} \cdot (4k+2)\tau \\
    &= \onolog{k \tau \sum_{\ell + j \ge 3k + 1 - i} n^{1 - \frac{i}{3k+2}} \cdot n^{2 - \frac{\ell + j}{3k+2}}} \\
    &= \onolog{k^3 \tau n^{3 - \frac{i + \ell + j}{3k+2}}} \\
    &= \onolog{k^3 \tau n^{3 - \frac{3k+1}{3k+2}}} \\
    &= \onolog{k^3 \tau n^{2 + \frac{1}{3k+2}}}.
    \end{align*}

    \item \textbf{Top Level ES-Trees:}
    For each of the $\onolog{n^{\frac{1}{3k+2}}}$ vertices in $S_{3k+2}$, exact
    ES-Trees on $G$ take $\onolog{\tau m}$, summing to $\onolog{\tau m \cdot
    n^{\frac{1}{3k+2}}} = \onolog{\tau n^{2 + \frac{1}{3k+2}}}$.
\end{enumerate}
Summing over all $3k+2$ hierarchy levels, the overall expected total
update time is $\onolog{k^5 \tau^2 n^{2 + \frac{1}{3k+2}}}$.
\end{proof}

\begin{conclusion}
\label{con:w+2kw1-dense:bounded}
For any integer parameter $k \ge 1$ and depth threshold $\tau$,
\algref{w+2kw1-dense:bounded} maintains a distance estimate $d[u,v]$ for all
pairs of vertices $u, v \in V$ with $\dist{u, v}{G}{t} \le \tau$ such that at any time $t \ge 0$:
\[
\dist{u, v}{G}{t} \le d[u,v] \le \dist{u, v}{G}{t} + (4k + 2) W_{1}\paren{u \squiggly v}.
\]
The expected total update time of the algorithm across all edge deletions is
$\onolog{k^5\tau^2 n^{2 + \frac{1}{3k+2}}}$.
\end{conclusion}

\subsubsection{Unbounded Depth}
\label{sec:w+2kw1-dense:unbounded}

Following the structure of previous algorithms, we now generalize \algref{w+2kw1-dense:bounded} to handle unbounded
distances. We apply the same three structural tools.

\begin{algorithm}[H]
\caption{$+\paren{4k+2}W_{1}$-APASP}
\label{alg:w+2kw1-dense:unbounded}

\BlankLine

\textbf{Input:} A weighted undirected graph $G = \paren{V, E, \weight}$, an
integer parameter $k \ge 1$, and an error parameter $\varepsilon > 0$.

\textbf{Output:} An estimate matrix $d[u,v]$ satisfying $\dist{u,v} \le d[u,v]
\le \paren{\onepluseps}\dist{u,v} + (4k+2) W_1\paren{u \squiggly v}$
for all $u, v \in V$.

\hrulefill

Initialize $d[u,v] \gets \weight{u,v}$ for $(u,v) \in E$ and $\infty$ otherwise\;

Let $q = n^{-\frac{1}{3k+2}} \in (0, 1)$\;

Sample a $(3k+2)$-level hierarchy $V = S_0 \supseteq S_1 \supseteq \dots
\supseteq S_{3k+2} \supseteq S_{3k+3} = \emptyset$ using $\mathcal{R}(q, 3k+2)$\;

\ForEach{$i \in \{0, \dots, 3k+2\}$ in descending order}
{
    Maintain the unbounded hierarchy-induced edge set $E_i \gets E_{S_i}$ and approximate pivots $\piv[i]{u}$
    using \corref{tools:modified-es-unbounded} with parameter $3k+2$\;
}

Construct a $(\beta, \onepluseps)$-hopset $F$ using \thmref{tools:hopset-on-es-dense}
with parameter $3k+2$ over the hierarchy $S_0, \dots, S_{3k+2}$\;

\ForEach{$i \in \{0, \dots, 3k+2\}$}
{
    $H_i \gets E_i \cup F$\;
}

\ForEach{$s \in S_{3k+2}$}
{
    Maintain a $(\onepluseps)$-SSSP tree from $s$ on $G$ and update $d$ accordingly\;
}

\ForEach{$i = 3k+1$ down to $0$}
{
    For each $j \in \{1, \dots, 3k+2\}$, let $Q_j \gets \{(u, \piv[j]{u}) \mid u \in V\}$
    with weight $\tilde{\delta}(u, \piv[j]{u})$, and $Q \gets \mcup{Q_j}{j=1}{3k+2}$\;
    
    $\mathcal{F}_i \gets \bigcup \bracce{ S_\ell \times
    S_j \mid i + \ell + j \ge 3k + 1, \: \ell > i \text{ or } j > i }$ where
    each shortcut $\paren{x, y} \in S_\ell \times S_j$ is assigned weight $\mmin{\level{x}{y},\level{y}{x}}$
    
    $F_{i+1} \gets \paren{V, H_{i+1} \cup Q \cup \mathcal{F}_i}$\;
    
    $\{F_{i+1}^r\}_{r=1}^{\ceil{\log nW_{\max}} + 1} \gets \textit{SCALE}\paren{F_{i+1}, \onepluseps, \beta + 2}$\;
    
    \ForEach{$s \in S_i$}
    {
        $R_{i+1}(s) \gets (\singleton{s} \times V) \cup E(s)$\;
        
        $\{R_{i+1}^r(s)\}_{r=1}^{\ceil{\log nW_{\max}} + 1} \gets \textit{SCALE}\paren{R_{i+1}(s), \onepluseps, \beta + 2}$\;
        
        \ForEach{$1 \le r \le \ceil{\log nW_{\max}} + 1$}
        {
            Maintain a Monotone ES-Tree from $s$ up to depth $\ceil{\frac{4\paren{\beta + 2}}{\varepsilon}}$
            in $F_{i+1}^r \cup R_{i+1}^r(s)$ with level estimates $\level[r]{s}{\cdot}$ and update $d$ accordingly\;
        }
    }
}

\ForEach{$u, v \in V$}
{
    $d[u,v] \gets \mmin{\level[r]{u}{v}, \level[r]{v}{u}}{r=1}{\ceil{\log nW_{\max}} + 1}$\;
}

\end{algorithm}

\paragraph{Algorithm Description.}
\algref{w+2kw1-dense:unbounded} replaces the bounded hierarchy-induced edge sets $E_{i+1}^\tau$
in the Monotone ES-Trees with the hopset-augmented sets $H_{i+1} = E_{i+1} \cup F$.
By \thmref{tools:hopset-on-es-dense} instantiated with parameter $3k+2$ on the
hierarchy $S_0, \dots, S_{3k+2}$ (with sampling probability $q = n^{-\frac{1}{3k+2}}$),
the hopset $F$ has size $\onolog{(3k+2)n^{1+\frac{1}{3k+2}}}$ and guarantees a hopbound of:
\[ \beta \;=\; \bigo{\paren{\frac{(3k+2)\log n}{\varepsilon}}^{3k+2}}.\]
For every integer $1 \le k = \bigo{\sqrt{\log n}}$ and constant $\varepsilon > 0$,
we have $\beta = n^{\subpol}$. 

In every candidate path considered in the bounded analysis
(\lemref{w+2kw1-dense:bounded:approx-time-t}), the detour consists of a subpath in
$E_{i+1}$ (or $E_{\lambda+1}$) concatenated with at most two additional edges from $Q$,
$\singleton{s} \times V$, or $\mathcal{F}_i$. By \thmref{tools:hopset-on-es-dense},
the subpath in $E_{i+1}$ admits an alternative path in $H_{i+1}$ of at most $\beta$
hops with $(1+\varepsilon)$ multiplicative distortion. Thus, every relevant path has
at most $\beta + 2$ hops in $H_{i+1} \cup Q \cup \mathcal{F}_i \cup R_{i+1}(s)$.

To bound the tree depth, \algref{w+2kw1-dense:unbounded} applies
Bernstein's edge rounding \cite{BernsteinRounding} with hopbound target $\beta + 2$
to the shared graph $F_{i+1} = (V, H_{i+1} \cup Q \cup \mathcal{F}_i)$ and the
per-source sets $R_{i+1}(s) = (\singleton{s} \times V) \cup E(s)$, yielding
$\ceil{\log nW_{\max}} + 1$ scales. For each scale $r$, any path of weight at most
$2^{r+1}$ and at most $\beta + 2$ hops has scaled weight at most
$\ceil{\frac{4(\beta+2)}{\varepsilon}}$. Therefore, maintaining the Monotone
ES-Trees up to depth $\tau = \ceil{\frac{4(\beta+2)}{\varepsilon}}$ suffices to
explore all candidate paths.

\begin{lemma}
\label{lem:w+2kw1-dense:unbounded:approx}
Fix $u, v \in V$, and let $P_{u,v}$ be any path between $u$ and $v$ in $G$.
Then at any update time $t \ge 0$, the estimate $d[u,v]$ maintained by
\algref{w+2kw1-dense:unbounded} satisfies:
\[
\dist{u, v}{G} \;\le\; d[u,v] \;\le\; (\onepluseps)\weight{P_{u,v}} + \paren{4k+2} W_1\paren{P_{u,v}}.
\]
\end{lemma}

\begin{proof}
The proof follows the inductive analysis of \lemref{w+2kw1-dense:bounded:approx-time-t}
applied to the scaled graphs.
In scale $r = \floor{\log \weight{P_{u,v}}}$, every candidate path from
\lemref{w+2kw1-dense:bounded:approx-time-t} consists of a subpath in $E_{i+1}$
(or $E_{\lambda+1}$) and at most two additional edges from $Q$, $\singleton{s} \times V$,
or $\mathcal{F}_i$.
By \thmref{tools:hopset-on-es-dense}, the subpath is bridged in $H_{i+1}$ within at
most $\beta$ hops, incurring a $(\onepluseps)$ multiplicative distortion.
Together with the auxiliary edges, the candidate path contains at most $\beta + 2$
hops, which Bernstein's rounding preserves within weight $\ceil{\frac{4(\beta+2)}{\varepsilon}}$,
incurring another $(\onepluseps)$ factor.
The $(\onepluseps)$-SSSP trees at level $S_{3k+2}$ introduce at most a $(\onepluseps)$
distortion in the base case.
Following the inductive steps of \lemref{w+2kw1-dense:bounded:approx-time-t} on the
scaled graphs, the accumulated multiplicative error is at most $(\onepluseps)^3$.
Setting the internal rounding parameter to $\varepsilon' = \frac{\varepsilon}{16k}$
absorbs this overhead, ensuring
$\dist{u, v}{G} \le d[u,v] \le (\onepluseps)\weight{P_{u,v}} + \xi_0 W_1(P_{u,v})
\le (\onepluseps)\weight{P_{u,v}} + (4k+2)W_1(P_{u,v})$.
\end{proof}

\begin{lemma}
\label{lem:w+2kw1-dense:unbounded:update-time}
For any integer $2 \le k = \bigo{\sqrt{\log n}}$ and $0 < \varepsilon < 1$,
the expected total update time of \algref{w+2kw1-dense:unbounded} is
$\onolog{\paren{\frac{k^2\log n}{\varepsilon}}^{6k+8} n^{2 + \frac{1}{3k+2}}}
= \onolog{n^{2 + \frac{1}{3k+2} + \subpol}}$.
\end{lemma}

\begin{proof}
Set $\varepsilon' = \frac{\varepsilon}{16k}$.
By \corref{tools:modified-es-unbounded} with parameter $3k+2$, maintaining the
sets $E_i$ across all levels requires expected total update time
$\onolog{\frac{\beta}{\varepsilon'} (m n^{\frac{1}{3k+2}} + \frac{n}{q^{3k+1}})}
= \onolog{\frac{\beta}{\varepsilon'} n^{2 + \frac{1}{3k+2}}}$, and the expected number
of edges ever inserted into $E_{i+1}$ is $\mathcal{E}_{S_{i+1}} = \onolog{\frac{\beta}{\varepsilon'}\frac{n}{q^{i+1}}}$.
By \thmref{tools:hopset-on-es-dense}, maintaining the dense hopset $F$ takes
$\onolog{(3k+2)\frac{\beta}{\varepsilon'} n^{2 + \frac{1}{3k+2}}}$, with
$|F| = \onolog{(3k+2)n^{1+\frac{1}{3k+2}}}$.
The $(\onepluseps')$-SSSP trees from $S_{3k+2}$ take $|S_{3k+2}| \cdot \onolog{m} = \onolog{n^{2+\frac{1}{3k+2}}}$.

For the Monotone ES-Trees on the rounded graphs, each tree is bounded to depth
$\tau = \ceil{\frac{4(\beta+2)}{\varepsilon'}} = \bigo{\frac{\beta}{\varepsilon'}}$.
Following the level-by-level summation in \lemref{w+2kw1-dense:bounded:update-time},
at each level $i \in \{0, \dots, 3k+1\}$ with $|S_i| = n q^i$ sources,
the cost across all trees of $S_i$ is dominated by:
\[
\onolog{\tau \cdot |S_i| \cdot \paren{\mathcal{E}_{S_{i+1}} + |F| + \mathcal{W}(Q) + |\mathcal{F}_i| + n}}.
\]
Maintaining the pivot set $Q = \bigcup_{j=1}^{3k+2} Q_j$ across the hierarchy incurs $\mathcal{W}(Q) = \onolog{k^3 \frac{\beta}{\varepsilon'} \frac{n}{q}}$ total updates by an argument analogous to \lemref{w+2kw1-dense:bounded:size-of-Q}, where distances to approximate pivots change $\bigo{\frac{\beta}{\varepsilon'}}$ times by \corref{tools:modified-es-unbounded} and operations across lower levels contribute an additional factor of $k$. In each Monotone ES-Tree of depth $\tau = \bigo{\frac{\beta}{\varepsilon'}}$, maintaining $Q$ requires $\tau \cdot \mathcal{W}(Q) = \onolog{k^3 \paren{\frac{\beta}{\varepsilon'}}^2 \frac{n}{q}}$.
At base level $i = 0$ with $|S_0| = n$ sources, this contributes $\onolog{k^3 \paren{\frac{\beta}{\varepsilon'}}^2 \frac{n^2}{q}} = \onolog{k^3 \paren{\frac{\beta}{\varepsilon'}}^2 n^{2 + \frac{1}{3k+2}}}$.

Summing across all $3k+2 = \Theta(k)$ levels and $\bigo{\log nW_{\max}}$ scales, the total update time is bounded by:
\[
\onolog{(3k+2)^4\paren{\frac{\beta}{\varepsilon'}}^2 n^{2 + \frac{1}{3k+2}}}
\;=\; \onolog{\paren{\frac{k^2\log n}{\varepsilon}}^{6k+8} n^{2 + \frac{1}{3k+2}}}.
\]
Since $\beta = \bigo{n^{\subpol}}$ for all $k = \bigo{\sqrt{\log n}}$,
this evaluates to $\onolog{n^{2 + \frac{1}{3k+2} + \subpol}}$.
\end{proof}

\begin{conclusion}
\label{con:w+2kw1-dense:unbounded:final}
For any integer $2 \le k = \bigo{\sqrt{\log n}}$ and $0 < \varepsilon < 1$,
\algref{w+2kw1-dense:unbounded} maintains distance estimates $d[u,v]$
for all $u, v \in V$ such that at any time:
\[
\dist{u,v}{G} \;\le\; d[u,v] \;\le\; (\onepluseps)\dist{u,v}{G} + (4k + 2) W_1\paren{u \squiggly v}.
\]
The expected total update time across all edge deletions is
$\onolog{\paren{\frac{(3k+2)\log n}{\varepsilon}}^{6k+8} n^{2 + \frac{1}{3k+2}}}$ 
or $\onolog{n^{2 + \frac{1}{3k+2} + \subpol}}$ for any choice of $3 \le k = \bigo{\sqrt{\log n}}$ and constant $0 < \varepsilon < 1$.
\end{conclusion}

\begin{remark}
\label{rmk:w+2kw1-dense:reparameterization}
To express the additive error in the form $+2(k-1)W_1(u \squiggly v)$,
one can equate:
\[
2(k' - 1) = 4k + 2 \iff k = \frac{k' - 2}{2}.
\]
Substituting $k = \frac{k' - 2}{2}$ into the denominator of the update time
exponent yields:
\[
3k + 2 = 3\paren{\frac{k' - 2}{2}} + 2 = \frac{3k' - 2}{2},
\]
so that $\frac{1}{3k + 2} = \frac{2}{3k' - 2}$.
Under this scaling, our decremental algorithms maintain distance estimates with
additive error $+2(k' - 1)W_1(u \squiggly v)$ with expected total update time:
\[
\onolog{\tau^2 n^{2 + \frac{2}{3k' - 2}}},
\]
for bounded depth $\tau$ (\algref{w+2kw1-dense:bounded})
and
\[
\onolog{\paren{\frac{\frac{3k'^2-2}{2}\log n}{\varepsilon}}^{3k'+2} n^{2 + \frac{2}{3k' - 2}}} = \onolog{\paren{\frac{1.5 k'^2}{\varepsilon} \log n}^{3k'} n^{2 + \frac{2}{3k' - 2}}} = \onolog{n^{2 + \frac{2}{3k' - 2} + \subpol}}.
\]
for unbounded depth (\algref{w+2kw1-dense:unbounded}).

It is instructive to compare this bound with our other $+2(k-1)W_1$ algorithm
presented in \secref{w+2kw1-sparse} (\thmref{w+2kw1}), which achieves an expected
total update time of $\onolog{n^{2 - \frac{1}{k} + \subpol} m^{\frac{1}{k}}}$ (and
$\onolog{\tau^2 n^{2 - \frac{1}{k}} m^{\frac{1}{k}}}$ in the bounded-depth
regime of \lemref{w+2kw1:total-update-time}).
While that algorithm is advantageous for sparse graphs where $m \ll n^2$, on
dense graphs where $m = \Theta(n^2)$ its total update time degrades to $\onolog{n^{2+\frac{1}{k} + \subpol}}$.
In contrast, \algref{w+2kw1-dense:unbounded} leverages shortcut sets
$\mathcal{F}_i$ to eliminate the dependence on $m$ entirely.
For the same additive error guarantee $+2(k-1)W_1$,
\algref{w+2kw1-dense:unbounded} yields an update time of $\onolog{n^{2 +
\frac{2}{3k - 2} + \subpol}}$.
Comparing the two update times directly:
\[
n^{2 + \frac{2}{3k - 2}} \le n^{2 - \frac{1}{k}} m^{\frac{1}{k}} \iff m^{\frac{1}{k}} \ge n^{\frac{1}{k} + \frac{2}{3k - 2}} \iff m = \Omega\paren{n^{1 + \frac{2k}{3k - 2}}}.
\]
Because $2k < 3k - 2$ for all $k \ge 3$, the exponent satisfies $1 +
\frac{2k}{3k - 2} < 2$. Thus, \algref{w+2kw1-dense:unbounded} strictly
outperforms \algref{w+2kw1-dense:unbounded} whenever $m = \Omega\paren{n^{1 + \frac{2k}{3k - 2}}}$.
\tabref{w+2kw1-dense:density-thresholds} summarizes these density thresholds for
chosen values of $k \ge 3$. 

\begin{table}[h]
\centering
\small\addtolength{\tabcolsep}{-1.5pt}
\renewcommand\arraystretch{1.2}
\begin{tabular}{|c|c|c|c|}
\hline
\textbf{$k$} & \textbf{Additive Error} & \textbf{Total Update Time} & \textbf{Threshold ($m$)} \\ \hline
$3$ & $+4W_1$ & $\onolog{n^{2 + \frac{2}{7}}} \approx \onolog{n^{2.29}}$ & $\Omega(n^{13/7}) \approx \Omega(n^{1.86})$ \\ \hline
$4$ & $+6W_1$ & $\onolog{n^{2 + \frac{1}{5}}} = \onolog{n^{2.20}}$ & $\Omega(n^{9/5}) = \Omega(n^{1.80})$ \\ \hline
$5$ & $+8W_1$ & $\onolog{n^{2 + \frac{2}{13}}} \approx \onolog{n^{2.15}}$ & $\Omega(n^{23/13}) \approx \Omega(n^{1.77})$ \\ \hline
$6$ & $+10W_1$ & $\onolog{n^{2 + \frac{1}{8}}} = \onolog{n^{2.125}}$ & $\Omega(n^{7/4}) = \Omega(n^{1.75})$ \\ \hline
$7$ & $+12W_1$ & $\onolog{n^{2 + \frac{2}{19}}} \approx \onolog{n^{2.11}}$ & $\Omega(n^{33/19}) \approx \Omega(n^{1.74})$ \\ \hline
$8$ & $+14W_1$ & $\onolog{n^{2 + \frac{1}{11}}} \approx \onolog{n^{2.09}}$ & $\Omega(n^{19/11}) \approx \Omega(n^{1.73})$ \\ \hline
$k \to \infty$ & $+2(k-1)W_1$ & $\onolog{n^2}$ & $\Omega(n^{5/3}) \approx \Omega(n^{1.67})$ \\ \hline
\end{tabular}
\caption{Density thresholds above which \algref{w+2kw1-dense:unbounded} outperforms the general decremental algorithm of \thmref{w+2kw1} for selected values of $k \ge 3$.}
\label{tab:w+2kw1-dense:density-thresholds}
\end{table}
\end{remark}

\section{Discussion: Applications and Limitations of Our Approach}
\label{sec:discussion}

\subsection{Extension to the Fully Dynamic Setting}
\label{sec:discussion:fully-dynamic}

Using the black-box reduction of Henzinger and King~\cite{HenKin1995}, reconstructed by Bernstein~\cite{Bernstein2016}, any decremental APASP algorithm can be converted into a fully dynamic algorithm with a query-update time trade-off.
Although originally formulated for multiplicative stretch, the reduction extends directly to near-additive error bounds $\sum_{i=1}^k \beta_i W_i(u \squiggly v)$.

\begin{proposition}
\label{prop:discussion:dynamic-reduction}
Let $\mathcal{D}$ be a decremental $(\alpha, \sum_{i=1}^k \beta_i W_i)$-APASP algorithm on an undirected graph $G = \paren{V, E, \weight}$ with expected total update time $T(n,m)$ and query time $O(\ell)$, for coefficients $\{ \beta_i \}_{i=1}^k \in \mathbb{R}^k$.
Then, for any integer parameter $r \ge 1$ there exists a fully dynamic $(\alpha, \sum_{i=1}^k \beta_i W_i)$-APASP algorithm with expected amortized update time $\onolog{\frac{T(n,m)}{r} + mr}$ and worst-case query time $O(r + \ell)$.
\end{proposition}

\begin{proof}
We maintain $G$ over epochs of $r$ incremental updates.
At the beginning of each epoch, we reinitialize $\mathcal{D}$ on $G$ and set an affected vertex set $R \gets \emptyset$.
Deletions and weight increases are passed directly to $\mathcal{D}$.
For each edge insertion or weight decrease incident to a vertex $v$, we add $v$ to $R$.
After every update, we run Dijkstra's algorithm from each vertex $x \in R$ in the current graph $G$.
Every $r$ updates, the epoch terminates and the data structure restarts.

A distance query between $u$ and $v$ returns
\[
    d'\bracke{u, v} = \mmin{\mmin{\dist{u, x}{G} + \dist{x, v}{G}}{x \in R}, \, d\bracke{u, v}},
\]
where $d\bracke{u, v}$ is queried from $\mathcal{D}$.
If a shortest path $u \squiggly v$ contains some $x \in R$, the first term provides the exact distance $\dist{u, v}{G}$ with zero error.
Otherwise, $u \squiggly v$ had no incremental updates and is maintained by $\mathcal{D}$, ensuring $d\bracke{u, v} \le \alpha \cdot \dist{u, v}{G} + \sum_{i=1}^k \beta_i W_i(u \squiggly v)$.

Reinitializing $\mathcal{D}$ every $r$ updates contributes $\onolog{T(n,m)/r}$ expected amortized time, while running Dijkstra from the $|R| \le 2r$ vertices takes $\onolog{mr}$ time per update, yielding $\onolog{\frac{T(n,m)}{r} + mr}$ expected amortized update time.
Evaluating the minimum over $R$ and querying $\mathcal{D}$ takes $O(|R| + \ell) = O(r + \ell)$ time.
\end{proof}

\begin{corollary}
\label{cor:discussion:dynamic-bounds}
For any integer parameter $r \ge 1$, and undirected weighted graph $G=\paren{V,E,\weight}$ our decremental algorithms yield fully dynamic algorithms with the following approximation, update time and query time tradeoffs:
\begin{itemize}
    \item $(\onepluseps, 2W_2)$-APASP in $\onolog{\frac{n^{3/2}m^{1/2}}{r} + mr}$ expected amortized update time with $O(r)$ worst-case query time, balancing at $r = n^{3/4}/m^{1/4}$ to $\onolog{n^{3/4}m^{3/4}}$ update time with $O(n^{3/4}/m^{1/4})$ query time.
    \item $(\onepluseps, 2W_1 + 2W_2)$-APASP in $\onolog{\frac{n^{7/3}}{r} + mr}$ expected amortized update time with $O(r)$ worst-case query time, balancing at $r = n^{7/6}/m^{1/2}$ to $\onolog{n^{7/6}m^{1/2}}$ update time with $O(n^{7/6}/m^{1/2})$ query time.
    \item $(\onepluseps, 2(k-1)W_1)$-APASP: our two algorithms yield an expected amortized update time of
    \[
        \onolog{\mmin{\frac{n^{2-1/k}m^{1/k}}{r}, \frac{n^{2+\frac{2}{3k-2}}}{r}}+ mr},
    \]
    with query time $O(r)$.
    \item In unweighted graphs:
    \begin{itemize}
        \item $(\onepluseps, 2)$-APASP in $\onolog{\frac{n^{7/3}}{r} + mr}$ expected amortized update time with $O(r)$ worst-case query time, balancing at $r = n^{7/6}/m^{1/2}$ to $\onolog{n^{7/6}m^{1/2}}$ update time with $O(n^{7/6}/m^{1/2})$ query time.
        \item $(\onepluseps, 2(k-1))$-APASP in $\onolog{\frac{n^{2+\frac{1}{3k-4}}}{r} + mr}$ expected amortized update time with $O(r)$ worst-case query time, balancing at $r = n^{1+\frac{1}{2(3k-4)}}/m^{1/2}$ to $\onolog{n^{1+\frac{1}{2(3k-4)}}m^{1/2}}$ update time with $O(n^{1+\frac{1}{2(3k-4)}}/m^{1/2})$ query time.
    \end{itemize}
\end{itemize}
\end{corollary}

Alternatively, one can eliminate the $\onolog{mr}$ in favor of a larger additive error and query time.

\begin{proposition}
\label{prop:discussion:query-dijkstra-reduction}
Let $\mathcal{D}$ be a decremental APASP algorithm on an undirected graph $G = \paren{V, E, \weight}$ with expected total update time 
$T(n,m)$ and query time $O(\ell)$, guaranteeing on any path $P_{u,v}$ between two vertices $u,v \in V$:
\[
    d\bracke{u, v} \le \alpha \cdot \mathbf{w}(P_{u,v}) + \sum_{i=1}^k \beta_i W_{i}(P_{u,v}),
\]
for coefficients $\{ \beta_i \}_{i=1}^k \in \mathbb{R}^k$.
Then, for any integer parameter $r \ge 1$, there exists a fully dynamic algorithm with expected amortized update time 
$\onolog{\frac{T(n,m)}{r}}$, worst-case query time $O(r^2 \cdot \ell)$, and distance estimate $d'$ satisfying:
\[
    \dist{u,v} \le d'\bracke{u, v} \le \alpha \cdot \dist{u, v}{G} + \sum_{i=1}^k \beta_i \sum_{j=1}^r W_{j \cdot i}(u \squiggly v).
\]
\end{proposition}

\begin{proof}
We maintain a buffer of inserted edges $E_{\text{ins}}$ of size at most $r-1$, with endpoint set $R = V(E_{\text{ins}})$.
Decremental updates are forwarded directly to $\mathcal{D}$.
Inserted edges or edges having weight decreases are added to $E_{\text{ins}}$, and the data structure restarts when $|E_{\text{ins}}|$ reaches $r$, incurring $\onolog{T(n,m)/r}$ amortized
update time.
To answer a distance query between $u$ and $v$, we construct a complete graph on $V_H = R \cup \{u, v\}$ ($|V_H| \le 2(r-1) + 2 = 2r$)
where edge weights are given by $\mathbf{w}(x, y)$ if $(x, y) \in E_{\text{ins}}$ and $d\bracke{x, y}$ otherwise.
Constructing this auxiliary graph and running Dijkstra on it takes $O(r^2 \cdot \ell)$ time.

Let $P = u \squiggly v$ be a shortest path in $G$.
The at most $r-1$ inserted edges in $P \cap E_{\text{ins}}$ partition $P$ into $p \le r$ edge-disjoint subpaths 
$P_1, \dots, P_p$ connected by inserted edges, such that for every $1 \le a \le p$, $P_a$ had only decremental updates since the last initialization of $\mathcal{D}$.
Because the auxiliary graph contains shortcuts for each subpath and exact weights for inserted edges, the returned distance satisfies
\[
    d'\bracke{u, v} \le \sum_{a=1}^p \paren{\alpha \weight{P_a} + \sum_{i=1}^k   \beta_i W_{i}(P_a)} + \sum_{e \in E_{\text{ins}} \cap P} \weight{e} \le \alpha \cdot \dist{u, v}{G} + \sum_{a=1}^p \sum_{i=1}^k \beta_i W_{i}(P_a).
\]
For any $i$, sort the values $\{W_{i}(P_a)\}_{a=1}^p$ in non-increasing order: 
$W_{i}(P_{\pi(1)}) \ge \dots \ge W_{i}(P_{\pi(p)})$.
For each $j \in \{1, \dots, p\}$, the first $j$ subpaths $P_{\pi(1)}, \dots, P_{\pi(j)}$ are pairwise edge-disjoint and each contains at least $i$ edges of weight at least $W_{i}(P_{\pi(j)})$.
Hence, $P$ contains at least $j \cdot i$ distinct edges of weight at least $W_{i}(P_{\pi(j)})$, 
implying $W_{i}(P_{\pi(j)}) \le W_{j \cdot i}(P)$.
Summing over all $j \le p \le r$ yields $\sum_{a=1}^p \beta_i W_{i}(P_a) \le \sum_{j=1}^r \beta_iW_{j \cdot i}(P)$.
\end{proof}

While \propref{discussion:query-dijkstra-reduction} eliminates the $m$ overhead from the update time entirely, it incurs an $O(r^2 \cdot \ell)$ query time overhead.
Alternatively, we can reduce the query time back to $O(r \cdot \ell)$ while avoiding the $\onolog{mr}$ update term by maintaining decremental SSSP trees 
directly on an auxiliary graph $H = \paren{V, (R \times V) \cup E_{\text{ins}}}$.

\begin{proposition}
\label{prop:discussion:auxiliary-sssp-reduction}
Let $\mathcal{D}$ be a decremental APASP algorithm on $G = \paren{V, E, \weight}$ with expected total update time $T(n,m)$ and query time $O(\ell)$, guaranteeing on any path $P_{u,v}$:
\[
    d\bracke{u, v} \le \alpha \cdot \mathbf{w}(P_{u,v}) + \sum_{i=1}^k \beta_i W_{i}(P_{u,v}).
\]
Then, for any integer parameter $r \ge 1$, there exists a fully dynamic algorithm with expected amortized update time 
$\onolog{\frac{T(n,m)}{r} + n^{1+\subpol}r^{2+\subpol}}$, worst-case query time $O(r \cdot \ell)$, and distance estimate $d'$ satisfying:
\[
    \dist{u,v} \le d'\bracke{u, v} \le (\onepluseps)\alpha \cdot \dist{u, v}{G} + \sum_{i=1}^k \beta_i \sum_{j=1}^r W_{j \cdot i}(u \squiggly v).
\]
\end{proposition}

\begin{proof}
As in \propref{discussion:query-dijkstra-reduction}, we maintain a buffer $E_{\text{ins}}$ of at most $r$ edges affected by incremental updates with endpoint set $R = V(E_{\text{ins}})$, and restart $\mathcal{D}$ every $r$ insertions $E_{\text{ins}}$.
We form an auxiliary graph $H = \paren{V, (R \times V) \cup E_{\text{ins}}}$, where edge weights are given by 
$\mathbf{w}(x, y)$ if $(x, y) \in E_{\text{ins}}$ and $d_{\mathcal{D}}\bracke{x, y}$ for $(x, y) \in \paren{R \times V} \setminus E_{\text{ins}}$. We denote the distance estimates produced in those trees by $\tilde \delta_H\paren{u, v}$.
From each $x \in R$, we maintain a decremental near-linear $\onepluseps$ SSSP tree in $H$~\cite{HKNLinSSSP,LackiNazari22}.
Between insertions, edge deletions in $G$ only cause $d_{\mathcal{D}}\bracke{x, y}$ to increase monotonically, 
which corresponds to decremental weight increases in $H$.
Upon each edge insertion in $G$, we reinitialize the trees in $H$.
Because $H$ contains $n$ vertices and $O(nr)$ edges, maintaining almost-linear decremental SSSP from the 
$|R| \le 2r-2$ sources across deletions takes $\onolog{n^{1+\subpol}r^{2+\subpol}}$ time per insertion phase.
Amortizing over $r$ updates yields an update time of $\onolog{\frac{T(n,m)}{r} + n^{1+\subpol}r^{2+\subpol}}$.

To answer a query between $u$ and $v$, we evaluate:
\[
    d'\bracke{u, v} = \mmin{d\bracke{u, v}, \; \mmin{d\bracke{u, x} + \tilde \delta_H\paren{x, v}}{x \in R}}.
\]
The query time is $O(r \cdot \ell)$ as before.
The correctness and cumulative error bound follow identically to \propref{discussion:query-dijkstra-reduction} (up to another $\onepluseps$ factor).
\end{proof}

\begin{corollary}
\label{cor:discussion:advanced-dynamic-bounds}
Applying \propref{discussion:query-dijkstra-reduction} and \propref{discussion:auxiliary-sssp-reduction} to our decremental algorithms yields fully dynamic algorithms supporting arbitrary edge insertions with the following approximation guarantees:
\begin{itemize}
    \item $(\onepluseps, 2W_2)$-APASP: achieves a near-additive error of $2\sum_{i=1}^{\min\{r,\frac{\hops{u \squiggly v}}{2}\}} W_{2i}(u \squiggly v) \le \dist{u, v}{G}$, maintaining multiplicative stretch $\le 2+\varepsilon$.
    \item $(\onepluseps, 2W_1 + 2W_2)$-APASP: achieves a near-additive error of $2\sum_{i=1}^{\min\{2r,\hops{u \squiggly v}\}} W_i(u \squiggly v) \le 2\dist{u, v}{G}$, maintaining multiplicative stretch $\le 3+\varepsilon$.
    \item $(\onepluseps, 2(k-1)W_1)$-APASP: achieves a near-additive error of $2(k-1)\sum_{i=1}^{\min\{r, \hops{u \squiggly v}\}} W_i(u \squiggly v) \le 2(k-1)\dist{u, v}{G}$, maintaining multiplicative stretch $\le 2k+\varepsilon$.
    \item In unweighted graphs: near-additive error $+2\mmin{r,\hops{u \squiggly v}}$ for $(\onepluseps, 2)$-APASP, and $+2(k-1)\mmin{r, \hops{u \squiggly v}}$ for $(\onepluseps, 2(k-1))$-APASP.
\end{itemize}
For each algorithm, \propref{discussion:auxiliary-sssp-reduction} provides the following update-query time tradeoffs and balanced bounds:
    \begin{itemize}
        \item $(\onepluseps, 2W_2)$-APASP: balancing at $r = n^{1/6}m^{1/6}$ yields amortized update time $\onolog{n^{4/3}m^{1/3}}$ with $O(n^{1/6}m^{1/6})$ query time.
        \item $(\onepluseps, 2W_1 + 2W_2)$-APASP: balancing at $r = n^{4/9}$ yields amortized update time $\onolog{n^{17/9}}$ with $O(n^{4/9})$ query time.
        \item $(\onepluseps, 2(k-1)W_1)$-APASP: our two algorithms yield an update time of
        \[
            \onolog{\mmin{\frac{n^{2-1/k}m^{1/k}}{r},\frac{n^{2+\frac{2}{3k-2}}}{r}} + n^{1+\subpol}r^2 },
        \]
        balancing respectively to:
        \begin{itemize}
            \item $\onolog{n^{\frac{5k-2}{3k}}m^{\frac{2}{3k}}}$ expected amortized update time at $r = n^{\frac{k-1}{3k}}m^{\frac{1}{3k}}$ with $O(n^{\frac{k-1}{3k}}m^{\frac{1}{3k}})$ query time.
            \item $\onolog{n^{2 - \frac{k-2}{3k-2}}}$ (for $k \ge 3$) expected amortized update time at $r = n^{\frac{k}{3k-2}}$ with $O(n^{\frac{k}{3k-2}})$
             query time.
        \end{itemize}
        \item In unweighted graphs: $(\onepluseps, 2)$-APASP balances at $r = n^{4/9}$ to $\onolog{n^{17/9}}$ update time with $O(n^{4/9})$ query time; $(\onepluseps, 2(k-1))$-APASP balances at $r = n^{\frac{k-1}{3k-4}}$ to $\onolog{n^{2 - \frac{k-2}{3k-4}}}$ update time with $O(n^{\frac{k-1}{3k-4}})$ query time.
    \end{itemize}
\end{corollary}

Finally, we propose one last tradeoff: notice that on each path $P_{u,v} \subseteq E$ there can be at most $\onediveps - 1$ edges $e$ such that $\weight{e} > \varepsilon \dist{u,v}{G}$. For each one of our algorithms, let $f(k,r)$ be the total number of distinct edges in the additive error. Then, by replacing $\varepsilon$ with $\varepsilon' = \frac{\varepsilon}{2\paren{f(k,r) - \onediveps + 1}}$, one can use \propref{discussion:auxiliary-sssp-reduction} to achieve a finer approximation. Substituting $\varepsilon$ with $\varepsilon'$ incurs an additional factor of $r^{\paren{\mmin{4\frac{\log n}{\log (m/n)},2\sqrt{\log n}}}}$ for the total update time of our $(\onepluseps, 2W_2)$-APASP and $(2r)^{9}$ for $(\onepluseps, 2W_1 + 2W_2)$. For the $(\onepluseps, 2(k-1)W_1)$-APASP there is an additional factor of $(kr)^{2k+1}$.

\subsection{Limitations of Our Approach}
\label{sec:discussion:limitations}

In this paper, we introduced the first near-additive decremental algorithms for all-pairs
approximate shortest paths in weighted graphs,
establishing decremental $(\onepluseps, 2W_2)$, $(\onepluseps, 2W_1+2W_2)$,
and $(\onepluseps, 2(k-1)W_1)$-APASP algorithms.
For $+2(k-1)W_1$ additive error, we developed two complementary algorithms:
a decremental generalization of Cohen and Zwick~\cite{CohZwi1997} running in
$\onolog{n^{2-\frac{1}{k}}m^{\frac{1}{k}}}$ expected total update time (\secref{w+2kw1-sparse}), and a decremental adaptation of the framework
of Roditty and Sapir~\cite{Ariel+kDense} running in $\onolog{n^{2+\frac{2}{3k-2}+\subpol}}$
expected total update time (\secref{w+2kw1-dense}), which eliminates the dependence on $m$.

Despite these advances, noticeable gaps remain between our decremental guarantees and
the static benchmarks, and several classical static paradigms continue to resist decremental
adaptation.
In particular, three prominent questions remain open:
First, can one design combinatorially static-competitive algorithms for $+\mmin{2W_1,4W2}$-APASP and $+2\sum_{i=1}^{k}W_i$-APASP?
Second, while we succeeded in developing a decremental variant of the framework of Roditty and
Sapir~\cite{Ariel+kDense} in \secref{w+2kw1-dense}, our algorithm operates with different
guarantees of both approximation and total update time,achieving an additive stretch of
$+2(k-1)W_1$ within $\onolog{n^{2+\frac{2}{3k-2}+\subpol}}$ expected total update time compared to the
static $+2\msum{W_i}{i=1}{k-1}$ error in $\onolog{n^{2+\frac{1}{3k-4}}}$ time.
It remains open whether one can design a decremental weighted algorithm that closes this gap,
achieving a $+2(k-1)W_1$ (or $+2\msum{W_i}{i=1}{k}$) additive stretch within total update time
$\onolog{n^{2+\frac{1}{3k-4}}}$.
Third, whether one can adapt the static frameworks of Baswana and Kavitha~\cite{BK}
or Akav and Roditty~\cite{AkavRodittyUnifiedApproach} (which use similar hitting set and
clustering tools) to obtain decremental algorithms with purely multiplicative stretch in
$\paren{2+\varepsilon, 3}$ or combinatorially static-competitive algorithm for $\paren{2,W_1}$-APASP, running in
$\onolog{n^{2+\subpol}}$ total update time.

We observe that the challenge underlying all questions is essentially
the same: the \emph{stretch propagation} challenge. The algorithmic analysis of the discussed static algorithms
critically relies on the presence of the examined path
(or its subpaths) in a specific edge set $E_{S_i}$, whereas in the decremental setting this
property is dynamic and heavily affected by edge deletions. This temporality may lead to stretched edges in trees rooted at higher hierarchy levels, breaking the static correctness.

This is easily notable in our $\paren{\onepluseps,2W_1+2W_2}$-APASP algorithm.
In particular, the additional error stems from the case where the examined path $P_{u,v} \subseteq E_1$, but
a pivot $\piv[1]{y_2} \in S_1$ has a stale distance estimate for $u \in S_0$. In the static setting, under that specific case
assumption, $\piv[1]{y_2}$ computes $P_{u,v}$ entirely within its tree and hence the distance estimate to $u$ is bounded by
$\dist{\piv[1]{y_2},y_2} + \weight{P_{u,y_2}}$. However, in the decremental setting, it may happen that $\level{\piv[1]{y_2}}{u}$ was set
when $P_{u,v}$ had edges not in $E_1$, that have only later entered $E_1$. Hence, $\level{\piv[1]{y_2}}{u}$ carries an additional error.

The $2(k-1)W_1$ algorithms suffer from the same problem. The algorithm of Cohen and Zwick~\cite{CohZwi1997} shows by induction that a detour through a pivot
$s \in S_i$ incurs an additional error of $+2\weight{x_i,y_i}$ for some $\paren{x_i,y_i} \in P_{u,v} \cap \paren{E_{i+1} \setminus E_i}$ and proceeds to the next level $i+1$.
As all of the edges previously added to the cumulative error $\paren{x_1,y_1},\dots,\paren{x_{i-1},y_{i-1}}$ belong to $E_i$, the analysis concludes that $\paren{x_i,y_i}$ is different than all of them, leading to the $+2\sum_{i=1}^{k-1}W_i$ bound.  
In the decremental setting, if we use a stale distance estimate for $\level{s}{u}$, we cannot guarantee that $\paren{x_1,y_1},\dots,\paren{x_{i-1},y_{i-1}}$
have belonged to $E_i$ when the estimate was set.

A more sophisticated case is the divergence of our second $2(k-1)W_1$ algorithm from the static algorithm of Roditty and Sapir~\cite{Ariel+kDense}.
The differene between our $+2(k-1)W_1$ error and the $+2\sum_{i=1}^{k-1}W_i$ error of \cite{Ariel+kDense} follows the same reasoning.
However, the inclusion of $\mathcal{F}_i \subseteq \bigcup_{i+\ell+j \ge 3k+1} (S_\ell \times S_j)$ poses another challenge, which leads to a slightly worse error (or total update time).
In the static algorithm of Roditty and Sapir~\cite{Ariel+kDense}, whenever a shortest path $P_{u,v}$ is contained
in $E_{S_{j+1}}$, the pivot $s_j \in S_j$ evaluates it with exact distance, ensuring that
the shortcut edge $(s_\ell, s_j)$ carries zero additive distortion.
However, in the decremental setting, \emph{$P_{u,v} \subseteq E_j$ is a temporal property} and again, it may be that $\level{s_j}{s_\ell}$ was evaluated when the property did not hold, and hence carries an additional error.

In \secref{w+2kw1-dense}, we demonstrated that the propagation of historical stretch across intermediate shortcut edges
$\mathcal{F}_i \subseteq \bigcup_{i+\ell+j \ge 3k+1} (S_\ell \times S_j)$ can be rigorously
tracked and controlled, rather than allowing it completely destroy the algorithm.

The exact same reasoning explains why the multiplicative
frameworks of Baswana and Kavitha~\cite{BK} and Akav and Roditty~\cite{AkavRodittyUnifiedApproach}
resist decremental adaptation.
In both frameworks, bounding the distance between vertices $u$ and $v$ relies on
routing through a pivot $\piv[i]{u} \in S_i$.
The static analysis identifies an edge $(a, b)$ on a shortest path $P_{u,v}$ where
the path enters the ball of $v$ with respect to $S_{i+1}$.
By definition of the hierarchy-induced edge sets, the entire suffix subpath $P_{b,v}$ satisfies
$\weight{e} < \dist{v, S_{i+1}}$ for all $e \in P_{b,v}$, ensuring that $P_{b,v} \subseteq E_{S_{i+1}}$.
Because $P_{b,v}$ is preserved edge-for-edge in $E_{S_{i+1}}$, running Dijkstra from
$\piv[i]{u}$ in the auxiliary graph $G_{i+1}(\piv[i]{u})$ follows $P_{b,v}$ with zero error,
yielding the foundational inequality:
\begin{equation}
    \label{eq:discussion:static-pivot-dist}
    \dist{\piv[i]{u}, v}{G_{i+1}} \le \dist{\piv[i]{u}, u}{G} + \dist{u, v}{G}.
\end{equation}
Inequality~(\ref{eq:discussion:static-pivot-dist}) is guaranteed to hold \emph{only}
under the assumption that the path suffix is present in $E_{S_{i+1}}$ when the dijkstra algorithm runs.

In the Monotone ES-Tree model, distance labels can never decrease.
To guarantee the approximation bound, we need a valid bound for the maintained level
$\level{\piv[i]{u}}{v}$ at \emph{any} time $t$ for the stretched case.
When an edge $(z', z) \in P_{u,v}$ becomes stretched in $\piv[i]{u}$'s tree, the tree
freezes the label $\level[t]{\piv[i]{u}}{z}$ to the value set at some earlier time $t' < t$.
To bound $\level[t]{\piv[i]{u}}{z}$, one requires a valid upper bound on the historical
label $\level[t']{\piv[i]{u}}{z}$.
However, the presence of the path in $E_{S_{i+1}}$ holds only at the current time $t$.
Because the hierarchy-induced edge sets expand monotonically over time, at the earlier time $t'$ the path $P_{u,v}$ was not necessarily
contained in $E_{S_{i+1}}$.
Consequently, at time $t'$, the graph $G_{i+1}$ had no guarantee of containing
the edges of $P_{u,v}$, and the historical distance $\dist{\piv[i]{u}, z}{G_{i+1}}{t'}$
could have been arbitrarily larger than $\dist{\piv[i]{u}, u}{G}{t'} + \dist{u, z}{G}{t'}$.
The static invariant that the path is present in $E_{S_{i+1}}$ therefore provides no
bound for $\level{\piv[i]{u}}{v}$ in the stretched case.
While additive algorithms can absorb historical stretch into a cumulative additive term (as demonstrated in our algorithms), the delicate balance equations that yield sub-3 multiplicative stretches in~\cite{BK} and~\cite{AkavRodittyUnifiedApproach} rely on exact algebraic cancellations that completely collapse once stretched edges introduce historical slack.

\subsection{Concluding Remarks}
Despite the obstacles outlined above, our work introduces the \emph{first} decremental maintenance of nested,
hierarchy-induced edge sets $E_{S_1} \subseteq E_{S_2} \subseteq \dots$, a central tool in weighted additive algorithms.
Our work proves the edge set can be maintained with bounded
total edge insertions and hop-bounded paths, and establishes an analysis framework for developing near-additive APASP in decremental graphs.
Moreover, we introduce the first near-additive algorithms for decremental weighted graphs, with non-trivial tools to maintain 
sparsity and bounded approximations under decremental updates. 

One should also notice that the depth-bounded algorithms are truly additive, extended to arbitrary depths with a solid framework. Then, if one improves any of the framework's building blocks - e.g., constructs a suitable hopset for the hierarchy-induced edge sets with a finer dependence on $\varepsilon$, or removes the $\varepsilon$ or $n^{\subpol}$ dependence completely -  that would lead to an immediate improvement in our algorithms.

Progressing beyond these results requires techniques that transcend
the assumption that shortest paths exist within the hierarchy-induced edge sets since the beginning of the algorithm.
The central open challenge is to develop decremental data structures whose approximation
guarantees remain robust even when paths are not initially contained within a specific set $E_{S_i}$,
or alternatively find an efficient way to decrease distance estimates when the path fully enters $E_{S_i}$.
This motivates several questions for future research:

\begin{question}\label{que:dynamic-shortcuts}
Can one design decremental data structures that use multi-level hierarchies without accumulating historical stretch
in the pivots' estimates? or, alternatively, can one maintain valid distance bounds from pivots without relying on the assumption that paths are contained in a specific hierarchy-induced edge set from the beginning?
\end{question}

\begin{question}\label{que:multiplicative-sub-three}
Can one design a decremental APASP algorithm for weighted graphs with a purely multiplicative approximation factor $\alpha \in (2+\varepsilon,3)$ 
(such as $\frac{7}{3}$ or $\frac{5}{2}$), even without maintaining an explicit additive bound?
\end{question}

Finally, an intriguing open direction is whether our framework can be derandomized. 
All of our algorithms are randomized and operate against an oblivious adversary, 
inheriting their randomness from the sampled pivot sets $S$ used to bound 
the cumulative edge insertions into $E_{S}$. 
In the static setting, Cohen and Zwick~\cite{CohZwi1997} achieved a deterministic algorithm 
by populating the sparse edge sets with the lightest incident edges of each vertex. 
In the decremental regime, however, an adversary that systematically deletes the lightest 
edges causes $\Omega(m)$ updates in the set. 
Because our construction establishes the first decremental maintenance of hierarchy-induced edge sets 
in weighted graphs, a natural and challenging question is whether this reliance on randomization 
can be eliminated:

\begin{question}\label{que:derandomized-hierarchy}
Can the decremental maintenance of the hierarchy-induced edge sets be derandomized efficiently? 
That is, can one deterministically maintain vertex sets $S_0,S_1,\dots,S_k$ and hierarchy-induced edge sets $E_{S_1},\dots,E_{S_{k-1}},E_{S_k}$ that provide local edge weight errors up to $\onepluseps$ factor, such that for each $1 \le i < k$, the total number of edges in $E_{S_i}$ across all updates is $\onolog{\frac{n^{2+\subpol}}{\abs{S_i}}}$?
\end{question}

\printbibliography

@article{HKNMonESTree,
    author  = "Monika Henzinger and Sebastian Krinninger and Danupon Nanongkai",
    title   = "Dynamic Approximate All-Pairs Shortest Paths: Breaking the $O(mn)$ Barrier and Derandomization",
    year    = "2015",
    journal = " SIAM Journal on Computing 45.3",
    volume  = "45",
    number  = "3",
    pages   = "947–1006",
    doi     = "http://dx.doi.org/10.1137/140957299"
}

@article{Doryetal,
    author  = "Michal Dory and Sebastian Forester and Yasamin Nazari and Tjin De Vos",
    title   = "New Tradeoffs for Decremental Approximate All-Pairs
 Shortest Paths",
    year    = "2024",
    doi     = "https://doi.org/10.48550/arXiv.2211.01152",
    journal = "ICALP"
}

@article{BK,
    author = {Baswana, Surender and Kavitha, Telikepalli},
    title = {Faster Algorithms for All-pairs Approximate Shortest Paths in Undirected Graphs},
    journal = {SIAM Journal on Computing},
    volume = {39},
    number = {7},
    pages = {2865-2896},
    year = {2006},
    doi = {10.1137/080737174}
}

@article{CohZwi1997,
    author = {Cohen, Edith and Zwick, Uri},
    title = {All-Pairs Small-Stretch Paths},
    year = {1997},
    journal = {Journal of Algorithms},
    volume = {38},
    pages = {335-353},
    number = {2},
    issn = {0196-6774},
    doi = {http://doi.org/10.1006/jagm.2000.1117},
    url = {http://dl.acm.org/doi/pdf/10.5555/314161.314190}
}

@article{DorHalZwi2000,
    author = {Dor, Dorit and Halperin, Shay and Zwick, Uri},
    title = {All-Pairs Almost Shortest Paths},
    journal = {SIAM Journal on Computing},
    volume = {29},
    number = {5},
    pages = {1740-1759},
    year = {2000},
    doi = {10.1137/S0097539797327908},
    URL = {https://doi.org/10.1137/S0097539797327908}
}

@article{HKNLinSSSP,
    author  = "Monika Henzinger and Sebastian Krinninger and Danupon Nanongkai",
    title   = "Decremental Single-Source Shortest Paths on Undirected Graphs in Near-Linear Total Update Time",
    year    = "2016",
    journal = "Journal of the ACM",
    doi     = "https://doi.org/10.48550/arXiv.1512.08148"
}

@article{LackiNazari22,
    author = "Jakub Łącki and Yasamin Nazari",
    title = "Near-Optimal Decremental Hopsets with Applications",
    year = "2022",
    journal = "In proceedings of ICALP",
    doi = "https://arxiv.org/abs/2009.08416"}

@article{HKN14,
    author  = "Monika Henzinger and Sebastian Krinninger and Danupon Nanongkai",
    title   = "A subquadratic time algorithm for decremental single-source shortest paths",
    year    = "2014",
    journal = "In Proceedings of the Twenty-Fifth Annual ACM-SIAM Symposium on Discrete Algorithms, SODA 2014"
}

@article{Rod23,
    author = "Liam Roditty",
    title = "New Algorithms for All Pairs Approximate Shortest Paths",
    year = "2023",
    journal = "STOC 2023: Proceedings of the 55th Annual ACM Symposium on Theory of Computing",
    doi = "https://doi.org/10.1145/3564246.3585197"}

@article{ThoZwick,
    title={Approximate distance oracles},
    author={Thorup, Mikkel and Zwick, Uri},
    journal={Journal of the ACM},
    year={2005},
    volume={52},
    pages = {1-24},
    url={https://www.cs.jhu.edu/~baruch/teaching/600.427/Papers/oracle-STOC-try.pdf}
}

@article{BernsteinRounding,
        author       = {Aaron Bernstein},
        title        = {Fully Dynamic {(2} + epsilon) Approximate All-Pairs Shortest Paths
                      with Fast Query and Close to Linear Update Time},
        booktitle    = {50th Annual {IEEE} Symposium on Foundations of Computer Science, {FOCS}
                      2009, Atlanta, Georgia, USA, October 25-27, 2009},
        pages        = {693--702},
        publisher    = {{IEEE} Computer Society},
        year         = {2009},
        url          = {https://doi.org/10.1109/FOCS.2009.16},
        doi          = {10.1109/FOCS.2009.16},
        bibsource    = {dblp computer science bibliography, https://dblp.org}
    }

@article{ChechikADO,
    author       = {Shiri Chechik},
      editor       = {Mikkel Thorup},
      title        = {Near-Optimal Approximate Decremental All Pairs Shortest Paths},
      journal    = {59th {IEEE} Annual Symposium on Foundations of Computer Science, {FOCS}                2018, Paris, France, October 7-9, 2018},
      pages        = {170--181},
      publisher    = {{IEEE} Computer Society},
      year         = {2018},
      url          = {https://doi.org/10.1109/FOCS.2018.00025},
      doi          = {10.1109/FOCS.2018.00025},
      bibsource    = {dblp computer science bibliography, https://dblp.org}
    }

@article{Bernstein2021DetSSSP,
        author       = {Aaron Bernstein and
                  Maximilian Probst Gutenberg and
                  Thatchaphol Saranurak},
        title        = {Deterministic Decremental {SSSP} and Approximate Min-Cost Flow in
                  Almost-Linear Time},
        booktitle    = {62nd {IEEE} Annual Symposium on Foundations of Computer Science, {FOCS}
                  2021, Denver, CO, USA, February 7-10, 2022},
        pages        = {1000--1008},
        publisher    = {{IEEE}},
        year         = {2021},
        url          = {https://doi.org/10.1109/FOCS52979.2021.00100},
        doi          = {10.1109/FOCS52979.2021.00100},
        bibsource    = {dblp computer science bibliography, https://dblp.org}
}

@article{FasterApproxBermanKavisi,
    author       = {Piotr Berman and
                      Shiva Prasad Kasiviswanathan},
    editor       = {Frank K. H. A. Dehne and
                  J{\"{o}}rg{-}R{\"{u}}diger Sack and
                  Norbert Zeh},
    title        = {Faster Approximation of Distances in Graphs},
    booktitle    = {Algorithms and Data Structures, 10th International Workshop, {WADS}
                  2007, Halifax, Canada, August 15-17, 2007, Proceedings},
    series       = {Lecture Notes in Computer Science},
    volume       = {4619},
    pages        = {541--552},
    publisher    = {Springer},
    year         = {2007},
    url          = {https://doi.org/10.1007/978-3-540-73951-7\_47},
    doi          = {10.1007/978-3-540-73951-7\_47},
    bibsource    = {dblp computer science bibliography, https://dblp.org}
}

@article{Ariel+kDense,
    author       = {Liam Roditty and
                  Ariel Sapir},
      editor       = {C. Aiswarya and
                      Ruta Mehta and
                      Subhajit Roy},
      title        = {Additive, Near-Additive, and Multiplicative Approximations for {APSP}
                      in Weighted Undirected Graphs: Trade-Offs and Algorithms},
      booktitle    = {45th {IARCS} Annual Conference on Foundations of Software Technology
                      and Theoretical Computer Science, {FSTTCS} 2025, {BITS} Pilani, {K}
                      {K} Birla Goa Campus, India, December 17-19, 2025},
      series       = {LIPIcs},
      volume       = {360},
      pages        = {50:1--50:22},
      publisher    = {Schloss Dagstuhl - Leibniz-Zentrum f{\"{u}}r Informatik},
      year         = {2025},
      url          = {https://doi.org/10.4230/LIPIcs.FSTTCS.2025.50},
      doi          = {10.4230/LIPICS.FSTTCS.2025.50},
      bibsource    = {dblp computer science bibliography, https://dblp.org}
}

@article{AkavRodittyUnifiedApproach,
  author       = {Maor Akav and
                  Liam Roditty},
  editor       = {Petra Mutzel and
                  Rasmus Pagh and
                  Grzegorz Herman},
  title        = {A Unified Approach for All Pairs Approximate Shortest Paths in Weighted
                  Undirected Graphs},
  booktitle    = {29th Annual European Symposium on Algorithms, {ESA} 2021, Lisbon,
                  Portugal (Virtual Conference), September 6-8, 2021},
  series       = {LIPIcs},
  volume       = {204},
  pages        = {4:1--4:18},
  publisher    = {Schloss Dagstuhl - Leibniz-Zentrum f{\"{u}}r Informatik},
  year         = {2021},
  url          = {https://doi.org/10.4230/LIPIcs.ESA.2021.4},
  doi          = {10.4230/LIPICS.ESA.2021.4},
  bibsource    = {dblp computer science bibliography, https://dblp.org}
}

@article{DorForKirNazVasVos2023,
      author={Dory, Michal and Forster, Sebastian and Kirkpatrick, Yael and Nazari, Yasamin and Vassilevska Williams, Virginia and De Vos, Tijn},
      title={Fast 2-Approximate All-Pairs Shortest Paths}, 
      year={2023},
      doi = {10.1137/1.9781611977912.169},
      eprint={2307.09258},
      archivePrefix={arXiv},
      primaryClass={cs.DS},
      url={https://arxiv.org/abs/2307.09258}, 
      journal={arXiv}
}

@inbook{SahYe2023,
    author = {Saha, Barna and Ye, Christopher},
    title = {Faster Approximate All Pairs Shortest Paths},
    booktitle = {Proceedings of the 2024 Annual ACM-SIAM Symposium on Discrete Algorithms (SODA)},
    chapter = {},
    pages = {4758-4827},
    doi = {10.1137/1.9781611977912.170},
    URL = {https://epubs.siam.org/doi/abs/10.1137/1.9781611977912.170},
    eprint = {https://epubs.siam.org/doi/pdf/10.1137/1.9781611977912.170},
    year = 2023,
    publisher = {siam}
}

@article{RodZwi2011,
    author={Roditty, Liam
    and Zwick, Uri},
    title={On Dynamic Shortest Paths Problems},
    journal={Algorithmica},
    year={2011},
    month={Oct},
    day={01},
    volume={61},
    number={2},
    pages={389-401},
    issn={1432-0541},
    doi={10.1007/s00453-010-9401-5},
    url={https://doi.org/10.1007/s00453-010-9401-5}
}

@InProceedings{Vas2015,
  author =	{Vassilevska Williams, Virginia},
  title =	{{Hardness of Easy Problems: Basing Hardness on Popular Conjectures such as the Strong Exponential Time Hypothesis}},
  booktitle =	{10th International Symposium on Parameterized and Exact Computation (IPEC 2015)},
  pages =	{17--29},
  series =	{Leibniz International Proceedings in Informatics (LIPIcs)},
  ISBN =	{978-3-939897-92-7},
  ISSN =	{1868-8969},
  year =	{2015},
  volume =	{43},
  editor =	{Husfeldt, Thore and Kanj, Iyad},
  publisher =	{Schloss Dagstuhl -- Leibniz-Zentrum f{\"u}r Informatik},
  address =	{Dagstuhl, Germany},
  URL =		{https://drops.dagstuhl.de/entities/document/10.4230/LIPIcs.IPEC.2015.17},
  URN =		{urn:nbn:de:0030-drops-55683},
  doi =		{10.4230/LIPIcs.IPEC.2015.17},
}

@article{AbbVas2023,
    author = {Abboud, Amir and Grandoni, Fabrizio and Vassilevska Williams, Virginia},
    title = {Subcubic Equivalences between Graph Centrality Problems, APSP, and Diameter},
    year = {2023},
    issue_date = {January 2023},
    publisher = {Association for Computing Machinery},
    address = {New York, NY, USA},
    volume = {19},
    number = {1},
    issn = {1549-6325},
    url = {https://doi.org/10.1145/3563393},
    doi = {10.1145/3563393},
    journal = {ACM Trans. Algorithms},
    month = mar,
    articleno = {3},
    numpages = {30}
}

@inbook{Vas2018,
    author = {Virginia Vassilevska Williams},
    title = {On Some Fine-Grained Questions in Algorithms and Complexity},
    booktitle = {Proceedings of the International Congress of Mathematicians (ICM 2018)},
    chapter = {},
    pages = {3447-3487},
    doi = {10.1142/9789813272880_0188},
    URL = {https://www.worldscientific.com/doi/abs/10.1142/9789813272880_0188},
    eprint = {https://www.worldscientific.com/doi/pdf/10.1142/9789813272880_0188},
}

@inproceedings{HenKriNanSara2015,
    author = {Henzinger, Monika and Krinninger, Sebastian and Nanongkai, Danupon and Saranurak, Thatchaphol},
    title = {Unifying and Strengthening Hardness for Dynamic Problems via the Online Matrix-Vector Multiplication Conjecture},
    year = {2015},
    isbn = {9781450335362},
    publisher = {Association for Computing Machinery},
    address = {New York, NY, USA},
    url = {https://doi.org/10.1145/2746539.2746609},
    doi = {10.1145/2746539.2746609},
    booktitle = {Proceedings of the Forty-Seventh Annual ACM Symposium on Theory of Computing},
    pages = {21–30},
    numpages = {10},
    location = {Portland, Oregon, USA},
    series = {STOC '15} 
}

@inproceedings{Thorup2005,
    author = {Thorup, Mikkel},
    title = {Worst-case update times for fully-dynamic all-pairs shortest paths},
    year = {2005},
    isbn = {1581139608},
    publisher = {Association for Computing Machinery},
    address = {New York, NY, USA},
    url = {https://doi.org/10.1145/1060590.1060607},
    doi = {10.1145/1060590.1060607},
    booktitle = {Proceedings of the Thirty-Seventh Annual ACM Symposium on Theory of Computing},
    pages = {112–119},
    numpages = {8},
    location = {Baltimore, MD, USA},
    series = {STOC '05}
}

@inproceedings{BerRod2011,
    author = {Bernstein, Aaron and Roditty, Liam},
    title = {Improved dynamic algorithms for maintaining approximate shortest paths under deletions},
    year = {2011},
    publisher = {Society for Industrial and Applied Mathematics},
    address = {USA},
    booktitle = {Proceedings of the Twenty-Second Annual ACM-SIAM Symposium on Discrete Algorithms},
    pages = {1355–1365},
    numpages = {11},
    location = {San Francisco, California},
    series = {SODA '11}
}

@article{GasJanLevLin2021,
    title = {Pushing the Online Boolean Matrix-vector Multiplication conjecture off-line and identifying its easy cases},
    journal = {Journal of Computer and System Sciences},
    volume = {118},
    pages = {108-118},
    year = {2021},
    issn = {0022-0000},
    doi = {https://doi.org/10.1016/j.jcss.2020.12.004},
    url = {https://www.sciencedirect.com/science/article/pii/S0022000021000027},
    author = {Leszek Gąsieniec and Jesper Jansson and Christos Levcopoulos and Andrzej Lingas and Mia Persson},
}

@inbook{LarRWill2017,
    author = {Kasper Green Larsen and Ryan Williams},
    title = {Faster Online Matrix-Vector Multiplication},
    booktitle = {Proceedings of the 2017 Annual ACM-SIAM Symposium on Discrete Algorithms (SODA)},
    chapter = {},
    pages = {2182-2189},
    doi = {10.1137/1.9781611974782.142},
    URL = {https://epubs.siam.org/doi/abs/10.1137/1.9781611974782.142},
    eprint = {https://epubs.siam.org/doi/pdf/10.1137/1.9781611974782.142},
    year = {2017}
}

@article{Kavitha2012,
    author = {Kavitha, Telikepalli},
    title = {Faster Algorithms for All-Pairs Small Stretch Distances in Weighted Graphs},
    journal = {Algorithmica},
    volume = {63},
    pages = {224-245},
    year = {2012},
    doi = {10.1007/s00453-011-9529-y},
    publisher   = {SPRINGER},
    URL = {https://doi.org/10.1007/s00453-011-9529-y}
}

@article{ESTree,
    author = {Shiloach, Yossi and Even, Shimon},
    title = {An On-Line Edge-Deletion Problem},
    year = {1981},
    issue_date = {Jan. 1981},
    publisher = {Association for Computing Machinery},
    address = {New York, NY, USA},
    volume = {28},
    number = {1},
    issn = {0004-5411},
    url = {https://doi.org/10.1145/322234.322235},
    doi = {10.1145/322234.322235},
    journal = {J. ACM},
    month = jan,
    pages = {1–4},
    numpages = {4}
}

@INPROCEEDINGS{KingWeightedESTree,
  author={King, V.},
  booktitle={40th Annual Symposium on Foundations of Computer Science (Cat. No.99CB37039)}, 
  title={Fully dynamic algorithms for maintaining all-pairs shortest paths and transitive closure in digraphs}, 
  year={1999},
  volume={},
  number={},
  pages={81-89},
  doi={10.1109/SFFCS.1999.814580}
  }

@article{Tzalik2026,
    title={Greedy Completion for Weighted $(\alpha,\beta)$-Spanners}, 
    author={Elad Tzalik},
    journal={arXiv},
    year={2026},
    eprint={2603.17047},
    primaryClass={cs.DS},
    volume={abs/2603.17047},
    url={https://arxiv.org/abs/2603.17047}, 
}

@InProceedings{ElkGitNei2021,
  author =	{Elkin, Michael and Gitlitz, Yuval and Neiman, Ofer},
  title =	{Improved Weighted Additive Spanners},
  booktitle =	{35th International Symposium on Distributed Computing (DISC 2021)},
  pages =	{21:1--21:15},
  series =	{Leibniz International Proceedings in Informatics (LIPIcs)},
  ISBN =	{978-3-95977-210-5},
  ISSN =	{1868-8969},
  year =	{2021},
  volume =	{209},
  publisher =	{Schloss Dagstuhl -- Leibniz-Zentrum f{\"u}r Informatik},
  address =	{Dagstuhl, Germany},
  URL =		{http://drops.dagstuhl.de/entities/document/10.4230/LIPIcs.DISC.2021.21},
  doi =		{10.4230/LIPIcs.DISC.2021.21}
}

@inproceedings{ChuzhoyAdapativeAdversaryAPASP,
author = {Chuzhoy, Julia},
title = {Decremental all-pairs shortest paths in deterministic near-linear time},
year = {2021},
isbn = {9781450380539},
publisher = {Association for Computing Machinery},
address = {New York, NY, USA},
url = {https://doi.org/10.1145/3406325.3451025},
doi = {10.1145/3406325.3451025},
booktitle = {Proceedings of the 53rd Annual ACM SIGACT Symposium on Theory of Computing},
pages = {626–639},
numpages = {14},
location = {Virtual, Italy},
series = {STOC 2021}
}

@inproceedings{ElkinGN2022SWAT,
  author    = {Michael Elkin and Yuval Gitlitz and Ofer Neiman},
  title     = {Almost Shortest Paths with Near-Additive Error in Weighted Graphs},
  booktitle = {18th Scandinavian Symposium and Workshops on Algorithm Theory (SWAT 2022)},
  pages     = {23:1--23:16},
  year      = {2022},
  eprint    = {1907.11422},
  archivePrefix = {arXiv}
}

@article{LaLe2024,
  author    = {An La and Hung Le},
  title     = {New Weighted Additive Spanners},
  journal   = {arXiv preprint arXiv:2408.14638},
  year      = {2024},
  eprint    = {2408.14638},
  archivePrefix = {arXiv}
}

@article{RodittySapir2026Emulators,
  author    = {Liam Roditty and Ariel Sapir},
  title     = {Weighted Emulators with Local Heaviest Edges Stretch for Undirected Graphs},
  journal   = {arXiv preprint arXiv:2604.26831},
  year      = {2026},
  eprint    = {2604.26831},
  archivePrefix = {arXiv}
}

@article{JinKSW25,
  author    = {Ce Jin and Yael Kirkpatrick and Micha{\l} Stawarz and Virginia Vassilevska Williams},
  title     = {Improved Additive Approximation Algorithms for {APSP}},
  journal   = {arXiv preprint arXiv:2511.04775},
  year      = {2025},
  eprint    = {2511.04775},
  archivePrefix = {arXiv}
}

@inproceedings{HenKin1995,
  author    = {Monika R. Henzinger and Valerie King},
  title     = {Fully Dynamic Biconnectivity and Transitive Closure},
  booktitle = {36th Annual Symposium on Foundations of Computer Science (FOCS 1995)},
  pages     = {664--672},
  year      = {1995},
  publisher = {{IEEE} Computer Society},
  doi       = {10.1109/SFCS.1995.492541}
}

@article{Bernstein2016,
  author    = {Aaron Bernstein},
  title     = {Maintaining Shortest Paths Under Deletions in Weighted Directed Graphs},
  journal   = {{SIAM} Journal on Computing},
  volume    = {45},
  number    = {2},
  pages     = {548--574},
  year      = {2016},
  doi       = {10.1137/130938670}
}

@article{AlmVas2026,
    author={Alman, Josh and Vassilevska Williams, Virginia},
    title={Truly Subquadratic {3SUM} and Truly Subcubic {APSP} via Triangles in Sparse Lopsided Graphs},
    journal={arXiv},
    year={2026},
    eprint={2610.06783},
    archivePrefix={arXiv},
    primaryClass={cs.DS},
    volume={abs/2610.06783},
    doi={10.48550/arXiv.2610.06783},
    url={https://arxiv.org/abs/2610.06783},
}

@article{SahYe2026,
    author={Saha, Barna and Ye, Christopher},
    title={Simple, Faster $(\min,+)$ Matrix Multiplication and {APSP}},
    journal={arXiv},
    year={2026},
    eprint={2610.20297},
    archivePrefix={arXiv},
    primaryClass={cs.DS},
    volume={abs/2610.20297},
    doi={10.48550/arXiv.2610.20297},
    url={https://arxiv.org/abs/2610.20297},
}

@article{BriKunWeg2019,
    author={Bringmann, Karl and K{\"u}nnemann, Marvin and W{\k{e}}grzycki, Karol},
    title={Approximating {APSP} without Scaling: Equivalence of Approximate Min-Plus and Exact Min-Max},
    journal={arXiv},
    year={2019},
    doi={10.1145/3313276.3316373},
    eprint={1907.11078},
    archivePrefix={arXiv},
    primaryClass={cs.DS},
    url={https://arxiv.org/abs/1907.11078},
}

@article{Durr2023,
  title = {Improved bounds for rectangular monotone Min-Plus Product and applications},
  author = {Dürr, Anita},
  journal = {Information Processing Letters},
  volume = {181},
  pages = {106358},
  year = {2023},
  issn = {0020-0190},
  doi = {https://doi.org/10.1016/j.ipl.2023.106358},
  url = {https://www.sciencedirect.com/science/article/pii/S0020019023000017}
}

@InProceedings{DenKirVicVasZho2022,
  author = {Deng, Mingyang and Kirkpatrick, Yael and Rong, Victor and Vassilevska Williams, Virginia and Zhong, Ziqian},
  title = {New Additive Approximations for Shortest Paths and Cycles},
  booktitle = {49th International Colloquium on Automata, Languages, and Programming (ICALP 2022)},
  pages = {50:1--50:10},
  series = {Leibniz International Proceedings in Informatics (LIPIcs)},
  ISBN = {978-3-95977-235-8},
  ISSN = {1868-8969},
  year = {2022},
  volume = {229},
  publisher = {Schloss Dagstuhl -- Leibniz-Zentrum f{\"u}r Informatik},
  address = {Dagstuhl, Germany},
  URL = {https://drops.dagstuhl.de/entities/document/10.4230/LIPIcs.ICALP.2022.50},
  doi = {10.4230/LIPIcs.ICALP.2022.50}
}

\begin{appendices}

\section{Related Algorithms}~\label{app:known}
The following tables summarize known algorithms for static and decremental APASP, as mentioned in \secref{intro} and \secref{intro:known}.

\begin{table}[H]
  \begin{center}
\small\addtolength{\tabcolsep}{-3pt}
    \renewcommand\arraystretch{1.5}
    \begin{tabular}{|c|c|c|c|} 
      \hline
      \textbf{\small Weights} & \textbf{\small Approximation}   & \textbf{\small Expected Runtime}
       & \textbf{\small Reference}\\
       \hline
      Unweighted & $\paren{1,2}$ &  $\onolog{\mmin{m^{\frac{1}{2}}n^{\frac{3}{2}},n^{\frac{7}{3}}}}$  & \cite{DorHalZwi2000}\\
      \hline
      
      Unweighted & $\paren{1,2\cdot\paren{k-1}}$ &  $\onolog{\mmin{m^{\frac{1}{k}}n^{2-\frac{1}{k}},n^{2+\frac{1}{3k-4}}}}$  & \cite{DorHalZwi2000}\\
      \hline

       Unweighted & $\paren{2,0}$ &  $\onolog{\mmin{n^{\frac{9}{4}},m^{\frac{1}{3}}n^{\frac{5}{3}}}}$  & \cite{Rod23}\\
       \hline

       Unweighted & $\paren{2+\varepsilon,0}$ &  $\onolog{n^{2.03106234}}$  & \cite{DorForKirNazVasVos2023}\\
       \hline

       Unweighted & $\paren{\frac{7}{3},0}$ &  $\onolog{n^{2}}$  & \cite{SahYe2023}\\
       \hline

       Weighted & $\paren{3,0}$ &  $\onolog{mn^{\frac{1}{2}}}$  & \cite{ThoZwick}\\
       \hline

       Weighted & $\paren{1,2W_2} \le \paren{2,0}$ &  $\onolog{n^{\frac{3}{2}}m^{\frac{1}{2}}}$  & \cite{CohZwi1997}\\
       \hline
       
       Weighted & $\paren{1,\mmin{2W_1,4W_2}} \le \paren{\frac{7}{3},0} $ &  $\onolog{n^{\frac{7}{3}}}$  & \cite{CohZwi1997}\\
      \hline

      \multirow{2}{*}{Weighted} & \multirow{2}{*}{$\paren{1,2\msum{W_i}{i=1}{k-1}} \leq \paren{3,0} $} &  \multirow{2}{*}{$\onolog{\mmin{m^{\frac{1}{k}} n^{2-\frac{1}{k}},n^{2+\frac{1}{3k-4}}}}$}  & \multirow{2}{*}{\cite{CohZwi1997}, \cite{Ariel+kDense}}\\
      &&&\\
      \hline

      Weighted & $\paren{2,W_1} \le \paren{3,0} $ &  $\onolog{n^{2}}$  & \cite{BK}, \cite{FasterApproxBermanKavisi}\\
      \hline

      Weighted & $\paren{2,0} $ &  $\onolog{n^{2} + m n^{\frac{1}{2}}}$  & \cite{BK}\\
      \hline
    \end{tabular}
    \renewcommand\arraystretch{1}
    \caption{Known $\paren{\alpha,\beta}$-APASP algorithms for undirected (weighted and unweighted) graphs in the static setting.}~\label{tab:known_static}
  \end{center}
\end{table}

\begin{table}[H]
  \begin{center}
\small\addtolength{\tabcolsep}{-3pt}
    \renewcommand\arraystretch{1.5}
    \begin{tabular}{|c|c|c|c|c|} 
      \hline
      \textbf{\small Weights} & \textbf{\small Approximation}   & \textbf{\small Total Update Time}
       & \textbf{\small Adversary} &  \textbf{\small Reference}\\
       \hline
      \multirow{2}{*}{Unweighted} & \multirow{2}{*}{$\paren{\onepluseps,0}$} &  \multirow{2}{*}{$\onolog{\frac{mn}{\varepsilon}}$} & \multirow{2}{*}{Oblivious} & \multirow{2}{*}{\cite{RodZwi2011}}\\
       &&&&\\ \hline

       \multirow{2}{*}{Unweighted} & \multirow{2}{*}{$\paren{\onepluseps,0}$}&  \multirow{2}{*}{$\onolog{\frac{mn}{\varepsilon}}$} & \multirow{2}{*}{Adaptive} & \cite{HKNMonESTree}\\
       &&&&\\
       \hline

       Unweighted & $\paren{\onepluseps,2}$ &  $\onolog{n^{\frac{5}{2}}}$ & Adaptive & \cite{HKNMonESTree}\\
       \hline
      
       Unweighted & $\paren{\onepluseps,2}$ &  $\onolog{m^{\frac{1}{2}}n^{\frac{3}{2}+\subpol}}$  &  Oblivious & \cite{Doryetal}\\
      \hline
      Unweighted & $\paren{\onepluseps,2\cdot\paren{k-1}}$ &  $\onolog{m^{\frac{1}{k}}n^{2-\frac{1}{k}+o\paren{1}}}$  & Oblivious & \cite{Doryetal}\\
      \hline

      Weighted & $\paren{\onepluseps,0}$ &  $\onolog{mn^{1+\subpol}}$ & Oblivious & \cite{HKNLinSSSP,LackiNazari22}\\
       \hline
      
      Weighted & $\paren{\twopluseps,0}$ &  $\onolog{n^{\frac{3}{2}+o\paren{1}}m^{\frac{1}{2}}}$ & Oblivious & \cite{Doryetal}\\
       \hline

       Weighted & $\paren{\twopluseps,W_1}  $ &  $\onolog{n^{1+o\paren{1}}m^{\frac{3}{4}}}$  & Oblivious & \cite{Doryetal} \\
      \hline

       Weighted & $\paren{2k-1 + \varepsilon,0}$ &  $\onolog{mn^{\frac{1}{k}+o\paren{1}}}$  & Oblivious & \cite{ChechikADO}\\
       \hline

        Weighted & $\paren{\paren{\log m}^{2^{\bigo{\onediveps}}},0}$ &  $\onolog{m^{1+\bigo{\varepsilon}}\cdot \paren{\log m}^{\bigo{\onediveps[2]}}}$  & Adaptive & \cite{ChuzhoyAdapativeAdversaryAPASP}\\
       \hline

      \hline

    \end{tabular}
    \renewcommand\arraystretch{1}
    \caption{Known $\paren{\alpha,\beta}$-APASP algorithms for undirected (weighted and unweighted) graphs in the decremental setting, operating against an oblivious adversary. The dependence on $\varepsilon$ is \qoute{hidden} within the $n^{o\paren{1}}$ component.}~\label{tab:known_decremental}
  \end{center}
\end{table}

\section{Bernstein's \cite{BernsteinRounding} Edge Rounding Method} 
\label{app:bernstein-rounding-process}
\begin{definition}
    For any two vertices $u,v \in V$, the $h-$distance $\dist^h(u,v)$ is defined as the weight of the shortest path between $u$ and $v$ of all paths that consist of at most $h$ edges.
\end{definition}

\begin{theorem}[Theorem 4 from \cite{BernsteinRounding}]
Given a source $s$ and a hop distance $h$, we can decrementally maintain distances $\delta'(s,v)$ to every vertex $v$ such that we always have $\delta(s,v) \le \delta'(s,v) \le (\onepluseps)\delta^h(s,v)$.

The total update time over all deletions and weight-increases is $O(mh\log(n)\log(nR)/\varepsilon+\Delta)$ in weighted graphs and $O(mh\log(n)\log\log(n))$ in unweighted ones.
\end{theorem}

\begin{lemma}
Assuming $0 < \varepsilon < 1$, we can maintain any $h\text{-SSSP}_k$ in total update time $O(mh/\varepsilon + \Delta)$.
\end{lemma}

\begin{proof}
Recall that here we are only concerned with approximating distances $\delta^h(s,v)$ for which $2^k \le \delta^h(s,v) \le 2^{k+1}$. For the rest of this proof, let $\alpha = \frac{\varepsilon 2^k}{h}$. We start by scaling the edge-weights of graph $G$ to obtain a new graph $G_k$ in the following way:

\begin{itemize}
    \item Delete all edges of weight $> 2^{k+1}$ from $G$.
    \item Round all remaining edge weights up to the nearest integer multiple of $\alpha$.
    \item Divide all edge weights by $\alpha$.
\end{itemize}

Note that the scaled weights in $G_k$ are positive and integral.

Our algorithm maintains a shortest path tree from $s$ in $G_k$ by simply running King's $O(md)$ decremental SSSP algorithm up to distance $d = \lceil 4h/\varepsilon \rceil$. More precisely, if an update deletes $(u, v)$ in the original graph, we simply delete $(u, v)$ in the scaled graph; if the update is increase-weight $(u, v)$, we first scale the new weight according to the three steps above and then change the weight of $w(u, v)$ to this new scaled weight.

We then output $\delta_k'(s, v) = \alpha \cdot \delta_{G_k}(s, v)$. Then, the total update time of King's algorithm is just $O(md) = O(mh/\varepsilon)$, as desired. The final $O(\Delta)$ term arises from weight increases in $G$ that do not change the scaled weights in $G_k$ (i.e., the old weight and the new weight scale up to the same nearest multiple of $\alpha$) and so are discarded in $O(1)$ time.

We now do an approximation analysis. Let $G_k^*$ be the graph $G_k$ before dividing the edge weights by $\alpha$ (but after scaling up to a multiple of $\alpha$), and note that since $G_k$ and $G_k^*$ are the same up to a scaling factor, our output is precisely $\delta_k'(s, v) = \alpha \cdot \delta_{G_k}(s, v) = \delta_{G_k^*}(s, v)$. All edge weights in $G_k^*$ are greater than those in $G$, so it is clear that $\delta_k'(s, v) = \delta_{G_k^*}(s, v) \ge \delta(s, v)$.

We now need to show that if $2^k \le \delta^h(s, v) \le 2^{k+1}$, then $\delta_{G_k^*}(s, v) \le (\onepluseps)\delta^h(s, v)$. To see this, let us examine how the weight changes $G \rightarrow G_k^*$ affect $\pi^h(s, v)$. 

Since $\delta^h(s, v) \le 2^{k+1}$, $\pi^h(s, v)$ does not contain any edges of weight $> 2^{k+1}$, so the first set of changes does not affect it at all. The second set of changes adds up to $\alpha$ weight to every edge on $\pi^h(s, v)$, so the weight of the path in $G_k^*$ is at most $\delta^h(s, v) + h\alpha = \delta^h(s, v) + \varepsilon 2^k \le (\onepluseps)\delta^h(s, v)$ (the last inequality follows from $2^k \le \delta^h(s, v)$). 

Thus, we have exhibited an $s-v$ path in $G_k^*$ of weight $\le (\onepluseps)\delta^h(s, v)$, so certainly the shortest $s-v$ path in $G_k^*$ will have weight $\le (\onepluseps)\delta^h(s, v)$, as desired. Finally, note that the weight of this path in $G_k$ is at most $(\onepluseps)\delta^h(s, v)/\alpha \le \frac{2 \cdot 2^{k+1}}{\varepsilon 2^k / h} = 4h/\varepsilon$, so running King's algorithm up to $d = \lceil 4h/\varepsilon \rceil$ in $G_k$ will in fact find this path.
\end{proof}

\end{appendices}

\end{document}